%% file: paper.tex
\documentclass[a4paper,USenglish,numberwithinsect]{lipics-v2021}

\EventEditors{Guillermo A. P\'{e}rez and Jean-Fran\c{c}ois Raskin}
\EventNoEds{2}
\EventLongTitle{34th International Conference on Concurrency Theory (CONCUR 2023)}
\EventShortTitle{CONCUR 2023}
\EventAcronym{CONCUR}
\EventYear{2023}
\EventDate{September 18--23, 2023}
\EventLocation{Antwerp, Belgium}
\EventLogo{}
\SeriesVolume{279}
\ArticleNo{30}

\nolinenumbers

\usepackage{amsmath} 
\usepackage{amssymb}
\usepackage{latexsym}
\usepackage{graphicx}
\usepackage{textcomp}
\usepackage{verbatim} 
\usepackage{xspace} 
\usepackage{array, booktabs, longtable, makecell} 
\usepackage{afterpage}
\usepackage{cancel}
\usepackage{tikz}
\usetikzlibrary{arrows,automata,calc,decorations.text,shapes,cd}
\usepackage{xxcolor}
\usepackage{multirow}
\usepackage{stmaryrd}
\usepackage{algorithm}
\usepackage{algpseudocode}
\usepackage{multicol}
\usepackage{xr-hyper}

\usepackage{ifoddpage}
\usepackage{marginnote}
\usepackage{skull}

\usepackage{csquotes}

\usepackage{subcaption}

\usepackage{booktabs}   
\usepackage{dashbox}
\usepackage{xargs}

\nolinenumbers

\input{defs.tex}

\title{Visibility and Separability for a Declarative Linearizability Proof of the Timestamped Stack:~Extended Version}

\author{Jes\'{u}s Dom\'{i}nguez}{IMDEA Software Institute, Spain \and Universidad Polit\'{e}cnica de Madrid, Spain}{jesus.dominguez@imdea.org}{https://orcid.org/0000-0002-5436-1384}{}

\author{Aleksandar Nanevski}{IMDEA Software Institute, Spain}{aleks.nanevski@imdea.org}{https://orcid.org/0000-0002-4851-1075}{} 

\authorrunning{J. Dom\'{i}nguez and A. Nanevski}

\titlerunning{Visibility and Separability for a Declarative Proof of the Timestamped Stack}

\Copyright{Jes\'{u}s Dom\'{i}nguez and Aleksandar Nanevski}



\funding{This work was partially supported by the ERC project MATHADOR
  (ERC2016-COG-724464), the Madrid regional government and EU project
  BLOQUES (S2018/TCS-4339), and the Spanish MCIN and EU project
  PRODIGY (TED2021-132464B-I00).}

\newcommand{\citet}[1]{\cite{#1}}

\begin{document}

\maketitle

\begin{abstract}
\input{abstract}
\end{abstract}


\input{intro}
\input{programs}

\input{visibility}
\input{spans}
\input{related}

\bibliography{references}

\newpage

\appendix

\input{linproof}
\input{visproof}
\input{structures}


\end{document}

%% file: defs.tex
\newcommand{\etal}{et~al.\xspace}

\newcounter{WarnCounts}

\algrenewcommand\alglinenumber[1]{\scriptsize #1:}

\algblockdefx[Record]{Record}{EndRecord}%
[1]{\textbf{record} {#1}:}{\textbf{end record}}

\algblockdefx[Proc]{Proc}{EndProc}%
[2]{\textbf{proc} {#1}(#2)}{\textbf{end proc}}

\algblockdefx[Type]{Type}{EndType}%
[1]{\textbf{type} {#1}:}{\textbf{end type}}

\algblockdefx[Enum]{Enum}{EndEnum}%
[1]{\textbf{enum} {#1}:}{\textbf{end enum}}

\algblockdefx[EnumLine]{EnumLine}{EndEnumLine}%
[2]{\textbf{enum} {#1}: {#2}}{\textbf{end enum}}

\algblockdefx[ForEach]{ForEach}{EndForEach}%
[2]{\textbf{for each} {#1} \textbf{in} {#2}:}{\textbf{end for each}}

\algblockdefx[ForIter]{ForIter}{EndForIter}%
[3]{\textbf{for} {#1} \textbf{from} {#2} \textbf{to} {#3} \textbf{do}}{\textbf{end for}}

\algblockdefx{Match}{EndMatch}%
[1]{\textbf{match} {#1} \textbf{with}}{\textbf{end match}}

\algcblockdefx[Match]{Match}{MatchCase}{EndMatch}%
[1]{$\mid$ {#1} $\Rightarrow$}{\textbf{end match}}

\algblockdefx[Ghost]{Ghost}{EndGhost}[1]{}{}

\algrenewcommand\algorithmicprocedure{\textbf{proc}}
\newcommand{\returnCmd}{\textbf{return\ }}

\floatname{algorithm}{Listing}
\algtext*{EndRecord}
\algtext*{EndType}
\algtext*{EndEnum}
\algtext*{EndEnumLine}
\algtext*{EndForEach}
\algtext*{EndIf}
\algtext*{EndProcedure}
\algtext*{EndProc}
\algtext*{EndWhile}
\algtext*{EndForIter}

\theoremstyle{remark}

\newtheorem*{notation}{Notation}

\newenvironment{defn}{\begin{definition}}{\end{definition}}
\newenvironment{lem}{\begin{lemma}}{\end{lemma}}

\newenvironment{prf}{\begin{proof}}{\end{proof}}

\newenvironment{thm}{\begin{theorem}}{\end{theorem}}

\NewDocumentEnvironment{subfigwrap}{mmO{\textwidth}}
   {\begin{subfigure}{#3}}
   {\caption{#1}
    \label{#2}
    \end{subfigure}
   }

\newcommand{\rdcssDesc}{\textsc{rdesc}} 
\newcommand{\mcasDesc}{\textsc{mdesc}} 

\newcommand{\rdcssAlg}{\textit{rdcss}}
\newcommand{\rdcssLoopAlg}{\textit{rdcss}'} 
\newcommand{\mcasAlg}{\textit{mcas}}
\newcommand{\mcasHelpAlg}{\textit{mcas}'} 
\newcommand{\completeAlg}{\textit{complete}} 
\newcommand{\rdcssReadAlg}{\textit{rread}}
\newcommand{\rdcssWriteAlg}{\textit{rwrite}}
\newcommand{\rdcssCasAlg}{\textit{rCAS}}
\newcommand{\rdcssReadCtlAlg}{\textit{rread-c}}
\newcommand{\rdcssWriteCtlAlg}{\textit{rwrite-c}}
\newcommand{\rdcssCasCtlAlg}{\textit{rCAS-c}}
\newcommand{\rdcssAllocAlg}{\textit{ralloc}}
\newcommand{\mcasReadAlg}{\textit{mread}}
\newcommand{\mcasWriteAlg}{\textit{mwrite}}
\newcommand{\mcasCasAlg}{\textit{mCAS}}
\newcommand{\mcasAllocAlg}{\textit{malloc}}
\newcommand{\writeAllDescsAlg}{\textit{writeall}} 
\newcommand{\pushAlg}{push}
\newcommand{\popAlg}{pop}
\newcommand{\enqAlg}{enq}
\newcommand{\deqAlg}{deq}

\newcommand{\lockAlg}{lock}
\newcommand{\unlockAlg}{unlock}
\newcommand{\regAlg}{reg}
\newcommand{\deregAlg}{dereg}

\newcommand{\readRepAlg}{!}
\newcommand{\writeRepAlg}{{:=}}
\newcommand{\allocRepAlg}{\textit{Alloc}}
\newcommand{\casRepAlg}{\textit{CAS}}
\newcommand{\isRdcssDescRepAlg}{\textit{is\_rdesc}}
\newcommand{\isMcasDescRepAlg}{\textit{is\_mdesc}}
\newcommand{\poolInsertAlg}{\textit{insert}}
\newcommand{\newStampAlg}{\textit{newTimestamp}}
\newcommand{\tryRemoveAlg}{\textit{tryRem}}
\newcommand{\getYoungAlg}{\textit{getYoungest}}
\newcommand{\removeNodeAlg}{\textit{remove}}

\newcommand{\initPoolAlg}{\textit{init}}

\newcommand{\UNDECIDED}{\textsc{Undec}}
\newcommand{\SUCCEEDED}{\textsc{Succ}}
\newcommand{\FAILED}{\textsc{Fail}}
\newcommand{\CONTROL}{\textsc{cptr}}
\newcommand{\DATA}{\textsc{dptr}}
\newcommand{\EMPTY}{\textsc{EMPTY}}
\newcommand{\rdcssNamespace}[1]{#1} 

\newcommand{\defini}{\eqdef}
\newcommand{\natorderSymbol}{<_{\mathbb{N}}}
\newcommand{\natorderEqSymbol}{\leq_{\mathbb{N}}}
\newcommand{\natorderSymbolRight}{>_{\mathbb{N}}}
\newcommand{\natorderEqSymbolRight}{\geq_{\mathbb{N}}}
\newcommand{\stamporderSymbol}{<_{\textsc{T}}}
\newcommand{\stamporderEqSymbol}{\leq_{\textsc{T}}}
\newcommand{\stamporderSymbolPadding}{\;\stamporderSymbol\;}
\newcommand{\stamporderEqSymbolPadding}{\;\stamporderEqSymbol\;}

\newcommand{\GstamporderSymbol}{\ll}
\newcommand{\GstamporderEqSymbol}{\llEq}

\newcommand{\ValType}{\textsc{Val}}

\newcommand{\BoolType}{\textsc{Bool}}
\newcommand{\IntType}{\textsc{Int}}
\newcommand{\OptionalType}[1]{\textsc{{#1}?}}
\newcommand{\NodeType}{\textsc{Node}}
\newcommand{\StampType}{\textsc{Stamp}}

\newcommand{\PoolType}{\textsc{SPPool}}

\newcommand{\projFirst}[1]{\pi_1({#1})}
\newcommand{\projSecond}[1]{\pi_2({#1})}
\newcommand{\oneSpan}[1]{start\,({#1})}
\newcommand{\twoSpan}[1]{end\,({#1})}
\newcommand{\ControlPtType}{\textsc{cptr}}
\newcommand{\ptKind}{\textsc{ptKind}}
\newcommand{\DataPtType}{\textsc{dptr}}
\newcommand{\mapEntry}[2]{{#1} \mapsto {#2}}
\newcommand{\mapExt}[2]{{#1}[{#2}]}
\newcommand{\mapExtThree}[3]{{#1}[{#2}]_{#3}}
\newcommand{\unitValue}{tt}
\newcommand{\nullPointer}{\textit{null}}
\newcommand{\absEvent}{\textsc{Ev}}
\newcommand{\repEvent}{\textsc{Rep}}

\newcommand{\ldot}{.\,}
\newcommand{\eqdef}{\mathrel{\:\widehat{=}\:}}
\newcommand{\inputValName}{\textsc{writer}}
\newcommand{\inputVal}[3]{\inputValName\ {#1}\ {#2}\ {#3}}

\newcommand{\refleTransCl}[1]{\mathrel{{#1}^{*}}}
\newcommand{\transCl}[1]{\mathrel{{#1}^{+}}}

\newcommand{\listvar}[1]{\overline{#1}}
\newcommand{\concat}{%
	\mathbin{{+}\mspace{-8mu}{+}}%
}

\newcommand{\stackFull}[2]{{#1} :: {#2}}
\newcommand{\stackEmpty}{[\,]}

\newcommand{\queueFull}[2]{{#1} :: {#2}}
\newcommand{\queueEmpty}{[\,]}
\newcommand{\queueList}[1]{[{#1}]}
\newcommand{\UNDEF}[1]{\textit{undefined}\,({#1})}
\newcommand{\DEF}[1]{\textit{defined}\,({#1})}

\newcommand{\PENDName}{\textsc{Miss}}
\newcommand{\PEND}[2]{\PENDName\ {#1}\ {#2}}

\newcommand{\TBName}{\textsc{Tb}}
\newcommand{\TB}[2]{\TBName\ {#1}\ {#2}}

\newcommand{\terminatedEvent}{T}
\newcommand{\closedEventSymbol}{\overline{\terminatedEvent}}

\newcommand{\closedEvent}{\closedEventSymbol}
\newcommand{\postPredSymbol}{\mathcal{Q}}
\newcommand{\postPred}[2]{\postPredSymbol_{{#1},{#2}}}

\newcommand*{\sepEq}{%
\mathrel{\vcenter{\offinterlineskip
\hbox{\phantom{$\ltimes$}}\vskip-.35ex\hbox{$\ltimes$}\vskip-.35ex\hbox{$-$}}}}

\newcommand*{\llEq}{%
	\mathrel{\vcenter{\offinterlineskip
			\hbox{\phantom{$\ll$}}\vskip-.35ex\hbox{$\ll$}\vskip-.35ex\hbox{$-$}}}}

\newcommand{\visObsSYMBOL}{\lessdot}
\newcommand{\visSepSYMBOL}{\ltimes}

\newcommand{\visSepEqSYMBOL}{\sepEq}
\newcommand{\nVisSepEqSYMBOL}{\not\sepEq}

\newcommand{\visObsSymbol}[1]{\mathrel{\visObsSYMBOL}_{#1}}
\newcommand{\visSepSymbol}[1]{\mathrel{\visSepSYMBOL}_{#1}}

\newcommand{\visSepEqSymbol}[1]{\mathrel{\visSepEqSYMBOL}_{#1}}
\newcommand{\nVisSepEqSymbol}[1]{\mathrel{\nVisSepEqSYMBOL}_{#1}}
\newcommand{\visObsIndxSymbol}[2]{\mathrel{\visObsSYMBOL}_{#1}^{#2}}
\newcommand{\visSepIndxSymbol}[2]{\mathrel{\visSepSYMBOL}_{#1}^{#2}}

\newcommand{\visSepEqIndxSymbol}[2]{\mathrel{\visSepEqSYMBOL}_{#1}^{#2}}

\newcommand{\visObs}[3]{{#2} \visObsSymbol {#1} {#3}}
\newcommand{\visSep}[3]{{#2} \visSepSymbol {#1} {#3}}

\newcommand{\visSepEq}[3]{{#2} \visSepEqSymbol {#1} {#3}}
\newcommand{\nVisSepEq}[3]{{#2} \nVisSepEqSymbol {#1} {#3}}
\newcommand{\visObsIndx}[4]{{#3} \visObsIndxSymbol {#1} {#2} {#4}}
\newcommand{\visSepIndx}[4]{{#3} \visSepIndxSymbol {#1} {#2} {#4}}

\newcommand{\genVisSymbol}{\prec}
\newcommand{\genVis}[2]{{#1} \genVisSymbol {#2}}
\newcommand{\genVisTransSymbol}{\genVisSymbol^{+}}
\newcommand{\genVisTrans}[2]{{#1} \genVisTransSymbol {#2}}

\newcommand{\triagVisSymbol}{\vartriangleleft}
\newcommand{\triagVis}[2]{{#1} \triagVisSymbol {#2}}

\newcommand{\triagVisEqSymbol}{\trianglelefteq}

\newcommand{\writesAbsSymbol}{\mathcal{W}}
\newcommand{\writesAbs}[1]{\writesAbsSymbol_{#1}}

\newcommand{\precedesAbsSymbol}{\sqsubset}
\newcommand{\precedesAbs}[2]{{#1} \precedesAbsSymbol {#2}}
\newcommand{\precedesAbsEqSymbol}{\sqsubseteq}
\newcommand{\precedesAbsEq}[2]{{#1} \precedesAbsEqSymbol {#2}}

\newcommand{\nprecedesAbsSymbol}{\not\sqsubset}
\newcommand{\nprecedesAbs}[2]{{#1} \nprecedesAbsSymbol {#2}}
\newcommand{\nprecedesAbsEqSymbol}{\not\sqsubseteq}
\newcommand{\nprecedesAbsEq}[2]{{#1} \nprecedesAbsEqSymbol {#2}}

\newcommand{\spansSymbol}{\mathcal{S}}
\newcommand{\spans}[1]{\spansSymbol_{#1}}

\newcommand{\writesSpansSymbol}{\mathcal{S}^W}
\newcommand{\writesSpans}[1]{\writesSpansSymbol_{#1}}

\newcommand{\hspans}[2]{\spans{#1}(#2)}

\newcommand{\spanEvSymbol}{\mathcal{S}}

\newcommand{\spanEv}[1]{\spanEvSymbol\ {#1}}
\newcommand{\spanEvOutSymbol}{\textsc{O}}
\newcommand{\spanEvOut}[1]{\spanEvOutSymbol({#1})}
\newcommand{\precedesSpansSymbol}{\sqsubset^S}
\newcommand{\precedesSpans}[2]{{#1} \precedesSpansSymbol {#2}}
\newcommand{\precedesSpansEqSymbol}{\sqsubseteq^S}

\newcommand{\precedesRepsSymbol}{<}
\newcommand{\precedesReps}[2]{{#1} \precedesRepsSymbol {#2}}
\newcommand{\precedesRepsEqSymbol}{\leq}
\newcommand{\precedesRepsEq}[2]{{#1} \precedesRepsEqSymbol {#2}}

\newcommand{\generatorEvent}[1]{\left\llbracket{#1}\right\rrbracket}
\newcommand{\gEv}[1]{\generatorEvent{#1}}

\newcommand{\outputPropName}{\textit{out}}
\newcommand{\outputProp}[1]{{#1}.\outputPropName}
\newcommand{\STimePropName}{\textit{start}}
\newcommand{\STimeProp}[1]{{#1}.\STimePropName}
\newcommand{\TimePropName}{\textit{time}}
\newcommand{\TimeProp}[1]{{#1}.\TimePropName}
\newcommand{\ETimePropName}{\textit{end}}
\newcommand{\ETimeProp}[1]{{#1}.\ETimePropName}
\newcommand{\inPropName}{\textit{in}}
\newcommand{\inProp}[1]{{#1}.\inPropName}

\newcommand{\idPropName}{\textit{id}}
\newcommand{\idProp}[1]{\idPropName\ {#1}}
\newcommand{\stampPropName}{t_s}
\newcommand{\stampProp}[1]{\stampPropName\ {#1}}

\newcommand{\AstampPropName}{at_s}
\newcommand{\AstampProp}[1]{\AstampPropName\ {#1}}

\newcommand{\elimRelName}{\textsc{Elim}}
\newcommand{\elimRel}[2]{{#1}\ \elimRelName\ {#2}}
\newcommand{\elimPairSetName}{P}
\newcommand{\elimPairSet}[1]{\elimPairSetName_{#1}}
\newcommand{\elimSetName}{E}
\newcommand{\elimSet}[1]{\elimSetName_{#1}}
\newcommand{\elimPairName}{\textsc{Elim}}
\newcommand{\elimPair}[1]{\elimPairName_{#1}}
\newcommand{\BEName}{\textsc{BE}}
\newcommand{\BE}[3]{\BEName_{#1}\ {#2}\ {#3}}
\newcommand{\cPairName}{C}
\newcommand{\cPair}[1]{\cPairName\ {#1}}

\newcommand{\pointOne}[1]{{#1}.pt_1}
\newcommand{\pointTwo}[1]{{#1}.pt_2}
\newcommand{\expOne}[1]{{#1}.exp_1}
\newcommand{\expTwo}[1]{{#1}.exp_2}
\newcommand{\newTwo}[1]{{#1}.new_2}
\newcommand{\pointGen}[1]{{#1}.pt}
\newcommand{\expGen}[1]{{#1}.exp}
\newcommand{\newGen}[1]{{#1}.new}
\newcommand{\statusProp}[1]{{#1}.\textit{status}}
\newcommand{\entriesProp}[1]{{#1}.\textit{entries}}

\newcommand{\updateEntry}[2]{{#1}_{#2}}
\newcommand{\pointGenEntry}[1]{\updateEntry{pt}{#1}}
\newcommand{\expGenEntry}[1]{\updateEntry{exp}{#1}}
\newcommand{\newGenEntry}[1]{\updateEntry{new}{#1}}

\newcommand{\pointerIndx}[1]{{#1}}

\newcounter{axiom}[figure]
\newcommand{\axiomHLabel}[1]{\refstepcounter{axiom}\label{#1}($H_{\theaxiom}$)}

\newcommand{\axiomSLabel}[1]{\refstepcounter{axiom}\label{#1}($C_{\theaxiom}$)}
\newcommand{\axiomQLabel}[1]{\refstepcounter{axiom}\label{#1}($Q_{\theaxiom}$)}
\newcommand{\axiomLLabel}[1]{\refstepcounter{axiom}\label{#1}($L_{\theaxiom}$)}
\newcommand{\axiomHRef}[1]{$H_{\ref{#1}}$}
\newcommand{\axiomSRef}[1]{$C_{\ref{#1}}$}
\newcommand{\axiomQRef}[1]{$Q_{\ref{#1}}$}
\newcommand{\axiomLRef}[1]{$L_{\ref{#1}}$}

\newcommand{\axiomILabel}[1]{\refstepcounter{axiom}\label{#1}($I_{\theaxiom}$)}
\newcommand{\axiomIRef}[1]{$I_{\ref{#1}}$}
\newcommand{\axiomPLabel}[1]{\refstepcounter{axiom}\label{#1}($P_{\theaxiom}$)}
\newcommand{\axiomPRef}[1]{$P_{\ref{#1}}$}



%% file: abstract.tex
Linearizability is a standard correctness criterion for concurrent
algorithms, typically proved by establishing the algorithms'
linearization points (LP). However, LPs often hinder abstraction, and
for some algorithms such as the timestamped stack, it is unclear how
to even identify their LPs.
In this paper, we show how to develop declarative proofs of
linearizability by foregoing LPs and instead employing axiomatization
of so-called visibility relations. While visibility relations have
been considered before for the timestamped stack, our study is the
first to show how to derive the axiomatization systematically and
intuitively from the sequential specification of the stack.
In addition to the visibility relation, a novel \emph{separability}
relation emerges to generalize real-time precedence of procedure
invocation. The visibility and separability relations have natural
definitions for the timestamped stack, and enable a novel proof that
reduces the algorithm to a simplified form where the timestamps are
generated atomically.

%% file: intro.tex
\section{Introduction}

A concurrent data structure is \emph{linearizable}~\cite{herlihy:90}
if in every concurrent execution history of the structure's exportable
methods, the method invocations can be ordered linearly just by
permuting \emph{overlapping} invocations, so that the obtained history
is \emph{sequentially sound}; that is, executing the methods
sequentially in the linear order produces the same outputs that the
methods had in the concurrent history. In other words, every
concurrent history is equivalent to a sequential one where methods
execute without interference, i.e., \emph{atomically}.

While linearizability is a standard correctness criterion, proving that
sophisticated data structures are linearizable is far from trivial.
The most common approach is to first describe the linearization points
(LPs) of the methods that the data structure exports. Given an
execution of a method (henceforth, event), its LP is the moment at
which the event’s effect can be considered to have occurred
abstractly, in the sense that the linearization order of the events is
determined by the real-time order of the chosen LPs.  LPs are
described operationally by indicating the line in the code together
with a run-time condition under which the line applies.
%
The proof then proceeds by a simulation argument to show that the
effect of the invocation abstractly occurs at the declared
line.

While LPs lead to a complete proof method~\cite{Schellhorn}, the
operational nature of the LP description leads to very low-level
proofs. Sometimes, it may even be unclear how to describe the position
of the LPs in the first place. An alternative, more \emph{declarative}
approach, that offers higher levels of abstraction, has been proposed
by Henzinger et al.~\citet{henzinger:concur13}. It advocates foregoing
LPs in favor of axiomatizing how the events of the structure depend on
each other. Such dependence relation has since been termed
\emph{visibility relation} in the literature~\cite{Viotti}, and has
been used to axiomatize concurrent queues~\cite{henzinger:concur13},
stacks~\cite{Dodds,Haas}, and snapshot algorithms~\cite{Joakim}. In
these cases, the higher abstraction capabilities of visibility
relations (compared to LPs) enabled that linearizability proofs of
different implementations of a data structure can share significant
proof components, or that a linearizability proof can be developed in
the first place where an LP-based proof did not exist. Nevertheless,
despite these recent successes, developing visibility-based proofs
remains an undeveloped area, with every proof approaching the
axiomatization in its own manner, without any specific
systematization.

This paper advances the visibility approach 
by proposing that the axiomatization of concurrent structures should
rely on a \emph{separability} relation between events, in addition to
the visibility relation. Separability relation partially characterizes
when two events are \emph{abstractly} non-overlapping, with one event
logically preceding the other. Thus, it is the abstract counterpart to
the ``returns-before'' relation, which is standard in the literature,
and holds between two events if, in real time, the first event
terminates before the second begins.

We employ the visibility and separability relations in tandem to derive
a new axiomatization and linearizability proof for the concurrent
structure of the \emph{timestamped stack}, initially designed and
proved linearizable by Dodds et al.~\cite{Dodds} and Haas~\cite{Haas}.
Since its inception, the timestamped stack has achieved some notoriety
for the difficulty of its linearizability proof, as it has so far
resisted an operational description of its LPs, and simulation-based
proof attempts. For example, Khyzha et al.~\citet{Khyzha} verified the
timestamped \emph{queue} by a simulation-based approach, but did not
scale to the stack. Bouajjani et al.~\citet{EnneaStack} employed
forward simulation on a simplified variant of the stack where
timestamps are allocated atomically, but did not attempt the general
variant, where the timestamp allocation is a more complex non-atomic
operation that produces behaviors not observed in the atomic case.
The original proof by Dodds et al.~is a large case analysis that mixes
visibility relations with \emph{approximate} LP descriptions, and then
adapts and corrects both as the proof advances.
However, the axioms and the definitions of the visibility relations
have been justified only technically, and have remained unconnected to
the intuition behind the structure's design.

By axiomatizing both separability and visibility relations, we derive
the following contributions: (1) We obtain a linearizability proof
that elides LPs, and is thus more declarative than the proof of Dodds
et al.; (2) The proof's declarative nature allows us to first consider
the simpler variant of the algorithm with atomic timestamp allocation,
and then show that the general variant reduces to the atomic case. The
staged proof is more intuitive than if we attempted the general case
directly, which is what Dodds et al. do.; (3) Our contribution goes
beyond a new proof for the timestamped stack, as it suggests a
\emph{systematic} way to axiomatize concurrent data structures in the
visibility style. More specifically, we show that the visibility
relation naturally emerges when one transforms an obvious state-based
sequential axiomatization of stacks to the concurrent setting with
histories. In the process, the separability relation also naturally
emerges, because one is immediately forced to generalize the
returns-before relation. So obtained axioms identify the abstractions
that are essential for understanding the algorithm, and strongly guide
the remaining proof. Finally, our approach to axiomatization applies
to other concurrent algorithms as well, and we comment in
Section~\ref{sec:related} how we did so for RDCSS and MCAS of Harris
\etal~\cite{Harris} and some other structures. Of course, the
generality of the approach remains to be evaluated on a wider set of
examples.

%% file: programs.tex
\section{The Timestamped Stack and its Timestamps}
\label{sect::stack::programs::ts-stack-algorithm}

\begin{figure}[t]
	\begin{multicols*}{2}
		
		\begin{algorithmic}[1]
			\scriptsize
			\State $pools$ : $\NodeType[maxThreads]$; \label{alg::stack::global::pool-array} 
			\State $TS$,\colorbox{lightgray}{\textit{ID}} : $\IntType = 0$; \label{alg::stack::global::global-counters} 
			\State
			%
			\Proc{$\pushAlg$\,}{$v: \ValType$}
			\State $\NodeType$ $n = $ \label{alg::stack::push::create-new-node-var-declaration}
			\State \qquad $\NodeType\{ v, \infty, pools[\textit{TID}], false,$\colorbox{lightgray}{$\textit{ID}$++}$\}$; \label{alg::stack::push::create-new-node}
			\State $pools[\textit{TID}] = n$; \label{alg::stack::push::new-node-as-top}
			\State $\StampType$ $ts = \newStampAlg()$; \label{alg::stack::push::create-stamp}
			\State $n.stamp = ts$; \label{alg::stack::push::assign-stamp}
			\EndProc
			\State
			\Proc{$\newStampAlg$\,}{}
			\State $\IntType$ $ts_1 = TS$; \label{alg::stack::newStamp::read-ts-counter-one}
			\State $pause()$; \label{alg::stack::newStamp::pause}
			\State $\IntType$ $ts_2 = TS$; \label{alg::stack::newStamp::read-ts-counter-two}
			\If {$ts_1 \neq ts_2$} \label{alg::stack::newStamp::compare-two-reads}
				\State \returnCmd{$[ts_1,ts_2-1]$}; 
			\ElsIf {$\textit{CAS}(TS,ts_1,ts_1+1)$} \label{alg::stack::newStamp::attempt-counter-increase}
				\State \returnCmd{$[ts_1,ts_1]$};
			\Else
				\State \returnCmd{$[ts_1,TS-1]$};
			\EndIf
			\EndProc
			
			\columnbreak
			\Record{$\NodeType$}
			\State $val$ : $\ValType$;
			       $stamp$ : $\StampType$;
			       $next$ : $\NodeType$; \label{alg::stack::global::node-record-start}
			\State $taken$ : $\BoolType$;
			       \colorbox{lightgray}{$id : \IntType$}; \label{alg::stack::global::node-record-end}
			\EndRecord 
			\State
			\Proc{$\popAlg$\,}{}
			\State $\BoolType$ $suc = false$; $\NodeType$ $chosen$;
			\While{\textbf{not} $suc$} \label{alg::stack::pop::full-attempt-loop}
				\State $\StampType$ $maxT = -\infty$; \label{alg::stack::pop::initialize-max-ts}
				\State $chosen = \nullPointer$;
				\ForIter{$i$}{0}{$maxThreads-1$} \label{alg::stack::pop::pools-loop}
					\State $\NodeType$ $n = pools[i]$; \label{alg::stack::pop::find-first-untaken-init}
					\While {$n.taken$ \textbf{and} $n.next \neq n$} \label{alg::stack::pop::find-first-untaken-loop}
						\State $n = n.next$;
					\EndWhile
					\State $\StampType$ $ts = n.stamp$; \label{alg::stack::pop::read-candidate-timestamp}
					\If{$maxT \stamporderSymbol ts$} \label{alg::stack::pop::compare-with-max-stamp}
						\State $chosen = n$; $maxT = ts$; \label{alg::stack::pop::update-max-variables}
					\EndIf
				\EndForIter
				\If {$chosen \neq \nullPointer$}
					\State $suc = \textit{CAS}(chosen.taken, false, true)$; \label{alg::stack::pop::try-take-chosen-node}
				\EndIf
			\EndWhile
			\State \returnCmd{$chosen.val$}; \label{alg::stack::pop::return-pop-val}
			\EndProc
		\end{algorithmic}
	\end{multicols*}
	\caption{Pseudocode of a simplified TS-stack.}
	\label{alg::stack::simple-stack}
\end{figure}

The timestamped stack (TS-stack) keeps an array of \emph{pools}, indexed
by thread IDs; one pool for each thread. A pool is a linked list of
nodes. The array index identifying the pool
(line~\ref{alg::stack::global::pool-array} in
Figure~\ref{alg::stack::simple-stack}) stores the head node of the pool
list, and each node
(lines~\ref{alg::stack::global::node-record-start}-\ref{alg::stack::global::node-record-end})
stores a value $val$, timestamp $stamp$, the $next$ node in the list, a
boolean $taken$ indicating if the value has been taken by some pop,
and a unique identifier $id$ for the node.\footnote{Unique identifiers
  $id$ are ghost code (gray color in
  Figure~\ref{alg::stack::simple-stack}), introduced solely for use in proofs.}
Each thread can only insert values in its own pool by allocating a
node at the head of the list.  A value is logically removed from the
pool once its $taken$ flag is set to $true$.\footnote{For presentation
  purposes, we simplified the original algorithm, but treat a more
  general form in 
  Appendix~\ref{subsect::stack::appendix::vis-proof::proof-of-invariants}. 
  The two versions exhibit the same
  challenges, and use the sames axiomatization and definitions of
  visibility and separability. The differences between them are
  discussed in Section~\ref{sec:related}.}

The $\pushAlg$ procedure inserts a new node containing the pushed
value $v$ into the pool of the executing thread with thread id
$\textit{TID}$.  More specifically,
in line~\ref{alg::stack::push::create-new-node}, $\pushAlg$ allocates a
new node with the value $v$, infinite timestamp, $next$ pointing
to the current head of the pool, taken flag set to
\textit{false}, and fresh unique identifier, where $\textit{ID}$++
denotes an atomic fetch and increment on the global counter
$\textit{ID}$. Then, the new node is set as the new head of
the $\textit{TID}$ pool
(line~\ref{alg::stack::push::new-node-as-top}), a new timestamp is
generated (line~\ref{alg::stack::push::create-stamp}) and assigned to
the node (line~\ref{alg::stack::push::assign-stamp}) as a replacement
for the original infinity timestamp. We will discuss 
infinity timestamps and the $\newStampAlg$ procedure further below.

The $\popAlg$ procedure traverses the pools (loop at
line~\ref{alg::stack::pop::pools-loop}), searching for an
untaken node with a maximal timestamp in the partial order
$\stamporderSymbol$, updating the current maximum in the variable
$maxT$
(lines~\ref{alg::stack::pop::compare-with-max-stamp}-\ref{alg::stack::pop::update-max-variables}).
Once a maximal node is found, $\popAlg$ attempts to remove it by
CAS-ing on its $taken$ flag at
line~\ref{alg::stack::pop::try-take-chosen-node}. The $\popAlg$ procedure
restarts (loop at line~\ref{alg::stack::pop::full-attempt-loop}) 
if it was not able to take a maximal node
at line~\ref{alg::stack::pop::try-take-chosen-node}.

The role of $\stamporderSymbol$ is to endow TS-stack with a 
LIFO discipline whereby an element with a larger timestamp (i.e, the
more recently pushed element), is popped first. In the concurrent
setting, however, the meaning of ``more recent'' is not as
straightforward as in the sequential setting, as the definition of
linearizability allows that overlapping operations can be linearized
in either order. In particular, if two invocations of $\pushAlg$
overlapped, they can actually be popped in either order. To reflect
this property of linearizability, the order $\stamporderSymbol$ is
\emph{partial} as opposed to total. However, to be sequentially sound,
it is of essence that if two pushes did \emph{not} overlap, then the
more recent push is indeed popped first.\footnote{Further assuming
  that the pops also did not overlap among themselves or with the
  pushes.}  This is why the implementation of $\newStampAlg$ should
satisfy the property that two non-overlapping calls to $\newStampAlg$
produce timestamps that actually \emph{are} ordered by
$\stamporderSymbol$.

There are several ways in which one can implement $\newStampAlg$ to
satisfy this property, and Figure~\ref{alg::stack::simple-stack} shows
the particularly efficient variant proposed by Dodds et
al.~\citet{Dodds}. We will return to this variant promptly. However,
for purposes of understanding and proving the algorithm linearizable,
one may consider a simpler version whereby timestamps are integers,
and $\newStampAlg$ is implemented to keep a global counter that is
\emph{atomically} fetch-and-incremented on each call, returning the
current count as the fresh timestamp. Such an atomic implementation
results in $\stamporderSymbol$ that is actually a total order, and
much simpler to analyze than the efficient variant in
Figure~\ref{alg::stack::simple-stack}. We will use the atomic
implementation as a stepping stone in our proof; we will prove it
linearizable first, and then show that the linearizability argument
for the efficient variant \emph{reduces} to the atomic case.

The reason to consider a non-atomic implementation at all is that the
atomic one suffers from a performance issue that threads contend on
the global timestamp counter. The efficient variant from
Figure~\ref{alg::stack::simple-stack} improves on this by introducing
\emph{interval} timestamps of the form $[a,b]$ for integers
$a \leq b$, where $[a,b] \stamporderSymbol [c,d]$ holds if $b < c$ in
the standard integer order. Obviously, so defined $\stamporderSymbol$
is only a partial order, as it does not order \emph{every} two
interval timestamps.  Nevertheless, it still suffices for
linearizability, because if two push events are assigned overlapping
interval timestamps, such events must overlap as well, and thus do not
constrain the order in which they are popped in a
linearization.

The $\newStampAlg$ from Figure~\ref{alg::stack::simple-stack} still
keeps a global counter $\textit{TS}$, as the atomic variant would, but
it does not always synchronize accesses to it. In particular,
$\textit{TS}$ is first read twice into $ts_1$ and $ts_2$
(lines~\ref{alg::stack::newStamp::read-ts-counter-one} and
\ref{alg::stack::newStamp::read-ts-counter-two}, respectively). In the
common case when some thread interfered on $\textit{TS}$ (i.e.,
$ts_1 \neq ts_2$), the method generates an interval with endpoint
$ts_2-1$, and terminates without having performed any synchronization.
Some synchronization is required only when no interference is detected
(i.e., $ts_1 = ts_2$). In that case, $\newStampAlg$ $\textit{CAS}$-es
over $\textit{TS}$
(line~\ref{alg::stack::newStamp::attempt-counter-increase}), to
atomically increment $\textit{TS}$. As $\textit{CAS}$ is an expensive
operation, invoking $\textit{pause}()$ in
line~\ref{alg::stack::newStamp::pause} increases the probability of
interference, and thus decreases the need for $\textit{CAS}$.
If the $\textit{CAS}$ succeeds, an interval with endpoint $ts_1$ is
returned.
If the $\textit{CAS}$ fails, some other thread increased $\textit{TS}$, and the
method returns an endpoint $\textit{TS}-1$.
In all cases, when $\newStampAlg$ terminates, $\textit{TS}$ has been
increased either by the executing thread or by another thread, and the
generated interval's endpoint is strictly smaller than the current
value of $\textit{TS}$. Thus, a subsequent non-overlapping invocation
of $\newStampAlg$ will produce an interval that is strictly larger in
$\stamporderSymbol$. This ensures that two sequentially
non-overlapping pushes generate non-overlapping interval
timestamps.

Note that $\newStampAlg$ could return the \emph{same} interval
timestamp for two different overlapping invocations. For example, with
initial $TS = 0$, a thread $T_1$, after reading $TS$ the first time at
line~\ref{alg::stack::newStamp::read-ts-counter-one} (returning $0$),
waits at line~\ref{alg::stack::newStamp::pause} while another thread
$T_2$ fully executes $\newStampAlg$, meaning that $T_2$ increased $TS$
at line~\ref{alg::stack::newStamp::attempt-counter-increase} and
returned timestamp $[0,0]$. When $T_1$ resumes, it again reads $TS$ at
line~\ref{alg::stack::newStamp::read-ts-counter-two} (returning $1$),
and so returns $[0,0]$.

Finally, $\stamporderSymbol$ is formally augmented with infinite
timestamps $-\infty$ and $\infty$, so that
$-\infty \stamporderSymbol t \stamporderSymbol \infty$ for any
timestamp $t$ generated by $\newStampAlg$. This enables $\popAlg$ to
start its search with minimum timestamp $-\infty$
(line~\ref{alg::stack::pop::initialize-max-ts}). Similarly, $\pushAlg$
can assign maximum timestamp $\infty$ to a fresh node
(line~\ref{alg::stack::push::create-new-node}) before assigning it a
finite timestamp; an intervening pop could take such a fresh node
immediately, as the node is the most recent.

%% file: visibility.tex
\section{Axiomatizing Visibility and Separability}

\begin{figure}[t]
	\centering 
	\begin{subfigwrap}{State-based sequential specification.}{subfig::stack::atomic-spec-state}
		\centering
		\begin{tabular}{l}
			$(A_1)$ Non-empty $\popAlg$ \\
			\quad $\stackFull v S \xrightarrow{\popAlg()\ \langle v \rangle} S$ \\
		\end{tabular}
		\quad
		\begin{tabular}{l}
			$(A_2)$ Empty $\popAlg$ \\
			\quad $\stackEmpty \xrightarrow{\popAlg()\ \langle \EMPTY \rangle} \stackEmpty$ \\
		\end{tabular}
		\quad
		\begin{tabular}{l}
			$(A_3)$ $\pushAlg$ \\
			\quad $S \xrightarrow{\pushAlg(v)\ \langle \unitValue \rangle} \stackFull v S$ \\
		\end{tabular}
	\end{subfigwrap}
	
	\begin{subfigwrap}{History-based sequential specification. Relation $\visObsSYMBOL : \absEvent \times \absEvent$ is
			abstract, and $\inProp{u}$ is event $u$'s input.}{subfig::stack::atomic-spec-history}
		\centering
		\begin{tabular}{c}
			\begin{tabular}{l}
				$(B_1)$ LIFO \\
				\quad $\visObs {} {u_1} {o_1} \wedge u_1 \precedesAbsSymbol u_2 \precedesAbsSymbol o_1 \implies 
				\exists o_2.\ \visObs {} {u_2} {o_2} \wedge \precedesAbs {o_2} {o_1}$ \\
				$(B_2)$ Pop uniqueness \\
				\quad $\visObs {} u {o_1} \wedge \visObs {} u {o_2} \implies o_1 = o_2$ \\
				$(B_3)$ Dependences occur in the past \\
				\quad $\visObs {} u o \implies \precedesAbs u o$\\
				$(B_{4.1})$ Non-empty $\popAlg$ \\
				\quad $o = \popAlg()\left<v\right> \wedge v \neq \EMPTY \implies 
				\exists u.\ \visObs {} u o \wedge v = \inProp{u}$ \\
				$(B_{4.2})$ Empty $\popAlg$ \\
				\quad $o_1 = \popAlg()\left<\EMPTY\right> \implies 
				\forall u.\ \precedesAbs u {o_1} \implies \exists o_2.\ \visObs {} u {o_2} \wedge \precedesAbs {o_2} {o_1}$ \\
				$(B_{4.3})$ $\pushAlg$ \\
				\quad $u = \pushAlg(\_)\left<v\right> \implies v = \unitValue$ \\
			\end{tabular}
		\end{tabular}
	\end{subfigwrap}
	\caption{State-based and history-based sequential specifications for stacks. Variables $u$, $o$, and their 
		indexed variants, range over pushes and pops, respectively.}
	\label{fig::stack::atomic-specs-stack}
\end{figure}

\subsection{Sequential History Specifications and Visibility Relations}
\label{subsect::stack::rels::visibility-relations-and-seq-specs}

Following Henzinger et al.~\citet{henzinger:concur13}, we start the
development of visibility relations by introducing \emph{history-based
  specifications} for our data structure.
History-based specifications describe relationships between the data
structure's procedures in an execution history. They are significantly
different from the perhaps more customary state-based specifications
that describe the actions of a procedure in terms of input and output
state. However, history-based specifications scale better to the
concurrent setting, which is why concurrent consistency criteria such
as linearizability are invariably defined in terms of execution
histories.

In this section we focus on \emph{sequential} histories in order to
introduce the idea of \emph{visibility relation} in a simple way,
before generalizing to concurrent histories in
Section~\ref{subsect::stack::rels::separability-relations-and-concur-specs}.
A sequential history is a sequence of the form
$[proc(in_1) \langle out_1 \rangle,\ldots, proc(in_n) \langle out_n
\rangle]$, where $proc(in_i) \langle out_i \rangle$ means that
$proc(in_i)$ executed \emph{atomically} and produced output
$out_i$. We term \emph{event} each element in a sequential history $h$,
and $\absEvent$ denotes the set of all events in $h$.

Figure~\ref{fig::stack::atomic-specs-stack} illustrates the
distinction between sequential state-based and history-based
specifications for stacks. For the state-based specification in
Figure~\ref{subfig::stack::atomic-spec-state}, let us denote by
$S \xrightarrow{proc(in)\ \langle out \rangle} S'$ the statement that
event $proc$ with input $in$ executes atomically on stack $S$,
produces output $out$ and modifies the stack into $S'$.  Axiom $A_1$
says that a $\popAlg$ removes the top element $v$ from a non-empty
stack and returns $v$.  Axiom $A_2$ says that $\popAlg$ returns
$\EMPTY$ when the stack is empty, leaving the stack unchanged. Axiom
$A_3$ says $\pushAlg(v)$ inserts $v$ into the stack as the new top
element, returning the trivial value $\unitValue$.

Figure~\ref{subfig::stack::atomic-spec-history} shows the
history-based sequential specification for stacks.  The specification
utilizes the \emph{visibility relation} $\visObsSYMBOL$ to capture a
push-pop causal dependence between events. In particular,
$\visObs {} u o$ means that ``event $o$ pops a value that event $u$
pushed onto the stack''. We usually say that $u$ is \emph{visible} to
$o$, or that $o$ \emph{observes} $u$. Under this interpretation,
axioms $B_1,...,B_{4.3}$ state the following expected
properties.\footnote{Our paper will make heavy use of several
  different relations. To help the reader keep track of them, we
  denote the relations by symbols that graphically associate to the
  relation's meaning. For example, we use $\visObsSYMBOL$ for the
  visibility relation, because the symbol graphically resembles an
  eye.}

Axiom $B_1$ (LIFO) states that more recent pushes are popped first.
More specifically, if $o_1$ observes $u_1$ (i.e,
$\visObs {} {u_1} {o_1}$) and $u_2$ is a later push executing between
$u_1$ and $o_1$ (i.e.,
$u_1 \precedesAbsSymbol u_2 \precedesAbsSymbol o_1$), then $u_2$ must
be popped before $o_1$ pops $u_1$, otherwise the value pushed by $u_1$
would not be at the top of the stack for $o_1$ to take.
Relation $\precedesAbsSymbol$ is the \emph{returns-before} relation
(with $\precedesAbsEqSymbol$ its reflexive closure), where
$\precedesAbs x y$ means that $x$ terminated before $y$ started.  Note
that $\precedesAbsSymbol$ is a total order on events, as in a
sequential execution, different events cannot overlap.

Axiom $B_2$ (Pop uniqueness) says that a push is observed by at most
one pop. 

Axiom $B_3$ (Dependences occur in the past) says that if a
pop depends on a push, then the push executes before the pop.

Axioms $B_{4.1}$ (Non-empty $\popAlg$), $B_{4.2}$ (Empty $\popAlg$),
and $B_{4.3}$ essentially are the counterparts of the state-based
sequential axioms $A_1$-$A_3$, respectively, as we show next.

Axiom $B_{4.1}$ says that a $\popAlg()\left<v\right>$ event $o$
observes a push $u$ that pushed $v$.
This axiom, along with $B_1$-$B_3$, ensures that $o$ relates to $u$ as
in the following diagram.
\vspace{-1mm}
\begin{center}
\begin{tikzcd}[ampersand replacement=\&, column sep=small, row sep=small] 
	S 
	\arrow[r, "{u}"{name=U,yshift=2pt}] \&
	\stackFull{v}{S}
	\arrow[r, "{u_{i}}"{name=Ui,yshift=2pt}] \& 
	\dots 
	\arrow[r, "{o_{j}}"{name=Oj,yshift=2pt}] \& 
	\stackFull{v}{S}
	\arrow[r, "{u_{k}}"{name=Uk,yshift=2pt}] \& 
	\ldots 
	\arrow[r, "{o_{l}}"{name=Ol,yshift=2pt}] \& 
	\stackFull{v}{S}
	\arrow[r, "{o = \popAlg()\left< v \right>}"{name=O}] \&[30pt] 
	S 
	\\
	\arrow[from=O, to=U, bend right, dashed, no head, "{\visObsSYMBOL}"{below}]
	\arrow[from=Oj, to=Ui, bend right, dashed, no head, "{\visObsSYMBOL}"{above}]
	\arrow[from=Ol, to=Uk, bend right, dashed, no head, "{\visObsSYMBOL}"{above}]
\end{tikzcd}
\end{center}
\vspace{-5mm}
In particular: (i) $u$ executes before $o$ (by axiom $B_3$, because
$\visObs{} {u}{o}$), (ii) every push between $u$ and $o$ is popped
before $o$ (by axiom $B_1$), and each push is popped exactly once (by
axiom $B_2$). Thus, once $o$ executes, the value $v$ pushed by the
observed $u$, is actually the most recent unpopped value, i.e., it is
on the top of the stack. Subsequent pops cannot observe this value anymore
either (again by axioms $B_1$ and $B_2$), thus the stack is modified
from $v :: S$ to $S$. This explains that $B_{4.1}$ is essentially a
history-based version of $A_1$.

Similarly, axiom $B_{4.2}$ states that if a
$\popAlg()\left<\EMPTY\right>$ event $o$ occurs, then every push
before $o$ must have been popped before $o$, as this ensures that the
stack is empty when $o$ is reached. Hence, the axiom is counterpart to
$A_2$.

Finally, axiom $B_{4.3}$ says that the output of a push event is the
trivial value $\unitValue$. The axiom imposes no conditions on the
stack, as a value can always be pushed. In this, $B_{4.3}$ is the
counterpart to $A_3$ which also imposes no conditions on the input
stack, and posits that push's output value is trivial. However, unlike
$A_3$, $B_{4.3}$ does not directly says that the value is pushed on
the top of the stack, as that aspect is captured by the relationships
between pushes and pops described by $B_{4.1}$.

\begin{figure}[t]
	\begin{subfigwrap}{Concurrent specification. Relations $\visObsSYMBOL, \visSepSYMBOL : \absEvent \times \absEvent$ are abstract.}{subfig::stack::concurrent-spec-history}
		\centering
		\begin{tabular}{l}
			\axiomSLabel{vis-ax::stack::cc-concurrent-lifo} Concurrent LIFO \\
			\quad $\visObs {} {u_1} {o_1} \wedge o_1 \nVisSepEqSYMBOL u_2 \nVisSepEqSYMBOL u_1 \implies 
			\exists o_2.\ \visObs {} {u_2} {o_2} \wedge \visSep {} {o_2} {o_1}$ \\
			\axiomSLabel{vis-ax::stack::cc-pop-uniqueness} Pop uniqueness \\
			\quad $\visObs {} u {o_1} \wedge \visObs {} u {o_2} \implies o_1 = o_2$ \\
		\end{tabular}
		\begin{tabular}{l}
			\axiomSLabel{vis-ax::stack::cc-no-future-dependence} No future dependences \\
			\quad $x \transCl{\genVisSymbol} y \implies \nprecedesAbsEq y x$\\
			\axiomSLabel{vis-ax::stack::cc-return-completion} Return value completion \\
			\quad $\exists v.\ \postPred x v \wedge (x \in \terminatedEvent \implies v = \outputProp x)$ \\
		\end{tabular} 
	\end{subfigwrap}
	
	\begin{subfigwrap}{Defined notions.}{subfig::stack::defined-notions-concurrent-spec-history}
		\centering
		\begin{tabular}{c}
			\begin{tabular}{l}
				Constraint relation \\
				\quad ${\genVisSymbol} \defini \visObsSYMBOL \cup \visSepSYMBOL$\\
				Returns-before relation\\
				\quad $\precedesAbs {e_1} {e_2} \defini \ETimeProp {e_1} \natorderSymbol \STimeProp {e_2}$\\
			\end{tabular}
			\begin{tabular}{l}
				Set of terminated events\\
				\quad $\terminatedEvent \defini \{ e \mid \ETimeProp e \neq \bot \}$\\
				Closure of terminated events\\
				\quad $\closedEvent \defini \{ e \mid \exists t \in \terminatedEvent.\ e \refleTransCl{\genVisSymbol} t \} $\\
			\end{tabular}
			\\
			\begin{tabular}{c}
				\\
				$\postPred {o_1} v \defini
				\begin{cases}
				\exists u.\ \visObs {} u {o_1} \wedge v = \inProp{u} &
				\text{if $v \neq \EMPTY$} \\
				\forall u.\ \nVisSepEq {} {o_1} u \implies \exists o_2.\ \visObs {} u {o_2} \wedge \visSep {} {o_2} {o_1} & \text{if $v = \EMPTY$} \\
				\end{cases}$
				\qquad $\postPred {u} v \defini v = \unitValue$
			\end{tabular}
		\end{tabular}
	\end{subfigwrap}
	\caption{Concurrent history-based specification for stacks. Variables $u$, $o$, and their 
		indexed variations, range over pushes and pops in $\closedEvent$, respectively.
		Variables $x$, $y$ range over $\closedEvent$.
		Variable $e$, and its indexed variations, range over $\absEvent$. $\outputProp x$ denotes $x$'s output.}
	\label{fig::stack::concurrent-spec-stack}
\end{figure}

\subsection{Concurrent Specifications and Separability Relations}
\label{subsect::stack::rels::separability-relations-and-concur-specs}

Concurrent execution histories do not satisfy the sequential axioms in
Figure~\ref{subfig::stack::atomic-spec-history} for two related
reasons.  First, concurrent events can \emph{overlap} in real time. As
a consequence, the axioms $B_1$ (LIFO), $B_3$ (Dependencies occur in
the past), and $B_{4.2}$ (Empty $\popAlg$) are too restrictive, as
they force events to be non-overlapping (i.e., disjoint in time) due
to the use of the returns-before relation $\precedesAbsSymbol$.
Second, events can no longer be treated as atomic; thus event's start
and end times (if the event terminated) must be taken into account. As
a consequence, axioms $B_{4.1}$, $B_{4.2}$, and $B_{4.3}$ must be
modified to account for the output of an unfinished event not being
available yet.  We continue using $\absEvent$ for the set of events in
the concurrent history. We denote by $\STimeProp {e}$ and
$\ETimeProp e$ the start and end time of event $e$, respectively; for
example, for the implementation in
Figure~\ref{alg::stack::simple-stack}, a $\pushAlg$ event starts when
line~\ref{alg::stack::push::create-new-node-var-declaration} executes,
and ends when line~\ref{alg::stack::push::assign-stamp} executes. We
use the standard order relation on natural numbers $\natorderSymbol$
to compare start and end times.

Figure~\ref{fig::stack::concurrent-spec-stack} shows the modified
axioms that address the above issues. Importantly, in addition to the
visibility relation, the axioms utilize the \emph{separability
  relation} $\visSep {} x y$ to capture that ``event $x$ is separable
before $y$'', i.e., $x$ should be linearized before $y$.\footnote{The
  symbol $\visSepSYMBOL$ twists $\precedesAbsSymbol$, suggesting that
  $\visSepSYMBOL$ relaxes (i.e., is a twist on) returns-before
  relation $\precedesAbsSymbol$.} The reason for the separation
depends on the particular stack implementation, but is kept abstract
in the axioms. Correspondingly, the relation $\visSepSYMBOL$ is
also kept abstract.
We now explain how the concurrent axioms in
Figure~\ref{fig::stack::concurrent-spec-stack} are
\emph{systematically} obtained from the sequential ones in
Figure~\ref{subfig::stack::atomic-spec-history}.

Axiom \axiomSRef{vis-ax::stack::cc-concurrent-lifo} is obtained from
$B_1$ by replacing $\precedesAbsSymbol$ with $\visSepSYMBOL$ or with
the (negation of the) reflexive closure $\visSepEqSYMBOL$, following
the rules below.  The goal is to relax the real-time strong separation
imposed by $\precedesAbsSymbol$ with a more permissive separation of
$\visSepSYMBOL$. 
\begin{itemize}
\item If subformula $\precedesAbs a b$ occurs in a condition of an
  implication (negative occurrence), it is replaced with
  $\nVisSepEq {} b a$. Notice the flip in the arguments and the
  negation.
\item If subformula $\precedesAbs a b$ occurs in the conclusion of an
  implication (positive occurrence), it is replaced with
  $\visSep {} a b$.
\end{itemize}
These rules have the following justification.  Let us suppose we have
a formula $\phi \defini \precedesAbs a b \implies \precedesAbs c d$ in
some sequential axiom.
In the sequential case, $\precedesAbsSymbol$ is a total order, which
means that $\phi$ is equivalent to
$\precedesAbsEq b a \vee \precedesAbs c d$.  After directly replacing
$\precedesAbsSymbol$ for $\visSepSYMBOL$, we obtain
$\visSepEq {} b a \vee \visSep {} c d$, which is further equivalent to
$\psi \defini \nVisSepEq {} b a \implies \visSep {} c d$.  Comparing
$\phi$ and $\psi$, we see that $\psi$'s condition is flipped,
replaced, and negated, while its conclusion is only replaced. An
important aspect of our procedure is that negative occurrences of
$\visSepSYMBOL$ in $\psi$ are themselves negated. Thus, intuitively,
$\psi$ as a whole remains positive with respect to~$\visSepSYMBOL$.
Positive formulas remain true under extensions of $\visSepSYMBOL$,
which is crucial, as the linearizability proof will involve extending
$\visSepSYMBOL$ until reaching a total order.

Axiom \axiomSRef{vis-ax::stack::cc-pop-uniqueness} is unchanged compared to $B_2$.

Axiom \axiomSRef{vis-ax::stack::cc-no-future-dependence} is obtained
from $B_3$ as follows. In the sequential specification,
$\visObsSYMBOL$ was the only relation encoding dependences between
events, but now we have two relations encoding dependences,
$\visObsSYMBOL$ and $\visSepSYMBOL$.  To collect them, we define a new
relation
${\genVisSymbol} \defini {{\visObsSYMBOL} \cup {\visSepSYMBOL}}$ which
we call \emph{constraint} relation.\footnote{The symbol
  $\genVisSymbol$ is like an eye with no iris; thus, ``blinder'' than
  $\visObsSYMBOL$, reflecting that $\genVisSymbol$ is a superset of
  $\visObsSYMBOL$.}
We can consider modifying Axiom $B_3$ into
$\genVis x y \implies \precedesAbs x y$ to say that any dependence $x$
of $y$ must terminate before $y$ starts.
However, such a modification of $B_3$ is too stringent, as it
does not allow $x$ to overlap with $y$. Instead, we relax the
conclusion to say that an event cannot depend on itself or events from
the future, i.e., $\genVis x y \implies \nprecedesAbsEq y x$. Finally,
we get axiom \axiomSRef{vis-ax::stack::cc-no-future-dependence} by replacing $\genVisSymbol$ with its transitive
closure $\transCl{\genVisSymbol}$ to account for \emph{indirect}
dependences of $y$; e.g., in $x_1 \genVisSymbol x_2 \genVisSymbol y$,
event $x_1$ is an indirect dependence of $y$.
Hence, \axiomSRef{vis-ax::stack::cc-no-future-dependence} reads ``any direct
or indirect dependence does not execute in the future, and events do
not depend on themselves''.

To understand Axiom \axiomSRef{vis-ax::stack::cc-return-completion},
we need to consider the set $T$ of all \emph{terminated} events and
its closure under the constraint relation
$\closedEvent \defini \{ e \in \absEvent \mid \exists t \in
\terminatedEvent.\ e \refleTransCl{\genVisSymbol} t \}$. As usual,
$\refleTransCl{\genVisSymbol}$ is the reflexive-transitive closure of
$\genVisSymbol$.
The reason for considering this set is that the variable $x$ over
which the axiom implicitly quantifies ranges over $\closedEvent$.

It is standard in linearizability that the linearization order
contains all the terminated events, plus selected unterminated events
with fictitious, but suitable, outputs. The selected unterminated
events are typically those that executed their effect, which then
influenced others, and must thus be included for sequential
soundness. The set $\closedEvent$ precisely determines the events to
be included by saturating the set of terminated events $T$ under
$\genVisSymbol$.

Axiom \axiomSRef{vis-ax::stack::cc-return-completion} then codifies
when an output $v$ is suitable for an event $x$ by means of the
\emph{postcondition predicate} $\postPred x v$. In particular,
\axiomSRef{vis-ax::stack::cc-return-completion} says that $v$ exists
such that $\postPred x v$. If $x \in \closedEvent$ is unterminated, we
use that $v$ as the fictitious output. If $x$ is terminated
($x \in T$), then $v$ must be $x$'s actual output.
The postcondition predicate $\postPred x v$ describes how $x$ and $v$
relate in the case of stacks. It is obtained by coalescing the axioms
$B_{4.1}$, $B_{4.2}$ and $B_{4.3}$, which themselves describe the
outputs of stack events in the sequential setting, and which we first
modify according to the systematic transformation outlined above.

We henceforth call the axioms in
Figure~\ref{fig::stack::concurrent-spec-stack}, \emph{visibility-style
  axioms}. These axioms imply linearizability of any stack
implementation satisfying them,

\begin{thm}
	\label{thm::stack::vis-rels::main-linearizable}
	Let $D$ be an arbitrary implementation of a concurrent stack.
        If there are relations $\visSepSYMBOL$ and $\visObsSYMBOL$
        definable using $D$ such that the visibility-style axioms
        hold, then $D$ is linearizable.
\end{thm}

The proof starts with the relation
${\triagVisSymbol} \defini (\genVisSymbol \cup \precedesAbsSymbol)^+$,
i.e, the transitive closure of the union of $\genVisSymbol$ and
$\precedesAbsSymbol$.  Then, it shows that $\triagVisSymbol$ is a
partial order that can be extended to a sequentially sound total order
$\leq$ by using the visibility-style axioms. Since $\leq$ contains
$\genVisSymbol$, this means that relations $\visObsSYMBOL$ and
$\visSepSYMBOL$ define ordering constraints that
linearization respects.

%% file: spans.tex
\section{Visibility and Separability for the TS-stack}
\label{sect::stack::spans::visi-sep-defs}

By Theorem~\ref{thm::stack::vis-rels::main-linearizable}, to prove
linearizability for the TS-stack, it suffices to define the relations
$\visObsSYMBOL$ and $\visSepSYMBOL$ and show that they satisfy the
axioms from Figure~\ref{fig::stack::concurrent-spec-stack}.
We carry out this proof in two stages: In
Section~\ref{subsect::spans::stack::atomic-case} we prove
linearizability when the $\newStampAlg$ procedure is implemented by an
atomic fetch-and-increment operation on a global counter
$\textit{TS}$, as discussed in
Section~\ref{sect::stack::programs::ts-stack-algorithm}. In
Section~\ref{subsect::spans::stack::interval-case} we show how the
general case of interval timestamps reduces to the atomic case.  The
lifting exploits that the difference between the atomic and interval
cases is only in the implementation of $\newStampAlg$.

For simplicity, in both cases we explicitly exclude \emph{elimination
  pairs} from the discussion. An elimination pair consists of a push
and an overlapping pop event that takes the value pushed.  The elision
allows the discussion to only consider pushes with \emph{finite}
timestamps. Indeed, every push is first assigned an infinite timestamp
(line~\ref{alg::stack::push::create-new-node} in
Figure~\ref{alg::stack::simple-stack}), which is then refined into a
finite one in line~\ref{alg::stack::push::assign-stamp}. If a push
$u$, having not yet reached line~\ref{alg::stack::push::assign-stamp},
is taken by some pop $o$, then $u$ and $o$ overlap, and hence form an
elimination pair.\footnote{Eliding elimination pairs when dealing with
  stacks is justified because such pairs can be linearized simply as a
  push that is immediately followed by a pop. The idea was originated
  by Hendler et al.~\citet{hsy:spaa04} and was also employed in Haas'
  PhD dissertation~\citet{Haas}, though with a different motivation
  from us and with a different soundness proof. For example, to prove
  the elimination sound, Haas shows how elimination pairs could be put
  back into the histories from which they have been removed. In
  contrast, we define visibility and separability relations that
  exclude elimination pairs, and show in
  Appendix~\ref{subsect::stack::appendix::vis-proof::axioms-from-invariants},
  how to extend the relations iteratively, one elimination pair at a
  time. The extension adds some bulk, but does not change the
  structure of the proof that we illustrate in this section.}

Also, in both cases, we utilize the abstraction we call \emph{spans},
to define the visibility and separability relations. A span of an
event is the interval in which the event accesses the shared state of
the stack. We could trivially take the span to be the whole interval
of the event, but in the case of TS-stack we can tighten it as
discussed below. In this sense, a span is a generalization of LPs;
being an interval, rather than a single point, it \emph{approximates}
where the LP of an event lies, but allows for some uncertainty as to
the LPs exact position.

The span of the $\pushAlg$ procedure starts when the new node is
linked as the first node of the pool
(line~\ref{alg::stack::push::new-node-as-top}), as this is the moment
when the new node becomes available for other events to see.  The span
ends when a finite timestamp is assigned to the new node
(line~\ref{alg::stack::push::assign-stamp}).
Notice how the span encompasses all the commands of $\pushAlg$ that
change the pool or the new node.

The span of the $\popAlg$ procedure starts at the infinity stamp
assignment (line~\ref{alg::stack::pop::initialize-max-ts}) of the last
iteration of the pools scan. The span ends at the successful CAS at
line~\ref{alg::stack::pop::try-take-chosen-node} which takes the node
for the pop to return.
Again, the span covers all the commands of $\popAlg$ that change the
pools or the taken node. These are all included in the last iteration of
the pools scan, because in all the prior iterations, the CAS modifying
the pools must have failed.

We formalize spans as pairs of rep events $(a,b)$, where $a$ and $b$
are the initial and final rep event in the span, respectively.  We
denote by $\oneSpan b$ and $\twoSpan{b}$ the standard projection
functions for span $b$.  Rep events are generated by the invocation of
a code line inside a procedure. For example, invoking
line~\ref{alg::stack::push::new-node-as-top} in
Figure~\ref{alg::stack::simple-stack} produces a rep event.  The set
of all rep events in an execution history is denoted as
$\repEvent$. The distinction between events ($\absEvent$) and rep
events is standard in linearizability~\cite{herlihy:90}.  We
denote by $\precedesRepsSymbol$ the real-time order between rep
events.
We consider only fully-formed spans; for example, if $\popAlg$ has not
executed its successful CAS, then it has no span.

We also extend our notion of timestamp into \emph{abstract
  timestamp}. An abstract timestamp is a pair $(i, t)$, where $i$ is
the (ghost) id of a node, and $t$ is a (plain) timestamp. The
extension is motivated by the observation explained in
Section~\ref{sect::stack::programs::ts-stack-algorithm} that two
pushes may actually generate the same (plain) timestamp. By attaching
the node id $i$ to the timestamp $t$, we differentiate such cases.  We
use ``timestamp'' to refer to abstract timestamps or plain timestamps
when the adjective can be inferred from the context.

We also utilize the following notation.
\begin{itemize}
\item Given event $e$, $\spanEv e$ is the unique span executed by
  $e$. The function is undefined if the argument event has not
  completed its span.

\item Given spans $a$, $b$, the relation $\precedesSpans{a}{b}$ means
  that $a$ finished before $b$ started. $\precedesSpansEqSymbol$
  denotes its reflexive closure.
      
\item For a push $u$, $\idProp u$ is the unique id of the node that
  $u$ inserted into the pool in
  line~\ref{alg::stack::push::new-node-as-top}.  Similarly, for a pop
  $o$, $\idProp o$ is the unique id of the node that $o$ took at the
  successful CAS in line~\ref{alg::stack::pop::try-take-chosen-node}.
  If $u$ and $o$ have not executed the mentioned lines, $\idPropName$
  is undefined.
\item For a push $u$, $\stampProp u$ is the abstract timestamp
  $(\idProp {u},t)$, combining $\idProp u$
      with the timestamp
      $t$ that $u$ assigned at
      line~\ref{alg::stack::push::assign-stamp}. In particular, $t$ is
      always \emph{finite}, because $\newStampAlg$ only generates
      finite timestamps.
      Similarly, for a pop $o$, $\stampProp o$ is the abstract
      timestamp $(\idProp {o},t)$, combining $\idProp {o}$
      with the timestamp $t$ of the taken node that $o$ read in
      line~\ref{alg::stack::pop::read-candidate-timestamp}. Generally,
      $\stampProp o$ may return an infinite plain timestamp; however,
      if elimination pairs are excluded, then timestamps are finite,
      as explained before.
      If an event $x$ has not executed its span, $\stampProp x$
      is undefined.

    \item Abstract timestamps admit the following partial order
      defined out of $\stamporderSymbol$ on plain timestamps, where we
      overload the symbol $\stamporderSymbol$ without confusion.
\[
(i_1,t_1) \stamporderSymbol (i_2,t_2) \defini t_1 \stamporderSymbol t_2
\]
\item We define when push $u$ and pop $o$ form an elimination pair.
\[
\elimRel{u}{o} \defini \idProp{u} = \idProp{o} \wedge u \not\precedesAbsSymbol o
\]
In English: (1) $o$ pops the node that $u$ pushed
($\idProp{u} = \idProp{o}$), and (2) $u$ and $o$ overlap.  Events $u$
and $o$ overlap if $u \not\precedesAbsSymbol o$ and
$o \not\precedesAbsSymbol u$, but it is not necessary to explicitly
check $o \not\precedesAbsSymbol u$, as that follows from
$\idProp{u} = \idProp{o}$ and a structural invariant that $o$ cannot
pop a node that has not been pushed yet 
(Appendix~\ref{subsect::stack::appendix::vis-proof::axioms-from-invariants}).

\item The set of events that occur in elimination pairs is
  $\elimSetName \defini \{ x \mid \exists y.\ \elimRel{x}{y} \vee
  \elimRel{y}{x}\}$.  As we explicitly exclude elimination pairs from
  the presentation, we assume that each event variable $x$ occurring
  in the forthcoming definitions is such that $x \notin
  \elimSetName$. In Section~\ref{sec:related} we comment how
  elimination pairs are placed back into consideration.

\end{itemize}

\subsection{Key Abstractions and Invariants}
\label{subsect::spans::stack::key-abstractions-and-invariants}

\subparagraph*{When pop misses a push} The key for understanding
TS-stacks is explaining what it means for a pop $o$ to have
\emph{missed} a push $u$. Informally, a miss occurs when $o$, in its
scan of the pools, takes a push $u'$ with a smaller timestamp than
that of $u$ (hence, $u'$ is less recent than $u$).
This is critical, because $o$ taking a less recent push than available
is seemingly a violation of the LIFO order. However, this does not
actually have to be so in the case of TS-stacks, where, for example,
it is fine for $u$ to insert into the pool after $o$ has already
scanned past the point of insertion. We can say that $u$ occurred too
late to really be available for $o$ to pop, and we simply linearize
$u$ after $o$.
The following definition formalizes when $o$ misses $u$ (i.e., when
$u$ occurs too late for $o$), focusing on the atomic timestamp case.
\begin{align}
	\label{eq::stack::miss-rel}
	\begin{aligned}
		\PEND {o} {u} \defini\ & \stampProp o \stamporderSymbolPadding \stampProp{u}\ \wedge\  \\
		& \forall o'.\ \stampProp{u} = \stampProp{o'} \implies 
                   \precedesReps {\twoSpan{\spanEv {o}}} {\twoSpan{\spanEv {o'}}} 
 	\end{aligned}
\end{align}

The first conjunct directly says that for $o$ to miss $u$, it must be
that $o$ takes a node with a smaller timestamp than that of $u$. The
second conjunct adds that, intuitively, $u$ \emph{remains untaken}
during the execution of $o$.
Indeed, if $u$ is taken by $o'$ ($\stampProp{u} = \stampProp{o'}$), the
definition requires that the span of $o'$ finishes after the span of
$o$
($\precedesReps {\twoSpan{\spanEv {o}}} {\twoSpan{\spanEv
    {o'}}}$). That is, the CAS that sets the $taken$ flag in the node
of $u$ executes after the span of $o$. In other words, if $u$ is taken
at all, then it is taken \emph{after} the span of $o$.

\begin{figure}[t]
	\includegraphics[scale=0.7]{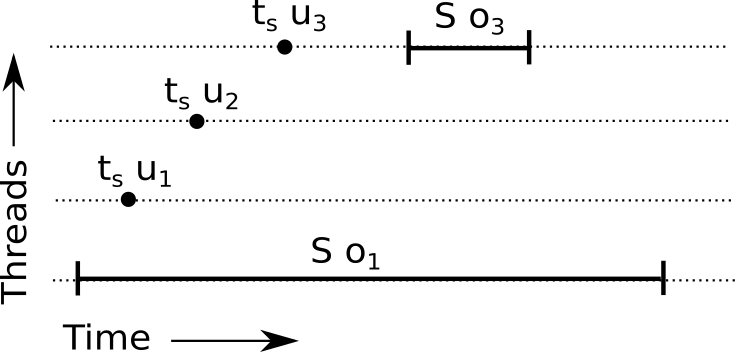}
	\centering
	\caption{Possible execution showing three atomic timestamp
          generation rep events for push events $u_1$, $u_2$, $u_3$,
          labeled by their generated timestamps; and two pop spans for
          pop events $o_1$ and $o_3$. Spans are shown as line segments
          and rep events (being atomic) as dots.  The timestamps are
          strictly increasing
          $\stampProp {u_1} \stamporderSymbol \stampProp {u_2}
          \stamporderSymbol \stampProp {u_3}$.  Event $o_1$ took
          $u_1$, while $o_3$ took $u_3$.  The events will be
          linearized as $u_1$, $o_1$, $u_2$, $u_3$, $o_3$. In
          particular, $o_1$ must be linearized before $u_2$.}
	\label{fig::stack::stack-example-diagram}
\end{figure}

Figure~\ref{fig::stack::stack-example-diagram} shows a push $u_2$ that
overlaps with a pop $o_1$, but $o_1$ takes a push $u_1$ whose
timestamp is smaller than that of $u_2$. In our definition,
$\PEND{o_1}{u_2}$ holds because $u_2$ remains untaken on the stack
after $o_1$ terminates.
$\PEND{o_1}{u_2}$ indicates that we must linearize $o_1$ before
$u_2$. And indeed, this is consistent with the situation in the
figure, as any order where $u_2$ appears before $o_1$ violates some
linearizability requirement. For example, the order $u_1$, $u_2$,
$o_1$ is sequentially unsound because $o_1$ pops $u_1$ while $u_2$ is
the top of the stack, while $u_2$, $u_1$, $o_1$ does not respect the
ordering of the timestamps of $u_1$ and $u_2$.\footnote{We 
  linearize pushes by the order of their timestamps.}

Continuing with Figure~\ref{fig::stack::stack-example-diagram}, $o_1$
\emph{does not miss} $u_3$, even though $u_3$ also overlaps with
$o_1$, and $o_1$ takes $u_1$ whose timestamp is smaller than that of
$u_3$. In our definition, $\neg\PEND{o_1}{u_3}$ because the span of
$o_3$ ends before the span of $o_1$.
$\neg\PEND{o_1}{u_3}$ indicates no restrictions on the ordering
between $o_1$ and $u_3$. For example, the only linearization order of
Figure~\ref{fig::stack::stack-example-diagram} is $u_1$, $o_1$, $u_2$,
$u_3$, $o_3$, but this is forced by the existence of $u_2$. Removing
$u_2$, the orders $u_1$, $u_3$, $o_3$, $o_1$ (where $u_3$ appears
before $o_1$) and $u_1$, $o_1$, $u_3$, $o_3$ (where $u_3$ appears
after $o_1$) are both valid.

\subparagraph*{Misses start late} 
Having defined $\PEND{o}{u}$, we can now explain the most important
invariants of the TS-stack, again focused on the atomic timestamps.
The first invariant says that a push $u$ missed by a pop $o$ has a span that
starts \emph{after} the pop's span starts. In other words, a missed
push starts after the pop that missed it.
\begin{align}
		\label{eq::stack::inv-misses-late}
	\PEND o u \implies \precedesReps {\oneSpan{\spanEv{o}}} {\oneSpan{\spanEv{u}}}
\end{align}

To intuit why \eqref{eq::stack::inv-misses-late} is an invariant,
consider a situation when $o$ misses $u$ but $u$'s span starts before
$o$'s. In that case, $u$'s pool contains $u$'s node \emph{before} $o$
even starts its scan. Thus $o$'s scan will encounter $u$ and proceed
to either take $u$, or take an even more recent push. At any rate, $o$
will not take a push with a timestamp below that of $u$; thus,
$\neg(\PEND o u)$.

\subparagraph*{Disjoint pushes order timestamps}

The next invariant is that pushes with disjoint spans, produce ordered
timestamps. Intuitively, this is so because disjoint push spans make
disjoint calls to $\newStampAlg$, which in turn generate ordered
timestamps as explained in
Section~\ref{sect::stack::programs::ts-stack-algorithm}.
\begin{align}
	\label{eq::stack::inv-disj-stamp-gen}
	\precedesSpans{\spanEv{u_1}}{\spanEv{u_2}} \implies \stampProp{u_1} \stamporderSymbol \stampProp{u_2}
\end{align}

\subsection{Case: Atomic Timestamps}
\label{subsect::spans::stack::atomic-case}

We next define the visibility $\visObsSYMBOL$ and separability
$\visSepSYMBOL$ relations for atomic timestamps.
\begin{align}
\label{eq::stack::vis-rel}
\visObs {} u o & \defini \stampProp{u} = \stampProp{o} \\
\label{eq::stack::sep-rel-push-push}
\visSep {} {u_1} {u_2} & \defini \stampProp{u_1} \stamporderSymbolPadding \stampProp{u_2} \\
\label{eq::stack::sep-rel-pop-push}
\visSep {} {o} {u} & \defini \exists u'.\ \PEND {o} {u'}\ \wedge\ \stampProp {u'} \stamporderEqSymbolPadding \stampProp u \\	
\label{eq::stack::sep-rel-pop-pop}
\visSep {} {o_2} {o_1} & \defini \stampProp{o_1} \stamporderSymbolPadding \stampProp{o_2}\ \wedge\  
\neg \exists u'.\ \PEND {o_1} {u'}\ \wedge\ \stampProp {u'} \stamporderEqSymbolPadding \stampProp {o_2}
\end{align}

The definition of $\visObs{}{}{}$ relates $u$ and $o$ if they have the
same timestamp (i.e., $o$ took $u$).

The definition of $\visSepSYMBOL$ comes with three clauses, motivated
by the form of the axioms from
Figure~\ref{fig::stack::concurrent-spec-stack}. In particular, we need
to separate a push from a push ($\visSep {} {u_1} {u_2}$), a pop from
a push ($\visSep {} {o} {u}$), and a pop from a pop
($\visSep {} {o_2} {o_1}$), but not a push from a pop, as only the
first three clauses of $\visSepSYMBOL$ appear in the axioms. 

The clause $\visSep {} {u_1} {u_2}$ naturally orders push events
according to their timestamps.

The clause $\visSep {} {o} {u}$ extends $\PEND o u'$ to account for
pushes being ordered by their timestamps, as per the previous
clause. It says that pop $o$ is separated before push $u$, if there is
a push $u'$ that was missed by $o$, and the timestamp of $u'$ is below
(or equals) that of $u$.
For example, in Figure~\ref{fig::stack::stack-example-diagram}
we have $\visSep {} {o_1} {u_2}$ and $\visSep {} {o_1} {u_3}$.

The clause $\visSep{} {o_2} {o_1}$ separates pops inversely to the
order of the taken timestamps, or equivalently, inversely to the order
of the taken pushes, but under the condition that \emph{$o_1$ did not
  miss any push with a timestamp below $o_2$}.  The last requirement
is important. For example, if we ignored it in
Figure~\ref{fig::stack::stack-example-diagram}, we would obtain
$\visSep {} {o_3} {o_1}$ since
$\stampProp {u_1} \stamporderSymbol \stampProp{u_3}$. But this order
is sequentially unsound; the pushes being ordered as $u_1$, $u_2$,
$u_3$, after $o_3$ takes $u_3$, the value pushed by $u_2$ is at the
top of the stack. But then $o_1$ cannot execute next, as we need an
intervening pop to remove $u_2$.

It is worth mentioning that we arrived at the definition of the clause
$\visSep{}{o_2}{o_1}$ by formal symbol manipulation aimed at
fulfilling axiom \axiomSRef{vis-ax::stack::cc-concurrent-lifo}
(Concurrent LIFO) after the definitions of the other clauses have been
unfolded in \axiomSRef{vis-ax::stack::cc-concurrent-lifo}. In
hindsight, this may have been expected, as the clauses
$\visSep {} {u_1} {u_2}$ and $\visSep {} {o} {u}$ are hypotheses of
\axiomSRef{vis-ax::stack::cc-concurrent-lifo}, while
$\visSep {} {o_2} {o_1}$ is in the conclusion.

The engineering of the (uniquely determined) definition of the clause
$\visSep{}{o_2}{o_1}$ thus makes the proof of axiom
\axiomSRef{vis-ax::stack::cc-concurrent-lifo} out of definitions
\eqref{eq::stack::vis-rel}-\eqref{eq::stack::sep-rel-pop-pop} quite
straightforward, but for one important observation.
Because the axiom contains negations of several clauses of
$\visSepSYMBOL$, unfolding the definitions of these clauses reveals
comparisons of the form
$\stampProp x \not\stamporderSymbol \stampProp y$, where the relation
$\stamporderSymbol$ appears negated. The proof then crucially relies
on $\stamporderSymbol$ being total, so that we can \emph{flip} the
negated comparisons into the form
$\stampProp y \stamporderEqSymbolPadding \stampProp x$. It is the
requirement of totality of $\stamporderSymbol$ that makes the
described development specific to atomic timestamps. In
Section~\ref{subsect::spans::stack::interval-case}, we shall see how
to adapt to interval timestamps where $\stamporderSymbol$ is
\emph{not} total.

\begin{thm}\label{thm::atomic}
  Given $\visObsSYMBOL$ and $\visSepSYMBOL$ as in
  \eqref{eq::stack::vis-rel}-\eqref{eq::stack::sep-rel-pop-pop},
  the TS-stack with atomic timestamps satisfies the invariants in
  Section~\ref{subsect::spans::stack::key-abstractions-and-invariants}
  and the axioms in Figure~\ref{fig::stack::concurrent-spec-stack},
  and is thus linearizable by
  Theorem~\ref{thm::stack::vis-rels::main-linearizable}.
\end{thm}

The characteristic part of the proof is showing that the axiom
\axiomSRef{vis-ax::stack::cc-no-future-dependence} (no future
dependences) holds, which is where we rely on the invariants
\eqref{eq::stack::inv-misses-late} and
\eqref{eq::stack::inv-disj-stamp-gen}. This proof generates
obligations, one of which is that
$o \visSepSYMBOL u \precedesAbsSymbol o$ for some push $u$ and pop $o$
is impossible, as such $u$ depends on $o$ which is in $u$'s future.
The proof proceeds by contradiction: suppose
$o \visSepSYMBOL u \precedesAbsSymbol o$. By definition of
$o \visSepSYMBOL u$, there exists a push $u'$ missed by $o$ such that
$\stampProp{u'} \stamporderEqSymbol \stampProp{u}$. By invariant
\eqref{eq::stack::inv-misses-late}, $u'$ starts after $o$ starts, and
since $u \precedesAbsSymbol o$, it must also be
$u \precedesAbsSymbol u'$. But then, by invariant
\eqref{eq::stack::inv-disj-stamp-gen}, it is also
$\stampProp{u} \stamporderSymbol \stampProp{u'}$.  In other words,
$\stampProp{u} \stamporderSymbol \stampProp{u'} \stamporderEqSymbol
\stampProp{u}$, a contradiction.

\subsection{Case: Interval Timestamps}
\label{subsect::spans::stack::interval-case}

The proof from Section~\ref{subsect::spans::stack::atomic-case} does
not directly apply to the interval timestamps because proving axiom
\axiomSRef{vis-ax::stack::cc-concurrent-lifo} (Concurrent LIFO) relies
on the totality of $\stamporderSymbol$ in order to flip the negated
inequalities $\stampProp x \not\stamporderSymbol \stampProp y$ into
positive facts $\stampProp y \stamporderEqSymbolPadding \stampProp x$.
The relation $\stamporderSymbol$ is total in the atomic case, but not
in the interval case.

The key observation that allows us to recover the argument is that
\emph{whenever $\stamporderSymbol$ is used to compare the timestamps
  of two push events in the proofs of the atomic case, at least one of
  the push events is invariably popped}. In other words, the proof
does not actually require totality, but only the following weaker
property of \emph{pop-totality}. Formally, if $R$ is a partial order
on abstract timestamps, then $R$ is pop-total if:
\begin{align}
	\label{eq::stack::inv-pop-totality}
\begin{aligned}
\forall u_1\ u_2\ o.\, &
	(\stampProp{u_1} = \stampProp{o}) \vee (\stampProp{u_2} = \stampProp{o}) \implies \\
&	(\stampProp{u_1})\ R\ (\stampProp{u_2}) \vee 
	(\stampProp{u_2})\ R\ (\stampProp{u_1}) \vee 
	(\stampProp{u_1} = \stampProp{u_2})
	\end{aligned}
\end{align}
In English: if at least one of the pushes is taken, 
then the timestamps generated by the pushes are totally 
comparable under $R$.

As an illustration why the weaker property suffices, consider the
hypotheses of the axiom \axiomSRef{vis-ax::stack::cc-concurrent-lifo}:
these are $\visObs {} {u_1} {o_1}$, $o_1 \nVisSepEqSYMBOL u_2$ and
$u_2 \nVisSepEqSYMBOL u_1$. Let us further assume that
$\stamporderSymbol$ in all the definitions is replaced by an arbitrary
pop-total $R$.
A common pattern throughout the proof of Theorem~\ref{thm::atomic} is
that three conjuncts of the above form appear together. Such
combination entails that $u_1$ and $u_2$ are both popped, thus
allowing us to flip any negated relation $R$ in which
$\stampProp{u_1}$ or $\stampProp{u_2}$ may appear.  

Indeed, that $u_1$ is popped, and by $o_1$, follows from
$\visObs{}{u_1}{o_1}$, which is defined as
$\stampProp{o_1} = \stampProp{u_1}$. To see that $u_2$ must also be
popped consider the following.
First, note that $o_1 \nVisSepEqSYMBOL u_2$ implies
$\neg\PEND{o_1}{u_2}$, by an easy derivation. Pushing the negation
inside the definition of $\PENDName$ and substituting
$\stampProp{o_1} = \stampProp{u_1}$ derives
$ \neg (\stampProp{u_1})\ R\ (\stampProp{u_2}) \vee \exists
o'.\,\stampProp{u_2} = \stampProp{o'} \wedge \ldots $.  The second
disjunct directly says that $u_2$ is popped by some $o'$.  By
pop-totality of $R$, the first disjunct implies
$(\stampProp{u_2})\ R\ (\stampProp{u_1}) \vee (\stampProp{u_2}) =
(\stampProp{u_1})$, and thus $\stampProp{u_2} = \stampProp{u_1}$,
because $\neg(\stampProp{u_2})\ R\ (\stampProp{u_1})$ by
$u_2 \nVisSepEqSYMBOL u_1$. Thus, $u_1$ and $u_2$ are the same push,
and $u_2$ is popped as well.

It follows that we could replicate the atomic case proof to the
interval case, if we could replace $\stamporderSymbol$ with some
pop-total relation over \emph{interval timestamps} throughout the
definitions and proofs in
Sections~\ref{subsect::spans::stack::key-abstractions-and-invariants}
and \ref{subsect::spans::stack::atomic-case}.
We next define such a relation $\GstamporderSymbol$ that includes
$\stamporderSymbol$. 
\begin{align*}
t_2 \GstamporderSymbol t_1 & \defini t_2 \stamporderSymbol t_1\ \vee\  
\exists u_1, u_2.\ \stampProp {u_1}  \not\stamporderSymbol \stampProp {u_2}\ \wedge\ \TB {u_1} {u_2}\ \wedge\\ 
& \hspace{38mm} 
 t_2 \stamporderEqSymbol \stampProp {u_2}\ \wedge\ \stampProp {u_1} \stamporderEqSymbol t_1 \\
\TB {u_1} {u_2} & \defini \exists o_1.\ \stampProp{u_1} = \stampProp{o_1}\ \wedge\
\forall o_2.\ \stampProp{u_2} = \stampProp{o_2} \implies \precedesReps {\twoSpan{\spanEv{o_1}}} {\twoSpan{\spanEv{o_2}}}
\end{align*}

The key insight of the definition is that if two pushes $u_1$ and
$u_2$ are not already ordered by $\stamporderSymbol$, i.e.,
$\stampProp {u_1} \not\stamporderSymbol \stampProp {u_2}$, we could
order their timestamps in $\GstamporderSymbol$ in the order in which
the pushes are popped. Indeed, if $u_1$ is \emph{taken before} $u_2$
($\TB {u_1} {u_2}$), then LIFO warrants that $u_2$ is linearized
before $u_1$. We thus order $u_2$'s timestamp before $u_1$'s timestamp
in $\GstamporderSymbol$. It follows that $\visSep {} {u_2} {u_1}$
(assuming $\GstamporderSymbol$ substitutes $\stamporderSymbol$ in the
definition of $\visSep{}{u_2}{u_1}$), and consequently that $u_2$ is
linearized before $u_1$. The definition of $\GstamporderSymbol$
further saturates the relation to include any
$t_2 \GstamporderSymbol t_1$ where
$t_2 \stamporderEqSymbol \stampProp {u_2}$ and
$\stampProp {u_1} \stamporderEqSymbol t_1$, as then
$t_2 \GstamporderSymbol t_1$ is forced by
$\stampProp{u_2} \GstamporderSymbol \stampProp{u_1}$.

Returning to taken-before, we define $\TB {u_1} {u_2}$ to hold of two
pops $u_1$ and $u_2$ if: (1) $u_1$ is taken and $u_2$ is not, or (2)
both are taken by pops $o_1$ and $o_2$, respectively. In the case (2)
we require that the span of $o_1$ ends before the span of $o_2$, i.e.,
$o_1$ took its push before $o_2$ did. 

One can now proceed to prove that $\GstamporderSymbol$ is a strict
partial order that is pop-total, that the invariants ``Misses start
late'' and ``Disjoint pushes order timestamps'' from
Section~\ref{subsect::spans::stack::key-abstractions-and-invariants},
continue to hold for the TS-stack with interval timestamps, after
substituting $\stamporderSymbol\ :=\ \GstamporderSymbol$ in the
definition of $\PENDName$ \eqref{eq::stack::miss-rel}, and definitions
\eqref{eq::stack::inv-misses-late},
\eqref{eq::stack::inv-disj-stamp-gen} of the invariants. The
visibility and separability relations for the TS-stack with interval
timestamps are exactly as in
\eqref{eq::stack::vis-rel}-\eqref{eq::stack::sep-rel-pop-pop} but with
substitution $\stamporderSymbol\ :=\ \GstamporderSymbol$, and our
final theorem about the correctness of TS-stack is obtained simply by
retracing the proof of Theorem~\ref{thm::atomic}.

\begin{thm}
  Let $\visObsSYMBOL$ and $\visSepSYMBOL$ defined as in
  \eqref{eq::stack::vis-rel}-\eqref{eq::stack::sep-rel-pop-pop} but
  under the substitution $\stamporderSymbol\ :=\
  \GstamporderSymbol$. The TS-stack with interval timestamps satisfies
  the invariants in
  Section~\ref{subsect::spans::stack::key-abstractions-and-invariants}
  under the substitution, and the axioms in
  Figure~\ref{fig::stack::concurrent-spec-stack}, and is thus
  linearizable by
  Theorem~\ref{thm::stack::vis-rels::main-linearizable}.
\end{thm}

%% file: related.tex
\section{Discussion, Related and Future Work}
\label{sec:related}

\subparagraph*{Dealing with elimination pairs} 

To handle elimination pairs that were excluded in
Section~\ref{sect::stack::spans::visi-sep-defs}, we recursively define
indexed families of visibility and separability relations, where
$\visObsSymbol{i}$ and $\visSepSymbol{i}$ means that the first $i$
elimination pairs have been added
(Appendix~\ref{subsect::stack::appendix::vis-proof::axioms-from-invariants}).
At level 0, $\visObsIndxSymbol{0}{}$ and $\visSepIndxSymbol{0}{}$ are
the relations from Section~\ref{sect::stack::spans::visi-sep-defs}. At
some limit level $n$, where $n$ is the number of elimination pairs,
we have the final
relations $\visObsIndxSymbol{n}{}$ and $\visSepIndxSymbol{n}{}$
that consider all the events.

The theorems in Section~\ref{sect::stack::spans::visi-sep-defs} show
that the visibility-style axioms in
Figure~\ref{fig::stack::concurrent-spec-stack} hold for events in
$\closedEvent \setminus E$, i.e., $\closedEvent$ without elimination
pairs. They are the base case of our proof in
Appendix~\ref{subsect::stack::appendix::vis-proof::axioms-from-invariants},
which proceeds to inductively show that if the visibility-style axioms
hold for the first $i$ pairs, they continue to hold when the pair
$i+1$ is added.

\subparagraph*{Differences with the original algorithm}
Figure~\ref{alg::stack::simple-stack} is a simplified version of the
algorithm from Appendix~\ref{subsect::stack::appendix::vis-proof::proof-of-invariants}. 
The latter further treats elimination
pair detection and node unlinking (i.e., node deallocation from
memory). We consider the simplified version solely for presentation
reasons, as the simplification still presents the same verification
challenges and suffices to motivate the visibility and separability
relations in Section~\ref{sect::stack::spans::visi-sep-defs}.  The
definitions of these relations transfers to 
Appendix~\ref{subsect::stack::appendix::vis-proof::axioms-from-invariants}, where
they serve as a basis for defining a family of augmented relations
that deal with elimination pairs, as described above.

Having said this, the program that we treat in 
Appendix~\ref{subsect::stack::appendix::vis-proof::proof-of-invariants} still
differs in a relatively minor way from the original program of Dodds
\etal~\cite{Dodds} in that we elide empty stack detection (i.e. pops
returning $\EMPTY$). This can be treated separately as an extra
independent step in the proof~\cite{Haas},
which means that considering empty pops changes neither the analysis
we already presented in
Section~\ref{sect::stack::spans::visi-sep-defs} nor the proof for
elimination pairs in 
Appendix~\ref{subsubsect::stack::appendix::vis-proof::with-elim-pairs}.  
Nevertheless, we plan to
augment the proof with an extra step that considers empty pops.
 
\subparagraph*{Related proofs} Dodds \etal~\citet{Dodds} proof is also
based on a visibility relation (their \textbf{val}), in addition to
several other relations. However, our two axiomatizations and proofs
differ significantly.
Our axiomatization arises from a systematic transformation of a
state-based sequential specification of stacks into a history-based
concurrent specification, while that of Dodds \etal does not seem to
derive from such prior principles, though it does suffice for the
linearizability proof.
The different axiomatizations give rise to different relations on
histories as well. For example, their insert-remove (\textbf{ir})
relation is defined in terms of LPs of submodules. The objective in
using LPs of submodules is to start with a definition for that may
have linearizability violations, which then gets adjusted along the
proof to remove such violations.  In contrast, the definitions of our
relations in 
Section~\ref{sect::stack::spans::visi-sep-defs} require no adjustments since they
already lead to a correct linearization, albeit by eliding LPs. As a
result, our relations are quite a bit more direct, and support better
proof decomposition. In particular, our proof transfers from the
easier atomic timestamp case to the more difficult interval timestamp
case, whereas Dodds \etal immediately consider the interval case.

Bouajjani et al.~\citet{EnneaStack} employs forward simulation on the
atomic timestamp variant of the TS-stack, but do not attempt the
interval timestamp variant. Our proof 
(Appendix~\ref{subsubsect::stack::appendix::vis-proof::interval-case}) 
does not
employ simulations, and also lifts the atomic timestamp case to the
interval timestamp case. The lifting exploits that the difference
between the atomic and interval timestamp cases is not in the program
structure, but only in the implementation of $\newStampAlg$.

\subparagraph*{Visibility relations in other contexts} Our
approach uses visibility and separability relations to model ordering dependencies
between events.  A general survey of the use of visibility relations in
concurrency and distributed systems is given by Viotti and
Vukoli\'c~\citet{Viotti}. Visibility relations and declarative proofs
have also been utilized to specify consistency criteria weaker than
linearizability (Emmi and Enea~\cite{EmmiEnea}), to introduce a
specification framework for weak memory models (Raad et
al.~\cite{RaadAzalea}), and to specify the RC11 memory model (Lahav et
al.~\cite{LahavRC11}).  

In contrast to the above papers that focus on the semantics of
consistency criteria, our use of visibility relations focuses on
verifying specific algorithms and data structures, and is thus closer
to the following work where visibility relations are applied to
concurrent queues (Henzinger et
al.~\cite{henzinger:concur13,aspects}), concurrent stacks (Dodds et
al.~\cite{Dodds} and Haas~\cite{Haas}), and memory snapshot
algorithms~(\"Ohman and Nanevski~\cite{Joakim}).  We differ from these
in the addressed structures, or in the case of Dodds \etal in the
structure of the proof and its components.

Our key innovation compared to these works is the introduction of the
separability relation and its utilization to systematically axiomatize
the stack structure in a novel way.

\subparagraph*{Visibility and separability as a general methodology}
The pattern suggested by
Sections~\ref{subsect::stack::rels::visibility-relations-and-seq-specs}
and
\ref{subsect::stack::rels::separability-relations-and-concur-specs},
whereby one transforms a history-based sequential specification into a
concurrent specification, by replacing the returns-before relation
$\precedesAbsSymbol$ with a separability relation $\visSepSYMBOL$,
points towards a general methodology for axiomatizing concurrent
structures.

To test the generality of the approach, we have applied
it---successfully~(\cite{arxiv:mcas2307.04653} and
Appendix~\ref{subsect::stack::appendix::structures::rdcss-and-mcas}---to
the RDCSS and MCAS algorithms of Harris \etal~\citet{Harris}. These
algorithms write \emph{descriptors} (a record with information about
the task that a thread requires help with) into pointers, so that a
thread that reads a descriptor can provide help by executing the
described task. These algorithms implicitly ``bunch'' their help
requests into related groups, and the separability relation models
gaps between such bunches. On the other hand, the visibility relation
models a writer-reader dependency, similarly to the push-pop
dependency in this paper.
We have also applied the approach to queues, where it derived a mildly
streamlined variant of the queue axioms of Henzinger
\etal~\citet{henzinger:concur13,aspects}, and to locks, including
readers-writers locks.
In the future, we plan to study if this pattern applies to other
concurrent data structures (e.g., memory snapshots, trees, lists,
sets, etc.).

%% file: linproof.tex
\section{Proof of Linearizability from the Visibility-Style Axioms}

We first define a couple of base definitions. Section~\ref{sect::stack::appendix::lin::generic} then 
shows the existence of a total order, and Section~\ref{sect::stack::appendix::lin::sequential-soundness} shows 
that the total order is sequentially sound, concluding with
linearizability from the visibility-style axioms of Figure~\ref{fig::stack::concurrent-spec-stack}.

\begin{defn}[Events in the Stack Data Structure]
	The following are the events in the Stack Data Structure,
	\begin{center}
		\begin{tabular}{ll}
			$\pushAlg(v: \ValType)$ & (Push of value $v$) \\
			$\popAlg()$ & (Pop) \\
		\end{tabular}
	\end{center}
\end{defn}

Towards defining the notion of sequential soundness for a total order over 
events, we need to describe what it means for events to execute.
For that matter, we define the following state-based operational semantics,
so that intuitively, events execute by building 
a path in the operational semantics.

\begin{defn}[Operational semantics for events in the Stack Data Structure]
	Let $Op^S$ denote the operational semantics generated by 
	the following base steps, where states are lists (representing abstract stacks) 
	and labels
	are of the form ``$proc(in)\ \langle out \rangle$''
	where $proc$ is the event name, $in$ the event's input, and $out$ the event's output.
	\begin{itemize}
		\item $S \xrightarrow{\pushAlg(v)\ \langle \unitValue \rangle} v::S$, 
		\item $v::S \xrightarrow{\popAlg()\ \langle v \rangle} S$, 
		\item $[\ ] \xrightarrow{\popAlg()\ \langle \EMPTY \rangle} [\ ]$
	\end{itemize}
\end{defn}

We now define the notion of sequential soundness of a total order $\leq$.

\begin{defn}[Sequential soundness of a total order]
	Given a linear order $\leq$ over a set of events, and a path $P$ in the operational semantics $Op^S$, 
	we say that $P$ \emph{matches} $\leq$ if for every step $i$ in $P$, 
	\begin{itemize}
		\item The $i$-th procedure's name in $P$ equals the procedure's name of the $i$-th event in $\leq$.
		\item The $i$-th procedure's input in $P$ equals the procedure's input of the $i$-th event in $\leq$.
		\item If the $i$-th event in $\leq$ is terminated, then the 
		$i$-th procedure's output in $P$ equals the $i$-th event output in $\leq$.
	\end{itemize}
	We say that $\leq$ is \emph{sequentially sound} if there is a matching path in $Op^S$, starting from the empty stack and
	having as many steps as the number of events in the domain of $\leq$.
\end{defn}

\begin{notation}
	If path $P$ matches $\leq$, and $S \xrightarrow{proc(in)\ \langle out \rangle} S'$ is a step in $P$,
	we denote by $\generatorEvent {proc}$ the corresponding event in $\leq$. 
\end{notation}

And we define linearizability as follows.

\begin{defn}[Linearizability] 
	\label{defn::stack::appendix::lin::defn-linearizability}
	We say that a stack's implementation $D$ is
	\emph{linearizable} if for any set of events $\absEvent$
	generated from an arbitrary execution history in $D$, there is a
	binary relation
	${\genVisSymbol} \subseteq {\absEvent \times \absEvent}$
	(\emph{constraint relation}) and a linear order
	${\leq} \subseteq {\closedEvent \times
		\closedEvent}$ (\emph{linearization})\footnote{$\closedEvent$ is the closure of the terminated events $\terminatedEvent$ under $\genVisSymbol$, as in Figure~\ref{fig::stack::concurrent-spec-stack}.} s.t.,
	\begin{itemize}
		\item $\leq$ respects the real-time ordering of events, 
		i.e., $\precedesAbsSymbol$ restricted to $\closedEvent$ is contained in $\leq$.
		\item $\leq$ respects the ordering constraints in $\genVisSymbol$,
		i.e., $\genVisSymbol$ restricted to $\closedEvent$ is 
		contained in $\leq$.
		\item $\leq$ is \emph{sequentially sound}.
	\end{itemize}
\end{defn}

In the rest of this section, we assume that there are two abstract relations
$\visObsSYMBOL$ and $\visSepSYMBOL$ satisfying the visibility-style axioms 
in Figure~\ref{fig::stack::concurrent-spec-stack}.

\subsection{Existence of Total Order}
\label{sect::stack::appendix::lin::generic}

In this section we prove that there is a total order built
from the visibility-style axioms.

\begin{lem}
	\label{lem::stack::appendix::lin::vis-acyclic}
	Relation $\genVisSymbol$ is acyclic on $\closedEvent$.
\end{lem}

\begin{prf}
	Suppose there is a cycle for some $x \in \closedEvent$.
	Hence, $\genVisTrans x x$ holds. But, by 
	Axiom \axiomHRef{vis-ax::stack::cc-no-future-dependence}, 
	$\nprecedesAbsEq x x$ must hold, which means 
	$x \neq x$ (Contradiction).
\end{prf}

\begin{lem}
	\label{lem::stack::appendix::lin::precedes-poset}
	Relation $\precedesAbsEqSymbol$ is a partial order on $\absEvent$. In addition, $\precedesAbsSymbol$ satisfies the interval 
	order property,
	\[
	\forall w,x,y,z \in \absEvent.\ 
	(\precedesAbs {w} {x} \wedge \precedesAbs {y} {z})
	\rightarrow
	(\precedesAbs {w} {z} \vee \precedesAbs {y} {x})
	\]
\end{lem}

\begin{prf}
	To show that $\precedesAbsEqSymbol$ is a partial order, 
	it is enough to show that $\precedesAbsSymbol$
	is irreflexive and transitive, because it is 
	a standard result that the reflexive closure of
	an irreflexive and transitive relation is a partial order.
	
	\begin{itemize}
		\item Irreflexivity. If $\precedesAbs x x$, then we would
		have $\ETimeProp x \natorderSymbol \STimeProp x 
		\natorderEqSymbol \ETimeProp x$, which is a contradiction.
		
		\item Transitivity. If $\precedesAbs {x} {y}$ and
		$\precedesAbs {y} {z}$, then 
		$\ETimeProp x \natorderSymbol \STimeProp y 
		\natorderEqSymbol \ETimeProp y 
		\natorderSymbol \STimeProp z$, which means
		$\ETimeProp x \natorderSymbol \STimeProp z$.
		Hence, $\precedesAbs {x} {z}$.
	\end{itemize}
	
	We now show that $\precedesAbsSymbol$ satisfies the
	interval order property.
	
	Let $\precedesAbs {w} {x}$ and $\precedesAbs {y} {z}$.
	Hence, $\ETimeProp w \natorderSymbol \STimeProp x$ 
	and $\ETimeProp y \natorderSymbol \STimeProp z$.
	
	Either $\ETimeProp w \natorderSymbol \STimeProp z$ or
	$\STimeProp z \natorderEqSymbol \ETimeProp w$. The first
	case leads to $\precedesAbs {w} {z}$.
	
	For the second case, by using the hypotheses, we have
	$\ETimeProp y \natorderSymbol \STimeProp z
	\natorderEqSymbol \ETimeProp w 
	\natorderSymbol \STimeProp x$.
	Therefore, $\precedesAbs {y} {x}$.
\end{prf}

\begin{lem}
	\label{lem::stack::appendix::lin::no-cycles}
	Let $n \natorderSymbolRight 0$. If 
	$a_0 R_1 a_1 R_2 \ldots R_{n-1} a_{n-1} R_n a_{n}$
	is a sequence of $n$ steps where each $R_i$ is either 
	$\genVisSymbol$ or $\precedesAbsSymbol$, and each $a_i \in \closedEvent$, 
	then $\nprecedesAbsEq {a_n} {a_0}$.
\end{lem}

\begin{prf}
	By strong induction on $n$.
	\begin{itemize}
		\item Case $n = 1$. So, we have $a_0 R_1 a_1$.
		
		If $R_1 = {\genVisSymbol}$, then from
		$\genVis {a_0} {a_1}$ and Axiom 
		\axiomHRef{vis-ax::stack::cc-no-future-dependence},
		we get $\nprecedesAbsEq {a_1} {a_0}$.
		
		If $R_1 = {\precedesAbsSymbol}$, then from
		$\precedesAbs {a_0} {a_1}$, we cannot have
		$\precedesAbsEq {a_1} {a_0}$, because we would get
		$a_0 \precedesAbsSymbol a_1 \precedesAbsEqSymbol
		a_0$ (Contradiction by Lemma \ref{lem::stack::appendix::lin::precedes-poset}).
		
		\item Inductive case. Let 
		$a_0 R_1 a_1 R_2 \ldots R_n a_n R_{n+1} a_{n+1}$ be a 
		sequence of $n+1$ steps.
		
		If all $R_i$ are $\genVisSymbol$, then 
		$\nprecedesAbsEq {a_{n+1}} {a_0}$ follows by Axiom
		\axiomHRef{vis-ax::stack::cc-no-future-dependence}.
		
		Therefore, we can assume that for some 
		$1 \natorderEqSymbol i \natorderEqSymbol n+1$,
		we have $R_i = {\precedesAbsSymbol}$, i.e., 
		
		\begin{align}
		\label{eqn::stack::appendix::lin::triag-irreflexive-1}
		\overbrace{a_0 R_1 \ldots R_{i-1} a_{i-1}}^{i-1 \text{ steps}} 
		\precedesAbsSymbol 
		\overbrace{a_{i} R_{i+1} \ldots R_{n+1} a_{n+1}}^{n-i+1 \text{ steps}}
		\end{align}
		
		Suppose for a contradiction that 
		$\precedesAbsEq {a_{n+1}} {a_0}$.
		
		If $a_{n+1} = a_0$. Then,
		\[
		a_{i} R_{i+1} \ldots R_n a_n R_{n+1} a_0 R_1 \ldots R_{i-1} a_{i-1}
		\]
		is a sequence of $(n-i+1) + (i-1) = n$ steps. Therefore,
		by the inductive hypothesis, 
		$\nprecedesAbsEq {a_{i-1}} {a_{i}}$, which contradicts
		\eqref{eqn::stack::appendix::lin::triag-irreflexive-1}.
		
		It remains to check the case 
		$\precedesAbs {a_{n+1}} {a_0}$.
		
		\begin{itemize}
			\item Subcase $i = 1$. Hence, 
			${a_{n+1}} \precedesAbsSymbol {a_0} 
			\precedesAbsSymbol a_1$.
			
			But ${a_{1}} R_2 \ldots R_{n+1} {a_{n+1}}$ is a sequence
			of $n$ steps. Therefore, by 
			the inductive hypothesis, 
			$\nprecedesAbsEq {a_{n+1}} {a_{1}}$ (Contradiction).
			
			\item Subcase $i = n+1$. Hence, 
			${a_{n}} \precedesAbsSymbol {a_{n+1}} 
			\precedesAbsSymbol a_0$.
			
			But ${a_{0}} R_1 \ldots R_n {a_{n}}$ is a sequence
			of $n$ steps. Therefore, by 
			the inductive hypothesis, 
			$\nprecedesAbsEq {a_{n}} {a_{0}}$ (Contradiction).
			
			\item Subcase $1 \natorderSymbol i \natorderSymbol n+1$.
			
			Since $\precedesAbs {a_{i-1}} {a_{i}}$ and 
			$\precedesAbs {a_{n+1}} {a_0}$, by Lemma 
			\ref{lem::stack::appendix::lin::precedes-poset}, 
			either $\precedesAbs {a_{i-1}} {a_0}$ or
			$\precedesAbs {a_{n+1}} {a_{i}}$.
			
			But $a_0 R_1 \ldots R_{i-1} a_{i-1}$ is a sequence with 
			$0 \natorderSymbol i-1 \natorderSymbol n$
			steps and 
			$a_{i} R_{i+1} \ldots R_{n+1} a_{n+1}$ 
			is a sequence with 
			$0 \natorderSymbol n-i+1 \natorderSymbol n$ steps. 
			Therefore, by the inductive hypothesis,
			$\nprecedesAbsEq {a_{i-1}} {a_0}$ and
			$\nprecedesAbsEq {a_{n+1}} {a_{i}}$ 
			(Contradiction).
		\end{itemize}
	\end{itemize}
\end{prf}

\begin{defn}
	Define the \emph{happens-before} relation ${\triagVisSymbol} \defini {(\genVisSymbol \cup \precedesAbsSymbol)^{+}}$, i.e.,
	the transitive closure of the union of the constraint and returns-before relations.
\end{defn}

\begin{lem}
	\label{lem::stack::appendix::lin::triag-irreflexive}
	The relation $\triagVisSymbol$ is irreflexive on
	$\closedEvent$.
\end{lem}

\begin{prf}
	If $\triagVis x x$, then for some 
	$n \natorderSymbolRight 0$, there is a sequence
	$a_0 R_1 \ldots R_n a_{n}$ of $n$ steps, where 
	each $R_i$ is either 
	$\genVisSymbol$ or $\precedesAbsSymbol$, each $a_i \in \closedEvent$ and
	$a_0 = x$, and $a_n = x$.
	
	But by Lemma 
	\ref{lem::stack::appendix::lin::no-cycles},
	we would have $\nprecedesAbsEq x x$ 
	which is a contradiction.
\end{prf}

\begin{lem}
	\label{lem::stack::appendix::lin::triag-poset}
	The relation $\triagVisEqSymbol$ is a partial order on $\closedEvent$.
\end{lem}

\begin{prf}
	It is a standard result that the reflexive closure of
	an irreflexive and transitive relation is a partial order.
	Hence, $\triagVisEqSymbol$ is a partial order on
	$\closedEvent$ by Lemma 
	\ref{lem::stack::appendix::lin::triag-irreflexive} and
	the fact that $\triagVisSymbol$ is transitive by definition.
\end{prf}

\begin{lem}
	\label{eq::stack::appendix::lin::committed-downward-closed}
	Set $\closedEvent$ is $\genVisSymbol$-downward closed. In other words,
	if $x \in \closedEvent$ and $\genVis y x$, then $y \in \closedEvent$.
\end{lem}

\begin{prf}
	Since $x \in \closedEvent$, there is $z \in \terminatedEvent$ such that 
	$x \refleTransCl{\genVisSymbol} z$. So $y \genVisSymbol x \refleTransCl{\genVisSymbol} z$
	which means $y \in \closedEvent$.
\end{prf}

\begin{figure}[t]
	\begin{subfigwrap}{Properties of the total order $<$.}{subfig::stack::appendix::lin::properties}
		\centering
		\begin{tabular}{l}
			\axiomPLabel{vis-ax::stack::appendix::lin::cc-concurrent-lifo} LIFO \\
			\quad $\visObs {} {u_1} {o_1} \wedge u_1 < u_2 < o_1 \implies 
			\exists o_2.\ \visObs {} {u_2} {o_2} \wedge {o_2} < {o_1}$ \\
			\axiomPLabel{vis-ax::stack::appendix::lin::cc-pop-uniqueness} Pop uniqueness \\
			\quad $\visObs {} u {o_1} \wedge \visObs {} u {o_2} \implies o_1 = o_2$ \\
		\end{tabular}
		\begin{tabular}{l}
			\axiomPLabel{vis-ax::stack::appendix::lin::cc-push-uniqueness} Push uniqueness \\
			\quad $\visObs {} {u_1} {o} \wedge \visObs {} {u_2} {o} \implies u_1 = u_2$\\
			\axiomPLabel{vis-ax::stack::appendix::lin::cc-return-completion} Return value completion \\
			\quad $\exists v.\ \postPred x v \wedge (x \in \terminatedEvent \implies v = \outputProp x)$ \\
		\end{tabular} 
	\end{subfigwrap}
	
	\begin{subfigwrap}{Defined notions.}{subfig::stack::appendix::lin::defined-notions}
		\centering
		\begin{tabular}{c}
			\begin{tabular}{l}
				Constraint relation \\
				\quad ${\genVisSymbol} \defini \visObsSYMBOL \cup \visSepSYMBOL$\\
				Returns-before relation\\
				\quad $\precedesAbs {e_1} {e_2} \defini \ETimeProp {e_1} \natorderSymbol \STimeProp {e_2}$\\
			\end{tabular}
			\begin{tabular}{l}
				Set of terminated events\\
				\quad $\terminatedEvent \defini \{ e \mid \ETimeProp e \neq \bot \}$\\
				Closure of terminated events\\
				\quad $\closedEvent \defini \{ e \mid \exists t \in \terminatedEvent.\ e \refleTransCl{\genVisSymbol} t \} $\\
			\end{tabular}
			\\
			\begin{tabular}{c}
				\\
				$\postPred {o_1} v \defini
				\begin{cases}
				\exists u.\ \visObs {} u {o_1} \wedge v = \inProp{u} &
				\text{if $v \neq \EMPTY$} \\
				\forall u.\ u < {o_1} \implies \exists o_2.\ \visObs {} u {o_2} \wedge {o_2} < {o_1} & \text{if $v = \EMPTY$} \\
				\end{cases}$
				\qquad $\postPred {u} v \defini v = \unitValue$
			\end{tabular}
		\end{tabular}
	\end{subfigwrap}
	\caption{Properties of the total order $<$ over $\closedEvent$. These will be used to prove sequential soundness.}
	\label{fig::stack::appendix::lin::concurrent-spec-stack}
\end{figure}

\begin{lem}
	\label{lem::stack::appendix::lin::recency-holds-in-any-extension}
	Given a total order $\leq$ over $\closedEvent$ 
	such that ${\triagVisEqSymbol} \subseteq {\leq}$, the properties in Figure~\ref{fig::stack::appendix::lin::concurrent-spec-stack} hold.
\end{lem}

\begin{prf}
	\begin{itemize}
		\item Case \axiomPRef{vis-ax::stack::appendix::lin::cc-concurrent-lifo}.
		
		Suppose $\visObs {} {u_1} {o_1}$ and $u_1 < u_2 < o_1$. 
		
		If $o_1 \visSepEqSYMBOL u_2$ holds, we would get $u_2 < o_1 \leq u_2$ (Contradiction), since $\leq$ contains $\triagVisEqSymbol$.
		Hence, we can assume $o_1 \nVisSepEqSYMBOL u_2$.
		
		If $u_2 \visSepEqSYMBOL u_1$ holds, we would get $u_2 \leq u_1 < u_2$ (Contradiction). Hence, we can assume $u_2 \nVisSepEqSYMBOL u_1$.
		
		Therefore, we have $o_1 \nVisSepEqSYMBOL u_2 \nVisSepEqSYMBOL u_1$, and by \axiomSRef{vis-ax::stack::cc-concurrent-lifo},
		we get $\visObs {} {u_2} {o_2}$ and $\visSep {} {o_2} {o_1}$ for some $o_2$,  or equivalently $o_2 < o_1$.
		
		\item Case \axiomPRef{vis-ax::stack::appendix::lin::cc-pop-uniqueness}.
		
		This is identical to axiom \axiomSRef{vis-ax::stack::cc-pop-uniqueness}.
		
		\item Case \axiomPRef{vis-ax::stack::appendix::lin::cc-push-uniqueness}.
		
		Suppose $\visObs {} {u_1} {o}$ and $\visObs {} {u_2} {o}$. By totality of $<$, we have the cases,
		
		\begin{itemize}
			\item Case $u_1 < u_2$. 
			
			Since $\visObs {} {u_2} {o}$, we also have $u_2 < o$. Therefore, $\visObs {} {u_1} {o}$ and $u_1 < u_2 < o$.
			
			Hence, by \axiomPRef{vis-ax::stack::appendix::lin::cc-concurrent-lifo} (which is already proved), 
			$\visObs {} {u_2} {o_2}$ and $o_2 < o$ for some $o_2$. But $\visObs {} {u_2} {o}$ and $\visObs {} {u_2} {o_2}$, therefore
			from \axiomSRef{vis-ax::stack::cc-pop-uniqueness}, $o = o_2$, and so $o_2 < o_2$ (Contradiction).
			
			\item Case $u_2 < u_1$. Similar to the previous case.
			
			\item $u_1 = u_2$. This is the desired conclusion.
		\end{itemize}
		
		\item Case \axiomPRef{vis-ax::stack::appendix::lin::cc-return-completion}.
		
		Follows directly from axiom \axiomSRef{vis-ax::stack::cc-return-completion}. We only need to check the case
		for empty pops. 
		
		Suppose $u < {o_1}$ holds. If $o_1 \visSepEqSYMBOL u$ also holds, we would obtain the contradiction $u < u$. So,
		$o_1 \nVisSepEqSYMBOL u$. But then by axiom \axiomSRef{vis-ax::stack::cc-return-completion}, 
		$\visObs {} u {o_2}$  and ${o_2} \visSepSYMBOL {o_1}$, so $o_2 < o_1$.
	\end{itemize}
\end{prf}

\begin{lem}
	\label{lem::stack::appendix::lin::existence-basic}
	There is a total order $\leq$ over $\closedEvent$ containing 
	$\genVisSymbol$ and $\precedesAbsSymbol$, such that the properties in Figure~\ref{fig::stack::appendix::lin::concurrent-spec-stack} 
	hold.
\end{lem}

\begin{prf}
	Any finite partial order is contained in a total order.
	Pick some total order $\leq$ containing 
	the partial order $\triagVisEqSymbol$ over $\closedEvent$.
	
	By Lemma \ref{lem::stack::appendix::lin::recency-holds-in-any-extension},
	the properties in Figure~\ref{fig::stack::appendix::lin::concurrent-spec-stack} hold.
\end{prf}

\subsection{Sequential Soundness Proof}
\label{sect::stack::appendix::lin::sequential-soundness}

This section focuses on proving sequential soundness 
of the total order built in Lemma \ref{lem::stack::appendix::lin::existence-basic}.

\begin{lem}[Stack lemma]
	\label{lem::stack::appendix::lin::stack-lemma}
	Let $\leq$ be the total order of Lemma \ref{lem::stack::appendix::lin::existence-basic}.
	Suppose $\mathcal P$ is a path of length $1 \natorderEqSymbol n \natorderEqSymbol \vert \closedEvent \vert$ in $Op^S$ that matches
	$\leq$,\footnote{Sets $U$ and $O$ collect all the push and pop events in $\absEvent$, respectively.}
	\[
	S_0 = [\ ] \xrightarrow{p_1(in_1)\ \langle out_1 \rangle} S_1 \xrightarrow{p_2(in_2)\ \langle out_2 \rangle} 
	\ldots \xrightarrow{p_n(in_n)\ \langle out_n \rangle} S_n 
	\]
	
	\begin{enumerate}
		\item Let $1 \natorderEqSymbol i \natorderEqSymbol n$. If we have,
		 \begin{itemize}
		 	\item $\gEv{p_i} \in U$,
		 	\item $\neg \exists k \natorderEqSymbol n.\ \gEv{p_k} \in O \wedge \visObs{}{\gEv{p_i}}{\gEv{p_k}}$,
		 	\item $\forall k \natorderEqSymbol n.\ \gEv{p_k} \in U \wedge \gEv{p_i} < \gEv{p_k} \implies \exists j \natorderEqSymbol n.\ \gEv{p_j} \in O \wedge \visObs{}{\gEv{p_k}}{\gEv{p_j}}$
		 \end{itemize}
		then 
		\[
		S_n = \inProp{\gEv{p_i}} :: \left[\ \inProp{\gEv{p_j}} \mid \gEv{p_j} < \gEv{p_i} \wedge \gEv{p_j} \in U \wedge \neg \exists k \natorderEqSymbol n.\ \gEv{p_k} \in O \wedge \visObs {} {\gEv{p_j}} {\gEv{p_k}}\ \right]^{-1}
		\]
		where $[a,\ldots]^{-1}$ denotes the reverse of list $[a,\ldots]$.
		
		\item If $(\forall i \natorderEqSymbol n.\ \gEv{p_i} \in U \implies \exists j \natorderEqSymbol n.\ \gEv{p_j} \in O \wedge \visObs{}{\gEv{p_i}}{\gEv{p_j}})$, then $S_n = [\ ]$. 
	\end{enumerate}
\end{lem}

\begin{prf}
	By induction on $n$.
	
	\begin{itemize}
		\item Case $n = 1$. 
		
		For part 1. If $p_1$ is a push, we have that the conclusion holds trivially, i.e., $S_1 = \inProp{\gEv{p_1}} :: [\ ]$
		because there are no pushes before $p_1$, as the path matches $\leq$. Now, $p_1$ cannot be a pop by the hypothesis $\gEv{p_i} \in U$.
		
		For part 2. $p_1$ cannot be a push by the hypothesis, so $p_1$ is a pop. By \axiomPRef{vis-ax::stack::appendix::lin::cc-return-completion}
		$p_1$ needs to be an empty pop, otherwise there would exists a push before $p_1$. So, $S_1 = [\ ]$.
		
		\item Inductive case. Let $n \natorderEqSymbolRight 1$ and 
		suppose both parts hold for any path of length $n$.
		Let $\mathcal P$ be a matching path of length $n+1$,
		\[
		S_0 = \emptyset \xrightarrow{p_1(in_1)\ \langle out_1 \rangle} 
		\ldots \xrightarrow{p_n(in_n)\ \langle out_n \rangle} S_n \xrightarrow{p_{n+1}(in_{n+1})\ \langle out_{n+1} \rangle} S_{n+1}
		\]
		
		We need to show that both parts hold for this path.
		
		\subparagraph*{Proof of Part 1.}
		
		Let $1 \natorderEqSymbol i \natorderEqSymbol n+1$. We do a case analysis on $p_{n+1}$.
		
			\begin{itemize}
				\item Case $p_{n+1} = \popAlg()$. 
				
				By \axiomPRef{vis-ax::stack::appendix::lin::cc-return-completion}, 
				the post-condition predicate holds for $\gEv{p_{n+1}}$ at some $v$.
				
				\begin{itemize}
					\item Case $v = \EMPTY$. Hence, 
					\[
					\forall u.\ u < \gEv{p_{n+1}} \implies \exists o.\ \visObs{}{u}{o} \wedge o < \gEv{p_{n+1}}
					\]
					
					In particular for $\gEv{p_i}$ (because $i \natorderEqSymbol n$ as $p_i$ is a push by hypothesis $\gEv{p_i} \in U$,
					while $p_{n+1}$ is a pop),
					\[
					\visObs{}{\gEv{p_i}}{o} \wedge o < \gEv{p_{n+1}}
					\]
					for some $o \in O$.
					
					Since the path matches $\leq$, $o = \gEv{p_k}$ for some $k \natorderEqSymbol n$, but then
					$\visObs{} {\gEv{p_i}} {\gEv{p_k}}$ 
					(Contradicts hypothesis $\neg \exists k \natorderEqSymbol n+1.\ \gEv{p_k} \in O \wedge \visObs{}{\gEv{p_i}}{\gEv{p_k}}$).
					
					\item Case $v \neq \EMPTY$. 
					
					Then, $\visObs{}{u}{\gEv{p_{n+1}}}$ and $v = \inProp{u}$ for some $u \in U$. Either $\gEv{p_i} < u$ or $u < \gEv{p_i}$ or $u = \gEv{p_i}$.
					
					\begin{itemize}
						\item The case $u = \gEv{p_i}$ contradicts hypothesis $\neg \exists k \natorderEqSymbol n+1.\ \gEv{p_k} \in O \wedge \visObs{}{\gEv{p_i}}{\gEv{p_k}}$
						because $\visObs{}{u}{\gEv{p_{n+1}}}$.
						
						\item Case $u < \gEv{p_i}$. 
						
						Since $\gEv{p_i} < \gEv{p_{n+1}}$ (because $p_{n+1}$ is a pop, while $p_i$ is a push), from
						\axiomPRef{vis-ax::stack::appendix::lin::cc-concurrent-lifo} 
						with $\visObs{}{u}{\gEv{p_{n+1}}}$, we get $\visObs{}{\gEv{p_i}}{o'}$ and $o' < \gEv{p_{n+1}}$ for some pop $o'$.
						But again, $o' = \gEv{p_k}$ for some $k \natorderEqSymbol n$, since the path matches $\leq$, which contradicts the second hypothesis.
					\end{itemize}
					
					Hence, we can assume the case $\gEv{p_i} < u$. 
					
					Since $\visObs{}{u}{\gEv{p_{n+1}}}$, we have $u = \gEv{p_j}$ for some $j \natorderEqSymbol n$
					since the path matches $\leq$.
					
					Now, we prove the conditions to use the IH on $\gEv{p_j}$. Let us consider the subpath $[p_1,\ldots,p_n]$.
					
					\begin{itemize}
						\item Suppose for a contradiction that $\visObs {} {\gEv{p_j}} {\gEv{p_k}}$ for some $k \natorderEqSymbol n$.
						But $\visObs {} {\gEv{p_j}} {\gEv{p_{n+1}}}$ also, which implies $\gEv{p_k} = \gEv{p_{n+1}}$ by \axiomPRef{vis-ax::stack::appendix::lin::cc-pop-uniqueness} (Contradiction).
						
						\item Let $\gEv{p_l} \in U$ with $\gEv{p_j} < \gEv{p_l}$ for $l \natorderEqSymbol n$. 
						
						Since $\gEv{p_i} < u = \gEv{p_j} < \gEv{p_l}$, by the second hypothesis on the $n+1$ path instantiated on $\gEv{p_l}$,
						we have $\visObs{}{\gEv{p_l}}{\gEv{p_k}}$ for some $k \natorderEqSymbol n+1$.
						
						If $k = n+1$, since $\visObs{}{\gEv{p_l}}{\gEv{p_{n+1}}}$ also, by \axiomPRef{vis-ax::stack::appendix::lin::cc-push-uniqueness}
						we have $\gEv{p_j} = \gEv{p_l}$ (Contradiction).
						
						So, $k \natorderEqSymbol n$.
					\end{itemize}
					
					By IH on the subpath,
					\[
					S_n = \inProp{\gEv{p_j}} :: \left[\ \inProp{\gEv{p_l}} \mid \gEv{p_l} < \gEv{p_j} \wedge \gEv{p_l} \in U \wedge \neg \exists k \natorderEqSymbol n.\ \visObs {} {\gEv{p_l}} {\gEv{p_k}}\ \right]^{-1}
					\]
					
					where we can strengthen $\neg \exists k \natorderEqSymbol n.\ \visObs {} {\gEv{p_l}} {\gEv{p_k}}$ to $k \natorderEqSymbol n+1$, 
					because $\gEv{p_{n+1}}$ is already observing $\gEv{p_j}$, i.e.,
					\[
					S_n = \inProp{\gEv{p_j}} :: \left[\ \inProp{\gEv{p_l}} \mid \gEv{p_l} < \gEv{p_j} \wedge \gEv{p_l} \in U \wedge \neg \exists k \natorderEqSymbol n+1.\ \visObs {} {\gEv{p_l}} {\gEv{p_k}}\ \right]^{-1}
					\]
					
					But $\gEv{p_i} < \gEv{p_j}$ and by the third hypothesis on the $n+1$ path, every push after $\gEv{p_i}$ is observed, meaning
					that every push between $\gEv{p_i}$ and $\gEv{p_j}$ is observed, hence,
					\begin{align*}
					\left[\ \inProp{\gEv{p_l}} \mid \gEv{p_l} < \gEv{p_j} \wedge \gEv{p_l} \in U \wedge \neg \exists k \natorderEqSymbol n+1.\ \visObs {} {\gEv{p_l}} {\gEv{p_k}}\ \right] = \\
					\left[\ \inProp{\gEv{p_l}} \mid \gEv{p_l} < \gEv{p_i} \wedge \gEv{p_l} \in U \wedge \neg \exists k \natorderEqSymbol n+1.\ \visObs {} {\gEv{p_l}} {\gEv{p_k}}\ \right] \concat [\ \inProp{\gEv{p_i}}\ ]
					\end{align*}
					and so,					
					\[
					S_n = \inProp{\gEv{p_j}} :: \inProp{\gEv{p_i}} :: \left[\ \inProp{\gEv{p_l}} \mid \gEv{p_l} < \gEv{p_j} \wedge \gEv{p_l} \in U \wedge \neg \exists k \natorderEqSymbol n+1.\ \visObs {} {\gEv{p_l}} {\gEv{p_k}}\ \right]^{-1}
					\]
					when we reverse the lists.
					
					But since $p_{n+1}$ is a pop,
					
					\[
					S_{n+1} = \inProp{\gEv{p_i}} :: \left[\ \inProp{\gEv{p_l}} \mid \gEv{p_l} < \gEv{p_j} \wedge \gEv{p_l} \in U \wedge \neg \exists k \natorderEqSymbol n+1.\ \visObs {} {\gEv{p_l}} {\gEv{p_k}}\ \right]^{-1}
					\]
					
				\end{itemize}
				
				\item Case $p_{n+1} = \pushAlg(v')$ for some $v'$.
				
				Suppose for a contradiction that $i \natorderEqSymbol n$. 
				Since $\gEv{p_i} < \gEv{p_{n+1}}$, by the third hypothesis on the $n+1$ path, $\visObs{}{\gEv{p_{n+1}}}{\gEv{p_j}}$ for some $j \natorderEqSymbol n+1$ (Contradiction). 
				Therefore, $i = n+1$.
				
				In order to use IH on the subpath $[p_1,\ldots,p_n]$, we do a case analysis on $p_n$.
				
				\begin{itemize}
					\item Case $p_n = \popAlg()$.
					
					By \axiomPRef{vis-ax::stack::appendix::lin::cc-return-completion}, 
					the post-condition predicate holds for $\gEv{p_{n}}$ at some $v$.
					
					\begin{itemize}
						\item Case $v \neq \EMPTY$.
						
						Hence, $\visObs{}{u}{\gEv{p_n}}$ and $v = \inProp{u}$ for some push $u$. Which means $u = \gEv{p_j}$ for $j \natorderSymbol n$ (which means $1 \natorderSymbol n$).
						
						We now prove the conditions of IH restricted to the path $[p_1,\ldots,p_{n-1}]$.
						
						We cannot have $\visObs{}{\gEv{p_j}}{\gEv{p_k}}$ for some $k \natorderEqSymbol n-1$ because $\gEv{p_n}$ already
						observes $\gEv{p_j}$. 
						
						Now, let push $\gEv{p_k}$ such that $\gEv{p_j} < \gEv{p_k}$ for $k \natorderEqSymbol n-1$.
						Since $\gEv{p_k} < \gEv{p_n}$, by \axiomPRef{vis-ax::stack::appendix::lin::cc-concurrent-lifo}, 
						$\visObs{}{\gEv{p_k}}{o'}$ and $o' < \gEv{p_n}$ for some $o'$,
						which means $o' = \gEv{p_l}$ for some $l \natorderEqSymbol n-1$.
						
						Therefore, we can use IH on the $n-1$ subpath,
						\[
						S_n = \inProp{\gEv{p_j}} :: \left[\ \inProp{\gEv{p_l}} \mid \gEv{p_l} < \gEv{p_j} \wedge \gEv{p_l} \in U \wedge \neg \exists k \natorderEqSymbol n-1.\ \visObs {} {\gEv{p_l}} {\gEv{p_k}}\ \right]^{-1}
						\]
						
						But $p_n$ is a pop and $p_{n+1} = p_i$ is a push, so
						\[
						S_{n+1} = \inProp{\gEv{p_i}} :: \left[\ \inProp{\gEv{p_l}} \mid \gEv{p_l} < \gEv{p_j} \wedge \gEv{p_l} \in U \wedge \neg \exists k \natorderEqSymbol n-1.\ \visObs {} {\gEv{p_l}} {\gEv{p_k}}\ \right]^{-1}
						\]
						
						But we already proved that any push $p_k$ (for $k \natorderEqSymbol n-1$) after $p_j$ is observed, and since $p_j$ is
						observed by $p_n$, we have,
						\begin{align*}
						\left[\ \inProp{\gEv{p_l}} \mid \gEv{p_l} < \gEv{p_j} \wedge \gEv{p_l} \in U \wedge \neg \exists k \natorderEqSymbol n-1.\ \visObs {} {\gEv{p_l}} {\gEv{p_k}}\ \right] =\\
						\left[\ \inProp{\gEv{p_l}} \mid \gEv{p_l} < \gEv{p_i} \wedge \gEv{p_l} \in U \wedge \neg \exists k \natorderEqSymbol n+1.\ \visObs {} {\gEv{p_l}} {\gEv{p_k}}\ \right]
						\end{align*}
						So,
						\[
						S_{n+1} = \inProp{\gEv{p_i}} :: \left[\ \inProp{\gEv{p_l}} \mid \gEv{p_l} < \gEv{p_i} \wedge \gEv{p_l} \in U \wedge \neg \exists k \natorderEqSymbol n+1.\ \visObs {} {\gEv{p_l}} {\gEv{p_k}}\ \right]^{-1}
						\]
						
						\item Case $v = \EMPTY$.
						
						By \axiomPRef{vis-ax::stack::appendix::lin::cc-return-completion}, 
						\[
						\forall u.\ u < \gEv{p_n} \implies \exists o.\ \visObs{}{u}{o} \wedge o < \gEv{p_n}
						\]
						
						In particular, for any push $\gEv{p_j}$ (for $j \natorderEqSymbol n$), there is pop $\gEv{p_k}$ (with $k \natorderEqSymbol n$)
						such that $\visObs{}{\gEv{p_j}}{\gEv{p_k}} \wedge \gEv{p_k} < \gEv{p_n}$.
						
						So by part 2 of the IH on the subpath $[p_1,\ldots,p_n]$, $S_n = [\ ]$. But since $p_{n+1} = p_i$ is a push,
						we have,
						\[
						S_{n+1} = \inProp{\gEv{p_i}} :: [\ ]
						\]
						
						But every push before $p_n$ is observed (equivalently, every push before $p_{n+1} = p_i$ is observed, as $p_n$ is a pop), 
						hence,
						\[
						\left[\ \inProp{\gEv{p_l}} \mid \gEv{p_l} < \gEv{p_i} \wedge \gEv{p_l} \in U \wedge \neg \exists k \natorderEqSymbol n+1.\ \visObs {} {\gEv{p_l}} {\gEv{p_k}}\ \right] = [\ ]
						\]
						
						and so,
						
						\[
						S_{n+1} = \inProp{\gEv{p_i}} :: \left[\ \inProp{\gEv{p_l}} \mid \gEv{p_l} < \gEv{p_i} \wedge \gEv{p_l} \in U \wedge \neg \exists k \natorderEqSymbol n+1.\ \visObs {} {\gEv{p_l}} {\gEv{p_k}}\ \right]^{-1}
						\]
						
					\end{itemize}
					
					\item Case $p_n = \pushAlg(v'')$ for some $v''$.
					
					We have trivially that $p_n$ is not observed in the path $[p_1,\ldots,p_n]$, and every push $p_k$ (for $k \natorderEqSymbol n$)
					after $p_n$ is observed (satisfied vacuously).
					
					Therefore, by the IH,
					\[
					S_n = \inProp{\gEv{p_n}} :: \left[\ \inProp{\gEv{p_l}} \mid \gEv{p_l} < \gEv{p_n} \wedge \gEv{p_l} \in U \wedge \neg \exists k \natorderEqSymbol n.\ \visObs {} {\gEv{p_l}} {\gEv{p_k}}\ \right]^{-1}
					\]
					
					But $p_{n+1} = p_i$ is a push, so,
						\[
						S_{n+1} = \inProp{\gEv{p_i}} :: \inProp{\gEv{p_n}} :: \left[\ \inProp{\gEv{p_l}} \mid \gEv{p_l} < \gEv{p_n} \wedge \gEv{p_l} \in U \wedge \neg \exists k \natorderEqSymbol n.\ \visObs {} {\gEv{p_l}} {\gEv{p_k}}\ \right]^{-1}
						\]
					
					But $p_n$ is not observed in the $n+1$ path, so
					\begin{align*}
					\inProp{\gEv{p_n}} :: \left[\ \inProp{\gEv{p_l}} \mid \gEv{p_l} < \gEv{p_n} \wedge \gEv{p_l} \in U \wedge \neg \exists k \natorderEqSymbol n.\ \visObs {} {\gEv{p_l}} {\gEv{p_k}}\ \right]^{-1} = \\
					\left[\ \inProp{\gEv{p_l}} \mid \gEv{p_l} < \gEv{p_i} \wedge \gEv{p_l} \in U \wedge \neg \exists k \natorderEqSymbol n+1.\ \visObs {} {\gEv{p_l}} {\gEv{p_k}}\ \right]^{-1}
					\end{align*}
					which means,
						\[
						S_{n+1} = \inProp{\gEv{p_i}} :: \left[\ \inProp{\gEv{p_l}} \mid \gEv{p_l} < \gEv{p_i} \wedge \gEv{p_l} \in U \wedge \neg \exists k \natorderEqSymbol n+1.\ \visObs {} {\gEv{p_l}} {\gEv{p_k}}\ \right]^{-1}
						\]
						
				\end{itemize}
				
			\end{itemize}

			\subparagraph*{Proof of Part 2.} Suppose that for any push $\gEv{p_i}$ (for $i \natorderEqSymbol n+1$),
			there is a pop $\gEv{p_j}$ ($j \natorderEqSymbol n+1$) such that $\visObs{}{\gEv{p_i}}{\gEv{p_j}}$.
			We need to show $S_{n+1} = [\ ]$.
			
			If $p_{n+1}$ is a push, by the hypothesis there is a pop $\natorderEqSymbol n+1$ that observes it, which is impossible.
			Therefore, $p_{n+1}$ is a pop.
			
			By \axiomPRef{vis-ax::stack::appendix::lin::cc-return-completion}, 
			the post-condition predicate holds for $\gEv{p_{n+1}}$ at some $v$.
			
			\begin{itemize}
				\item Case $v \neq \EMPTY$. We have $\visObs{}{u}{\gEv{p_{n+1}}}$ and $v = \inProp{u}$ for some push $u$.
				Hence, $u = \gEv{p_j}$ for some $j \natorderEqSymbol n$.
				
				We now prove the conditions for Part 1 on the IH for the subpath $[p_1,\ldots,p_n]$.
				
				Suppose for a contradiction that $\visObs{}{\gEv{p_j}}{\gEv{p_k}}$ ($k \natorderEqSymbol n$). But
				$\visObs{}{\gEv{p_j}}{\gEv{p_{n+1}}}$ also, which means $\gEv{p_j} = \gEv{p_{n+1}}$ (Contradiction).
				
				Now, let push $\gEv{p_k}$ such that $\gEv{p_j} < \gEv{p_k}$ ($k \natorderEqSymbol n$).
				Since $\gEv{p_k} < \gEv{p_{n+1}}$, by \axiomPRef{vis-ax::stack::appendix::lin::cc-concurrent-lifo}, 
				$\visObs{}{\gEv{p_k}}{o'}$ and $o' < \gEv{p_{n+1}}$
				for some $o'$, which means that $p_k$ is observed in the $n$ subpath.
				
				Therefore, by the first part of the IH, we have $S_n = \inProp{\gEv{p_j}} :: L^{-1}$, where 
				\[
				L = \left[\ \inProp{\gEv{p_l}} \mid \gEv{p_l} < \gEv{p_j} \wedge \gEv{p_l} \in U \wedge \neg \exists k \natorderEqSymbol n.\ \visObs {} {\gEv{p_l}} {\gEv{p_k}}\ \right]
				\] 
				
				We claim that $L = [\ ]$. For suppose for a contradiction that $L$ contains some push $\gEv{p_k}$
				such that $\gEv{p_k} < \gEv{p_j}$ and $\gEv{p_k}$ is not observed in the $n$ subpath.
				
				But by hypothesis of part 2, $\visObs{}{\gEv{p_k}}{\gEv{p_l}}$ (for $l \natorderEqSymbol n+1$) for some $p_l$.
				But $p_k$ is not observed in the $n$ subpath, therefore $l = n+1$. But $\visObs{}{\gEv{p_j}}{\gEv{p_{n+1}}}$ also,
				so $\gEv{p_k} = \gEv{p_j}$ (Contradiction).
				
				Therefore $L = [\ ]$, and since $p_{n+1}$ is a pop, we have,
				\[
				S_{n+1} = [\ ]
				\]
				
				\item Case $v = \EMPTY$. By \axiomPRef{vis-ax::stack::appendix::lin::cc-return-completion},
				\[
				\forall u.\ u < \gEv{p_{n+1}} \implies \exists o.\ \visObs{}{u}{o} \wedge o < \gEv{p_{n+1}}
				\]
				We now prove the hypothesis of the second part in IH for the subpath $[p_1,\ldots,p_n]$.
				Let push $\gEv{p_j}$ ($j \natorderEqSymbol n$). By the the above, $\visObs{}{\gEv{p_j}}{o}$ and $o < \gEv{p_{n+1}}$ 
				for some pop $o$. Hence $o = \gEv{p_k}$ ($k \natorderEqSymbol n$).
				
				Hence, by the second part of IH on the $n$ subpath, $S_n = [\ ]$. But $p_{n+1}$ is an empty pop,
				and so $S_{n+1} = [\ ]$.
			\end{itemize}
			
	\end{itemize}
\end{prf}

\begin{lem}[Path Existence]
	\label{lem::stack::appendix::lin::stack-path-existence}
	Let $\leq$ be the total order of Lemma \ref{lem::stack::appendix::lin::existence-basic}.
	For any $1 \natorderEqSymbol n \natorderEqSymbol \lvert \closedEvent \rvert$, 
	there is a path in $Op^S$ of length $n$ that matches $\leq$ and starts from the empty stack.
\end{lem}

\begin{prf}
	By induction on $n$.
	
	\begin{itemize}
		\item Case $n = 1$. Denote by $x_1$ the first event in $\leq$. 
		
		If $x_1$ has terminated, then it cannot be a non-empty pop, 
		because \axiomPRef{vis-ax::stack::appendix::lin::cc-return-completion}
		and the definition of the postcondition predicate force $x_1$
		to observe another event. This observed event
		must be in $\closedEvent$ because $\closedEvent$ is $\genVisSymbol$-downward closed
		(Lemma \ref{eq::stack::appendix::lin::committed-downward-closed}).
		Therefore, there must exist an event occurring \emph{before}
		$x_1$ in $\leq$, which is impossible.
		
		Therefore, $x_1$ must be either a push or an unterminated pop, or 
		a terminated empty pop. 
		
		If $x_1$ is a push, 
		$[\ ] \xrightarrow{\pushAlg(v)\ \left\langle \unitValue \right\rangle} v::[\ ]$ is a path
		of length $1$ starting from the empty stack that matches $\leq$.
		
		If $x_1$ is an unterminated pop or a terminated empty pop, 
		$[\ ] \xrightarrow{\pushAlg(v)\ \left\langle \EMPTY \right\rangle} [\ ]$ is a path
		of length $1$ starting from the empty stack that matches $\leq$.
		
		\item Inductive case. Let $n \natorderEqSymbolRight 1$. Suppose 
		$n + 1 \natorderEqSymbol \vert \closedEvent \vert$. 
		We need to show that there is a matching path of length $n+1$.
		Since $1 \natorderEqSymbol n \natorderEqSymbol \vert \closedEvent \vert$, 
		the inductive hypothesis implies that there is a matching path of length $n$,
		\begin{align}
		S_0 = [\ ] \xrightarrow{p_1(in_1)\ \langle out_1 \rangle} S_1 
		\xrightarrow{p_2(in_2)\ \langle out_2 \rangle} 
		\ldots \xrightarrow{p_n(in_n)\ \langle out_n \rangle} S_n
		\end{align}
		
		We need to show that we can extend this path with a matching $n+1$ step 
		for the $n+1$ event in $\leq$. Denote the $n+1$ event in $\leq$ as $x_{n+1}$.
		
		We do a case analysis on $x_{n+1}$,
		\begin{itemize}
			\item Case $x_{n+1} = \popAlg()$. By \axiomPRef{vis-ax::stack::appendix::lin::cc-return-completion}
			the postcondition holds for some $v$.

			\begin{itemize} 
				\item Case $v = \EMPTY$. So,
				\[
				\forall u.\ u < x_{n+1} \implies \exists o.\ \visObs{}{u}{o} \wedge o < x_{n+1}
				\]
				
				From this and the fact that the path $[p_1,\ldots,p_n]$ matches $\leq$, we have
				that for any push $\gEv{p_i}$ ($i \natorderEqSymbol n$) there is a pop $\gEv{p_j}$
				such that $\visObs{}{\gEv{p_i}}{\gEv{p_j}}$.
				
				Therefore, by part 2 of Lemma~\ref{lem::stack::appendix::lin::stack-lemma},
				$S_n = [\ ]$, and we can augment the path with
				$S_n \xrightarrow{\popAlg()\ \left\langle \EMPTY \right\rangle} [\ ]$.
				
				In case $x_{n+1} \in \terminatedEvent$, we know from 
				\axiomPRef{vis-ax::stack::appendix::lin::cc-return-completion} that 
				$\EMPTY = v = \outputProp{x_{n+1}}$,
				and the step matches $\leq$.
				
				\item Case $v \neq \EMPTY$. So,
				\[
				\exists u.\ \visObs{}{u}{x_{n+1}} \wedge v = \inProp{u}
				\]
				
				From this and the fact that the path $[p_1,\ldots,p_n]$ matches $\leq$, we have
				$u = \gEv{p_i}$, for some $i \natorderEqSymbol n$.
				
				We know show that the conditions for Part 1 in Lemma~\ref{lem::stack::appendix::lin::stack-lemma} hold.
				
				\begin{itemize}
					\item Suppose for a contradiction that there is a pop $\gEv{p_k}$ ($k \natorderEqSymbol n$) 
					such that $\visObs{}{\gEv{p_i}}{\gEv{p_k}}$. 
					
					Since $\visObs{}{\gEv{p_i}}{x_{n+1}}$ also, from \axiomIRef{vis-ax::stack::appendix::lin::cc-pop-uniqueness}, it follows
					$\gEv{p_k} = x_{n+1}$ (Contradiction).
					
					\item Suppose a push $\gEv{p_k}$ ($k \natorderEqSymbol n$) with $\gEv{p_i} < \gEv{p_k}$.
					
					By \axiomPRef{vis-ax::stack::appendix::lin::cc-concurrent-lifo} with $\visObs{}{\gEv{p_i}}{x_{n+1}}$,
					there is pop $o'$ such that $\visObs{}{\gEv{p_k}}{o'}$ and $o' < x_{n+1}$.
					Therefore, $o' = \gEv{p_j}$, for $j \natorderEqSymbol n$.
				\end{itemize}
				
				There by Part 1 in Lemma~\ref{lem::stack::appendix::lin::stack-lemma},
				\[
				S_n = \inProp{\gEv{p_i}} :: L
				\]
				for some $L$.
				
				Hence, we make the step $S_n \xrightarrow{\popAlg()\ \left\langle v \right\rangle} L$,
				since $v = \inProp{u} = \inProp{\gEv{p_i}}$.
				
				In case $x_{n+1} \in \terminatedEvent$, we know from 
				\axiomPRef{vis-ax::stack::appendix::lin::cc-return-completion} that 
				$v = \outputProp{x_{n+1}}$,
				and the step matches $\leq$.
				
			\end{itemize}
			
			\item Case $x_{n+1} = \pushAlg(v)$.
			
			We make the step $S_n \xrightarrow{\pushAlg(v)\ \left\langle \unitValue \right\rangle} v::S_n$.
			In case $x_{n+1} \in \terminatedEvent$, we know from 
			\axiomPRef{vis-ax::stack::appendix::lin::cc-return-completion} that 
			$\outputProp{x_{n+1}} = \unitValue$,
			and the step matches $\leq$.

		\end{itemize}
	\end{itemize}
\end{prf}

\begin{lem}[Sequential Soundness]
	\label{lem::stack::appendix::lin::linear-order-stack-sequential-soundness}
	The total order $\leq$ of Lemma \ref{lem::stack::appendix::lin::existence-basic} is sequentially sound.
\end{lem}

\begin{prf}
	If $\closedEvent = \emptyset$, then the empty path matches $\leq$. 
	If $\closedEvent \neq \emptyset$, then Lemma \ref{lem::stack::appendix::lin::stack-path-existence}
	applied with $n = \vert \closedEvent \vert$ ensures the existence of a matching path
	for $\leq$.
\end{prf}

\begin{thm}[Stack Linearizability]
\label{thm::stack::appendix::lin-proof::vis-axioms-imply-linearizability}
Let $D$ be an arbitrary implementation of a concurrent stack. 
Suppose that for any set of
abstract events $\absEvent$ generated from an arbitrary execution
history in $D$, there are relations
$\visObsSYMBOL$,
$\visSepSYMBOL$ definable using $D$ such that 
the visibility-style axioms in Figure~\ref{fig::stack::concurrent-spec-stack} hold. Then, $D$ is linearizable.
\end{thm}

\begin{prf}
	Let $\absEvent$ be a set of events generated from an arbitrary
	execution history in the implementation. 
	From the hypothesis, the visibility-style axioms hold
	for the relations $\visObsSYMBOL$,
	$\visSepSYMBOL$.
	Let $\leq$ be the total order of Lemma
	\ref{lem::stack::appendix::lin::existence-basic} (which depends 
	on the visibility-style axioms).  Then, we take the
	constraint relation $\genVisSymbol$ (i.e., $\genVisSymbol \defini \visObsSYMBOL \cup \visSepSYMBOL$) 
	and $\leq$ to be the
	relations required by the definition of linearizability.  
	By construction, $\leq$ respects both $\genVisSymbol$ and
	$\precedesAbsSymbol$. Also,
	$\leq$ is sequentially sound by Lemma
	\ref{lem::stack::appendix::lin::linear-order-stack-sequential-soundness}.
\end{prf}

%% file: visproof.tex
\section{Proof of the Visibility-style Axioms}
\label{sect::stack::appendix::vis-proof::proof-of-axioms}

In this appendix we show that both versions of the TS-stack 
(atomic timestamps and interval timestamps)
satisfy the visibility-style axioms 
of Figure~\ref{fig::stack::concurrent-spec-stack}.

The proof is divided in two parts. First, we show that 
the visibility-style axioms follow from the TS-stack invariants in 
Figure~\ref{fig::stack::appendix::vis-proof::ts-stack-invariants} (Section~\ref{subsect::stack::appendix::vis-proof::axioms-from-invariants}), 
as this part of the proof is common to both versions of the TS-stack. 
Then we show that both versions of the TS-stack satisfy the 
TS-stack invariants (Section~\ref{subsect::stack::appendix::vis-proof::proof-of-invariants}).

\subsection{Proof of Axioms from TS-stack invariants}
\label{subsect::stack::appendix::vis-proof::axioms-from-invariants}

This step is divided into two. Section~\ref{subsubsect::stack::appendix::vis-proof::without-elim-pairs} 
proves that the visibility-style axioms
hold when elimination pairs are removed from $\closedEvent$.
Section~\ref{subsubsect::stack::appendix::vis-proof::with-elim-pairs}
then shows that the visibility-style axioms still hold when the elimination
pairs are put back into $\closedEvent$.
Section~\ref{subsubsect::stack::appendix::vis-proof::common-defs} provides
common definitions, the TS-stack algorithm and presents the TS-stack invariants.

\subsubsection{Common Definitions}
\label{subsubsect::stack::appendix::vis-proof::common-defs}

\begin{figure}[h]
	\begin{multicols*}{2}
		
		\begin{algorithmic}[1]
			\scriptsize
			\State $pools$ : $\PoolType[maxThreads]$ 
			\State
			%
			\Proc{$\pushAlg$\,}{$v: \ValType$}
			\State $\PoolType$ $pool = pools[TID]$; \label{alg::stack::appendix::vis-proof::push::read-thread-id} 
			\State $\NodeType$ $node = pool.\poolInsertAlg(v)$; \label{alg::stack::appendix::vis-proof::push::insert-new-node} 
			\State $\StampType$ $ts = \newStampAlg()$; \label{alg::stack::appendix::vis-proof::push::create-stamp}
			\State $node.stamp = ts$; \label{alg::stack::appendix::vis-proof::push::assign-stamp}
			\EndProc
			\State 
			\Proc{$\popAlg$\,}{}
			\State $\StampType$ $sTime = \newStampAlg()$; \label{alg::stack::appendix::vis-proof::pop::create-start-stamp}
			\State $\OptionalType{\ValType}$ $val = \bot$;
			\While{$val = \bot$}
			\State $val = \tryRemoveAlg(sTime)$; \label{alg::stack::appendix::vis-proof::pop::try-remove-call}
			\EndWhile
			\State \returnCmd{$\textit{val}$}; \label{alg::stack::appendix::vis-proof::pop::return-pop-val}
			\EndProc
			
			\columnbreak
			
			\Proc{$\tryRemoveAlg$\,}{$sTime: \StampType$}
			\State $\NodeType$ $chosen = \nullPointer$; \label{alg::stack::appendix::vis-proof::pop::initialize-chosen}
			\State $\StampType$ $maxT = -\infty$; \label{alg::stack::appendix::vis-proof::pop::initialize-max-ts}
			\State $\NodeType$ $top = \nullPointer$; \label{alg::stack::appendix::vis-proof::pop::initialize-top}
			\State $\PoolType$ $chPool = \nullPointer$; \label{alg::stack::appendix::vis-proof::pop::initialize-chPool}
			\ForEach{$pool$}{$pools$}
			\State $\NodeType$ $n$; $\NodeType$ $poolTop$;
			\State $(n,poolTop) = pool.\getYoungAlg()$; \label{alg::stack::appendix::vis-proof::pop::find-first-untaken-node}
			\If {$n \neq \nullPointer$}
			\State $\StampType$ $ts = n.stamp$; \label{alg::stack::appendix::vis-proof::pop::read-candidate-timestamp}
			\If {$sTime \stamporderSymbol ts$} \label{alg::stack::appendix::vis-proof::pop::compare-with-start-stamp}
			\State \returnCmd{$pool.\removeNodeAlg(poolTop, n)$}; \label{alg::stack::appendix::vis-proof::pop::try-take-elimination-pair}
			\ElsIf{$maxT \stamporderSymbol ts$} \label{alg::stack::appendix::vis-proof::pop::compare-with-max-stamp}
			\State $chosen = n$; $maxT = ts$; \label{alg::stack::appendix::vis-proof::pop::update-vars-one}
			\State $chPool = pool$; $top = poolTop$; \label{alg::stack::appendix::vis-proof::pop::update-vars-two}
			\EndIf
			\EndIf
			\EndForEach
			\If {$chosen \neq \nullPointer$}
			\State \returnCmd{$chPool.\removeNodeAlg(top, chosen)$}; \label{alg::stack::appendix::vis-proof::pop::try-take-chosen-node}
			\Else
			\State \returnCmd{$\bot$};
			\EndIf
			\EndProc
			\algstore{StackAlg}
		\end{algorithmic}
	\end{multicols*}
	\caption{Pseudo code of the TS-stack. $TID$ is the id of the executing thread. 
		The $\PoolType$ is shown in Figure~\ref{alg::stack::appendix::vis-proof::pool-type}.
	}
	\label{alg::stack::appendix::vis-proof::full-stack}
\end{figure}

\begin{figure}[h]
	\begin{multicols*}{2}
		\begin{algorithmic}[1]
			\scriptsize
			\algrestore{StackAlg}
			\Record{$\NodeType$}
			\State $val$ : $\ValType$ \label{alg::stack::appendix::vis-proof::push::node-record-start}
			\State $stamp$ : $\StampType$
			\State $next$ : $\NodeType$
			\State $taken$ : $\BoolType$
			\State \colorbox{lightgray}{$id : \IntType$} \label{alg::stack::appendix::vis-proof::push::node-record-end}
			\EndRecord 
			%
			\State 
			\State $top$ : $\NodeType$
			\State \colorbox{lightgray}{$ID$ : $\IntType = 0$}
			\State
			\Proc{$\getYoungAlg$\,}{}
			\State $\NodeType$ $oldTop = top$;
			\State $\NodeType$ $n = oldTop$;
			\While {$true$}
			\If {\textbf{not} $n.taken$} \label{alg::stack::appendix::vis-proof::pool::find-first-untaken}
			\State \returnCmd{($n$, $oldTop$)};
			\ElsIf {$n.next = n$}
			\State \returnCmd{($\nullPointer$, $oldTop$)};
			\Else 
			\State $n = n.next$;
			\EndIf
			\EndWhile
			\EndProc
			
			\State
			
			\Proc{$\initPoolAlg$\,}{}
			\State $\NodeType$ $sentinel = $
			\State \qquad $\NodeType\{ \nullPointer, \infty, \nullPointer, true,$\colorbox{lightgray}{-1}$\}$;
			\State $sentinel.next = sentinel$;
			\State $top = sentinel$;
			\EndProc
			\columnbreak
			\Proc{$\poolInsertAlg$\,}{$v: \ValType$}
			\State $\NodeType$ $n = $
			\State \qquad $\NodeType\{ v, \infty, top, false,$\colorbox{lightgray}{$ID$++}$\}$; \label{alg::stack::appendix::vis-proof::pool::create-new-node}
			\State $top = n$; \label{alg::stack::appendix::vis-proof::pool::new-node-as-top}
			\State $\NodeType$ $next = n.next$; // Unlinking starts
			\While {$next.next \neq next$ \textbf{and} $next.taken$} \label{alg::stack::appendix::vis-proof::pool::unlinking-insert-start}
			\State $next = next.next$; \label{alg::stack::appendix::vis-proof::pool::unlinking-insert-finish}
			\EndWhile
			\State $n.next = next$; 
			\State  \returnCmd{$n$};
			\EndProc
			
			\State
			
			\Proc{$\removeNodeAlg$\,}{$oldTop: \NodeType$, $n: \NodeType$}
			\If {$CAS(n.taken, false, true)$} \label{alg::stack::appendix::vis-proof::pool::try-take-node}
			\State $CAS(top, oldTop, n)$; // Unlinking starts \label{alg::stack::appendix::vis-proof::pool::unlinking-remove-cas}
			\If{$oldTop \neq n$} \label{alg::stack::appendix::vis-proof::pool::unlinking-remove-condition}
			\State $oldTop.next = n$; \label{alg::stack::appendix::vis-proof::pool::unlinking-remove-condition-true}
			\EndIf
			\State $\NodeType$ $next = n.next$;
			\While{$next.next \neq next$ \textbf{and} $next.taken$} \label{alg::stack::appendix::vis-proof::pool::unlinking-remove-start}
			\State $next = next.next$; 
			\EndWhile
			\State $n.next = next$; \label{alg::stack::appendix::vis-proof::pool::unlinking-remove-finish}
			\State \returnCmd{$n.val$}; \label{alg::stack::appendix::vis-proof::pool::read-val-to-return}
			\Else
			\State \returnCmd{$\bot$};
			\EndIf
			\EndProc
			\algstore{StackAlg}
		\end{algorithmic}
	\end{multicols*}
	\caption{Pseudo code of the $\PoolType$ type. $ID$++ is a fetch-and-increment atomic operation.
		Code in gray is ghost code.
	}
	\label{alg::stack::appendix::vis-proof::pool-type}
\end{figure}

\begin{figure}[h]
	\begin{subfigwrap}{Atomic timestamps.}{fig::sub::stack::appendix::vis-proof::atomic-timestamps-gen}[0.45\textwidth]
		\begin{algorithmic}[1]
			\algrestore{StackAlg}
			\scriptsize
			\Proc{$\newStampAlg$\,}{}
			\State \returnCmd{$TS$++};
			\EndProc
			\algstore{StackAlg}
		\end{algorithmic}
	\end{subfigwrap}
	\begin{subfigwrap}{Interval timestamps.}{fig::sub::stack::appendix::vis-proof::interval-timestamps-gen}[0.5\textwidth]
		\begin{algorithmic}[1]
			\algrestore{StackAlg}
			\scriptsize
			\Proc{$\newStampAlg$\,}{}
			\State $\IntType$ $ts_1 = TS$; \label{alg::stack::appendix::vis-proof::ts-gen::first-read}
			\State $pause()$;
			\State $\IntType$ $ts_2 = TS$; \label{alg::stack::appendix::vis-proof::ts-gen::second-read}
			\If {$ts_1 \neq ts_2$}
			\State \returnCmd{$[ts_1,ts_2-1]$}; \label{alg::stack::appendix::vis-proof::ts-gen::interval1}
			\ElsIf {$CAS(TS,ts_1,ts_1+1)$}
			\State \returnCmd{$[ts_1,ts_1]$}; \label{alg::stack::appendix::vis-proof::ts-gen::interval2}
			\Else
			\State \returnCmd{$[ts_1,TS-1]$}; \label{alg::stack::appendix::vis-proof::ts-gen::interval3}
			\EndIf
			\EndProc
		\end{algorithmic}
	\end{subfigwrap}
	\caption{Different ways of generating timestamps. In both cases, $TS$ is a global $\IntType$ variable initialized to $0$.
		$TS$++ is an atomic fetch-and-increment.
	}
	\label{alg::stack::appendix::vis-proof::stamp-type}
\end{figure}

Figures \ref{alg::stack::appendix::vis-proof::full-stack},
\ref{alg::stack::appendix::vis-proof::pool-type},
\ref{alg::stack::appendix::vis-proof::stamp-type} show the full algorithm.
The atomic TS-stack consists on 
Figures \ref{alg::stack::appendix::vis-proof::full-stack},
\ref{alg::stack::appendix::vis-proof::pool-type}, and
\ref{fig::sub::stack::appendix::vis-proof::atomic-timestamps-gen}.
The interval TS-stack consists on 
Figures \ref{alg::stack::appendix::vis-proof::full-stack},
\ref{alg::stack::appendix::vis-proof::pool-type}, and
\ref{fig::sub::stack::appendix::vis-proof::interval-timestamps-gen}.

We now define some concepts that are common to both versions of the TS-stack.

\begin{defn}
	\label{defn::stack::appendix::vis-proof::common-definitions}
	We have the following definitions, which refer to the code in
	Figures \ref{alg::stack::appendix::vis-proof::full-stack} and
	\ref{alg::stack::appendix::vis-proof::pool-type}.
	
	\begin{description}
		\item[Events and Rep Events] For event $e$, we denote by $\STimeProp{e}$, $\ETimeProp{e}$, 
		$\inProp{e}$, $\outputProp{e}$ the start time, end time, input, and output of event $e$, 
		respectively.
		$\ETimeProp{e}$ and $\outputProp{e}$ remain undefined if $e$ has not terminated.
		
		Similarly, for rep event $r$, we denote by $\TimeProp{r}$, $\inProp{r}$, $\outputProp{r}$
		the time, input, and output of rep event $r$, respectively. The three properties
		are always defined for rep events, because rep events are atomic.
		Additionally, we denote by $<$ the strict total order on rep events.
		
		\item[Spans] Spans are pairs of rep events $(a,b)$. For span $s$, we denote by $\oneSpan{s}$
		and $\twoSpan{s}$ the first and second projection functions.
		
		We define the order on spans as,
		\[
		s_1 \precedesSpansSymbol s_2 \defini \twoSpan{s_1} < \oneSpan{s_2}
		\]
		and denote by $\precedesSpansEqSymbol$ its reflexive closure.
		
		\item[Span Function ($\spanEvSymbol$)] Function $\spanEv e$ returns the unique 
		span executed by event $e$. More specifically,

		For a push event $u$, $\spanEv u$ returns the span starting when 
		the new node is
		linked as first node of the pool
		(line~\ref{alg::stack::appendix::vis-proof::pool::new-node-as-top}). The span
		ends when a finite timestamp is assigned to the new node
		(line~\ref{alg::stack::appendix::vis-proof::push::assign-stamp}).
		
		For a pop event $o$, $\spanEv o$ returns the span starting before 
		the pools loop (line~\ref{alg::stack::appendix::vis-proof::pop::initialize-chosen}) as long as this line
		executed before the \emph{last} pools loop.  The span ends on the successful
		CAS at line~\ref{alg::stack::appendix::vis-proof::pool::try-take-node} which takes a
		node for the pop to return. 
		
		If event $e$ has not completed its span, $\spanEv e$ is undefined.
		
		\item [Node Id Function ($\idPropName$)] Function $\idProp e$ returns the unique 
		node id generated by $e$ (if $e$ is a push) or the unique node id of the taken node
		(if $e$ is a pop). More specifically,
		
		For a push event $u$, $\idProp u$ returns the id of the generated node at 
		line~\ref{alg::stack::appendix::vis-proof::pool::create-new-node} as soon
		as the node is linked to the pool at line~\ref{alg::stack::appendix::vis-proof::pool::new-node-as-top}. 
		For pop event $o$, $\idProp o$ 
		returns the id of the taken node at line~\ref{alg::stack::appendix::vis-proof::pool::try-take-node}. 
		
		If event $e$ has not executed the mentioned lines, $\idProp e$ is undefined.
		
		\item[Timestamp Function ($\stampPropName$)] Function $\stampProp{e}$ returns the 
		timestamp generated by $e$ (if $e$ is a push) or the timestamp of the taken node
		(if $e$ is a pop). More specifically,
		
		For a push $u$, $\stampProp u$ returns the generated timestamp as soon as it is assigned to the node at line~\ref{alg::stack::appendix::vis-proof::push::assign-stamp}.
		
		For a pop $o$, $\stampProp o$ 
		returns the timestamp (as was read at line~\ref{alg::stack::appendix::vis-proof::pop::read-candidate-timestamp}) 
		of the taken node at line~\ref{alg::stack::appendix::vis-proof::pool::try-take-node}.
		
		If event $e$ has not executed the mentioned lines, $\stampProp e$ is undefined.
		Note that $\stampProp e$ is defined if and only if $\spanEv e$ is defined,
		because $e$ completes its span precisely when the mentioned lines in the 
		definition of $\stampProp{e}$ execute.
		
		\item[Timestamp Order ($\stamporderSymbol$)] The relation $\stamporderSymbol$
		denotes the strict partial order on timestamps. 
		
		For the atomic TS-stack, $\stamporderSymbol$
		is simply the strict total order on natural numbers $\natorderSymbol$, 
		extended with minimum value $-\infty$ and maximum value $\infty$.
		
		For the interval TS-stack, 
		$\stamporderSymbol$ is the strict partial order on interval timestamps defined as,
		\[
		[a,b] \stamporderSymbol [c,d] \defini b \natorderSymbol c
		\]
		and extended with a minimum value $-\infty$ and maximum value $\infty$.
		
		\item[Abstract Timestamp Function ($\AstampPropName$)] For event $e$, 
		$\AstampProp{e} \defini (\idProp e, \stampProp{e})$ returns a pair 
		combining the id and timestamp of $e$. If either $\idProp e$ or $\stampProp{e}$ is undefined,
		then $\AstampProp{e}$ is undefined.
		
		We call the pairs $(i,t)$, which combine a natural number $i$
		(representing an id) and a timestamp $t$, \emph{abstract timestamps},
		and we overload the timestamp order $\stamporderSymbol$ into abstract timestamps
		in the natural way, as there is no risk of confusion,
		\[
		(i_1,t_1) \stamporderSymbol (i_2,t_2) \defini t_1 \stamporderSymbol t_2
		\]
		
		\item[Output Function ($\spanEvOutSymbol$)] Function $\spanEvOut e$ returns the 
		output associated with the possibly not terminated event $e$. More specifically,
		
		For a push event $u$, $\spanEvOut u$ returns $\unitValue$ 
		if $u$ has generated an id and linked the new node at 
		line~\ref{alg::stack::appendix::vis-proof::pool::new-node-as-top}.
		
		For a pop event $o$, $\spanEvOut o$ returns the value stored in the taken node
		at line~\ref{alg::stack::appendix::vis-proof::pool::try-take-node}.
		
		If event $e$ has not executed the mentioned lines, $\spanEvOut e$ is undefined.
		
		Do not confuse $\spanEvOut e$ with $\outputProp{e}$. Function $\outputProp{e}$
		is defined only when $e$ has terminated, while $\spanEvOut e$ describes
		the output that is going to be chosen when $e$ is included in the completion $\closedEvent$,
		even if $e$ has not terminated. For terminated events $e$, we will have
		$\outputProp{e} = \spanEvOut{e}$.
		
		\item[Elimination pairs] We define when push $u$ and pop $o$ form an elimination pair.
		\[
		\elimRel{u}{o} \defini \idProp{u} = \idProp{o} \wedge u \not\precedesAbsSymbol o
		\]
		In English: (1) $o$ pops the node that $u$ pushed
		($\idProp{u} = \idProp{o}$), and (2) $u$ and $o$ overlap.  Events $u$
		and $o$ overlap if $u \not\precedesAbsSymbol o$ and
		$o \not\precedesAbsSymbol u$, but it is not necessary to explicitly
		check $o \not\precedesAbsSymbol u$, as that follows from
		$\idProp{u} = \idProp{o}$ and a structural invariant that $o$ cannot
		pop a node that has not been pushed yet. We shall present the invariants soon.
		
		\item[Elimination Pair Events Set ($\elimSetName$)] We define the set
		of events occurring in elimination pairs as,
		\[
		\elimSetName \defini \{ x \mid \exists y.\ \elimRel{x}{y} \vee \elimRel{y}{x}\}
		\]
		
		\item[Pop-Totality] We assume that we are given a strict partial order on 
		abstract timestamps $\GstamporderSymbol$, such that it is \emph{pop-total},
		\[
		\begin{aligned}
			\forall u_1\ u_2\ o \notin \elimSet{}.\, &
			(\AstampProp{u_1} = \AstampProp{o}) \vee (\AstampProp{u_2} = \AstampProp{o}) \implies \\
			& (\AstampProp{u_1})\ \GstamporderSymbol\ (\AstampProp{u_2}) \vee 
			(\AstampProp{u_2})\ \GstamporderSymbol\ (\AstampProp{u_1}) \vee 
			(\AstampProp{u_1} = \AstampProp{u_2})
		\end{aligned}
		\]
		As usual, we denote by $\GstamporderEqSymbol$ its reflexive closure.
		
		Pop-totality is a weaker form of totality for a partial order, in which two
		abstract timestamps are required to be totally comparable only when one of the pushes
		that generated them is taken by a pop.
		
		The property of pop-totality will be used to prove theorems where elimination pairs
		are excluded.  Hence, we explicitly exclude elimination pairs with the condition $\notin \elimSet{}$
		in the universally quantified variables of the property.
		
		Sections~\ref{subsubsect::stack::appendix::vis-proof::atomic-case}
		and \ref{subsubsect::stack::appendix::vis-proof::interval-case} 
		will define $\GstamporderSymbol$ for the atomic and interval
		cases, respectively.
	\end{description}	
\end{defn}

In what follows, we assume that variables $o$ and $u$, and their 
variations, range over pop and push events, respectively.
Variables $x$, $y$ range over arbitrary events.

We have these immediate lemmas from the definitions,

\begin{lem}
	\label{defn::stack::appendix::vis-proof::abstract-timestamps-order-is-strict}
	$\stamporderSymbol$ overloaded to abstract timestamps is a strict partial order.
\end{lem}

\begin{prf}
	We prove each property,
	
	Irreflexivity. If $(i,t) \stamporderSymbol (i,t)$, then $t \stamporderSymbol t$ (Contradiction).
	
	Transitivity. If $(i_1,t_1) \stamporderSymbol (i_2,t_2)$ and $(i_2,t_2) \stamporderSymbol (i_3,t_3)$, then
	$t_1 \stamporderSymbol t_2$ and $t_2 \stamporderSymbol t_3$, which means $t_1 \stamporderSymbol t_3$.
	Therefore $(i_1,t_1) \stamporderSymbol (i_3,t_3)$ by definition.
\end{prf}

\begin{lem}
	\label{lem::stack::appendix::vis-proof::elim-set-basic-disj-fact}
	The following holds,
	\begin{enumerate}
		\item If $u$ is a push such that $u \notin \elimSet{}$, then 
		 $\forall o.\ \idProp{u} = \idProp{o} \implies u \precedesAbsSymbol o$.
		\item If $o$ is a pop such that $o \notin \elimSet{}$, then 
		$\forall u.\ \idProp{u} = \idProp{o} \implies u \precedesAbsSymbol o$.
	\end{enumerate}
\end{lem}

\begin{prf}
	We prove each part,
	\begin{enumerate}
		\item Since $u \notin \elimSet {}$, then $\forall y.\ \neg(\elimRel{u}{y}) \wedge \neg(\elimRel{y}{u})$.
		In particular, $\forall y.\ \neg(\elimRel{u}{y})$.
		
		After some De Morgan manipulations on the definition of $\elimRelName$,
		\[
		\forall y.\ y \text{ is a pop} \wedge \idProp{u} = \idProp{y} \implies u \precedesAbsSymbol y
		\]
		and so, $\forall o.\ \idProp{u} = \idProp{o} \implies u \precedesAbsSymbol o$ directly.
		
		\item Similar to the previous case.
	\end{enumerate}
\end{prf}

We now define the visibility and separability relations. 

First, we define the base relations, which are those 
in Section~\ref{subsect::spans::stack::atomic-case},
but they have been abstracted to use
$\AstampPropName$ and $\GstamporderSymbol$, instead of 
$\stampPropName$ and $\stamporderSymbol$; also, the exclusion
of elimination pairs have been made explicit with the clause $\notin \elimSet{}$. 

\begin{defn}[Base Relations]
\label{defn::stack::appendix::vis-proof::base-relations}
\begin{align*}
\visObsIndx{}{B} u o \defini & u,o\notin \elimSet{}\ \wedge\ \AstampProp{u} = \AstampProp{o}\\
\visSepIndx {} {B} {u_1} {u_2} \defini & u_1,u_2 \notin \elimSet{}\ \wedge\ \AstampProp{u_1} \GstamporderSymbol \AstampProp{u_2}\\
\visSepIndx {} {B} {o} {u} \defini & o,u \notin \elimSet{}\ \wedge\ \exists u' \notin \elimSet{}.\ \PEND {o} {u'}\ \wedge\ \AstampProp {u'} \GstamporderEqSymbol \AstampProp u \\
\PEND {o} {u'} \defini & o,u' \notin \elimSet{}\ \wedge\ 
\AstampProp o \GstamporderSymbol \AstampProp{u'}\ \wedge \\ 
               & \forall o' \notin \elimSet{}.\ \AstampProp{u'} = \AstampProp{o'} \implies \precedesReps {\twoSpan{\spanEv {o}}} {\twoSpan{\spanEv {o'}}} \\
\visSepIndx {} {B} {o_2} {o_1} \defini & o_2,o_1 \notin \elimSet{}\ \wedge\ \AstampProp{o_1} \GstamporderSymbol \AstampProp{o_2}\ \wedge \\
               & \neg \exists u' \notin \elimSet{}.\ \PEND {o_1} {u'}\ \wedge\ \AstampProp {u'} \GstamporderEqSymbol \AstampProp {o_2}
\end{align*}
\end{defn}

We next define the relations that extend the base relations to include
elimination pairs. We define them as a recursively defined indexed family
$\visObsIndxSymbol{i}{}$ and
$\visSepIndxSymbol{i}{}$. 
The intuition behind the
index $i$ in $\visObsIndxSymbol{i}{}$ and
$\visSepIndxSymbol{i}{}$ is that the $i$-th elimination
pair has been added to the relations
(the elimination pairs are enumerated from 0 to $|\elimRelName|-1$, where
$|\elimRelName|$ is the number of pairs in the relation $\elimRelName$). 
The zero indexed relations are simply the base relations, 
which exclude all elimination pairs.

\begin{defn}[Elimination Pair Relations]
\label{defn::stack::appendix::vis-proof::elim-pair-relations}
Given the definitions,
\begin{align*}
\elimPairSet i & \defini \{ \projFirst {\elimPair i}, \projSecond{\elimPair i}\} \\
\elimSet{0} & \defini \elimSet{} \\
\elimSet{i+1} & \defini \elimSet i \setminus \elimPairSet i \qquad \text{ for } i \natorderSymbol |\elimRelName|
\end{align*}
where $\elimPair i$ is an enumeration of the pairs in $\elimRelName$, from 0 to $|\elimRelName| - 1$.\footnote{Any enumeration will work.}
And $\projFirst p$, $\projSecond{p}$ are the standard projection functions for pair $p$.
Also, given $\elimRel u o$, we define $\cPair u = o$ and $\cPair{o} = u$, i.e., $\cPair e$ is
the event that is coupled with $e$ in the elimination pair.

We define the visibility and separability relations recursively, for $i \natorderSymbol |\elimRelName|$,
\begin{align*}
\visObsIndx{0}{} u o \defini & \visObsIndx{}{B} u o \\
\visSepIndx {0}{} {x} {y} \defini & \visSepIndx {} {B} {x} {y} \\
\visObsIndx{i+1}{} u o \defini & \visObsIndx{i}{} u o \vee \elimPair i = (u,o) \\
\visSepIndx {i+1}{} {x} {y} \defini & (x \notin \elimSet i \wedge y \in \elimPairSet i \wedge \BE i x y) \vee 
                                        (x \in \elimPairSet{i} \wedge y \notin \elimSet i \wedge \neg \BE i y x) \vee 
                                        \visSepIndx {i} {} {x} {y} \\
\BE i x y \defini & y \in \elimSet{} \wedge \exists z.\ x \refleTransCl{\genVisSymbol_i} z \wedge (\precedesAbs{z}{\cPair{y}} \vee \precedesAbs{z}{y}) \\
\genVisSymbol_i \defini & \visObsIndxSymbol{i}{} \cup \visSepIndxSymbol{i}{}
\end{align*}
where $\refleTransCl{\genVisSymbol_i}$ is the reflexive and transitive closure of $\genVisSymbol_i$.
The closure $\refleTransCl{\genVisSymbol_i}$ is computed over $\absEvent$.
\end{defn}

The indexed set $\elimPairSet{i}$ collects the events in the $i$-th elimination pair 
(remember that the pairs in relation $\elimRelName$ are enumerated by $\elimPair{i}$).

The family of sets $\elimSet{i}$ will be useful to state situations when
we want to express ``all elimination pairs but the first $i$ ones''.
Set $\elimSet{0}$ contains all events belonging to an elimination pair, since it is 
simply $\elimSet{}$.
Set $\elimSet{i+1}$ is simply $\elimSet{i}$ with the $i$-th elimination pair removed.
So, for example, $\elimSet{0}$ has all the elimination pairs; $\elimSet{1}$ removes from $\elimSet{0}$ the 
0-th elimination pair (remember that the pairs in relation $\elimRelName$ are enumerated 
by $\elimPair{i}$ starting from index 0); $\elimSet{2}$ removes from $\elimSet{1}$ the 
1-th elimination pair, or equivalently, $\elimSet{2}$ removes from $\elimSet{0}$ the 
0-th and 1-th elimination pairs; and so on.

The family of relations $\visObsIndxSymbol{i}{}$ and $\visSepIndxSymbol{i}{}$ are the visibility and
separability relations when the first $i$ elimination pairs have been included.
Relations $\visObsIndxSymbol{0}{}$ and $\visSepIndxSymbol{0}{}$ encode
the visibility and separability relations when \emph{no} elimination
pairs are included.

Relation $\visObsIndx{i+1}{}{u}{o}$ states that
$u$ is observed by $o$ if either they were already related 
in this way in the previous stage $\visObsIndx{i}{}{u}{o}$,
or $(u,o)$ is the $i$-th elimination pair in the enumeration,
since the push in an elimination pair is taken by the pop 
in the elimination pair,
and hence, the push is observed by the pop.

To understand $\visSepIndx{i+1}{}{x}{y}$, it is necessary
to explain how separation works in the presence of elimination
pairs.
The key property of an elimination pair is that it can be freely
ordered anywhere in a linearization, because it will not affect
the abstract state of the stack. The reason for this is that an elimination 
pair is linearized as a push immediately followed by a pop. So, if $S$ is the stack
before executing the elimination pair, after executing the push and
pop in the elimination pair, the stack is again $S$.
Since the push and pop in an elimination pair are linearized as
one immediately followed by the other, the push and pop
in an elimination pair can be treated as a single unit in any argument.

However, linearizability must respect the returns-before relation.
Therefore, elimination pairs can be freely ordered as long as returns-before
is respected.
This means that an event $x$ that returns-before 
$u$ \emph{or} $o$ in the elimination pair $(u,o)$,
should be ordered before \emph{both} events $u$ and $o$,
since we treat the push and pop in the elimination pair as a single unit.

\begin{figure}[t]
	\includegraphics[scale=0.7]{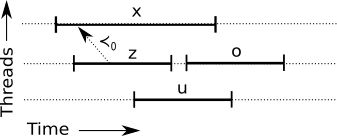}
	\centering
	\caption{Possible execution showing four events. Push $u$ and pop $o$ 
	form an elimination pair. Event $z$ returns-before $o$, and $x$
	overlaps with $z$, $u$, and $o$. 
	Event $x$ is a dependency of $z$ under the constraint relation 
	$\genVisSymbol_0$ at level 0,
indicated with a dashed arrow.}
	\label{fig::stack::appendix::vis-proof::be-example-diagram}
\end{figure}

However, the above intuition is still incomplete 
because it does not capture all the possible 
situations in which $x$ should be ordered before both $u$ and $o$. 
To complete the intuition, the key
insight is to also take into account dependencies under the constraint
relation $\genVisSymbol$.
For example, Figure~\ref{fig::stack::appendix::vis-proof::be-example-diagram} 
shows an execution in which $x$ is a dependency at level 0
of some event $z$ (i.e., $x \genVisSymbol_0 z$, where $\genVisSymbol_0 \defini \visObsIndxSymbol{0}{} \cup \visSepIndxSymbol{0}{}$)
and $z$ returns-before $o$. Even though $x$ \emph{does not} return-before
$u$ or $o$, event $x$ should still be ordered before $u$ and $o$ because 
(1) $x$ is a dependency of $z$, hence, $x$ should be ordered before $z$ and
(2) $z$ returns-before the elimination pair, hence, $z$ 
should be ordered before $u$ and $o$.

We can formally capture the above idea by saying that 
$x$ \emph{should be ordered before} elimination pair event $y$ at level 0,
denoted $\BE{0}{x}{y}$ (we will later generalize to an arbitrary level $i$) if,
\begin{align*}
\BE{0}{x}{y} \defini y \in \elimSet{} \wedge \exists z.\ x \refleTransCl{\genVisSymbol_0} z \wedge (\precedesAbs{z}{\cPair{y}} \vee \precedesAbs{z}{y})
\end{align*}
In English: $y$ is an elimination pair event and 
$x$ is a dependency at level 0 of some event $z$ that either returns-before 
the elimination couple of $y$ or returns-before $y$.
We use the reflexive-transitive closure $\refleTransCl{\genVisSymbol_0}$ to account
for the fact that $z$ could be $x$. Hence, in 
Figure~\ref{fig::stack::appendix::vis-proof::be-example-diagram}
we have both $\BE{0}{x}{u}$ and $\BE{0}{x}{o}$.

With the $\BEName$ predicate defined at level 0, we can now think how we should define the separability
relation at level 1 (i.e., $\visSepIndxSymbol{1}{}$).
At level 1 we take into consideration the two events in the first elimination
pair in the enumeration of $\elimRelName$, i.e., the two events in set 
$\elimPairSet{0}$,\footnote{Remember that indexes in the enumeration 
	of $\elimRelName$ start from 0.} 
together with all the events that were previously related under 
the visibility and separability relations at level 0, i.e. all non-elimination 
pair events, or equivalently, all events 
in the set $\absEvent \setminus \elimSet{0}$.
Therefore, we can either separate 
(1) an $x \notin \elimSet{0}$ before a $y \in \elimPairSet{0}$,
(2) an $x \in \elimPairSet{0}$ before a $y \notin \elimSet{0}$,
(3) an $x \notin \elimSet{0}$ before a $y \notin \elimSet{0}$.
Note that there is no need to separate an $x \in \elimPairSet{0}$
before a $y \in \elimPairSet{0}$ because $x$ and $y$ would
belong to the same elimination pair and we treat elimination
pairs as a single unit.

For the (1) case, we simply define $\visSepIndx{1}{}{x}{y}$ if 
$\BE{0}{x}{y}$, i.e., if $x$ should be ordered before the elimination pair event $y$.

For (2) we cannot use $\BE{0}{x}{y}$ because the $\BEName$ predicate requires $y$
to be an elimination pair event. Also, even though $x$ is an elimination pair event, 
we cannot define $\visSepIndx{1}{}{x}{y}$ simply as $\BE{0}{y}{x}$ because 
if $y$ actually returns-before $x$, the separability relation
will introduce a violation of axiom \axiomSRef{vis-ax::stack::cc-no-future-dependence}, i.e., 
$y \precedesAbsSymbol x \visSepIndxSymbol{1}{} y$.
The key insight is to define $\visSepIndx{1}{}{x}{y}$ if $\neg\BE{0}{y}{x}$, because the negation is actually
stating the following,
\[
x \in \elimSet{} \implies \forall z.\ y \refleTransCl{\genVisSymbol_0} z \implies (\nprecedesAbs{z}{\cPair{x}} \wedge \nprecedesAbs{z}{x})
\]
In particular, since $x \in \elimPairSet{0} \subseteq \elimSet{}$ and by instantiating $z$ with $y$, we obtain
$\nprecedesAbs{y}{\cPair{x}} \wedge \nprecedesAbs{y}{x}$, and more specifically 
$\nprecedesAbs{y}{x}$. In other words, either $y$ overlaps with $x$ or 
$y$ actually starts after $x$ finished. In both cases, it makes sense to separate $x$ 
before $y$.\footnote{In the case when $y$ overlaps with $x$, it is ok to arbitrarily 
	separate $x$ before $y$ because elimination pairs can be freely ordered 
as long as returns-before is respected. In this case, since $x$ and $y$ overlap, there is
no restriction between $x$ and $y$ regarding returns-before.}

For (3) we simply separate $x$ before $y$ if they were already separated 
at level 0, i.e., $x \visSepIndxSymbol{0}{} y$.

Each one of the above definitions for the cases (1), (2) and (3) corresponds to each
of the three disjuncts in $\visSepIndxSymbol{i+1}{}$ 
(Definition~\ref{defn::stack::appendix::vis-proof::elim-pair-relations}),
but specialized to $i = 0$.

Now that we have the definition of $\visSepIndxSymbol{1}{}$, we can then define 
$\BEName_1$, identically as in level 0 but using the constraint relation at level 1, 
$\genVisSymbol_{1} \defini \visSepIndxSymbol{1}{} \cup \visObsIndxSymbol{1}{}$.
Then, we define $\visSepIndxSymbol{2}{}$ identically as in level 1, but using $\BEName_1$ and 
sets $\elimPairSet{1}$ and $\absEvent\setminus\elimSet{1}$. We use $\elimPairSet{1}$ because it contains the two events
in the second elimination pair in the enumeration, while set $\absEvent\setminus\elimSet{1}$
contains all non-elimination pairs together with the first elimination pair, which has been
already related by the visibility and separability relations at the previous level 1.
We can continue this construction process for arbitrary higher levels, so that
the general step is described by $\visSepIndxSymbol{i+1}{}$ 
as in Definition~\ref{defn::stack::appendix::vis-proof::elim-pair-relations},
for arbitrary $i$.

This process finishes when $i+1$ reaches $|\elimRelName|$, i.e.,
the total number of elimination pairs. Hence,
we define the final visibility and separability relations
precisely at this index,
which include all the elimination pairs,

\begin{defn}[Final Visibility and Separability Relations]
\label{defn::stack::appendix::vis-proof::final-relations}
\begin{align*}
\visObsSYMBOL & \defini \visObsIndxSymbol {|\elimRelName|} {} \\
\visSepSYMBOL & \defini \visSepIndxSymbol {|\elimRelName|} {}
\end{align*}
\end{defn}

Using $\visObsSYMBOL$ and $\visSepSYMBOL$ as in Definition~\ref{defn::stack::appendix::vis-proof::final-relations},
the proof that the TS-stack satisfies the visibility-style axioms
works as follows.
We prove inductively that for any natural number $i \natorderEqSymbol |\elimRelName|$,
the visibility-style axioms hold when all the variables are restricted to the domain
$\closedEvent\setminus\elimSet{i}$. This means that when $i = |\elimRelName|$,
all the visibility-style axioms hold at domain $\closedEvent\setminus\elimSet{|\elimRelName|} = \closedEvent$,
since we will show that $\elimSet{|\elimRelName|} = \emptyset$.

Section~\ref{subsubsect::stack::appendix::vis-proof::without-elim-pairs} proves the base case ($i=0$), 
i.e., when all the elimination pairs are elided. 
Section~\ref{subsubsect::stack::appendix::vis-proof::with-elim-pairs} then shows the inductive step:
if the axioms hold at domain $\closedEvent\setminus\elimSet{i}$
then they also hold at domain 
$\closedEvent\setminus\elimSet{i+1}$.

\begin{figure}[t]
	\begin{subfigwrap}{Key TS-stack invariants.}{subfig::stack::appendix::vis-proof::key-ts-stack-invariants}
		\centering
		\begin{tabular}{l}
			\axiomILabel{vis-ax::stack::appendix::vis-proof::disjoint-push} Disjoint Push Timestamp Generation \\
			\quad $\precedesSpans{\spanEv{u_1}}{\spanEv{u_2}} \implies \AstampProp{u_1} \GstamporderSymbol \AstampProp{u_2}$ \\
			\axiomILabel{vis-ax::stack::vis-proof::misses-are-late} Misses start late \\
			\quad $\PEND{o}{u} \implies \precedesReps {\oneSpan{\spanEv{o}}} {\oneSpan{\spanEv{u}}}$ \\
		\end{tabular} 
	\end{subfigwrap}
	
	\begin{subfigwrap}{Structural TS-stack invariants.}{subfig::stack::appendix::vis-proof::structural-ts-stack-invariants}
		\centering
		\begin{tabular}{l}
			\axiomILabel{inv::stack::appendix::vis-proof::pop-not-precedes-push} 
			$\idProp{u} = \idProp{o} \implies \TimeProp{\twoSpan{\spanEv{o}}} \not\natorderSymbol \STimeProp{u}$ \\
			\axiomILabel{inv::stack::appendix::vis-proof::start-end-of-span} 
			$\DEF{\spanEv{x}} \implies \precedesRepsEq{\oneSpan{\spanEv{x}}}{\twoSpan{\spanEv{x}}}$ \\
			\axiomILabel{inv::stack::appendix::vis-proof::timestamps-equal}
			$\idProp{u} = \idProp{o} \wedge \precedesSpans{\spanEv{u}}{\spanEv{o}} \implies \stampProp{u} = \stampProp{o}$ \\
			\axiomILabel{inv::stack::appendix::vis-proof::pop-ids-imply-span}
			$\DEF{\idProp{o}} \implies \DEF{\spanEv{o}}$ \\
			\axiomILabel{inv::stack::appendix::vis-proof::timestamps-imply-span}
			$\DEF{\stampProp{x}} \implies \DEF{\spanEv{x}}$ \\
			\axiomILabel{inv::stack::appendix::vis-proof::terminated-implies-span}
			$x \in \terminatedEvent \implies \DEF{\spanEv{x}} \wedge \outputProp{x} = \spanEvOut{x}$ \\
			\axiomILabel{inv::stack::appendix::vis-proof::span-implies-all-properties}
			$\DEF{\spanEv{x}} \implies \DEF{\stampProp{x}} \wedge \DEF{\idProp{x}} \wedge \DEF{\spanEvOut{x}}$ \\
			\axiomILabel{inv::stack::appendix::vis-proof::spans-are-injective}
			$\spanEv{x} = \spanEv{y} \implies x = y$ \\
			\axiomILabel{inv::stack::appendix::vis-proof::reps-are-injective}
			$\twoSpan{\spanEv{x}} = \twoSpan{\spanEv{y}} \implies \spanEv{x} = \spanEv{y}$ \\
			\axiomILabel{inv::stack::appendix::vis-proof::prop-funcs-are-injective} \\
			\quad (i) $\idProp{u_1} = \idProp{u_2} \implies u_1 = u_2$ \\
			\quad (ii) $\idProp{o_1} = \idProp{o_2} \implies o_1 = o_2$ \\
			\axiomILabel{inv::stack::appendix::vis-proof::span-inside-event} \\
			\quad (i) $\DEF{\spanEv{x}} \implies \STimeProp{x}\ \natorderEqSymbol\ \TimeProp{\oneSpan{\spanEv{x}}}$ \\
			\quad (ii) $\DEF{\ETimeProp{x}} \implies \TimeProp{\twoSpan{\spanEv{x}}}\ \natorderEqSymbol\ \ETimeProp{x}$ \\
			\axiomILabel{inv::stack::appendix::vis-proof::pops-take-pushes} \\
			\quad (i) $\DEF{\idProp{o}} \implies \exists u.\ \idProp{u} = \idProp{o} \wedge \spanEvOut{o} = \inProp{u}$ \\
			\quad (ii) $\DEF{\idProp{u}} \implies \spanEvOut{u} = \unitValue$
		\end{tabular} 
	\end{subfigwrap}
	\caption{TS-stack invariants.}
	\label{fig::stack::appendix::vis-proof::ts-stack-invariants}
\end{figure}

During the entire proof, we assume that all the TS-stack invariants
in Figure~\ref{fig::stack::appendix::vis-proof::ts-stack-invariants} hold. 
In other words, we will prove the visibility-style axioms under the assumption
of these invariants. Later, Section~\ref{subsect::stack::appendix::vis-proof::proof-of-invariants}
will show that both versions of the TS-stack satisfy the invariants.

Before starting the main proof, we have the following lemmas which follow
directly from the definitions.

\begin{lem}
	\label{lem::stack::appendix::vis-proof::elim-unique-enumeration}
Let $i,j \in |\elimRelName|$. If $i \neq j$ and $\elimPair{i} = (u_i,o_i)$ and $\elimPair{j} = (u_j,o_j)$, then $u_i \neq u_j$ and $o_i \neq o_j$.
\end{lem}

\begin{prf}
	Suppose for a contradiction that $u_i = u_j \vee o_i = o_j$. We focus on the $u_i = u_j$ case, as the other one is similar.
	
	Since $\elimPair{i} = (u_i,o_i)$ and $\elimPair{j} = (u_j,o_j)$, we have by definition of the $\elimRelName$ relation that 
	$\idProp{u_i} = \idProp{o_i}$ and $\idProp{u_j} = \idProp{o_j}$. But $u_i = u_j$, which implies $\idProp{o_i} = \idProp{o_j}$. 
	So, $o_i = o_j$ by \axiomIRef{inv::stack::appendix::vis-proof::prop-funcs-are-injective}. 
	But then $\elimPair{i} = (u_i,o_i) = (u_j,o_j) = \elimPair{j}$, meaning that they are the same pair in 
	the enumeration, and so $i = j$ (Contradiction).
\end{prf}

\begin{lem}
\label{lem::stack::appendix::vis-proof::elim-set-basic-facts}
Let $i,j \natorderEqSymbol |\elimRelName|$.
\begin{enumerate}
	\item $\elimPairSet{i} \subseteq \elimSet{i}$.
	\item If $j \natorderEqSymbol i$, then $\elimSet{i} \subseteq \elimSet{j}$.
	\item If $j \natorderEqSymbol i$, then ${\visObsIndxSymbol{j}{}} \subseteq {\visObsIndxSymbol{i}{}}$ and ${\visSepIndxSymbol{j}{}} \subseteq {\visSepIndxSymbol{i}{}}$.
\end{enumerate}
\end{lem}

\begin{prf}
	We prove each part.
\begin{enumerate}
	\item Let $x \in \elimPairSet i$. Since $x$ is in the $i$-th elimination pair in the enumeration of $\elimRelName$, 
	we have $x \in \elimSet{0}$. Suppose for a contradiction $x \in \bigcup_{j \natorderSymbol i} \elimPairSet{j}$. Then, $x$ is in both
	the $i$-th elimination pair and the $j$-th elimination pair ($j \natorderSymbol i$). So, $x \neq x$ by Lemma~\ref{lem::stack::appendix::vis-proof::elim-unique-enumeration} (Contradiction). Hence, $x \notin \bigcup_{j \natorderSymbol i} \elimPairSet{j}$.
	
	The result follows from the following claim when $l = i$.
	\begin{claim*}
		Let $l \natorderEqSymbol |\elimRelName|$. If $x \in \elimSet{0}$ and $x \notin \bigcup_{j \natorderSymbol l} \elimPairSet{j}$, 
		then $x \in \elimSet{l}$.
	\end{claim*}
	
	The base case ($l = 0$) is immediate. For the inductive step, suppose 
	$x \in \elimSet{0}$ and $x \notin \bigcup_{j \natorderSymbol l+1} \elimPairSet{j}$. Hence,
	$x \notin \bigcup_{j \natorderSymbol l} \elimPairSet{j}$ and $x \notin \elimPairSet{l}$.
	
	By IH, $x \in \elimSet{l}$. But from $x \notin \elimPairSet{l}$, we get $x \in \elimSet{l+1}$ by
	definition.
	
	\item By induction on $i$. 
	
	For the base case ($i = 0$), $j \natorderEqSymbol 0$ implies $j = 0$, and $\elimSet{0} \subseteq \elimSet{0}$ follows trivially.
	
	For the inductive case, suppose $j \natorderEqSymbol i+1$. If $j = i+1$, we have trivially $\elimSet{i+1} \subseteq \elimSet{i+1} = \elimSet{j}$. 
	
	Hence, we can assume $j \natorderSymbol i+1$. So, $j \natorderEqSymbol i$ and by IH, $\elimSet{i} \subseteq \elimSet{j}$.
	But if $x \in \elimSet{i+1}$, then $x \in \elimSet{i}$ by definition. Therefore, $\elimSet{i+1} \subseteq \elimSet{j}$.
	
	\item By induction on $i$.
	
	For the base case ($i = 0$), $j \natorderEqSymbol 0$ implies $j = 0$, and ${\visObsIndxSymbol{0}{}} \subseteq {\visObsIndxSymbol{0}{}}$ and ${\visSepIndxSymbol{0}{}} \subseteq {\visSepIndxSymbol{0}{}}$ follow trivially.
	
	For the inductive case, suppose $j \natorderEqSymbol i+1$. If $j = i+1$, we have trivially $\visObsIndxSymbol{j}{} = {\visObsIndxSymbol{i+1}{}} \subseteq {\visObsIndxSymbol{i+1}{}}$ and $\visSepIndxSymbol{j}{} = {\visSepIndxSymbol{i+1}{}} \subseteq {\visSepIndxSymbol{i+1}{}}$.
	
	Hence, we can assume $j \natorderSymbol i+1$. So, $j \natorderEqSymbol i$ and by IH, ${\visObsIndxSymbol{j}{}} \subseteq {\visObsIndxSymbol{i}{}}$ and ${\visSepIndxSymbol{j}{}} \subseteq {\visSepIndxSymbol{i}{}}$.
	
	But if $\visObsIndx{i}{}{x}{y}$ holds, then $\visObsIndx{i+1}{}{x}{y}$ by definition. Similarly for $\visSepIndx{i}{}{x}{y}$. Therefore,
	${\visObsIndxSymbol{j}{}} \subseteq {\visObsIndxSymbol{i+1}{}}$ and ${\visSepIndxSymbol{j}{}} \subseteq {\visSepIndxSymbol{i+1}{}}$.
	
\end{enumerate}
\end{prf}

\begin{lem}
\label{lem::stack::appendix::vis-proof::gen-vis-restricted-to-elim-level}
Let $i \natorderEqSymbol |\elimRelName|$. Suppose $x,y \in \absEvent \setminus \elimSet{i}$.
\begin{itemize}
	\item If $\visObs{} x y$, then $x \visObsIndxSymbol{i}{} y$.
	\item If $\visSep{} x y$, then $x \visSepIndxSymbol{i}{} y$.
	\item If $\genVis x y$, then $x \genVisSymbol_i y$.
\end{itemize}
\end{lem}

\begin{prf}
	Suppose $x,y \in \absEvent \setminus \elimSet{i}$. If either $\visObs{} x y$, or $\visSep{} x y$, or $\genVis x y$, 
	then $x \visObsIndxSymbol{|\elimRelName|}{} y$ or $x \visSepIndxSymbol{|\elimRelName|}{} y$ by definition.
	
	The result follows from the following claim when $l = |\elimRelName|$. We prove the claim by induction on $l$.
	
		\begin{claim*}
			Let $l \natorderEqSymbol |\elimRelName|$. 
			\begin{itemize}
				\item If $x \visObsIndxSymbol{l}{} y$, then $x \visObsIndxSymbol{i}{} y$.
				\item If $x \visSepIndxSymbol{l}{} y$, then $x \visSepIndxSymbol{i}{} y$.
			\end{itemize}
		\end{claim*}
		
		The base case ($l = 0$) follows directly from part 3 of Lemma~\ref{lem::stack::appendix::vis-proof::elim-set-basic-facts}. 
		
		For the inductive case, suppose either $x \visObsIndxSymbol{l+1}{} y$ or $x \visSepIndxSymbol{l+1}{} y$. 
		If $l \natorderSymbol i$, then $l+1 \natorderEqSymbol i$, and by part 3 of Lemma~\ref{lem::stack::appendix::vis-proof::elim-set-basic-facts},
		$x \visObsIndxSymbol{i}{} y$ and $x \visSepIndxSymbol{i}{} y$ hold, respectively.
		
		Hence, we can assume $i \natorderEqSymbol l$. We now check each case in the definition of $\visObsIndxSymbol{l+1}{}$ and $\visSepIndxSymbol{l+1}{}$.
		
		\begin{itemize}
			\item Case $x$ is a push and $y$ is a pop, and $\visObsIndx{l}{} x y$. By IH, $x \visObsIndxSymbol{i}{} y$ holds.
			
			\item Case $x$ is a push and $y$ is a pop, and $\elimPair l = (x,y)$. Then, $x,y \in \elimPairSet{l}$ by definition. 
			
			By part 1 of Lemma~\ref{lem::stack::appendix::vis-proof::elim-set-basic-facts}, we have $x,y \in \elimSet{l}$. But by
			part 2 of Lemma~\ref{lem::stack::appendix::vis-proof::elim-set-basic-facts}, we have $x,y \in \elimSet{i}$ (since $i \natorderEqSymbol l$), 
			which contradicts hypothesis $x,y \in \absEvent \setminus \elimSet{i}$.
			
			\item Case $x \notin \elimSet l \wedge y \in \elimPairSet l \wedge \BE l x y$. But $y \in \elimPairSet l$ 
			and part 1 of Lemma~\ref{lem::stack::appendix::vis-proof::elim-set-basic-facts} imply $y \in \elimSet{l}$.
			Hence, part 2 of Lemma~\ref{lem::stack::appendix::vis-proof::elim-set-basic-facts} implies $y \in \elimSet{i}$ (since $i \natorderEqSymbol l$) 
			which contradicts hypothesis $y \in \absEvent \setminus \elimSet{i}$.
			
			\item Case $x \in \elimPairSet{l} \wedge y \notin \elimSet l \wedge \neg \BE l y x$. Similar to the previous case, but using $x$.
			
			\item Case $\visSepIndx {l} {} {x} {y}$. By IH, $x \visSepIndxSymbol{i}{} y$ holds.
		\end{itemize}
		This proves the claim.
\end{prf}

\begin{lem}
	\label{lem::stack::appendix::vis-proof::elim-pairs-go-together}
		Let $i \natorderEqSymbol |\elimRelName|$. Let $x$ a push and $y$ a pop. 
		If $\idProp{x} = \idProp{y}$, then $x \in \elimSet{i}$ if and only if $y \in \elimSet{i}$.
\end{lem}

\begin{prf}
By induction on $i$. 

\begin{itemize}
	\item Base case $i = 0$. We focus on the forward direction, since the other one is similar. 
	
	Since $x \in \elimSet{0}$, by definition of $\elimSet{0}$, there is a pop $o$ such that $\elimRel{x}{o}$. 
	But this implies by definition of $\elimRelName$ that $\idProp{x} = \idProp{o}$. And hence,
	$\idProp{o} = \idProp{y}$ since $\idProp{x} = \idProp{y}$ by hypothesis.
	Therefore, \axiomIRef{inv::stack::appendix::vis-proof::prop-funcs-are-injective} 
	implies $o = y$, meaning that $\elimRel{x}{y}$, and so $y \in \elimSet{0}$ by
	definition.
	
	\item Inductive case. We focus on the forward direction, since the other one is similar. 
	
	Since $x \in \elimSet{i+1}$, we have $x \in \elimSet{i}$ and $x \notin \elimPairSet{i}$ by definition. 
	By IH, $y \in \elimSet{i}$. Suppose for a contradiction that $y \in \elimPairSet{i}$. Then,
	$\elimPair{i} = (u,y)$ for some push $u$ by definition of $\elimPairSet{i}$. But this means $\elimRel{u}{y}$.
	
	So, $\idProp{u} = \idProp{y}$ by definition of $\elimRelName$. And hence,
	$\idProp{u} = \idProp{x}$ since $\idProp{x} = \idProp{y}$ by hypothesis.
	Therefore, \axiomIRef{inv::stack::appendix::vis-proof::prop-funcs-are-injective} 
	implies $u = x$, meaning that $\elimPair{i} = (x,y)$, and so $x \in \elimPairSet{i}$ (Contradiction).
	
	Hence, $y \notin \elimPairSet{i}$, and so, $y \in \elimSet{i+1}$ by definition.
\end{itemize}
\end{prf}

\begin{lem}
	\label{lem::stack::appendix::vis-proof::vis-level-inmplies-no-elim-pair-level}
	Let $i \natorderEqSymbol |\elimRelName|$. If $x \genVisSymbol_i y$, then $x,y \notin \elimSet{i}$.
\end{lem}

\begin{prf}
	By induction on $i$.
	
	The base case $(i = 0)$ follows directly from the definitions of $\visObsIndxSymbol{0}{}$ and $\visSepIndxSymbol{0}{}$. 
	Hence, we can focus on the inductive case. We consider each case in the definition of $x \genVisSymbol_{i+1} y$,
	
	\begin{itemize}
		\item Case $\visObsIndx{i}{}{x}{y}$. From the IH, we have $x,y \notin \elimSet{i}$. So, $x,y \notin \elimSet{i+1}$,
		since $\elimSet{i+1} \subseteq \elimSet{i}$ by Lemma~\ref{lem::stack::appendix::vis-proof::elim-set-basic-facts}.
		
		\item Case $\elimPair{i} = (x,y)$. Hence, $x,y \in \elimPairSet{i}$ by definition. This means that 
		$x,y \notin \elimSet{i+1}$, for otherwise we would have $x,y \notin \elimPairSet{i}$ by definition of $\elimSet{i+1}$.
		
		\item Case $x \notin \elimSet i \wedge y \in \elimPairSet i \wedge \BE i x y$. Since $y \in \elimPairSet i$, this means that 
		$y \notin \elimSet{i+1}$, for otherwise we would have $y \notin \elimPairSet{i}$ by definition of $\elimSet{i+1}$.
		
		Also, since $x \notin \elimSet i$, then $x \notin \elimSet{i+1}$,
		since $\elimSet{i+1} \subseteq \elimSet{i}$ by Lemma~\ref{lem::stack::appendix::vis-proof::elim-set-basic-facts}.
		
		\item Case $x \in \elimPairSet{i} \wedge y \notin \elimSet i \wedge \neg \BE i y x$. Similar to the previous case, but
		inverting the roles of $x$ and $y$.
		
		\item Case $\visSepIndx {i} {} {x} {y}$. From the IH, we have $x,y \notin \elimSet{i}$. So, $x,y \notin \elimSet{i+1}$,
		since $\elimSet{i+1} \subseteq \elimSet{i}$ by Lemma~\ref{lem::stack::appendix::vis-proof::elim-set-basic-facts}.
	\end{itemize}
\end{prf}

\begin{lem}
\label{lem::stack::appendix::vis-proof::cardinality-of-elim-pairs}
Let $i \natorderEqSymbol |\elimRelName|$. We have $|\elimSet{i}| = 2(|\elimRelName| - i)$.
\end{lem}

\begin{prf}
	By induction on $i$.
	
	For the base case $(i = 0)$, $\elimSet{0}$ consists, by definition, on the domain and codomain of the $\elimRelName$ relation. Lemma~\ref{lem::stack::appendix::vis-proof::elim-unique-enumeration} implies that any two distinct tuples in $\elimRelName$ have their 
	components mutually different. Therefore $|\elimSet{0}|$ counts all the events in the first component of tuples in $\elimRelName$ plus
	all the events in the second component of tuples in $\elimRelName$. In other words, $|\elimSet{0}|$ is twice the tuples in $\elimRelName$,
	i.e., $|\elimSet{0}| = 2|\elimRelName|$.
	
	For the inductive case, $\elimSet{i+1}$ consists, by definition, on $\elimSet{i}$ minus the two events in the $i$-th tuple in the enumeration
	of $\elimRelName$. In other words, $|\elimSet{i+1}| = |\elimSet{i}| - 2$. Therefore, from the IH,
	\begin{align*}
		|\elimSet{i+1}| & = |\elimSet{i}| - 2 \\
		                & = 2(|\elimRelName| - i) - 2 \\
		                & = 2(|\elimRelName| - i - 1) \\
		                & = 2(|\elimRelName| - (i + 1))
	\end{align*}
\end{prf}

\begin{lem}
	\label{lem::stack::appendix::event-prec-implies-span-prec}
	If $\precedesAbs{x}{y}$ and $\DEF{\spanEv{y}}$, then $\precedesSpans{\spanEv{x}}{\spanEv{y}}$.
\end{lem}

\begin{prf}
	Since $\precedesAbs{x}{y}$, event $x$ has terminated. But $\DEF{\spanEv{y}}$ by hypothesis, so 
	\axiomIRef{inv::stack::appendix::vis-proof::span-inside-event} implies,
	\[
	\TimeProp{\twoSpan{\spanEv{x}}}\ \natorderEqSymbol\ \ETimeProp{x}\ \natorderSymbol\ \STimeProp{y}\ \natorderEqSymbol\ \TimeProp{\oneSpan{\spanEv{y}}}
	\]
	i.e., $\TimeProp{\twoSpan{\spanEv{x}}}\ \natorderSymbol\ \TimeProp{\oneSpan{\spanEv{y}}}$, which means $\twoSpan{\spanEv{x}} < \oneSpan{\spanEv{y}}$. 
	Hence, $\spanEv{x} \precedesSpansSymbol \spanEv{y}$ by definition.
\end{prf}

\subsubsection{Eliding Elimination Pairs}
\label{subsubsect::stack::appendix::vis-proof::without-elim-pairs}

Given the relations in Definition~\ref{defn::stack::appendix::vis-proof::final-relations}, 
in this section we show that all the visibility-style axioms in Figure~\ref{fig::stack::concurrent-spec-stack} hold
when all the variables in the axioms range over $\closedEvent \setminus \elimSet{}$, i.e., 
when we elide the elimination pairs.\footnote{Axiom \axiomSRef{vis-ax::stack::cc-no-future-dependence}
	has as hypothesis a transitive closure of the constraint relation $x \transCl{\genVisSymbol} y$.
	We prove that the axiom holds when the transitive closure is computed over the domain 
	$\closedEvent\setminus\elimSet{}$.}

\begin{lem}
	\label{lem::stack::appendix::vis-proof::eq-id-implies-eq-timestamp}
	The following holds,
	\begin{enumerate}
		\item Let $x$ be a push and $y$ a pop. If $(x \notin \elimSet{} \vee y \notin \elimSet{})$ and $\idProp{x} = \idProp{y}$, then $\spanEv{x} \precedesSpansSymbol \spanEv{y}$ and $\AstampProp{x} = \AstampProp{y}$.
		\item Let $y$ be a pop such that $\DEF{\AstampProp{y}}$. If $y \notin \elimSet{}$, then there is a push $x \notin \elimSet{}$ such that 
		$\AstampProp{x} = \AstampProp{y}$ and $\spanEv{x} \precedesSpansSymbol \spanEv{y}$.
	\end{enumerate}
\end{lem}

\begin{prf}
We prove each part in turn.
\begin{enumerate}
	\item We focus on the $x \notin \elimSet{}$ case, since the case $y \notin \elimSet{}$ is similar.

		By part 1 of Lemma~\ref{lem::stack::appendix::vis-proof::elim-set-basic-disj-fact}, 
		$\precedesAbs{x}{y}$ after instantiating with $o \defini y$.
		Hypothesis $\idProp{x} = \idProp{y}$ implies $\DEF{\idProp{y}}$. 
		So, from \axiomIRef{inv::stack::appendix::vis-proof::pop-ids-imply-span}
		we have $\DEF{\spanEv{y}}$. Hence, Lemma~\ref{lem::stack::appendix::event-prec-implies-span-prec}
		implies $\spanEv{x} \precedesSpansSymbol \spanEv{y}$.
		
		Finally, \axiomIRef{inv::stack::appendix::vis-proof::timestamps-equal} implies
		$\stampProp{x} = \stampProp{y}$, which means $\AstampProp{x} = \AstampProp{y}$.
			
	\item We know $\DEF{\AstampProp{y}}$, hence $\DEF{\stampProp{y}}$.
	Applying \axiomIRef{inv::stack::appendix::vis-proof::timestamps-imply-span}, 
	\axiomIRef{inv::stack::appendix::vis-proof::span-implies-all-properties}, and
	\axiomIRef{inv::stack::appendix::vis-proof::pops-take-pushes} on $y$, there is a push $u_y$ such that $\idProp{u_y} = \idProp{y}$. 
	
	Since $y \notin \elimSet{}$, the first part of this lemma implies $\AstampProp{u_y} = \AstampProp{y}$
	and $\spanEv{u_y} \precedesSpansSymbol \spanEv{y}$.
	
	Also, $u_y \notin \elimSet{}$ by Lemma~\ref{lem::stack::appendix::vis-proof::elim-pairs-go-together},
	since $y \notin \elimSet{}$.
\end{enumerate}
\end{prf}

\begin{lem}
\label{lem::stack::appendix::vis-proof::pop-totality-for-pops}
	Let $x$ be any event and $y$ a pop. If $\DEF{\AstampProp{x}}$, $\DEF{\AstampProp{y}}$ and $x,y \notin \elimSet{}$, then
	$\AstampProp{x} \GstamporderSymbol \AstampProp{y}$ or $\AstampProp{y} \GstamporderSymbol \AstampProp{x}$ or $\AstampProp{x} = \AstampProp{y}$.
\end{lem}

\begin{prf}
		We need to check the case when $x$ is a push or a pop.
		\begin{itemize}
			\item Case $x$ is a push. 
			
			Since $y \notin \elimSet{}$, Part 2 of Lemma~\ref{lem::stack::appendix::vis-proof::eq-id-implies-eq-timestamp}
			implies that there is a push $u_y \notin \elimSet{}$ such that $\AstampProp{u_y} = \AstampProp{y}$.
			
			Therefore, by Pop-Totality, we obtain
			$\AstampProp {x} \GstamporderSymbol \AstampProp {u_y}$ or $\AstampProp {u_y} \GstamporderSymbol \AstampProp {x}$
			or $\AstampProp {x} = \AstampProp {u_y}$, which means,
			$\AstampProp {x} \GstamporderSymbol \AstampProp {y}$ or $\AstampProp {y} \GstamporderSymbol \AstampProp {x}$ or
			$\AstampProp {x} = \AstampProp {y}$, since $\AstampProp{u_y} = \AstampProp{y}$.
			
			\item Case $x$ is a pop. 
			
			Since $x,y \notin \elimSet{}$, Part 2 of Lemma~\ref{lem::stack::appendix::vis-proof::eq-id-implies-eq-timestamp}
			implies that there are pushes $u_x, u_y \notin \elimSet{}$ such that $\AstampProp{u_x} = \AstampProp{x}$, $\AstampProp{u_y} = \AstampProp{y}$.
			
			Therefore, by Pop-Totality, we obtain
			$\AstampProp {u_x} \GstamporderSymbol \AstampProp {u_y}$ or $\AstampProp {u_y} \GstamporderSymbol \AstampProp {u_x}$
			or $\AstampProp {u_x} = \AstampProp {u_y}$, which means,
			$\AstampProp {x} \GstamporderSymbol \AstampProp {y}$ or $\AstampProp {y} \GstamporderSymbol \AstampProp {x}$ or
			$\AstampProp {x} = \AstampProp {y}$, since $\AstampProp{u_x} = \AstampProp{x}$ and $\AstampProp{u_y} = \AstampProp{y}$.
		\end{itemize}
\end{prf}

\begin{lem}
	\label{lem::stack::appendix::vis-proof::negated-miss}
	Let $x$ be a pop. If the following conditions hold,
	\begin{itemize}
		\item $\neg \exists u' \notin \elimSet{}.\ \PEND {x} {u'}\ \wedge\ \AstampProp {u'} \GstamporderEqSymbol \AstampProp {y}$,
		\item $\AstampProp {u} \GstamporderEqSymbol \AstampProp {y}$,
		\item $\AstampProp {x} \GstamporderSymbol \AstampProp {u}$,
		\item $x,u \notin \elimSet{}$	
	\end{itemize}
	then, $\precedesReps{\twoSpan{\spanEv{o}}}{\twoSpan{\spanEv{x}}}$ for some pop $o \notin \elimSet{}$ such that
	$\AstampProp{u} = \AstampProp{o}$.
\end{lem}

\begin{prf}
	After de Morgan manipulations on hypothesis $\neg \exists u'\notin\elimSet{}.\ \PEND {x} {u'}\ \wedge\ \AstampProp {u'} \GstamporderEqSymbol \AstampProp {y}$,
	and expanding the definition of $\PEND {x} {u'}$, we obtain,\footnote{Recall that whenever we negate an atomic formula that contains functions 
	that could be undefined, like $\twoSpan{\spanEv {x}} \not\precedesRepsSymbol \twoSpan{\spanEv {o'}}$ in our case, we are implicitly negating the
    formula $\DEF{\spanEv {x}} \wedge \DEF{\spanEv {o'}} \wedge \twoSpan{\spanEv {x}} \precedesRepsSymbol \twoSpan{\spanEv {o'}}$.}
	\begin{align*}
	&\forall u'\notin\elimSet{}.\ \AstampProp {u'} \GstamporderEqSymbol \AstampProp {y}\ \wedge x,u' \notin \elimSet{} \wedge \AstampProp x \GstamporderSymbol \AstampProp{u'}
	\implies \\ 
	&\exists o'\notin \elimSet{}.\ \AstampProp{u'} = \AstampProp{o'} \wedge (\UNDEF{\spanEv {o'}} \vee \UNDEF{\spanEv {x}} \vee \precedesRepsEq {\twoSpan{\spanEv {o'}}} {\twoSpan{\spanEv {x}}})
	\end{align*}

Instantiating with $u' \defini u$, and using the second, third and fourth hypotheses (i.e. $\AstampProp {u} \GstamporderEqSymbol \AstampProp {y}$, $\AstampProp {x} \GstamporderSymbol \AstampProp {u}$ and $x,u \notin \elimSet{}$), we obtain that there is a pop $o\notin\elimSet{}$ such that 
$\AstampProp{u} = \AstampProp{o}$ and $\UNDEF{\spanEv {o}} \vee \UNDEF{\spanEv {x}} \vee \precedesRepsEq {\twoSpan{\spanEv {o}}} {\twoSpan{\spanEv {x}}}$.

The case $\UNDEF{\spanEv {o}} \vee \UNDEF{\spanEv {x}}$ is discarded by \axiomIRef{inv::stack::appendix::vis-proof::timestamps-imply-span}, 
since we know $\AstampProp{o}$ and $\AstampProp {x}$ are defined.

If $\twoSpan{\spanEv {o}} = \twoSpan{\spanEv {x}}$, then $\spanEv {o} = \spanEv {x}$ by \axiomIRef{inv::stack::appendix::vis-proof::reps-are-injective}, which implies $o = x$ by \axiomIRef{inv::stack::appendix::vis-proof::spans-are-injective}. So, 
$\AstampProp{u} = \AstampProp{o} = \AstampProp{x}$ (since $o = x$). But this contradicts hypothesis 
$\AstampProp {x} \GstamporderSymbol \AstampProp {u}$.

Therefore, $\precedesReps {\twoSpan{\spanEv {o}}} {\twoSpan{\spanEv {x}}}$.
\end{prf}

\begin{lem}
\label{lem::stack::appendix::vis-proof::transitive-closure}
Given the domain $\absEvent \setminus \elimSet{}$ for the relation $\genVisTransSymbol$, 
if $\genVisTrans{x}{y}$, then one of the following cases hold,
\begin{itemize}
	\item $x$ and $y$ are pushes and $\AstampProp{x} \GstamporderSymbol \AstampProp{y}$.
	\item $x$ is a push, $y$ is a pop, and, 
	\[
	\AstampProp{x} \GstamporderEqSymbol \AstampProp{y} \vee
	(\AstampProp{y} \GstamporderSymbol \AstampProp{x}\ \wedge\  
	\neg \exists u'\notin \elimSet{}.\ \PEND {y} {u'}\ \wedge\ \AstampProp {u'} \GstamporderEqSymbol \AstampProp {x})
	\]
	\item $x$ is a pop, $y$ is a push, and $\exists u'\notin \elimSet{}.\ \PEND {x} {u'}\ \wedge\ \AstampProp {u'} \GstamporderEqSymbol \AstampProp {y}$.
	\item $x$ and $y$ are pops, and,
	\begin{align*}
	& (\exists u'\notin\elimSet{}.\ \PEND {x} {u'}\ \wedge\ \AstampProp {u'} \GstamporderEqSymbol \AstampProp {y})\ \vee \\
	& (\AstampProp{y} \GstamporderSymbol \AstampProp{x}\ \wedge\  
	\neg \exists u'\notin\elimSet{}.\ \PEND {y} {u'}\ \wedge\ \AstampProp {u'} \GstamporderEqSymbol \AstampProp {x})
	\end{align*}
\end{itemize}
\end{lem}

\begin{prf}
By repeating Lemma~\ref{lem::stack::appendix::vis-proof::gen-vis-restricted-to-elim-level}
on hypothesis $\genVisTrans{x}{y}$, we have
$x \transCl{\genVisSymbol_0} y$ in the domain $\absEvent \setminus \elimSet{0} = \absEvent \setminus \elimSet{}$.

Denote by $P(x,y)$ the statement of the four cases in the lemma we want to prove. 

To prove $x \transCl{\genVisSymbol_0} y \implies P(x,y)$, it 
suffices to prove the two properties,
\begin{itemize}
	\item (Base case). $x \genVisSymbol_0 y \implies P(x,y)$.
	\item (Inductive case). $x \genVisSymbol_0 y \wedge P(y,z) \implies P(x,z)$.
\end{itemize}

The base case follows directly from the definitions of $\visObsIndx{}{B} x y$ and $\visSepIndx {} {B} {x} {y}$,
since they appear in some case of $P(x,y)$.
We now focus on the inductive case. We only need to consider the cases where the types of the events match.

\begin{itemize}
	\item Case $\visObsIndx{}{B} x y$ for push $x$ and pop $y$, and $\exists u' \notin \elimSet{}.\ \PEND {y} {u'}\ \wedge\ \AstampProp {u'} \GstamporderEqSymbol \AstampProp {z}$ for push $z$. 
	
	We have $\AstampProp{x} = \AstampProp{y}$ by definition of $\visObsIndx{}{B} x y$.
	
	But by definition of $\PEND {y} {u'}$, we have $\AstampProp y \GstamporderSymbol \AstampProp{u'}$. Hence,
	\[
	\AstampProp{x} = \AstampProp y \GstamporderSymbol \AstampProp{u'} \GstamporderEqSymbol \AstampProp {z}
	\]
	So, $\AstampProp{x} \GstamporderSymbol \AstampProp {z}$.
	
	\item Case $\visObsIndx{}{B} x y$ for push $x$ and pop $y$, and 
	$\exists u'\notin \elimSet{}.\ \PEND {y} {u'}\ \wedge\ \AstampProp {u'} \GstamporderEqSymbol \AstampProp {z}$ for pop $z$.
	
	Similar to the previous case, we will obtain $\AstampProp{x} \GstamporderEqSymbol \AstampProp {z}$.

	\item Case $\visObsIndx{}{B} x y$ for push $x$ and pop $y$, and
	$\AstampProp{z} \GstamporderSymbol \AstampProp{y}\ \wedge\  
	\neg \exists u'\notin \elimSet{}.\ \PEND {z} {u'}\ \wedge\ \AstampProp {u'} \GstamporderEqSymbol \AstampProp {y}$ for pop $z$.
	
	Again by definition of $\visObsIndx{}{B} x y$, we have $\AstampProp{x} = \AstampProp{y}$, so,
	\[
	\AstampProp{z} \GstamporderSymbol \AstampProp{x}\ \wedge\  
	\neg \exists u' \notin \elimSet{}.\ \PEND {z} {u'}\ \wedge\ \AstampProp {u'} \GstamporderEqSymbol \AstampProp {x}
	\]
	
	\item Case $\visSepIndx{}{B} x y$ for pushes $x$ and $y$, and $\AstampProp{y} \GstamporderSymbol \AstampProp{z}$ for push $z$.
	
	From definition of $\visSepIndx{}{B} x y$, we have $\AstampProp{x} \GstamporderSymbol \AstampProp{y}$.
	Hence, $\AstampProp{x} \GstamporderSymbol \AstampProp{z}$.
	
	\item Case $\visSepIndx{}{B} x y$ for pushes $x$ and $y$, and $\AstampProp{y} \GstamporderEqSymbol \AstampProp{z}$ for pop $z$.
	
	Similarly to the previous case, $\AstampProp{x} \GstamporderEqSymbol \AstampProp{z}$.
	
	\item Case $\visSepIndx{}{B} x y$ for pushes $x$ and $y$, and $\AstampProp{z} \GstamporderSymbol \AstampProp{y}\ \wedge\  
	\neg \exists u' \notin \elimSet{}.\ \PEND {z} {u'}\ \wedge\ \AstampProp {u'} \GstamporderEqSymbol \AstampProp {y}$ for pop $z$.
	
	By Lemma~\ref{lem::stack::appendix::vis-proof::pop-totality-for-pops}, we have
	$\AstampProp {x} \GstamporderEqSymbol \AstampProp {z}$ or $\AstampProp {z} \GstamporderSymbol \AstampProp {x}$.
	
	For the case $\AstampProp {x} \GstamporderEqSymbol \AstampProp {z}$ we are done. 
	For the case $\AstampProp {z} \GstamporderSymbol \AstampProp {x}$, we claim $\neg \exists u'\notin \elimSet{}.\ \PEND {z} {u'}\ \wedge\ \AstampProp {u'} \GstamporderEqSymbol \AstampProp {x}$.
	
	For suppose there is such $u'$, so $\AstampProp {u'} \GstamporderEqSymbol \AstampProp{x}$. 
	But $\AstampProp {x} \GstamporderSymbol \AstampProp {y}$ from the hypothesis $\visSepIndx{}{B} x y$. Therefore,
	$\AstampProp {u'} \GstamporderEqSymbol \AstampProp{y}$, which contradicts 
	the hypothesis $\neg \exists u'\notin \elimSet{}.\ \PEND {z} {u'}\ \wedge\ \AstampProp {u'} \GstamporderEqSymbol \AstampProp {y}$.
	
	\item Case $\visSepIndx{}{B} x y$ for pop $x$ and push $y$, and $\AstampProp{y} \GstamporderSymbol \AstampProp{z}$ for push $z$.
	
	From definition of $\visSepIndx{}{B} x y$, we have 
	$\exists u'\notin \elimSet{}.\ \PEND {x} {u'}\ \wedge\ \AstampProp {u'} \GstamporderEqSymbol \AstampProp {y}$. But since $\AstampProp{y} \GstamporderSymbol \AstampProp{z}$,
	we have $\exists u'\notin \elimSet{}.\ \PEND {x} {u'}\ \wedge\ \AstampProp {u'} \GstamporderEqSymbol \AstampProp {z}$.
	
	\item Case $\visSepIndx{}{B} x y$ for pop $x$ and push $y$, and $\AstampProp{y} \GstamporderEqSymbol \AstampProp{z}$ for pop $z$.
	
	By a similar reasoning as in the previous case, $\exists u'\notin \elimSet{}.\ \PEND {x} {u'}\ \wedge\ \AstampProp {u'} \GstamporderEqSymbol \AstampProp {z}$.
	
	\item Case $\visSepIndx{}{B} x y$ for pop $x$ and push $y$, and $\AstampProp{z} \GstamporderSymbol \AstampProp{y}\ \wedge\  
	\neg \exists u'\notin \elimSet{}.\ \PEND {z} {u'}\ \wedge\ \AstampProp {u'} \GstamporderEqSymbol \AstampProp {y}$ for pop $z$.
	
	From definition of $\visSepIndx{}{B} x y$, we have 
	$\exists u'\notin \elimSet{}.\ \PEND {x} {u'}\ \wedge\ \AstampProp {u'} \GstamporderEqSymbol \AstampProp {y}$.
	By Lemma~\ref{lem::stack::appendix::vis-proof::pop-totality-for-pops}, $\AstampProp {x} \GstamporderSymbol \AstampProp {z}$ or $\AstampProp {z} \GstamporderSymbol \AstampProp {x}$
	or $\AstampProp {x} = \AstampProp {z}$.
	
	\begin{itemize}
		\item Case $\AstampProp {x} \GstamporderSymbol \AstampProp {z}$. By  
		Lemma~\ref{lem::stack::appendix::vis-proof::pop-totality-for-pops}, $\AstampProp {u'} \GstamporderEqSymbol \AstampProp {z}$
		or $\AstampProp {z} \GstamporderSymbol \AstampProp {u'}$.
		\begin{itemize}
			\item Case $\AstampProp {u'} \GstamporderEqSymbol \AstampProp {z}$. Hence, we have $\exists u'\notin \elimSet{}.\ \PEND {x} {u'}\ \wedge\ \AstampProp {u'} \GstamporderEqSymbol \AstampProp {z}$.
			\item Case $\AstampProp {z} \GstamporderSymbol \AstampProp {u'}$. 
			
			From the hypotheses $\AstampProp {u'} \GstamporderEqSymbol \AstampProp {y}$ and
			$\neg \exists u'.\ \PEND {z} {u'}\ \wedge\ \AstampProp {u'} \GstamporderEqSymbol \AstampProp {y}$ and 
			Lemma~\ref{lem::stack::appendix::vis-proof::negated-miss}, we have
			$\precedesReps{\twoSpan{\spanEv{o'}}}{\twoSpan{\spanEv{z}}}$ for some pop $o'\notin \elimSet{}$ such that 
			$\AstampProp{u'} = \AstampProp{o'}$. 
			
			But from hypothesis $\PEND {x} {u'}$, 
			we also have $\precedesReps{\twoSpan{\spanEv{x}}}{\twoSpan{\spanEv{o'}}}$. Therefore, $\precedesReps{\twoSpan{\spanEv{x}}}{\twoSpan{\spanEv{z}}}$.
			
			Since $z \notin \elimSet{}$, Lemma~\ref{lem::stack::appendix::vis-proof::eq-id-implies-eq-timestamp} implies that there is a push $u_z \notin \elimSet{}$
			such that $\AstampProp{u_z} = \AstampProp{z}$. 
			
			So, $\AstampProp {x} \GstamporderSymbol \AstampProp {z} = \AstampProp{u_z}$. 
			
			We want to show $\PEND x {u_z}$, which means it remains 
			to show $\forall o_2 \notin \elimSet{}.\ \AstampProp{u_z} = \AstampProp{o_2} \implies \precedesReps{\twoSpan{\spanEv{x}}}{\twoSpan{\spanEv{o_2}}}$.
			So, let $o_2 \notin \elimSet{}$ such that $\AstampProp{u_z} = \AstampProp{o_2}$. Then, $\AstampProp{z} = \AstampProp{u_z} = \AstampProp{o_2}$.
			Therefore, $z = o_2$ by \axiomIRef{inv::stack::appendix::vis-proof::prop-funcs-are-injective}. 
			But we already know $\precedesReps{\twoSpan{\spanEv{x}}}{\twoSpan{\spanEv{z}}}$. So,
			$\precedesReps{\twoSpan{\spanEv{x}}}{\twoSpan{\spanEv{o_2}}}$.
			
			Since $\AstampProp{u_z} = \AstampProp{z}$, we have,
			\[
			\exists u_z \notin \elimSet{}.\ \PEND x {u_z} \wedge \AstampProp {u_z} \GstamporderEqSymbol \AstampProp {z}
			\]
		\end{itemize}
		
		\item Case $\AstampProp {z} \GstamporderSymbol \AstampProp {x}$. We claim that $\neg \exists u_2 \notin \elimSet{}.\ \PEND {z} {u_2}\ \wedge\ \AstampProp {u_2} \GstamporderEqSymbol \AstampProp {x}$. For suppose there is such $u_2$. But then from the hypothesis $\PEND {x} {u'}\ \wedge\ \AstampProp {u'} \GstamporderEqSymbol \AstampProp {y}$,
		\[
		\AstampProp{u_2} \GstamporderEqSymbol \AstampProp {x} \GstamporderSymbol \AstampProp{u'} \GstamporderEqSymbol \AstampProp {y}
		\]
		contradicting the hypothesis $\neg \exists u' \notin \elimSet{}.\ \PEND {z} {u'}\ \wedge\ \AstampProp {u'} \GstamporderEqSymbol \AstampProp {y}$.
		
		\item Case $\AstampProp {x} = \AstampProp {z}$. We get $x = z$ from \axiomIRef{inv::stack::appendix::vis-proof::prop-funcs-are-injective}, but then
		the two hypotheses $\exists u' \notin \elimSet{}.\ \PEND {x} {u'}\ \wedge\ \AstampProp {u'} \GstamporderEqSymbol \AstampProp {y}$
		and $\neg \exists u' \notin \elimSet{}.\ \PEND {z} {u'}\ \wedge\ \AstampProp {u'} \GstamporderEqSymbol \AstampProp {y}$ are contradictory.
	\end{itemize}
	
	\item Case $\visSepIndx{}{B} x y$ for pops $x$ and $y$, and $\exists u' \notin \elimSet{}.\ \PEND {y} {u'}\ \wedge\ \AstampProp {u'} \GstamporderEqSymbol \AstampProp {z}$ for push $z$.
	
	From definition of $\visSepIndx{}{B} x y$, we have $\AstampProp{y} \GstamporderSymbol \AstampProp{x}\ \wedge\  
	\neg \exists u' \notin \elimSet{}.\ \PEND {y} {u'}\ \wedge\ \AstampProp {u'} \GstamporderEqSymbol \AstampProp {x}$.
	
	Since $x \notin \elimSet{}$, Lemma~\ref{lem::stack::appendix::vis-proof::eq-id-implies-eq-timestamp} implies that there is a push $u_x \notin \elimSet{}$
	such that $\AstampProp{u_x} = \AstampProp{x}$.
			
	From $\neg \exists u' \notin \elimSet{}.\ \PEND {y} {u'}\ \wedge\ \AstampProp {u'} \GstamporderEqSymbol \AstampProp {x}$ and 
	$\AstampProp{y} \GstamporderSymbol \AstampProp{u_x} = \AstampProp{x}$ and Lemma~\ref{lem::stack::appendix::vis-proof::negated-miss}, 
	we obtain $\precedesReps{\twoSpan{\spanEv{o'}}}{\twoSpan{\spanEv{y}}}$ for some pop $o' \notin \elimSet{}$ such that $\AstampProp{u_x} = \AstampProp{o'}$. 
	Therefore, $o' = x$ by \axiomIRef{inv::stack::appendix::vis-proof::prop-funcs-are-injective}, meaning 
	$\precedesReps{\twoSpan{\spanEv{x}}}{\twoSpan{\spanEv{y}}}$.
	
	By Lemma~\ref{lem::stack::appendix::vis-proof::pop-totality-for-pops}, 
	$\AstampProp {x} \GstamporderSymbol \AstampProp {u'}$ or $\AstampProp {u'} \GstamporderEqSymbol \AstampProp {x}$.
	
	If $\AstampProp {x} \GstamporderSymbol \AstampProp {u'}$, we have $\precedesReps{\twoSpan{\spanEv{x}}}{\twoSpan{\spanEv{o_{u'}}}}$
	for any pop $o_{u'}$ taking $u'$, since $\precedesReps{\twoSpan{\spanEv{x}}}{\twoSpan{\spanEv{y}}}$ and $\PEND {y} {u'}$. 
	Therefore $\exists u' \notin \elimSet{}.\ \PEND {x} {u'}\ \wedge\ \AstampProp {u'} \GstamporderEqSymbol \AstampProp {z}$.
	
	If $\AstampProp {u'} \GstamporderEqSymbol \AstampProp{x}$, then
	$\exists u' \notin \elimSet{}.\ \PEND {y} {u'}\ \wedge\ \AstampProp {u'} \GstamporderEqSymbol \AstampProp {x}$ which contradicts
	hypothesis $\neg \exists u' \notin \elimSet{}.\ \PEND {y} {u'}\ \wedge\ \AstampProp {u'} \GstamporderEqSymbol \AstampProp {x}$.
	
	\item Case $\visSepIndx{}{B} x y$ for pops $x$ and $y$, and $\exists u' \notin \elimSet{}.\ \PEND {y} {u'}\ \wedge\ \AstampProp {u'} \GstamporderEqSymbol \AstampProp {z}$ for
	pop $z$. This is similar to the previous case.
	
	\item Case $\visSepIndx{}{B} x y$ for pops $x$ and $y$, and $\AstampProp{z} \GstamporderSymbol \AstampProp{y}\ \wedge\  
	\neg \exists u' \notin \elimSet{}.\ \PEND {z} {u'}\ \wedge\ \AstampProp {u'} \GstamporderEqSymbol \AstampProp {y}$ for
	pop $z$.
	
	From definition of $\visSepIndx{}{B} x y$, we have $\AstampProp{y} \GstamporderSymbol \AstampProp{x}\ \wedge\  
	\neg \exists u' \notin \elimSet{}.\ \PEND {y} {u'}\ \wedge\ \AstampProp {u'} \GstamporderEqSymbol \AstampProp {x}$.
	
	So, $\AstampProp {z} \GstamporderSymbol \AstampProp {x}$ and we claim that
	$\neg \exists u' \notin \elimSet{}.\ \PEND {z} {u'}\ \wedge\ \AstampProp {u'} \GstamporderEqSymbol \AstampProp {x}$. For suppose such $u'$ exists.
	
	By Lemma~\ref{lem::stack::appendix::vis-proof::pop-totality-for-pops}, 
	$\AstampProp {u'} \GstamporderEqSymbol \AstampProp {y}$ or $\AstampProp {y} \GstamporderSymbol \AstampProp {u'}$.
	
	If $\AstampProp {u'} \GstamporderEqSymbol \AstampProp {y}$, then this contradicts the hypothesis,
	$\neg \exists u' \notin \elimSet{}.\ \PEND {z} {u'}\ \wedge\ \AstampProp {u'} \GstamporderEqSymbol \AstampProp {y}$.
	
	If $\AstampProp {y} \GstamporderSymbol \AstampProp {u'}$, then 
	from hypothesis $\neg \exists u' \notin \elimSet{}.\ \PEND {y} {u'}\ \wedge\ \AstampProp {u'} \GstamporderEqSymbol \AstampProp {x}$ and 
	Lemma~\ref{lem::stack::appendix::vis-proof::negated-miss}, we 
	obtain $\precedesReps{\twoSpan{\spanEv{o_{u'}}}}{\twoSpan{\spanEv{y}}}$ for some pop $o_{u'} \notin \elimSet{}$ such that $\AstampProp{u'} = \AstampProp{o_{u'}}$.
	
	But from $\PEND {z} {u'}$ we have 
	$\precedesReps{\twoSpan{\spanEv{z}}}{\twoSpan{\spanEv{o_{u'}}}}$, which means $\precedesReps{\twoSpan{\spanEv{z}}}{\twoSpan{\spanEv{y}}}$.
	
	Since $y \notin \elimSet{}$, Lemma~\ref{lem::stack::appendix::vis-proof::eq-id-implies-eq-timestamp} implies that there is a push $u_y \notin \elimSet{}$
	such that $\AstampProp{u_y} = \AstampProp{y}$.
	
	We want to show $\PEND z {u_y}$. We already know $\AstampProp{z} \GstamporderSymbol \AstampProp{y} = \AstampProp{u_y}$, which means it remains 
	to show $\forall o_2 \notin \elimSet{}.\ \AstampProp{u_y} = \AstampProp{o_2} \implies \precedesReps{\twoSpan{\spanEv{z}}}{\twoSpan{\spanEv{o_2}}}$.
	So, let $o_2 \notin \elimSet{}$ such that $\AstampProp{u_y} = \AstampProp{o_2}$. Then, $\AstampProp{y} = \AstampProp{u_y} = \AstampProp{o_2}$.
	Therefore, $y = o_2$ by \axiomIRef{inv::stack::appendix::vis-proof::prop-funcs-are-injective}. 
	But we already know $\precedesReps{\twoSpan{\spanEv{z}}}{\twoSpan{\spanEv{y}}}$. So,
	$\precedesReps{\twoSpan{\spanEv{z}}}{\twoSpan{\spanEv{o_2}}}$.
	
	Since $\AstampProp {u_y} = \AstampProp{y}$, we have,
	\[
	\exists u_y \notin \elimSet{}.\ \PEND{z}{u_y} \wedge \AstampProp{u_y} \GstamporderEqSymbol \AstampProp {y}
	\]
	which contradicts hypothesis $\neg \exists u' \notin \elimSet{}.\ \PEND {z} {u'}\ \wedge\ \AstampProp {u'} \GstamporderEqSymbol \AstampProp {y}$.
\end{itemize}
\end{prf}

\begin{lem}
\label{lem::stack::appendix::vis-proof::transitive-closure-implies-defined-span}
If $\genVisTrans{x}{\_}$ and $x \notin \elimSet{}$, then $\DEF{\spanEv{x}}$.
\end{lem}

\begin{prf}
	Denote by $P(x,y)$ the statement,
	\[
	x \notin \elimSet{} \implies \DEF{\spanEv{x}}
	\]
	
	To prove $\genVisTrans{x}{y} \implies P(x,y)$ (which proves the lemma), it 
	suffices to prove the two properties,
	\begin{itemize}
		\item (Base case). $x \genVisSymbol y \implies P(x,y)$.
		\item (Inductive case). $P(x,y) \wedge P(y,z) \implies P(x,z)$.
	\end{itemize}
	
	We prove first the inductive case. By definition of $P(x,y)$ and $P(y,z)$, we have,
	\begin{align*}
	x \notin \elimSet{} \implies \DEF{\spanEv{x}} \\
	y \notin \elimSet{} \implies \DEF{\spanEv{y}}
	\end{align*}
	but then we trivially have $P(x,z)$.
	
	The base case follows from the following claim when $i = |\elimRelName|$. We prove the claim by induction on $i$.
	
	\begin{claim*}
		Let $i \natorderEqSymbol |\elimRelName|$. Let $x,y \in \absEvent$. 
		If $x \genVisSymbol_i y$ and $x \notin \elimSet{}$, then $\DEF{\spanEv{x}}$.
	\end{claim*}
	
	For the base case ($i = 0$), we check each possibility in $x \genVisSymbol_0 y$,
	
	\begin{itemize}
		\item Case $\visObsIndx{B}{}{x}{y}$. From definition of $\visObsIndxSymbol{B}{}$,
		$\AstampProp{x} = \AstampProp{y}$ for $x$ a push and $y$ a pop. 
		So, $\DEF{\stampProp{x}}$.
		
		Therefore, by \axiomIRef{inv::stack::appendix::vis-proof::timestamps-imply-span}, $\DEF{\spanEv{x}}$.
		
		\item Case $\visSepIndx{B}{}{x}{y}$. Notice that for each case in Definition~\ref{defn::stack::appendix::vis-proof::base-relations},
		we have $\DEF{\AstampProp{x}}$. Therefore, 
		by \axiomIRef{inv::stack::appendix::vis-proof::timestamps-imply-span}, $\DEF{\spanEv{x}}$.
	\end{itemize}
	
	For the inductive case, let $x \genVisSymbol_{i+1} y$ and $x \notin \elimSet{}$. 
	We check each case in the definition of $x \genVisSymbol_{i+1} y$,
	
	\begin{itemize}
		\item Case $\visObsIndx{i}{E} x y$. By IH, $\DEF{\spanEv{x}}$ holds.
		
		\item Case $\elimPair i = (x,y)$. Hence, $x \in \elimPairSet{i}$. 
		
		By Part 1 of Lemma~\ref{lem::stack::appendix::vis-proof::elim-set-basic-facts},
		$x \in \elimSet{i}$. But by Part 2 of Lemma~\ref{lem::stack::appendix::vis-proof::elim-set-basic-facts},
		$x \in \elimSet{0} = \elimSet{}$ (Contradiction).
		
		\item Case $x \notin \elimSet i \wedge y \in \elimPairSet i \wedge \BE i x y$.
		
		By definition of $\BE i x y$, there is a $z$ such that $x \refleTransCl{\genVisSymbol_i} z$ and $(\precedesAbs{z}{\cPair{y}} \vee \precedesAbs{z}{y})$.
		
		If $x = z$, then either $\precedesAbs{x}{\cPair{y}}$ or $\precedesAbs{x}{y}$. In both cases $x \in \terminatedEvent$. Therefore,
		by \axiomIRef{inv::stack::appendix::vis-proof::terminated-implies-span}, $\DEF{\spanEv{x}}$.
		
		If $x \neq z$, then $x \transCl{\genVisSymbol_i} z$. In particular, $x \genVisSymbol_i z' \refleTransCl{\genVisSymbol_i} z$ for some $z'$.
		By the IH on $x \genVisSymbol_i z'$, we have $\DEF{\spanEv{x}}$.
		
		\item Case $x \in \elimPairSet{i} \wedge y \notin \elimSet i \wedge \neg \BE i y x$.
		
		By Part 1 of Lemma~\ref{lem::stack::appendix::vis-proof::elim-set-basic-facts},
		$x \in \elimSet{i}$. But by Part 2 of Lemma~\ref{lem::stack::appendix::vis-proof::elim-set-basic-facts},
		$x \in \elimSet{0} = E$ (Contradiction).
		
		\item Case $\visSepIndx {i} {E} {x} {y}$. By IH, $\DEF{\spanEv{x}}$ holds.
	\end{itemize}
\end{prf}

\begin{lem}
\label{lem::stack::appendix::vis-proof::terminated-closure-implies-defined-span}
If $x \in \closedEvent \setminus \elimSet{}$, then $\DEF{\spanEv{x}}$.
\end{lem}

\begin{prf}
Since $x \in \closedEvent$, there is $y \in \terminatedEvent$ such that $x \refleTransCl{\genVisSymbol} y$.

If $x = y$, then $x \in \terminatedEvent$, and by \axiomIRef{inv::stack::appendix::vis-proof::terminated-implies-span}, $\DEF{\spanEv{x}}$.

If $x \neq y$, then $x \transCl{\genVisSymbol} y$. From hypothesis $x \notin \elimSet{}$ and 
Lemma~\ref{lem::stack::appendix::vis-proof::transitive-closure-implies-defined-span}, $\DEF{\spanEv{x}}$.
\end{prf}

\begin{lem}
\label{lem::stack::appendix::closed-event-prec-implies-span-prec}
Let $x,y \in \closedEvent \setminus \elimSet{}$. If $\precedesAbs{x}{y}$, then $\precedesSpans{\spanEv{x}}{\spanEv{y}}$.
\end{lem}

\begin{prf}
From Lemma~\ref{lem::stack::appendix::vis-proof::terminated-closure-implies-defined-span}, we have $\DEF{\spanEv{x}}$
and $\DEF{\spanEv{y}}$. Hence, by 
Lemma~\ref{lem::stack::appendix::event-prec-implies-span-prec}, $\precedesSpans{\spanEv{x}}{\spanEv{y}}$.
\end{prf}

\begin{lem}
	\label{lem::stack::appendix::vis-proof::lem-concurrent-lifo}
	Axiom \axiomSRef{vis-ax::stack::cc-concurrent-lifo} holds
	at domain $\closedEvent \setminus \elimSet{}$.
\end{lem}

\begin{prf}
	From the fact that $u_1, u_2, o_1 \in \closedEvent \setminus \elimSet {}$,
	we know that $\spanEvSymbol$ is defined for the three events 
	by Lemma~\ref{lem::stack::appendix::vis-proof::terminated-closure-implies-defined-span}.
	In addition, \axiomIRef{inv::stack::appendix::vis-proof::span-implies-all-properties} implies 
	that function $\AstampPropName$ is defined for these three events.
	
	From the hypothesis
	$o_1 \nVisSepEqSYMBOL u_2 \nVisSepEqSYMBOL u_1$ and Part 3 of Lemma~\ref{lem::stack::appendix::vis-proof::elim-set-basic-facts},
	we have $o_1 \not\visSepEqIndxSymbol{0}{} u_2 \not\visSepEqIndxSymbol{0}{} u_1$.
	In particular $o_1 \not\visSepEqIndxSymbol{}{B} u_2 \not\visSepEqIndxSymbol{}{B} u_1$.
	
	Since $\visObs{}{u_1}{o_1}$ and $u_1, o_1 \notin \elimSet 0$,  
	Lemma~\ref{lem::stack::appendix::vis-proof::gen-vis-restricted-to-elim-level} implies $\visObsIndx{0}{}{u_1}{o_1}$, which in turn implies
	$\AstampProp{u_1} = \AstampProp{o_1}$ by definition. 
	Hence, by Pop-Totality,
	$\AstampProp{u_1} \GstamporderSymbol \AstampProp{u_2}$ or $\AstampProp{u_2} \GstamporderSymbol \AstampProp{u_1}$ or
	$\AstampProp{u_1} = \AstampProp{u_2}$.
	
	The case $\AstampProp{u_2} \GstamporderSymbol \AstampProp{u_1}$ contradicts $u_2 \not\visSepEqIndxSymbol{}{B} u_1$.
	The case $\AstampProp{u_1} = \AstampProp{u_2}$ leads to $u_1 = u_2$ by 
	\axiomIRef{inv::stack::appendix::vis-proof::prop-funcs-are-injective}, which contradicts $u_2 \not\visSepEqIndxSymbol{}{B} u_1$.
	Hence, we can assume $\AstampProp{u_1} \GstamporderSymbol \AstampProp{u_2}$. 
	So, $\AstampProp{o_1} = \AstampProp{u_1} \GstamporderSymbol \AstampProp{u_2}$.
	
	Together with $o_1 \not\visSepEqIndxSymbol{}{B} u_2$, we have,
	\begin{align}
	\label{eq::stack::appendix::vis-proof::negated-formulas-1}
	\AstampProp{o_1} \GstamporderSymbol \AstampProp{u_2}\ \wedge\  
	\neg\exists u' \notin \elimSet{}.\ \PEND{o_1}{u'} \wedge \AstampProp{u'} \GstamporderEqSymbol \AstampProp {u_2}
	\end{align}
	
    From the second conjunct in \eqref{eq::stack::appendix::vis-proof::negated-formulas-1} and 
    Lemma~\ref{lem::stack::appendix::vis-proof::negated-miss} (instantiating $u \defini u_2$), there is a pop $o_2 \notin \elimSet{}$ 
	such that $\AstampProp{u_2} = \AstampProp{o_2}$.
	Therefore, $\visObsIndx{0}{}{u_2}{o_2}$, and hence $\visObs{}{u_2}{o_2}$ by Part 3 of 
	Lemma~\ref{lem::stack::appendix::vis-proof::elim-set-basic-facts}.
	
	Substituting $\AstampProp{u_2} = \AstampProp{o_2}$ in \eqref{eq::stack::appendix::vis-proof::negated-formulas-1},
	we obtain $\visSepIndx{0}{}{o_2}{o_1}$ by definition, and hence $\visSep{}{o_2}{o_1}$ by 
	Lemma~\ref{lem::stack::appendix::vis-proof::elim-set-basic-facts}.
	
	Since $\visSep{} {o_2} {o_1}$ holds, this also implies that 
	$o_2 \in \closedEvent$ since $\closedEvent$ is $\genVisSymbol$-downward closed 
	(Lemma~\ref{eq::stack::appendix::lin::committed-downward-closed}),
	and hence $o_2 \in \closedEvent \setminus \elimSet{}$.
\end{prf}

\begin{lem}
	Axiom \axiomSRef{vis-ax::stack::cc-pop-uniqueness} holds
	at domain $\closedEvent \setminus \elimSet{}$.
\end{lem}

\begin{prf}
From $\visObs {} u {o_1}$ and $\visObs {} u {o_2}$,  
Lemma~\ref{lem::stack::appendix::vis-proof::gen-vis-restricted-to-elim-level} implies 
$\visObsIndx{0}{}{u}{o_1}$ and $\visObsIndx{0}{}{u}{o_2}$ (since $u, o_1, o_2 \notin \elimSet 0$), which in turn imply
$\AstampProp{u} = \AstampProp{o_1}$ and $\AstampProp{u} = \AstampProp{o_2}$ by definition. 

Therefore $\AstampProp{o_1} = \AstampProp{o_2}$ (hence, $\idProp{o_1} = \idProp{o_2}$), which implies $o_1 = o_2$
by \axiomIRef{inv::stack::appendix::vis-proof::prop-funcs-are-injective}. 
\end{prf}

\begin{lem}
	\label{lem::stack::appendix::vis-proof::lem-no-future-dependence}
	Axiom \axiomSRef{vis-ax::stack::cc-no-future-dependence} holds
	at domain $\closedEvent \setminus \elimSet{}$.
\end{lem}

\begin{prf}
Since $\closedEvent \setminus \elimSet{} \subseteq \absEvent \setminus \elimSet{}$, 
by Lemma~\ref{lem::stack::appendix::vis-proof::transitive-closure}, we need to consider four cases,
\begin{itemize}
		\item Case $x$ and $y$ are pushes and $\AstampProp{x} \GstamporderSymbol \AstampProp{y}$.
		
		Suppose for a contradiction that $\precedesAbsEq y x$. If $\precedesAbs{y}{x}$, then, 
		$\spanEv{y} \precedesSpansSymbol \spanEv{x}$ by Lemma~\ref{lem::stack::appendix::closed-event-prec-implies-span-prec}.
		So, \axiomIRef{vis-ax::stack::appendix::vis-proof::disjoint-push} 
		(Disjoint Push Timestamp Generation) implies $\AstampProp{y} \GstamporderSymbol \AstampProp{x}$. 
		Hence, $\AstampProp{x} \GstamporderSymbol \AstampProp{x}$ (Contradiction).
		
		If $y = x$, then $\AstampProp{x} \GstamporderSymbol \AstampProp{x}$ (Contradiction).
		
		\item Case $x$ is a push, $y$ is a pop, and, 
		\[
		\AstampProp{x} \GstamporderEqSymbol \AstampProp{y} \vee
		(\AstampProp{y} \GstamporderSymbol \AstampProp{x}\ \wedge\  
		\neg \exists u' \notin \elimSet{}.\ \PEND {y} {u'}\ \wedge\ \AstampProp {u'} \GstamporderEqSymbol \AstampProp {x})
		\]
		
		Suppose for a contradiction that $\precedesAbsEq y x$. The case $y = x$ leads to a contradiction, since $x$ and $y$ are 
		of different type. Hence, we can assume $\precedesAbs{y}{x}$. So, $\spanEv{y} \precedesSpansSymbol \spanEv{x}$ by Lemma~\ref{lem::stack::appendix::closed-event-prec-implies-span-prec}.
		
		\begin{itemize}
			\item Case $\AstampProp{x} \GstamporderEqSymbol \AstampProp{y}$. Since $y \notin \elimSet{}$, 
			Lemma~\ref{lem::stack::appendix::vis-proof::eq-id-implies-eq-timestamp} implies that there is a push 
			$u_y \notin \elimSet{}$ such that $\AstampProp{u_y} = \AstampProp{y}$ and $\precedesSpans{\spanEv{u_y}} {\spanEv{y}}$.
						
			But $\precedesSpans{\spanEv{y}} {\spanEv{x}}$. Therefore, $\precedesSpans{\spanEv{u_y}} {\spanEv{x}}$. So, by \axiomIRef{vis-ax::stack::appendix::vis-proof::disjoint-push} (Disjoint Push Timestamp Generation) we have $\AstampProp{u_y} \GstamporderSymbol \AstampProp{x}$, and we get the contradiction,
			
			\[
			\AstampProp{y} = \AstampProp{u_y} \GstamporderSymbol \AstampProp{x} \GstamporderEqSymbol \AstampProp{y}
			\]
			
			\item Case $\AstampProp{y} \GstamporderSymbol \AstampProp{x}\ \wedge\  
			\neg \exists u' \notin \elimSet{}.\ \PEND {y} {u'}\ \wedge\ \AstampProp {u'} \GstamporderEqSymbol \AstampProp {x}$.
			
			From the case hypothesis and 
			Lemma~\ref{lem::stack::appendix::vis-proof::negated-miss} (instantiating with $u \defini x$), we get $\precedesReps{\twoSpan{\spanEv{o_x}}}{\twoSpan{\spanEv{y}}}$
			where $o_x \notin \elimSet{}$ is a pop such that $\AstampProp{o_x} = \AstampProp{x}$. 
			By Lemma~\ref{lem::stack::appendix::vis-proof::eq-id-implies-eq-timestamp}, we also 
			have $\precedesReps{\twoSpan{\spanEv{x}}}{\oneSpan{\spanEv{o_x}}}$. But then we have the contradiction 
			(using \axiomIRef{inv::stack::appendix::vis-proof::start-end-of-span}),
			
			\[
			\twoSpan{\spanEv{x}} \precedesRepsSymbol \oneSpan{\spanEv{o_x}} \precedesRepsEqSymbol \twoSpan{\spanEv{o_x}} \precedesRepsSymbol \twoSpan{\spanEv{y}}
			\precedesRepsSymbol \oneSpan{\spanEv x} \precedesRepsEqSymbol \twoSpan{\spanEv{x}} 
			\]
			
		\end{itemize}
		
		\item Case $x$ is a pop, $y$ is a push, and $\exists u' \notin \elimSet{}.\ \PEND {x} {u'}\ \wedge\ \AstampProp {u'} \GstamporderEqSymbol \AstampProp {y}$.
		
		Suppose for a contradiction that $\precedesAbsEq y x$. The case $y = x$ leads directly to a contradiction since $x$ and $y$ are 
		of different type. Hence, we can assume $\precedesAbs y x$. So, $\spanEv{y} \precedesSpansSymbol \spanEv{x}$ by Lemma~\ref{lem::stack::appendix::closed-event-prec-implies-span-prec}.
		
		By \axiomIRef{vis-ax::stack::vis-proof::misses-are-late} (Misses start late) with $\PEND {x} {u'}$, we have 
		$\oneSpan{\spanEv x} \precedesRepsSymbol \oneSpan{\spanEv{u'}}$. 
		
		Since $\spanEv{y} \precedesSpansSymbol \spanEv{x}$, we have $\twoSpan{\spanEv{y}} \precedesRepsSymbol \oneSpan{\spanEv{x}} \precedesRepsSymbol \oneSpan{\spanEv{u'}}$, which means $\spanEv{y} \precedesSpansSymbol \spanEv{u'}$. 
		
		So, by \axiomIRef{vis-ax::stack::appendix::vis-proof::disjoint-push} (Disjoint Push Timestamp Generation) we have $\AstampProp{y} \GstamporderSymbol \AstampProp{u'}$, and we get the contradiction from hypothesis $\AstampProp {u'} \GstamporderEqSymbol \AstampProp {y}$.
		
		\item Case $x$ and $y$ are pops, and,
		\begin{align*}
		& (\exists u' \notin \elimSet{}.\ \PEND {x} {u'}\ \wedge\ \AstampProp {u'} \GstamporderEqSymbol \AstampProp {y})\ \vee \\
		& (\AstampProp{y} \GstamporderSymbol \AstampProp{x}\ \wedge\  
		\neg \exists u' \notin \elimSet{}.\ \PEND {y} {u'}\ \wedge\ \AstampProp {u'} \GstamporderEqSymbol \AstampProp {x})
		\end{align*}
		
		Suppose for a contradiction that $\precedesAbsEq y x$. We have four cases two consider,
		
		\begin{itemize}
			\item Case $\precedesAbs{y}{x}$ and $\exists u' \notin \elimSet{}.\ \PEND {x} {u'}\ \wedge\ \AstampProp {u'} \GstamporderEqSymbol \AstampProp {y}$.
			
			So, $\spanEv{y} \precedesSpansSymbol \spanEv{x}$ by Lemma~\ref{lem::stack::appendix::closed-event-prec-implies-span-prec}.
			
			By \axiomIRef{vis-ax::stack::vis-proof::misses-are-late} (Misses start late) with $\PEND {x} {u'}$, we have 
			$\oneSpan{\spanEv x} \precedesRepsSymbol \oneSpan{\spanEv{u'}}$. 
			
			Since $\spanEv{y} \precedesSpansSymbol \spanEv{x}$, we have $\twoSpan{\spanEv{y}} \precedesRepsSymbol \oneSpan{\spanEv{x}} \precedesRepsSymbol \oneSpan{\spanEv{u'}}$. Hence, $\spanEv{y} \precedesSpansSymbol \spanEv{u'}$.
			
			Since $y \notin \elimSet{}$, Lemma~\ref{lem::stack::appendix::vis-proof::eq-id-implies-eq-timestamp} 
			implies that there is a push $u_y \notin \elimSet{}$ such that $\AstampProp{u_y} = \AstampProp{y}$ and 
			$\spanEv{u_y} \precedesSpansSymbol \spanEv{y}$. 
			Hence (using \axiomIRef{inv::stack::appendix::vis-proof::start-end-of-span}),
			\[
			 \twoSpan{\spanEv{u_y}} \precedesRepsSymbol \oneSpan{\spanEv{y}} \precedesRepsEqSymbol \twoSpan{\spanEv{y}} \precedesRepsSymbol \oneSpan{\spanEv{u'}}
			\] 
			which means $\spanEv{u_y} \precedesSpansSymbol \spanEv{u'}$.
			
			So, by \axiomIRef{vis-ax::stack::appendix::vis-proof::disjoint-push} (Disjoint Push Timestamp Generation) we have $\AstampProp{y} = \AstampProp{u_y} \GstamporderSymbol \AstampProp{u'}$, and we get the contradiction from hypothesis $\AstampProp {u'} \GstamporderEqSymbol \AstampProp {y}$.
			
			\item Case $y = x$ and $\exists u' \notin \elimSet{}.\ \PEND {x} {u'}\ \wedge\ \AstampProp {u'} \GstamporderEqSymbol \AstampProp {y}$.
			
			From $\PEND {x} {u'}$ we have $\AstampProp{y} = \AstampProp{x} \GstamporderSymbol \AstampProp{u'}$, and we get the contradiction
			from hypothesis $\AstampProp {u'} \GstamporderEqSymbol \AstampProp {y}$.
			
			\item Case $\precedesAbs{y}{x}$ and $\AstampProp{y} \GstamporderSymbol \AstampProp{x}\ \wedge\  
			\neg \exists u' \notin \elimSet{}.\ \PEND {y} {u'}\ \wedge\ \AstampProp {u'} \GstamporderEqSymbol \AstampProp {x}$.
			
			So, $\spanEv{y} \precedesSpansSymbol \spanEv{x}$ by Lemma~\ref{lem::stack::appendix::closed-event-prec-implies-span-prec}.
			
			Since $x \notin \elimSet{}$, Lemma~\ref{lem::stack::appendix::vis-proof::eq-id-implies-eq-timestamp} 
			implies that there is a push $u_x \notin \elimSet{}$ such that $\AstampProp{u_x} = \AstampProp{x}$.
			
			From the case hypothesis and Lemma~\ref{lem::stack::appendix::vis-proof::negated-miss} (instantiating with $u \defini u_x$), we get $\precedesReps{\twoSpan{\spanEv{o_x}}}{\twoSpan{\spanEv{y}}}$ for some pop $o_x \notin \elimSet{}$ such that 
			$\AstampProp{u_x} = \AstampProp{o_x}$. 
			Hence, $\AstampProp{x} = \AstampProp{u_x} = \AstampProp{o_x}$, which implies $x = o_x$ by 
			\axiomIRef{inv::stack::appendix::vis-proof::prop-funcs-are-injective}, meaning 
			$\precedesReps{\twoSpan{\spanEv{x}}}{\twoSpan{\spanEv{y}}}$.
			
			But then we have the contradiction (using $\spanEv{y} \precedesSpansSymbol \spanEv{x}$ and \axiomIRef{inv::stack::appendix::vis-proof::start-end-of-span}),
			\[
			\twoSpan{\spanEv{x}} \precedesRepsSymbol \twoSpan{\spanEv{y}}
			\precedesRepsSymbol \oneSpan{\spanEv x} \precedesRepsEqSymbol \twoSpan{\spanEv{x}} 
			\]
			
			\item Case $y = x$ and $\AstampProp{y} \GstamporderSymbol \AstampProp{x}\ \wedge\  
			\neg \exists u' \notin \elimSet{}.\ \PEND {y} {u'}\ \wedge\ \AstampProp {u'} \GstamporderEqSymbol \AstampProp {x}$.
			
			We get the contradiction directly from case hypothesis $\AstampProp{y} \GstamporderSymbol \AstampProp{x} = \AstampProp{y}$.
		\end{itemize}
\end{itemize}
\end{prf}
		
\begin{lem}
	Axiom \axiomSRef{vis-ax::stack::cc-return-completion} holds
	at domain $\closedEvent \setminus \elimSet{}$.
\end{lem}

\begin{prf}
	Since $x \in \closedEvent \setminus \elimSet{}$,
	Lemma~\ref{lem::stack::appendix::vis-proof::terminated-closure-implies-defined-span} 
	implies $\DEF{\spanEv x}$.
	Also, \axiomIRef{inv::stack::appendix::vis-proof::span-implies-all-properties} implies
	$\DEF{\spanEvOut{x}}$ and $\DEF{\AstampProp{x}}$.
	
	Define $v \defini \spanEvOut x$. To show $\postPred{x}{v}$, we need to consider the following cases,
	\begin{itemize}
		\item Case $x$ is a push. 
		
		\axiomIRef{inv::stack::appendix::vis-proof::pops-take-pushes} implies $v = \spanEvOut x = \unitValue$.
		
		\item Case $x$ is a pop and $v \neq \EMPTY$.
		
		\axiomIRef{inv::stack::appendix::vis-proof::pops-take-pushes}
		implies that there is a push $u$ such that $\idProp{u} = \idProp{x}$ 
		and $v = \spanEvOut x = \inProp{u}$.
		Lemma~\ref{lem::stack::appendix::vis-proof::eq-id-implies-eq-timestamp} then
		implies $\AstampProp{u} = \AstampProp{x}$.
		Additionally, Lemma~\ref{lem::stack::appendix::vis-proof::elim-pairs-go-together} implies 
		$u \notin \elimSet{}$ because of hypothesis $x \notin \elimSet{}$. 
		
		Hence, $\visObsIndx{0}{}{u}{x}$ by definition, which means $\visObs{}{u}{x}$
		by Lemma~\ref{lem::stack::appendix::vis-proof::elim-set-basic-facts}.
		
		Also, since $\visObs{}{u}{x}$, then $u \in \closedEvent$, since $\closedEvent$ is 
		$\genVisSymbol$-downward closed (Lemma~\ref{eq::stack::appendix::lin::committed-downward-closed}).
		
		\item Case $x$ is a pop and $v = \EMPTY$.
		
		\axiomIRef{inv::stack::appendix::vis-proof::pops-take-pushes}
		implies that there is a push $u$ such that $\idProp{u} = \idProp{x}$ 
		and $v = \spanEvOut x = \inProp{u} \in \ValType$. Therefore, $v \in \ValType$,
		which contradicts $v = \EMPTY$, i.e., this simply means that
		pops never return $\EMPTY$, which means that this case is impossible.
	\end{itemize} 
	
	Finally, if $x \in \terminatedEvent$, \axiomIRef{inv::stack::appendix::vis-proof::terminated-implies-span}
	implies $v = \spanEvOut{x} = \outputProp{x}$.
\end{prf}

\begin{thm}
	\label{thm::stack::appendix::vis-proof::vis-axioms-hold-at-base-domain}
	All the visibility-style axioms hold at domain $\closedEvent \setminus \elimSet{}$.
\end{thm}

\begin{prf}
Directly from the previous lemmas.
\end{prf}

\subsubsection{Putting Back Elimination Pairs}
\label{subsubsect::stack::appendix::vis-proof::with-elim-pairs}

Given the relations in Definition~\ref{defn::stack::appendix::vis-proof::final-relations},
this section shows the inductive step of the proof:
if the visibility-style axioms hold at domain $\closedEvent\setminus\elimSet{i}$
then they also hold at domain 
$\closedEvent\setminus\elimSet{i+1}$.\footnote{Axiom \axiomSRef{vis-ax::stack::cc-no-future-dependence}
has as hypothesis a transitive closure of the constraint relation $x \transCl{\genVisSymbol} y$.
We prove that if the axiom holds when the transitive closure is computed over the domain $\closedEvent\setminus\elimSet{i}$
then the axiom also holds when the transitive closure is computed over the domain 
$\closedEvent\setminus\elimSet{i+1}$.}
Finally, Theorem~\ref{thm::stack::appendix::vis-proof::vis-axioms-hold-at-max-domain}
shows that all the visibility-style axioms hold at domain $\closedEvent$.

\begin{lem}
	\label{subsubsect::stack::appendix::BE-with-elim-couples}
If $o = \cPair{u}$, then $(\BE{i}{x}{u} \Longleftrightarrow \BE{i}{x}{o})$.
\end{lem}

\begin{prf}
We will do the forward implication. The other direction is similar.
From definition of $\BE{i}{x}{u}$, there is a $z$ such that $x \genVisSymbol_{i}^* z$ and $(z \precedesAbsSymbol u \vee 
z \precedesAbsSymbol \cPair{u})$. 

But $\cPair{u} = o$ (so, $\cPair{o} = u$ by definition). Hence, 
$(z \precedesAbsSymbol \cPair{o} \vee 
z \precedesAbsSymbol o)$ holds as well, which is $\BE{i}{x}{o}$ by definition.
\end{prf}

\begin{lem}
	\label{subsubsect::stack::appendix::BE-with-left-vis}
	If $x,y \notin \elimSet i$, $z \in \elimPairSet{i}$, $y \genVisSymbol_i^* x$, and $\BE{i}{x}{z}$, then $\BE{i}{y}{z}$.
\end{lem}

\begin{prf}
By definition of $\BE{i}{x}{z}$, there is $z'$ such that $x \genVisSymbol_i^* z'$ and $(z' \precedesAbsSymbol z \vee z' \precedesAbsSymbol \cPair{z})$.
So, from $y \genVisSymbol_i^* x$ we obtain $y \genVisSymbol_i^* z'$, and therefore $\BE{i}{y}{z}$ by definition.
\end{prf}

\begin{lem}
\label{lem::stack::appendix::vis-proof::elim-pair-transitive-closure}
Let $i \natorderSymbol |\elimRelName|$.
Given the domain $\absEvent \setminus \elimSet{i+1}$ for relation $\genVisSymbol$, 
if $\genVisTrans{x}{y}$, then one of the following cases hold,
\begin{itemize}
	\item $x$ is a push and $y$ is a pop, and $\elimPair{i} = (x,y)$.
	\item $x \genVisSymbol_{i}^+ y$.
	\item $x \notin \elimSet{i}$, $y \in \elimPairSet{i}$, and $\BE{i}{x}{y}$.
	\item $x, y \notin \elimSet{i}$ and $\exists z \in \elimPairSet{i}.\ \BE{i}{x}{z} \wedge \neg\BE{i}{y}{z}$.
	\item $x \in \elimPairSet{i}$, $y \notin \elimSet{i}$, and $\neg\BE{i}{y}{x}$.
\end{itemize}
\end{lem}

\begin{prf}
	By repeating Lemma~\ref{lem::stack::appendix::vis-proof::gen-vis-restricted-to-elim-level}
	on hypothesis $\genVisTrans{x}{y}$, we have
	$x \transCl{\genVisSymbol_{i+1}} y$ in the domain $\absEvent \setminus \elimSet{i+1}$.
	
	Denote by $P(x,y)$ the statement of the five cases in the lemma we want to prove. 
	
	To prove $x \transCl{\genVisSymbol_{i+1}} y \implies P(x,y)$, it 
	suffices to prove the two properties,
	\begin{itemize}
		\item (Base case). $x \genVisSymbol_{i+1} y \implies P(x,y)$.
		\item (Inductive case). $P(x,y) \wedge P(y,z) \implies P(x,z)$.
	\end{itemize}
	
	For the base case, the cases in the definition of $x \genVisSymbol_{i+1} y$ are directly one of the conclusions 
	in the lemma. Hence, we focus on the inductive case. For this, we check each of the 25 cases in the hypothesis 
	$P(x,y) \wedge P(y,z)$.
\begin{itemize}
	\item Case $x$ is a push and $y$ is a pop, and $\elimPair{i} = (x,y)$, and $y$ is a push and $z$ is a pop, and $\elimPair{i} = (y,z)$.
	
	Then, $y$ is a pop and a push (Contradiction).
	
	\item Case $x$ is a push and $y$ is a pop, and $\elimPair{i} = (x,y)$, and $y \genVisSymbol_{i}^+ z$.
	
	By definition, $x,y \in \elimPairSet{i}$, so Lemma~\ref{lem::stack::appendix::vis-proof::elim-set-basic-facts} implies $x,y \in \elimSet{i}$. 
	From $y \genVisSymbol_{i}^+ z$, repeating Lemma~\ref{lem::stack::appendix::vis-proof::vis-level-inmplies-no-elim-pair-level} 
	implies $y \notin \elimSet{i}$ (Contradiction).
	
	\item Case $x$ is a push and $y$ is a pop, and $\elimPair{i} = (x,y)$, and $y \notin \elimSet{i}$, $z \in \elimPairSet{i}$, and $\BE{i}{y}{z}$.
	
	By definition, $x,y \in \elimPairSet{i}$, so Lemma~\ref{lem::stack::appendix::vis-proof::elim-set-basic-facts} implies $x,y \in \elimSet{i}$
	(Contradiction).
	
	\item Case $x$ is a push and $y$ is a pop, and $\elimPair{i} = (x,y)$, and $y,z \notin \elimSet{i}$, and $\exists z' \in \elimPairSet{i}.\ \BE{i}{y}{z'} \wedge \neg\BE{i}{z}{z'}$.
	
	By definition, $x,y \in \elimPairSet{i}$, so Lemma~\ref{lem::stack::appendix::vis-proof::elim-set-basic-facts} implies $x,y \in \elimSet{i}$
	(Contradiction).
	
	\item Case $x$ is a push and $y$ is a pop, and $\elimPair{i} = (x,y)$, and $y \in \elimPairSet{i}$, $z \notin \elimSet{i}$, and $\neg\BE{i}{z}{y}$.
	
	Since $\cPair{y} = x$, by Lemma~\ref{subsubsect::stack::appendix::BE-with-elim-couples} with $\neg\BE{i}{z}{y}$, we get $\neg\BE{i}{z}{x}$.
	Also, since $\elimPair{i} = (x,y)$, we have $x \in \elimPairSet{i}$.
	
	\item Case $x \genVisSymbol_{i}^+ y$ and $y$ is a push and $z$ is a pop, and $\elimPair{i} = (y,z)$.
	
	By definition, $y \in \elimPairSet{i}$, so Lemma~\ref{lem::stack::appendix::vis-proof::elim-set-basic-facts} implies $y \in \elimSet{i}$. 
	From $x \genVisSymbol_{i}^+ y$, repeating Lemma~\ref{lem::stack::appendix::vis-proof::vis-level-inmplies-no-elim-pair-level} 
	implies $y \notin \elimSet{i}$ (Contradiction).
	
	\item Case $x \genVisSymbol_{i}^+ y$, and $y \genVisSymbol_{i}^+ z$.
	
	Then trivially, $x \genVisSymbol_{i}^+ z$.
	
	\item Case $x \genVisSymbol_{i}^+ y$, and $y \notin \elimSet{i}$, $z \in \elimPairSet{i}$, and $\BE{i}{y}{z}$.
	
	By Lemma~\ref{subsubsect::stack::appendix::BE-with-left-vis}, we obtain $\BE{i}{x}{z}$. 
	Repeating Lemma~\ref{lem::stack::appendix::vis-proof::vis-level-inmplies-no-elim-pair-level} 
	implies $x \notin \elimSet{i}$. Also, $z \in \elimPairSet{i}$ by hypothesis.
	
	\item Case $x \genVisSymbol_{i}^+ y$, and $y, z \notin \elimSet{i}$ and $\exists z' \in \elimPairSet{i}.\ \BE{i}{y}{z'} \wedge \neg\BE{i}{z}{z'}$.
	
	By Lemma~\ref{subsubsect::stack::appendix::BE-with-left-vis}, we obtain $\BE{i}{x}{z'}$. But we also have $\neg\BE{i}{z}{z'}$.
	
	Repeating Lemma~\ref{lem::stack::appendix::vis-proof::vis-level-inmplies-no-elim-pair-level} 
	implies $x \notin \elimSet{i}$. Also, $z \notin \elimSet{i}$ by hypothesis.
	
	\item Case $x \genVisSymbol_{i}^+ y$, and $y \in \elimPairSet{i}$, $z \notin \elimSet{i}$, and $\neg\BE{i}{z}{y}$.
	
	From $y \in \elimPairSet{i}$, Lemma~\ref{lem::stack::appendix::vis-proof::elim-set-basic-facts} implies $y \in \elimSet{i}$. 
	From $x \genVisSymbol_{i}^+ y$, repeating Lemma~\ref{lem::stack::appendix::vis-proof::vis-level-inmplies-no-elim-pair-level} 
	implies $y \notin \elimSet{i}$ (Contradiction).
	
	\item Case $x \notin \elimSet{i}$, $y \in \elimPairSet{i}$, and $\BE{i}{x}{y}$, and $y$ is a push and $z$ is a pop, and $\elimPair{i} = (y,z)$.
	
	Since $\cPair{y} = z$, by Lemma~\ref{subsubsect::stack::appendix::BE-with-elim-couples} with $\BE{i}{x}{y}$, we get $\BE{i}{x}{z}$.
	Also, since $\elimPair{i} = (y,z)$, we have $z \in \elimPairSet{i}$.
	
	\item Case $x \notin \elimSet{i}$, $y \in \elimPairSet{i}$, and $\BE{i}{x}{y}$, and $y \genVisSymbol_{i}^+ z$.
	
	From $y \in \elimPairSet{i}$, Lemma~\ref{lem::stack::appendix::vis-proof::elim-set-basic-facts} implies $y \in \elimSet{i}$. 
	But from $y \genVisSymbol_{i}^+ z$, repeating Lemma~\ref{lem::stack::appendix::vis-proof::vis-level-inmplies-no-elim-pair-level} 
	implies $y \notin \elimSet{i}$ (Contradiction).
	
	\item Case $x \notin \elimSet{i}$, $y \in \elimPairSet{i}$, and $\BE{i}{x}{y}$, and $y \notin \elimSet{i}$, $z \in \elimPairSet{i}$, and $\BE{i}{y}{z}$.
	
	From $y \in \elimPairSet{i}$, Lemma~\ref{lem::stack::appendix::vis-proof::elim-set-basic-facts} implies $y \in \elimSet{i}$ (Contradiction).
	
	\item Case $x \notin \elimSet{i}$, $y \in \elimPairSet{i}$, and $\BE{i}{x}{y}$, and $y, z \notin \elimSet{i}$ and $\exists z' \in \elimPairSet{i}.\ \BE{i}{y}{z'} \wedge \neg\BE{i}{z}{z'}$.
	
	From $y \in \elimPairSet{i}$, Lemma~\ref{lem::stack::appendix::vis-proof::elim-set-basic-facts} implies $y \in \elimSet{i}$ (Contradiction).
	
	\item Case $x \notin \elimSet{i}$, $y \in \elimPairSet{i}$, and $\BE{i}{x}{y}$, and $y \in \elimPairSet{i}$, $z \notin \elimSet{i}$, and $\neg\BE{i}{z}{y}$.
	
	In  other words, $x,z \notin \elimSet{i}$, and $\exists y \in \elimPairSet{i}.\ \BE{i}{x}{y} \wedge \neg\BE{i}{z}{y}$.
	
	\item Case $x, y \notin \elimSet{i}$ and $\exists z' \in \elimPairSet{i}.\ \BE{i}{x}{z'} \wedge \neg\BE{i}{y}{z'}$, and $y$ is a push and $z$ is a pop, and $\elimPair{i} = (y,z)$.
	
	By definition, $y \in \elimPairSet{i}$, so Lemma~\ref{lem::stack::appendix::vis-proof::elim-set-basic-facts} implies $y \in \elimSet{i}$ (Contradiction). 
		
	\item Case $x, y \notin \elimSet{i}$ and $\exists z' \in \elimPairSet{i}.\ \BE{i}{x}{z'} \wedge \neg\BE{i}{y}{z'}$, and $y \genVisSymbol_{i}^+ z$.
	
	From $y \genVisSymbol_{i}^+ z$, repeating Lemma~\ref{lem::stack::appendix::vis-proof::vis-level-inmplies-no-elim-pair-level} 
	implies $z \notin \elimSet{i}$.
	From $\neg\BE{i}{y}{z'}$, Lemma~\ref{subsubsect::stack::appendix::BE-with-left-vis} implies $\neg\BE{i}{z}{z'}$.
	So, $\exists z' \in \elimPairSet{i}.\ \BE{i}{x}{z'} \wedge \neg\BE{i}{z}{z'}$.
	
	\item Case $x, y \notin \elimSet{i}$ and $\exists z' \in \elimPairSet{i}.\ \BE{i}{x}{z'} \wedge \neg\BE{i}{y}{z'}$, and $y \notin \elimSet{i}$, $z \in \elimPairSet{i}$, and $\BE{i}{y}{z}$.
	
	If $z = z'$, we would get $\neg\BE{i}{y}{z}$ and $\BE{i}{y}{z}$ (Contradiction).
	
	Hence, $z \neq z'$. But $z,z' \in \elimPairSet{i}$, and $\elimPairSet{i}$ is a two-element set, 
	we must have $\cPair{z} = z'$ by definition of $\elimPairSet{i}$. 
	But then by Lemma~\ref{subsubsect::stack::appendix::BE-with-elim-couples}, 
	$\neg\BE{i}{y}{z'}$ implies $\neg\BE{i}{y}{z}$ (Contradicts $\BE{i}{y}{z}$).
		
	\item Case $x, y \notin \elimSet{i}$ and $\exists z' \in \elimPairSet{i}.\ \BE{i}{x}{z'} \wedge \neg\BE{i}{y}{z'}$, and
	$y, z \notin \elimSet{i}$ and $\exists z'' \in \elimPairSet{i}.\ \BE{i}{y}{z''} \wedge \neg\BE{i}{z}{z''}$.
	
	If $z' = z''$, then $\exists z' \in \elimPairSet{i}.\ \BE{i}{x}{z'} \wedge \neg\BE{i}{z}{z'}$. 
	If $z' \neq z''$, since $z',z'' \in \elimPairSet{i}$, and $\elimPairSet{i}$ is a two-element set, 
	we must have $\cPair{z''} = z'$. But then by Lemma~\ref{subsubsect::stack::appendix::BE-with-elim-couples}, 
	$\neg\BE{i}{z}{z''}$ implies $\neg\BE{i}{z}{z'}$, and so $\exists z' \in \elimPairSet{i}.\ \BE{i}{x}{z'} \wedge \neg\BE{i}{z}{z'}$.
	
	\item Case $x, y \notin \elimSet{i}$ and $\exists z' \in \elimPairSet{i}.\ \BE{i}{x}{z'} \wedge \neg\BE{i}{y}{z'}$, and
	$y \in \elimPairSet{i}$, $z \notin \elimSet{i}$, and $\neg\BE{i}{z}{y}$.
	
	From $y \in \elimPairSet{i}$, Lemma~\ref{lem::stack::appendix::vis-proof::elim-set-basic-facts} implies $y \in \elimSet{i}$ (Contradiction).
	
	\item Case $x \in \elimPairSet{i}$, $y \notin \elimSet{i}$, and $\neg\BE{i}{y}{x}$, and $y$ is a push and $z$ is a pop, and $\elimPair{i} = (y,z)$.
	
	By definition, $y \in \elimPairSet{i}$. So, Lemma~\ref{lem::stack::appendix::vis-proof::elim-set-basic-facts} implies $y \in \elimSet{i}$ (Contradiction).

	\item Case $x \in \elimPairSet{i}$, $y \notin \elimSet{i}$, and $\neg\BE{i}{y}{x}$, and $y \genVisSymbol_{i}^+ z$.
	
	From $y \genVisSymbol_{i}^+ z$, repeating Lemma~\ref{lem::stack::appendix::vis-proof::vis-level-inmplies-no-elim-pair-level} 
	implies $z \notin \elimSet{i}$. From $\neg\BE{i}{y}{x}$, Lemma~\ref{subsubsect::stack::appendix::BE-with-left-vis} implies $\neg\BE{i}{z}{x}$.
	
	\item Case $x \in \elimPairSet{i}$, $y \notin \elimSet{i}$, and $\neg\BE{i}{y}{x}$, and $y \notin \elimSet{i}$, $z \in \elimPairSet{i}$, and $\BE{i}{y}{z}$.
	
	If $x = z$, then $\neg\BE{i}{y}{x}$ and $\BE{i}{y}{x}$ (Contradiction).
	
	Hence, $x \neq z$. But since $x,z \in \elimPairSet{i}$, and $\elimPairSet{i}$ is a two-element set, 
	we must have $\cPair{z} = x$. But then by Lemma~\ref{subsubsect::stack::appendix::BE-with-elim-couples}, 
	$\neg\BE{i}{y}{x}$ implies $\neg\BE{i}{y}{z}$, which contradicts $\BE{i}{y}{z}$.
	
	\item Case $x \in \elimPairSet{i}$, $y \notin \elimSet{i}$, and $\neg\BE{i}{y}{x}$, and $y, z \notin \elimSet{i}$ and $\exists z' \in \elimPairSet{i}.\ \BE{i}{y}{z'} \wedge \neg\BE{i}{z}{z'}$.
	
	If $x = z'$, then $\neg\BE{i}{y}{x}$ and $\BE{i}{y}{x}$ (Contradiction).
	
	Hence, $x \neq z'$. But since $x,z' \in \elimPairSet{i}$, and $\elimPairSet{i}$ is a two-element set, 
	we must have $\cPair{z'} = x$. But then by Lemma~\ref{subsubsect::stack::appendix::BE-with-elim-couples}, 
	$\neg\BE{i}{y}{x}$ implies $\neg\BE{i}{y}{z'}$, which contradicts $\BE{i}{y}{z'}$.
	
	\item Case $x \in \elimPairSet{i}$, $y \notin \elimSet{i}$, and $\neg\BE{i}{y}{x}$, and $y \in \elimPairSet{i}$, $z \notin \elimSet{i}$, and $\neg\BE{i}{z}{y}$.
	
	From $y \in \elimPairSet{i}$, Lemma~\ref{lem::stack::appendix::vis-proof::elim-set-basic-facts} implies $y \in \elimSet{i}$ (Contradiction).
\end{itemize}
\end{prf}

\begin{lem}
\label{lem::stack::appendix::vis-proof::elim-implies-returns-before-formulation}
We have the following facts,
\begin{enumerate}
	\item If $\elimRel{x}{y}$, then $\nprecedesAbs{x}{y}$ and $\nprecedesAbs{y}{x}$.
	\item If $x = \cPair{y}$, then $\nprecedesAbs{x}{y}$ and $\nprecedesAbs{y}{x}$
	(equivalently $\nprecedesAbs{\cPair{y}}{y}$ and $\nprecedesAbs{y}{\cPair{y}}$).
\end{enumerate}
\end{lem}

\begin{prf}
We prove each fact in turn,
\begin{enumerate}
	\item $\nprecedesAbs{x}{y}$ follows directly from the definition of $\elimRel{x}{y}$.
	
	We now show $\nprecedesAbs{y}{x}$.
	
	Suppose for a contradiction that $\precedesAbs{y}{x}$. 
	The definition of $\elimRel{x}{y}$ implies $\idProp{x} = \idProp{y}$. So,
	\axiomIRef{inv::stack::appendix::vis-proof::pop-not-precedes-push} implies 
	$\TimeProp{\twoSpan{\spanEv{y}}} \not\natorderSymbol \STimeProp{x}$.
	However, from \axiomIRef{inv::stack::appendix::vis-proof::span-inside-event}(ii) 
	and the fact $\precedesAbs{y}{x}$ we obtain, 
	$\TimeProp{\twoSpan{\spanEv{y}}} \natorderEqSymbol \ETimeProp{y} \natorderSymbol \STimeProp{x}$ (Contradiction).
	
	\item If $x$ is a push and $y$ is a pop, then we have $\elimRel{x}{y}$ (equivalently, $\elimRel{(\cPair{y})}{y}$) and the result follows
	by part 1. If $x$ is a pop and $y$ is a push, then we have $\elimRel{y}{x}$ (equivalently, $\elimRel{y}{(\cPair{y})}$) and the result follows
	by part 1.
\end{enumerate}
\end{prf}

\begin{lem}
	Let $i \natorderSymbol |\elimRelName|$.
	If Axiom \axiomSRef{vis-ax::stack::cc-concurrent-lifo} holds
	at domain $\closedEvent \setminus \elimSet{i}$, then 
	the axiom also holds at domain $\closedEvent \setminus \elimSet{i+1}$.
\end{lem}

\begin{prf}	
	Let $u_1, o_1, u_2 \in \closedEvent \setminus \elimSet{i+1}$. From hypothesis $\visObs{}{u_1}{o_1}$, 
	Lemma~\ref{lem::stack::appendix::vis-proof::gen-vis-restricted-to-elim-level} implies 
	$\visObsIndx{i+1}{}{u_1}{o_1}$. Hence, by definition, we have two cases,
	\begin{itemize}
		\item Case $\elimPair{i} = (u_1,o_1)$. So, $u_1,o_1 \in \elimPairSet{i}$ by definition. 
		
		We must have $u_2 \notin \elimPairSet{i}$ because $\elimPairSet{i}$ is a two element set
		by definition, containing 
		already $u_1$ and $o_2$, and we have $u_1 \neq u_2$ as hypothesis. Therefore, we must also have $u_2 \notin \elimSet{i}$,
		otherwise we would have $u_2 \in \elimSet{i+1}$ by definition (Contradicts hypothesis $u_2 \in \closedEvent \setminus \elimSet{i+1}$).
		
		From the hypothesis $o_1 \nVisSepEqSYMBOL u_2 \nVisSepEqSYMBOL u_1$, 
		Lemma~\ref{lem::stack::appendix::vis-proof::elim-set-basic-facts} implies 
		$o_1 \not\visSepIndxSymbol{i+1}{} u_2 \not\visSepIndxSymbol{i+1}{} u_1$ (since $i+1 \natorderEqSymbol |\elimRelName|$). 
		Expanding negations
		in the definition of $\visSepIndxSymbol{i+1}{}$, we have the following two implications,
		\begin{align*}
			o_1 \in \elimPairSet{i} \wedge u_2 \notin \elimSet{i} \implies \BE{i}{u_2}{o_1} \\
			u_2 \notin \elimSet{i} \wedge u_1 \in \elimPairSet{i} \implies \neg\BE{i}{u_2}{u_1}
		\end{align*}
		
		Hence, $\BE{i}{u_2}{o_1}$ and $\neg\BE{i}{u_2}{u_1}$ which is a 
		contradiction by Lemma~\ref{subsubsect::stack::appendix::BE-with-elim-couples} (since $\cPair{u_1} = o_1$).
		
		\item Case $\visObsIndx{i}{}{u_1}{o_1}$. 
		Lemma~\ref{lem::stack::appendix::vis-proof::vis-level-inmplies-no-elim-pair-level} 
		implies $u_1,o_1 \notin \elimSet{i}$. 
		
		We now consider two cases,
		\begin{itemize}
			\item Case $u_2 \in \elimPairSet{i}$. By definition, $\elimPairSet{i}$ contains a push and a pop. 
			Let $o_2$ be the second element in the set $\elimPairSet{i}$, which is a pop such that
			$\elimPair{i} = (u_2,o_2)$, and so $\visObsIndx{i+1}{}{u_2}{o_2}$ by definition,
			which means $\visObs{}{u_2}{o_2}$ by Lemma~\ref{lem::stack::appendix::vis-proof::elim-set-basic-facts}. 
			
			But from the hypothesis $o_1 \nVisSepEqSYMBOL u_2$, 
			Lemma~\ref{lem::stack::appendix::vis-proof::elim-set-basic-facts} implies  
			$o_1 \not\visSepIndxSymbol{i+1}{} u_2$ (since $i+1 \natorderEqSymbol |\elimRelName|$). 
			Expanding negations
			in the definition of $\visSepIndxSymbol{i+1}{}$, we have the following implication,
			\[
			o_1 \notin \elimSet{i} \wedge u_2 \in \elimPairSet{i} \implies \neg\BE{i}{o_1}{u_2}
			\]
			Therefore, $\neg\BE{i}{o_1}{u_2}$, but by Lemma~\ref{subsubsect::stack::appendix::BE-with-elim-couples},
			$\neg\BE{i}{o_1}{o_2}$ also (because $\elimPair{i} = (u_2,o_2)$ and so $\cPair{u_2} = o_2$). 
			Therefore, $\visSepIndx{i+1}{}{o_2}{o_1}$ by definition (because we also know 
			$o_1 \notin \elimSet{i}$ and $o_2 \in \elimPairSet{i}$).
			So, $\visSep{}{o_2}{o_1}$
			by Lemma~\ref{lem::stack::appendix::vis-proof::elim-set-basic-facts} since $i+1 \natorderEqSymbol |\elimRelName|$.
			
			We must also have $o_2 \in \closedEvent$, since $\visSep{}{o_2}{o_1}$, and $o_1 \in \closedEvent$, and
			$\closedEvent$ is $\genVisSymbol$-downward closed (Lemma~\ref{eq::stack::appendix::lin::committed-downward-closed}).
			In addition, $o_2 \notin \elimSet{i+1}$, for otherwise $o_2 \notin \elimPairSet{i}$, which is a contradiction.
			
			\item Case $u_2 \notin \elimPairSet{i}$. Therefore, $u_2 \notin \elimSet{i}$ as well, for otherwise
			we would have $u_2 \in \elimSet{i+1}$ by definition (contradicting the hypothesis $u_2 \notin \elimSet{i+1}$).
			
			Since $u_1, u_2, o_1 \notin \elimSet{i}$ and the axiom holds at domain $\closedEvent \setminus \elimSet{i}$,
			we obtain $\visObs{}{u_2}{o_2}$ and $\visSep{}{o_2}{o_1}$ for some pop 
			$o_2 \in \closedEvent \setminus \elimSet{i}$. But Lemma~\ref{lem::stack::appendix::vis-proof::elim-set-basic-facts} 
			implies $o_2 \notin \elimSet{i+1}$ also (since $i \natorderSymbol i+1$).
		\end{itemize}
	\end{itemize}
\end{prf}

\begin{lem}
	Let $i \natorderSymbol |\elimRelName|$.
	If Axiom \axiomSRef{vis-ax::stack::cc-pop-uniqueness} holds
	at domain $\closedEvent \setminus \elimSet{i}$, then 
	the axiom also holds at domain $\closedEvent \setminus \elimSet{i+1}$.
\end{lem}

\begin{prf}
	From the hypotheses $\visObs {} u {o_1}$ and $\visObs {} u {o_2}$, 
	Lemma~\ref{lem::stack::appendix::vis-proof::gen-vis-restricted-to-elim-level}
	implies $\visObsIndx{i+1}{}{u}{o_1}$ and $\visObsIndx{i+1}{}{u}{o_2}$.
	From the definition of $\visObsIndxSymbol{i+1}{}$, we consider four cases,
	
	\begin{itemize}
		\item Case $\elimPair{i} = (u, o_1)$ and $\elimPair{i} = (u, o_2)$.
		Hence $(u, o_1) = (u, o_2)$, which implies $o_1 = o_2$.
		
		\item Case $\elimPair{i} = (u, o_1)$ and $\visObsIndx{i}{}{u}{o_2}$.
		
		By definition $u \in \elimPairSet{i}$. Lemma~\ref{lem::stack::appendix::vis-proof::elim-set-basic-facts} 
		implies $u \in \elimSet{i}$. 
		But from $\visObsIndx{i}{}{u}{o_2}$, 
		Lemma~\ref{lem::stack::appendix::vis-proof::vis-level-inmplies-no-elim-pair-level} 
		implies $u \notin \elimSet{i}$
		(Contradiction).
		
		\item Case $\visObsIndx{i}{}{u}{o_1}$ and $\elimPair{i} = (u, o_2)$.
		Similar to the previous case.
		
		\item Case $\visObsIndx{i}{}{u}{o_1}$ and $\visObsIndx{i}{}{u}{o_2}$.
		
		Lemma~\ref{lem::stack::appendix::vis-proof::vis-level-inmplies-no-elim-pair-level} 
		implies $u, o_1, o_2 \notin \elimSet{i}$.
		Since the axiom holds
		at domain $\closedEvent \setminus \elimSet{i}$, it follows $o_1 = o_2$.
	\end{itemize}
\end{prf}

\begin{lem}
	Let $i \natorderSymbol |\elimRelName|$.
	If Axiom \axiomSRef{vis-ax::stack::cc-no-future-dependence} holds
	at domain $\closedEvent \setminus \elimSet{i}$, then 
	the axiom also holds at domain $\closedEvent \setminus \elimSet{i+1}$.
\end{lem}

\begin{prf}
	Since $\closedEvent \setminus \elimSet{i+1} \subseteq \absEvent \setminus \elimSet{i+1}$,
	Lemma~\ref{lem::stack::appendix::vis-proof::elim-pair-transitive-closure} implies that we need to consider five cases,
	\begin{itemize}
		\item Case $x$ is a push and $y$ is a pop, and $\elimPair{i} = (x,y)$.
		
		Directly from Lemma~\ref{lem::stack::appendix::vis-proof::elim-implies-returns-before-formulation}, 
		we have $\nprecedesAbs{y}{x}$. While $x \neq y$ holds because $x$ and $y$ have different types.
		
		\item Case $x \genVisSymbol_{i}^+ y$. 
		
		By repeating Lemma~\ref{lem::stack::appendix::vis-proof::vis-level-inmplies-no-elim-pair-level} we have $x \genVisSymbol_{i}^+ y$
		in domain $\absEvent \setminus \elimSet{i}$. Since $i \natorderSymbol |\elimRelName|$, repeating 
		Lemma~\ref{lem::stack::appendix::vis-proof::elim-set-basic-facts} 
		implies $x \genVisSymbol^+ y$ in domain $\absEvent \setminus \elimSet{i}$. But
		$\closedEvent$ is $\genVisSymbol$-downward closed (Lemma~\ref{eq::stack::appendix::lin::committed-downward-closed}), 
		and since $y \in \closedEvent$ by assumption, 
		$x \genVisSymbol^+ y$ holds in domain $\closedEvent \setminus \elimSet{i}$.
		Finally, since \axiomSRef{vis-ax::stack::cc-no-future-dependence} holds at domain $\closedEvent \setminus \elimSet{i}$ we have 
		$\nprecedesAbsEq{y}{x}$. 
		
		\item Case $x \notin \elimSet{i}$, $y \in \elimPairSet{i}$, and $\BE{i}{x}{y}$.
		
		Suppose for a contradiction that $x = y$. From $y \in \elimPairSet{i}$, 
		Lemma~\ref{lem::stack::appendix::vis-proof::elim-set-basic-facts} implies $y \in \elimSet{i}$, which
		means $x \in \elimSet{i}$ (Contradiction). Hence, $x \neq y$.
		
		Suppose for a contradiction that $\precedesAbs{y}{x}$.
		
		By definition of $\BE{i}{x}{y}$, there is $z$ such that $x \genVisSymbol_{i}^* z$ and $(z \precedesAbsSymbol y \vee 
		z \precedesAbsSymbol \cPair{y})$. From $x \genVisSymbol_{i}^* z$ we have either $x = z$ or $x \genVisSymbol_{i}^+ z$.
		Hence, we have four cases,
		
		\begin{itemize}
			\item Case $x = z$ and $z \precedesAbsSymbol y$. So, $x = z \precedesAbsSymbol y \precedesAbsSymbol x$ (Contradiction).
			
			\item Case $x = z$ and $z \precedesAbsSymbol \cPair{y}$. So, $y \precedesAbsSymbol x = z \precedesAbsSymbol \cPair{y}$.
			But $y$ and $\cPair{y}$ are elimination pair couples, so $y \nprecedesAbsSymbol \cPair{y}$ from Lemma~\ref{lem::stack::appendix::vis-proof::elim-implies-returns-before-formulation} (Contradiction).
			
			\item Case $x \genVisSymbol_{i}^+ z$ and $z \precedesAbsSymbol y$. By repeating 
			Lemma~\ref{lem::stack::appendix::vis-proof::vis-level-inmplies-no-elim-pair-level} we have $x \genVisSymbol_{i}^+ z$
			in domain $\absEvent \setminus \elimSet{i}$. Since $i \natorderSymbol |\elimRelName|$, repeating 
			Lemma~\ref{lem::stack::appendix::vis-proof::elim-set-basic-facts} 
			implies $x \genVisSymbol^+ z$ in domain $\absEvent \setminus \elimSet{i}$. But
			$\closedEvent$ is $\genVisSymbol$-downward closed (Lemma~\ref{eq::stack::appendix::lin::committed-downward-closed}), 
			and since $z \in \terminatedEvent \subseteq \closedEvent$, 
			$x \genVisSymbol^+ z$ holds in domain $\closedEvent \setminus \elimSet{i}$.
			
			But from $\precedesAbs{y}{x}$ and $z \precedesAbsSymbol y$ and the interval property of
			$\precedesAbsSymbol$ (Lemma~\ref{lem::stack::appendix::lin::precedes-poset}), we get $y \precedesAbsSymbol y$ (Contradiction) or 
			$z \precedesAbsSymbol x$ which contradicts
			\axiomSRef{vis-ax::stack::cc-no-future-dependence} on the domain $\closedEvent \setminus \elimSet{i}$
			(since $x \genVisSymbol^+ z$).
			
			\item Case $x \genVisSymbol_{i}^+ z$ and $z \precedesAbsSymbol \cPair{y}$. 
			By repeating 
			Lemma~\ref{lem::stack::appendix::vis-proof::vis-level-inmplies-no-elim-pair-level} we have $x \genVisSymbol_{i}^+ z$
			in domain $\absEvent \setminus \elimSet{i}$. Since $i \natorderSymbol |\elimRelName|$, repeating 
			Lemma~\ref{lem::stack::appendix::vis-proof::elim-set-basic-facts} 
			implies $x \genVisSymbol^+ z$ in domain $\absEvent \setminus \elimSet{i}$. But
			$\closedEvent$ is $\genVisSymbol$-downward closed (Lemma~\ref{eq::stack::appendix::lin::committed-downward-closed}), 
			and since $z \in \terminatedEvent \subseteq \closedEvent$, 
			$x \genVisSymbol^+ z$ holds in domain $\closedEvent \setminus \elimSet{i}$.
			
			But from $\precedesAbs{y}{x}$ and $z \precedesAbsSymbol \cPair{y}$ and the interval property of
			$\precedesAbsSymbol$ (Lemma~\ref{lem::stack::appendix::lin::precedes-poset}), we get 
			$y \precedesAbsSymbol \cPair{y}$ or $z \precedesAbsSymbol x$.
			
			The case $y \precedesAbsSymbol \cPair{y}$ is a contradiction because $y$ and $\cPair{y}$ 
			are elimination pair couples, so $y \nprecedesAbsSymbol \cPair{y}$ from Lemma~\ref{lem::stack::appendix::vis-proof::elim-implies-returns-before-formulation}. 
			
			The case $z \precedesAbsSymbol x$ contradicts
			\axiomSRef{vis-ax::stack::cc-no-future-dependence} on the domain $\closedEvent \setminus \elimSet{i}$
			(since $x \genVisSymbol^+ z$).
		\end{itemize}
		
		\item Case $x, y \notin \elimSet{i}$ and $\exists z \in \elimPairSet{i}.\ \BE{i}{x}{z} \wedge \neg\BE{i}{y}{z}$.
		
		Lemma~\ref{lem::stack::appendix::vis-proof::elim-set-basic-facts} 
		implies $\elimPairSet{i} \subseteq \elimSet{i} \subseteq \elimSet{0}$. So, 
		from $z \in \elimPairSet{i}$, we have $z \in \elimSet{0} = \elimSet{}$.
		
		We must have $x \neq y$ because otherwise $\BE{i}{x}{z}$ and $\neg\BE{i}{x}{z}$.
		
		Suppose for a contradiction that $y \precedesAbsSymbol x$.
		
		After unfolding the definition of $\BEName$ in $\neg\BE{i}{y}{z}$, doing some de Morgan manipulations and the fact that $z \in \elimSet{}$, we obtain,
		\begin{align}
		\label{eq::stack::appendix::vis-proof::neq-be-i-y-z}
		\forall w.\ y \refleTransCl{\genVisSymbol_i} w \implies (\nprecedesAbs{w}{\cPair{z}} \wedge \nprecedesAbs{w}{z})
		\end{align}
		
		By definition of $\BE{i}{x}{z}$, there is $z'$ such that $x \genVisSymbol_{i}^* z'$ and $(z' \precedesAbsSymbol z \vee 
		z' \precedesAbsSymbol \cPair{z})$. From $x \genVisSymbol_{i}^* z'$ we have either $x = z'$ or $x \genVisSymbol_{i}^+ z'$.
		So, we have four cases,
		
		\begin{itemize}
			\item Case $x = z'$ and $z' \precedesAbsSymbol z$. So, $y \precedesAbsSymbol x = z' \precedesAbsSymbol z$.
			But this contradicts \eqref{eq::stack::appendix::vis-proof::neq-be-i-y-z} when instantiated with $w \defini y$.
			
			\item Case $x = z'$ and $z' \precedesAbsSymbol \cPair{z}$. So, $y \precedesAbsSymbol x = z' \precedesAbsSymbol \cPair{z}$
			But this contradicts \eqref{eq::stack::appendix::vis-proof::neq-be-i-y-z} when instantiated with $w \defini y$.
			
			\item Case $x \genVisSymbol_{i}^+ z'$ and $z' \precedesAbsSymbol z$. By repeating 
			Lemma~\ref{lem::stack::appendix::vis-proof::vis-level-inmplies-no-elim-pair-level} we have $x \genVisSymbol_{i}^+ z'$
			in domain $\absEvent \setminus \elimSet{i}$. Since $i \natorderSymbol |\elimRelName|$, repeating 
			Lemma~\ref{lem::stack::appendix::vis-proof::elim-set-basic-facts} 
			implies $x \genVisSymbol^+ z'$ in domain $\absEvent \setminus \elimSet{i}$. But
			$\closedEvent$ is $\genVisSymbol$-downward closed (Lemma~\ref{eq::stack::appendix::lin::committed-downward-closed}), 
			and since $z' \in \terminatedEvent \subseteq \closedEvent$, 
			$x \genVisSymbol^+ z'$ holds in domain $\closedEvent \setminus \elimSet{i}$.
			
			But from $\precedesAbs{y}{x}$ and $z' \precedesAbsSymbol z$ and the interval property of
			$\precedesAbsSymbol$ (Lemma~\ref{lem::stack::appendix::lin::precedes-poset}), we get 
			$y \precedesAbsSymbol z$ or $z' \precedesAbsSymbol x$.
			
			The case $y \precedesAbsSymbol z$ contradicts \eqref{eq::stack::appendix::vis-proof::neq-be-i-y-z} 
			when instantiated with $w \defini y$.
			
			The case $z' \precedesAbsSymbol x$ contradicts 
			\axiomSRef{vis-ax::stack::cc-no-future-dependence} on the domain $\closedEvent \setminus \elimSet{i}$
			(since $x \genVisSymbol^+ z'$).
			
			\item Case $x \genVisSymbol_{i}^+ z'$ and $z' \precedesAbsSymbol \cPair{z}$. By repeating 
			Lemma~\ref{lem::stack::appendix::vis-proof::vis-level-inmplies-no-elim-pair-level} we have $x \genVisSymbol_{i}^+ z'$
			in domain $\absEvent \setminus \elimSet{i}$. Since $i \natorderSymbol |\elimRelName|$, repeating 
			Lemma~\ref{lem::stack::appendix::vis-proof::elim-set-basic-facts} 
			implies $x \genVisSymbol^+ z'$ in domain $\absEvent \setminus \elimSet{i}$. But
			$\closedEvent$ is $\genVisSymbol$-downward closed (Lemma~\ref{eq::stack::appendix::lin::committed-downward-closed}), 
			and since $z' \in \terminatedEvent \subseteq \closedEvent$, 
			$x \genVisSymbol^+ z'$ holds in domain $\closedEvent \setminus \elimSet{i}$.
			
			But from $\precedesAbs{y}{x}$ and $z' \precedesAbsSymbol \cPair{z}$ and the interval property of
			$\precedesAbsSymbol$ (Lemma~\ref{lem::stack::appendix::lin::precedes-poset}), we get 
			$y \precedesAbsSymbol \cPair{z}$ or $z' \precedesAbsSymbol x$.
			
			The case $y \precedesAbsSymbol \cPair{z}$ contradicts \eqref{eq::stack::appendix::vis-proof::neq-be-i-y-z} 
			when instantiated with $w \defini y$.
			
			The case $z' \precedesAbsSymbol x$ contradicts 
			\axiomSRef{vis-ax::stack::cc-no-future-dependence} on the domain $\closedEvent \setminus \elimSet{i}$
			(since $x \genVisSymbol^+ z'$).
		\end{itemize}
				
		\item Case $x \in \elimPairSet{i}$, $y \notin \elimSet{i}$, and $\neg\BE{i}{y}{x}$.
		
		Lemma~\ref{lem::stack::appendix::vis-proof::elim-set-basic-facts}
		implies $\elimPairSet{i} \subseteq \elimSet{i} \subseteq \elimSet{0}$. So, 
		from $x \in \elimPairSet{i}$, we have both $x \in \elimSet{i}$ and $x \in \elimSet{0} = \elimSet{}$.
		
		We must have $x \neq y$, otherwise, $x = y \notin \elimSet{i}$ (Contradicting $x \in \elimSet{i}$).
		
		After unfolding the definition of $\BEName$ in $\neg\BE{i}{y}{x}$, doing some de Morgan manipulations and the fact that $x \in \elimSet{}$, we obtain,
		\begin{align*}
			\forall w.\ y \refleTransCl{\genVisSymbol_i} w \implies (\nprecedesAbs{w}{\cPair{x}} \wedge \nprecedesAbs{w}{x})
		\end{align*}
	
	    So, instantiating with $w \defini y$, we obtain $\nprecedesAbs{y}{x}$ as required.
	\end{itemize}
\end{prf}

\begin{lem}
	Let $i \natorderSymbol |\elimRelName|$.
	If Axiom \axiomSRef{vis-ax::stack::cc-return-completion} holds
	at domain $\closedEvent \setminus \elimSet{i}$, then 
	the axiom also holds at domain $\closedEvent \setminus \elimSet{i+1}$.
\end{lem}

\begin{prf}
	Let $x \in \closedEvent \setminus \elimSet{i+1}$. Since $x \notin \elimSet{i+1}$, by definition of 
	$\elimSet{i+1}$, either $x \in \elimPairSet{i}$ or $x \notin \elimSet{i}$.
	\begin{itemize}
	\item Case $x \in \elimPairSet{i}$. 
	
	If $x$ is a push, then by definition of $\elimPairSet{i}$, 
	$\elimRel{x}{(\cPair{x})}$,
	and so $\idProp{x} = \idProp{(\cPair{x})}$. If $x$ is a pop, then 
	$\elimRel{(\cPair{x})}{x}$, and so $\idProp{x} = \idProp{(\cPair{x})}$.
	In both cases, $\idProp{x}$ is defined, and from 
	\axiomIRef{inv::stack::appendix::vis-proof::pops-take-pushes}, $\spanEvOut x$ 
	is also defined.
	
	Define $v \defini \spanEvOut x$. To show $\postPred{x}{v}$, we need to consider the following cases,
	\begin{itemize}
		\item Case $x$ is a push. 
		
		\axiomIRef{inv::stack::appendix::vis-proof::pops-take-pushes} implies $v = \spanEvOut x = \unitValue$.
		
		\item Case $x$ is a pop and $v \neq \EMPTY$.
		
		\axiomIRef{inv::stack::appendix::vis-proof::pops-take-pushes}
		implies that there is a push $u$ such that $\idProp{u} = \idProp{x}$ 
		and $v = \spanEvOut x = \inProp{u}$.
		
		But $x \in \elimPairSet{i}$, and by definition of $\elimPairSet{i}$,
		$\elimPair{i} = (u',x)$ for some push $u'$. Hence, 
		$\idProp{u'} = \idProp{x}$ by definition of $\elimRelName$.
		Therefore, $\idProp{u'} = \idProp{x} = \idProp{u}$, which means
		$u = u'$ by \axiomIRef{inv::stack::appendix::vis-proof::prop-funcs-are-injective}.
		
		Since $\elimPair{i} = (u',x) = (u,x)$, we have $\visObsIndx{i+1}{}{u}{x}$ by definition.
		So, Lemma~\ref{lem::stack::appendix::vis-proof::elim-set-basic-facts} implies $\visObs{}{u}{x}$
		(since $i+1 \natorderEqSymbol |\elimRelName|$).
		
		Also, since $\visObs{}{u}{x}$, then $u \in \closedEvent$, because $\closedEvent$ is 
		$\genVisSymbol$-downward closed (Lemma~\ref{eq::stack::appendix::lin::committed-downward-closed}).
		Finally, since $\elimPair{i} = (u,x)$, we have by definition $u \in \elimPairSet{i}$.
		Therefore, $u \notin \elimSet{i+1}$, for otherwise we would have $u \notin \elimPairSet{i}$
		by definition of $\elimSet{i+1}$. In other words, $u \in \closedEvent \setminus \elimSet{i+1}$.
		
		\item Case $x$ is a pop and $v = \EMPTY$.
		
		\axiomIRef{inv::stack::appendix::vis-proof::pops-take-pushes}
		implies that there is a push $u$ such that $\idProp{u} = \idProp{x}$ 
		and $v = \spanEvOut x = \inProp{u} \in \ValType$. Therefore, $v \in \ValType$,
		which contradicts $v = \EMPTY$, i.e., this simply means that
		pops never return $\EMPTY$, which means that this case is impossible.
	\end{itemize} 
	
	Finally, if $x \in \terminatedEvent$, \axiomIRef{inv::stack::appendix::vis-proof::terminated-implies-span}
	implies $v = \spanEvOut{x} = \outputProp{x}$.
	
	\item Case $x \notin \elimSet{i}$.
	We know the axiom holds on domain $\closedEvent \setminus \elimSet{i}$.
	In other words, there is a $v$ such that $\postPred{x}{v}$ and $x \in \terminatedEvent \implies v = \outputProp{x}$.
	
	Notice that the cases in $\postPred{x}{v}$ hold in domain $\closedEvent \setminus \elimSet{i}$, so
	we still need to show that the cases in $\postPred{x}{v}$ hold in domain $\closedEvent \setminus \elimSet{i+1}$.
	We consider the cases,
	
	\begin{itemize}
		\item Case $x$ is a push. 
		
		From $\postPred{x}{v}$ (in domain $\closedEvent \setminus \elimSet{i}$), we have $v = \unitValue$.
		
		\item Case $x$ is a pop and $v \neq \EMPTY$.
		
		From $\postPred{x}{v}$ (in domain $\closedEvent \setminus \elimSet{i}$),
		there is $u \in \closedEvent \setminus \elimSet{i}$ such that $\visObs{}{u}{x}$
		and $v = \inProp{u}$.
		But Lemma~\ref{lem::stack::appendix::vis-proof::elim-set-basic-facts} 
		implies $\elimSet{i+1} \subseteq \elimSet{i}$ (since $i \natorderSymbol i+1$),
		which means $u \notin \elimSet{i+1}$, i.e., $u \in \closedEvent \setminus \elimSet{i+1}$.
		
		\item Case $x$ is a pop and $v = \EMPTY$.
		
		Let $u \in \closedEvent \setminus \elimSet{i+1}$ such that $x \not\visSepEqSYMBOL u$. 
		By definition of $\elimSet{i+1}$, either $u \in \elimPairSet{i}$ or $u \notin \elimSet{i}$.
		
		\begin{itemize}
			\item Case $u \in \elimPairSet{i}$. By definition of $\elimPairSet{i}$,
			$\elimPair{i} = (u,o)$ for some pop $o$. In particular, 
			we have $\visObsIndx{i+1}{}{u}{o}$ by definition.
			So, Lemma~\ref{lem::stack::appendix::vis-proof::elim-set-basic-facts} implies $\visObs{}{u}{o}$
			(since $i+1 \natorderEqSymbol |\elimRelName|$).
			
			It remains to show that $\visSep{}{o}{x}$. From hypothesis $x \not\visSepEqSYMBOL u$,
			Lemma~\ref{lem::stack::appendix::vis-proof::elim-set-basic-facts} implies 
			$x \not\visSepIndxSymbol{i+1}{} u$
			(since $i+1 \natorderEqSymbol |\elimRelName|$). After some De Morgan manipulations on 
			$x \not\visSepIndxSymbol{i+1}{} u$,
			\[
			x \notin \elimSet{i} \wedge u \in \elimPairSet{i} \implies \neg \BE{i}{x}{u}
			\]
			So, $\neg \BE{i}{x}{u}$. But $o$ is the elimination couple of $u$, so 
			Lemma~\ref{subsubsect::stack::appendix::BE-with-elim-couples} implies $\neg \BE{i}{x}{o}$.
			And since $o \in \elimPairSet{i}$ and $x \notin \elimSet{i}$, we have 
			$\visSepIndx{i+1}{}{o}{x}$ by definition.
			Hence, Lemma~\ref{lem::stack::appendix::vis-proof::elim-set-basic-facts} implies $\visSep{}{o}{x}$
			(since $i+1 \natorderEqSymbol |\elimRelName|$).
			
			Also, since $\visSep{}{o}{x}$, then $o \in \closedEvent$, because $\closedEvent$ is 
			$\genVisSymbol$-downward closed (Lemma~\ref{eq::stack::appendix::lin::committed-downward-closed}).
			Finally, since $\elimPair{i} = (u,o)$, we have by definition $o \in \elimPairSet{i}$.
			Therefore, $o \notin \elimSet{i+1}$, for otherwise we would have $o \notin \elimPairSet{i}$
			by definition of $\elimSet{i+1}$. In other words, $o \in \closedEvent \setminus \elimSet{i+1}$.
			
			\item Case $u \notin \elimSet{i}$. 
			
			From $\postPred{x}{v}$ (in domain $\closedEvent \setminus \elimSet{i}$), there is a pop
			$o \in \closedEvent \setminus \elimSet{i}$ such that $\visObs{}{u}{o}$ and $\visSep{}{o}{x}$.
			But then, Lemma~\ref{lem::stack::appendix::vis-proof::elim-set-basic-facts} 
			implies that $o \notin \elimSet{i+1}$ (since $i \natorderSymbol i+1$ and 
			$o \notin \elimSet{i}$). In other words, $o \in \closedEvent \setminus \elimSet{i+1}$.
		\end{itemize}
		
	\end{itemize} 
	
    \end{itemize}
	
\end{prf}

\begin{thm}
\label{thm::stack::appendix::vis-proof::vis-axioms-hold-at-max-domain}
All the visibility-style axioms hold at domain $\closedEvent$.
\end{thm}

\begin{prf}
	We prove by induction on $i \natorderEqSymbol |\elimRelName|$ 
	that all the axioms hold at domain $\closedEvent \setminus \elimSet{i}$.
	
	The base case $i = 0$ is Theorem~\ref{thm::stack::appendix::vis-proof::vis-axioms-hold-at-base-domain}.
	
	The inductive case are all the previous lemmas that show that 
	if the axioms are true at domain
	$\closedEvent \setminus \elimSet{i}$ (for $i \natorderSymbol |\elimRelName|$, since
	$i+1 \natorderEqSymbol |\elimRelName|$ by hypothesis),
	then they also hold at domain $\closedEvent \setminus \elimSet{i+1}$.
	
Therefore, the axioms hold at domain $\closedEvent \setminus \elimSet{|\elimRelName|}$.
But Lemma~\ref{lem::stack::appendix::vis-proof::cardinality-of-elim-pairs} 
implies $|\elimSet{|\elimRelName|}| = 2(|\elimRelName| - |\elimRelName|) = 0$, i.e.,
$\elimSet{|\elimRelName|} = \emptyset$. 
Therefore, all axioms hold at $\closedEvent \setminus \elimSet{|\elimRelName|} = \closedEvent$.
\end{prf}

\begin{thm}
	\label{thm::stack::appendix::vis-proof::ts-invariants-imply-vis-axioms}
Given the relations in Definition~\ref{defn::stack::appendix::vis-proof::final-relations}. 
If all the TS-stack invariants of Figure~\ref{fig::stack::appendix::vis-proof::ts-stack-invariants} hold,
then the visibility-style axioms in Figure~\ref{fig::stack::concurrent-spec-stack} hold.
\end{thm}

\begin{prf}
The hypothesis states that the TS-stack invariants hold. Therefore, the result follows
by Theorem~\ref{thm::stack::appendix::vis-proof::vis-axioms-hold-at-max-domain}.
\end{prf}

\subsection{Proof of TS-Stack invariants}
\label{subsect::stack::appendix::vis-proof::proof-of-invariants}

In the following subsections
we will define the pop-total strict partial order $\GstamporderSymbol$
and show that the invariants in Figure~\ref{fig::stack::appendix::vis-proof::ts-stack-invariants} hold
for the atomic timestamp TS-stack (Figures \ref{alg::stack::appendix::vis-proof::full-stack},
\ref{alg::stack::appendix::vis-proof::pool-type} and \ref{fig::sub::stack::appendix::vis-proof::atomic-timestamps-gen}) 
and the interval timestamp TS-stack (Figures \ref{alg::stack::appendix::vis-proof::full-stack},
\ref{alg::stack::appendix::vis-proof::pool-type} and \ref{fig::sub::stack::appendix::vis-proof::interval-timestamps-gen}).
	
\subsubsection{Version: Atomic Timestamps}
\label{subsubsect::stack::appendix::vis-proof::atomic-case}

For the atomic TS-stack, we define $\GstamporderSymbol$ simply as,\footnote{Recall that we overloaded the symbol $\stamporderSymbol$
	to apply to both timestamps and abstract timestamps in Definition~\ref{defn::stack::appendix::vis-proof::common-definitions}.}
\[
\GstamporderSymbol\ \defini\ \stamporderSymbol
\]

First we prove the structural invariants, since they do not depend on $\GstamporderSymbol$.
In this way, we can freely use the structural invariants in the rest of the lemmas.

\begin{lem}
	\label{lem::stack::appendix::vis-proof::struct-invariants-hold-atomic}
	All the structural invariants in Figure~\ref{fig::stack::appendix::vis-proof::ts-stack-invariants} hold for the atomic timestamp TS-stack.
\end{lem}

\begin{prf}
	Most of them are trivial, and follow directly from Definition~\ref{defn::stack::appendix::vis-proof::common-definitions}.
	We focus only on the ones that require a little bit of more explanation. 
	
	For \axiomIRef{inv::stack::appendix::vis-proof::pop-not-precedes-push}, since a pop takes a node with some id,
	such node must have been inserted before by a push (no other procedure can link new nodes into the pool). 
	Since a push's span links the node
	into the pool as is first rep event, the push's span cannot start after the pop took the node
	(i.e., the node must be inserted into the pool before a pop can take such node, otherwise the pop
	would not have been able to discover the node in the first place).
	
	For \axiomIRef{inv::stack::appendix::vis-proof::timestamps-equal}, since the push finished its span before the pop started
	its last iteration, the push assigned the generated timestamp to the node as its last rep event, before the pop started its last
	iteration. 
	Therefore, when during its iteration, the pop reads at line~\ref{alg::stack::appendix::vis-proof::pop::read-candidate-timestamp} 
	the timestamp of the node, at that moment, the node already has the timestamp assigned by the push.
	
	For \axiomIRef{inv::stack::appendix::vis-proof::terminated-implies-span}, we need to check for pushes and pops.
	If a push $x$ terminated, then the executing thread completed the lines defining the push span, and the procedure 
	finished with the trivial output $\unitValue$ (there is an implicit return), and so, the definition 
	of $\spanEvOut{x}$ coincides with the output of the event $\outputProp{x}$. 
	If a pop $x$ terminated, the executing thread must have executed a successful 
	CAS at line~\ref{alg::stack::appendix::vis-proof::pool::try-take-node}, otherwise it would have looped infinitely,
	meaning that the executing thread had a last iteration and therefore, completed the lines defining its span. 
	After the successful CAS, the procedure reads the value stored in the taken node and returns, and so
	$\spanEvOut{x}$ coincides with the output of the event $\outputProp{x}$.
	
	For \axiomIRef{inv::stack::appendix::vis-proof::spans-are-injective}, if two events $x$ and $y$ generate the same span, 
	then they executed the same rep events inside the spans. This implies that both events must have been executed
	by the same thread (a rep event, being atomic, can only be executed by a single thread).
	Suppose for a contradiction that $x \neq y$. Since events executed by the same thread appear sequentially and 
	disjoint in the history, then either $x$
	executes completely before $y$ or $y$ executes completely before $x$. But this implies that the rep events
	inside the events cannot be the same (Contradiction). Therefore, $x = y$.
	A similar argument applies to \axiomIRef{inv::stack::appendix::vis-proof::reps-are-injective}.
	
	For \axiomIRef{inv::stack::appendix::vis-proof::prop-funcs-are-injective} since ids are atomically
	generated and assigned to fresh nodes, they are unique per push and also per node. 
	Hence, if two pushes have the same id, it means that they executed the same rep event,
	leading to conclude that the pushes must be the same by following an argument similar to
	\axiomIRef{inv::stack::appendix::vis-proof::spans-are-injective} above.
	If two pops have the same id, then they took the same node, since ids are also unique per node.
	This means that both pops executed the same CAS rep event that set the $taken$ flag at
	line~\ref{alg::stack::appendix::vis-proof::pool::try-take-node} 
	(once the $taken$ flag is set, it cannot be reset, since there are no instructions 
	that could change the boolean back to $false$). Therefore, both pops executed the same rep event,
	leading to conclude that the pops must be the same by following an argument similar to
	\axiomIRef{inv::stack::appendix::vis-proof::spans-are-injective} above.
	
	For \axiomIRef{inv::stack::appendix::vis-proof::span-inside-event} all the events execute rep events inside the duration 
	of the event because all the rep events inside the event are executed by the same thread that invoked the event, 
	as such, the unique span that events execute, does it so within the duration of the event.
	
	For \axiomIRef{inv::stack::appendix::vis-proof::pops-take-pushes}(i), when a pop finishes its span (i.e., takes a node),
	such node must have been inserted by a push and the output of the pop will be the value stored in the node
	(line~\ref{alg::stack::appendix::vis-proof::pool::read-val-to-return}).
\end{prf}

\begin{lem}
$\GstamporderSymbol$ is a pop-total strict partial order on abstract timestamps.
\end{lem}

\begin{prf}
By Lemma~\ref{defn::stack::appendix::vis-proof::abstract-timestamps-order-is-strict},\footnote{Note that 
	the proof of Lemma~\ref{defn::stack::appendix::vis-proof::abstract-timestamps-order-is-strict} 
	does not make use of any invariants, because the result
	follows directly from definitions. Therefore, we can use
	Lemma~\ref{defn::stack::appendix::vis-proof::abstract-timestamps-order-is-strict} safely in our proofs.}
$\stamporderSymbol$ is still a strict partial order when it is extended to abstract timestamps.
It remains to show that $\GstamporderSymbol$ is pop-total.

Let $\AstampProp{u_1} = (\idProp{u_1}, \stampProp{u_1})$ and $\AstampProp{u_2}=(\idProp{u_2}, \stampProp{u_2})$ be the generated abstract timestamps for pushes $u_1$ and $u_2$.
The generated standard timestamps $\stampProp{u_1}$ and $\stampProp{u_2}$ are natural numbers, because
the $\newStampAlg$ method in Figure~\ref{fig::sub::stack::appendix::vis-proof::atomic-timestamps-gen} 
returns a natural number. Therefore, we must have either $\stampProp{u_1} \stamporderSymbol \stampProp{u_2}$
or $\stampProp{u_2} \stamporderSymbol \stampProp{u_1}$ or $\stampProp{u_1} = \stampProp{u_2}$.

The case $\stampProp{u_1} \stamporderSymbol \stampProp{u_2}$
trivially leads to $\AstampProp{u_1} = (\idProp{u_1},\stampProp{u_1}) \stamporderSymbol (\idProp{u_2},\stampProp{u_2}) = \AstampProp{u_2}$
from the overloading of $\stamporderSymbol$ to abstract timestamps. 

The case $\stampProp{u_2} \stamporderSymbol \stampProp{u_1}$ leads to a similar result.

For the case $\stampProp{u_1} = \stampProp{u_2}$, 
since timestamps are atomic natural numbers, timestamps are unique per push. 
This means that pushes $u_1$, $u_2$ must have executed the same rep event that
created the timestamp.
The rest of the argument is as in the proof of \axiomIRef{inv::stack::appendix::vis-proof::spans-are-injective}
in Lemma~\ref{lem::stack::appendix::vis-proof::struct-invariants-hold-atomic}.
This means that $u_1 = u_2$, and hence, 
they must have the same id, i.e., $\idProp{u_1} = \idProp{u_2}$. In other words, we have,
\[
\AstampProp{u_1} = (\idProp{u_1},\stampProp{u_1}) = (\idProp{u_2},\stampProp{u_2}) = \AstampProp{u_2}
\]

Therefore, $\GstamporderSymbol$ is pop-total.
\end{prf}

\begin{lem}
\label{lem::stack::appendix::vis-proof::decreasing-stamps-in-pool}
In a pool, at any moment, all untaken nodes reachable from the top have strictly decreasing timestamps. In other words,
let $t_n$ and $t_m$ be the timestamps of untaken nodes $n$ and $m$, respectively. If $m$ occurs after $n$ when
traversing the linked list starting from the top, then $t_m \stamporderSymbol t_n$.
\end{lem}

\begin{prf}
Insertions to a pool are single-threaded, i.e., all nodes in a pool were inserted by the same thread. This implies that two
nodes in the pool were assigned timestamps generated by disjoint calls to the $\newStampAlg$ procedure, and the nodes were inserted
in the same order as the calls to the $\newStampAlg$ procedure. In other words, the most recently inserted node will have a bigger
timestamp. It remains to see if the unlinking code preserves the decreasing timestamp property.

The loop at lines~\ref{alg::stack::appendix::vis-proof::pool::unlinking-insert-start}-\ref{alg::stack::appendix::vis-proof::pool::unlinking-insert-finish} 
exits only when it reaches the sentinel node (stored in the $next$ variable) or
the  first untaken node (also stored in the $next$ variable). In either case, when the loop ends,
the procedure sets the new node's next $n.next$ to
the found node in $next$. In case the found node is the sentinel node, then the pool will only contain the new node, which trivially
satisfies the strictly decreasing timestamp property. In case the found node is an untaken node, then the pool still 
satisfies the decreasing timestamp property, since the new node $n$ has the freshest timestamp and $n$ is the new top.

Lines~\ref{alg::stack::appendix::vis-proof::pool::unlinking-remove-cas}-\ref{alg::stack::appendix::vis-proof::pool::unlinking-remove-finish} 
carry out unlinking of nodes around the node $n$ that was taken. Variable $oldTop$
stores the top that the pool had when $n$ was found. If the current top has not changed from $oldTop$
(meaning that no new nodes were inserted into the pool), the CAS at line~\ref{alg::stack::appendix::vis-proof::pool::unlinking-remove-cas} 
will set as the new top the taken node $n$. This action preserves the decreasing timestamp property, because the property was true
before changing the top to one of the nodes in the pool (even if the new top has been marked as taken already).
Lines~\ref{alg::stack::appendix::vis-proof::pool::unlinking-remove-condition}-\ref{alg::stack::appendix::vis-proof::pool::unlinking-remove-condition-true} 
unlink nodes between the old top and the taken node $n$, which preserves the decreasing timestamp property,
as it is only removing nodes.
Similarly, lines~\ref{alg::stack::appendix::vis-proof::pool::unlinking-remove-start}-\ref{alg::stack::appendix::vis-proof::pool::unlinking-remove-finish} 
unlink nodes between the taken node $n$ and the first untaken node after $n$ (or the sentinel node
if every node after $n$ is already taken). This also preserves the decreasing timestamp property,
as it is only removing nodes.
\end{prf}

\begin{lem}
\label{lem::stack::appendix::vis-proof::untaken-nodes-are-reachable}
Untaken nodes do not get unlinked. In other words, if $n$ is an untaken node that was inserted into a pool, then it remains reachable from the top
of the pool.
\end{lem}

\begin{prf}
When inserting a new node in the $\poolInsertAlg$ procedure in Figure~\ref{alg::stack::appendix::vis-proof::pool-type},
the new node is set as the top at line~\ref{alg::stack::appendix::vis-proof::pool::new-node-as-top}. 
It remains to check that the unlinking code does not remove untaken nodes.

The loop at lines~\ref{alg::stack::appendix::vis-proof::pool::unlinking-insert-start}-\ref{alg::stack::appendix::vis-proof::pool::unlinking-insert-finish} 
exits only when it reaches the sentinel node (stored in the $next$ variable) or
the  first untaken node (also stored in the $next$ variable). In either case, when the loop ends,
the procedure sets the new node's next $n.next$ to
the found node in $next$. Therefore, any node between the new node and the node in $next$ is already taken
and can be unlinked.

Lines~\ref{alg::stack::appendix::vis-proof::pool::unlinking-remove-cas}-\ref{alg::stack::appendix::vis-proof::pool::unlinking-remove-finish} 
carry out unlinking of nodes around the node $n$ that was taken. Variable $oldTop$
stores the top that the pool had when $n$ was found. If the current top has not changed from $oldTop$
(meaning that no new nodes were inserted into the pool), the CAS at line~\ref{alg::stack::appendix::vis-proof::pool::unlinking-remove-cas} 
will set as the new top the taken node $n$ (i.e., since $n$ was the first untaken node to be found and the top 
has not changed, then any node between the top and node $n$ is taken and can be unlinked).
If the CAS fails, it means that new nodes were inserted before $oldTop$ and so, we cannot unlink nodes
before $oldTop$. But nodes between $oldTop$ and $n$ are already taken (since $n$ was the first untaken
node when the pool was traversed starting from $oldTop$), and therefore, these nodes can be unlinked
(lines~\ref{alg::stack::appendix::vis-proof::pool::unlinking-remove-condition}-\ref{alg::stack::appendix::vis-proof::pool::unlinking-remove-condition-true}).
Finally, the loop in lines~\ref{alg::stack::appendix::vis-proof::pool::unlinking-remove-start}-\ref{alg::stack::appendix::vis-proof::pool::unlinking-remove-finish} 
unlink nodes between the taken node $n$ and the first untaken node after $n$ (or the sentinel node
if every node after $n$ is already taken). Hence, these nodes can be unlinked.
\end{prf}

\begin{lem}[Misses Lemma]
	\label{lem::stack::appendix::vis-proof::atomic-misses-lemma}
	If, 
	\begin{itemize}
		\item $o, u \notin \elimSet{}$,
		\item $\stampProp o \stamporderSymbol \stampProp{u}$, and
		\item $\forall o' \notin \elimSet{}.\ \idProp{u} = \idProp{o'} \implies \precedesReps {\twoSpan{\spanEv {o}}} {\twoSpan{\spanEv {o'}}}$,
	\end{itemize} 
	then $\precedesReps {\oneSpan{\spanEv{o}}} {\oneSpan{\spanEv{u}}}$.
\end{lem}

\begin{prf}
	Let $maxT_1$, $maxT_2$, ..., $maxT_n$
	be the strictly increasing sequence of timestamps assigned to the $maxT$ variable during the last iteration of pop $o$ (excluding
	the initialization assignment at line~\ref{alg::stack::appendix::vis-proof::pop::initialize-max-ts}).
	Similarly, let $chosen_1$, $chosen_2$, ..., $chosen_n$ and $chPool_1$, $chPool_2$, ..., $chPool_n$ be the sequences of assignments
	to variables $chosen$ and $chPool$ (excluding initializing assignments at 
	lines~\ref{alg::stack::appendix::vis-proof::pop::initialize-chosen} 
	and \ref{alg::stack::appendix::vis-proof::pop::initialize-chPool}). All three sequences have 
	the same length $1 \natorderEqSymbol n$ because variables $maxT$, $chosen$, and $chPool$ are assigned together at
	lines~\ref{alg::stack::appendix::vis-proof::pop::update-vars-one}-\ref{alg::stack::appendix::vis-proof::pop::update-vars-two}.
	
	Claim: Pop $o$ cannot reach line~\ref{alg::stack::appendix::vis-proof::pop::try-take-elimination-pair} because $o$ is not in an elimination pair
	by hypothesis (i.e., $o \notin \elimSet{}$).
	Indeed, if $o$ executed such line, then we must have $t_{start} \stamporderSymbol t_g$ 
	(the condition at line~\ref{alg::stack::appendix::vis-proof::pop::compare-with-start-stamp}), 
	where $t_g$ is the timestamp of the node $g$ taken by $o$ and $t_{start}$ is the timestamp generated by $o$ at 
	line~\ref{alg::stack::appendix::vis-proof::pop::create-start-stamp}. 
	Node $g$ was inserted by some push $u_g$, which
	means we have $\idProp{u_g} = \idProp{o}$. But since $o \notin \elimSet{}$, we also have $\precedesAbs{u_g}{o}$,
	meaning that the calls to $\newStampAlg$ executed by $u_g$ 
	(line~\ref{alg::stack::appendix::vis-proof::push::create-stamp}) and by $o$ 
	(line~\ref{alg::stack::appendix::vis-proof::pop::create-start-stamp}) are disjoint, and 
	so $t_g \stamporderSymbol t_{start}$ (Contradiction).
	This proves the claim.
	
	Hence, pop $o$ took node $chosen_n$ from pool $chPool_n$ and with timestamp $maxT_n$ after scanning all the pools
	(i.e., pop $o$ cannot prematurely end the scan of the pools by entering line 26).
	Additionally, hypothesis $o \notin \elimSet{}$ implies that the last assignment $maxT_n$ must be a finite timestamp, because 
	the push that inserted $chosen_n$ executed before $o$, meaning that $u$ assigned the finite timestamp 
	to the node before $o$ started the scan. This implies that every timestamp in the sequence 
	$maxT_i$ is finite, because it is a strictly increasing sequence and $maxT_n$ is the last assignment.
	
	The third hypothesis to the lemma implies that the node $m$ inserted by $u$ remains untaken during the last loop of $o$,
	while the second hypothesis states that $m$ was assigned a bigger timestamp $t_m$ than $maxT_n$. Denote by $p_m$ the pool where node $m$ was inserted. 
	Notice that, during the last loop of $o$, node $m$ could 
	have infinite timestamp or finite timestamp $t_m$, depending on whether or not the span of $u$ finished during the last loop of $o$.
	
	Assume for a contradiction that the span of $u$ started before the span of $o$. This means
	that node $m$ was in the pool before $o$ started its last loop. 
	
	Claim: Node $m$ must be different from each $chosen_i$. If $m$ is equal to $chosen_i$ for some $i$, then
	we have $t_m = maxT_i \stamporderEqSymbol maxT_n \stamporderSymbol t_m$ (contradiction). Notice that $m$ must have finite
	timestamp at the moment it is discovered (hence, equal to $t_m$), because $m$ is $chosen_i$, and each one of the $chosen_i$ nodes had finite timestamp
	when they were discovered. This proves the claim.
	
	Claim: Pool $p_m$ is different from each $chPool_i$. Suppose for a contradiction that $p_m$ is equal to $chPool_i$ for some $i$.
	Since $m$ remains untaken during $o$, when $o$ finds node $chosen_i$ at line~\ref{alg::stack::appendix::vis-proof::pool::find-first-untaken},
	nodes $m$ and $chosen_i$ were simultaneously untaken in the same pool at that moment. 
	Hence, Lemma~\ref{lem::stack::appendix::vis-proof::untaken-nodes-are-reachable} implies that nodes 
	$m$ and $chosen_i$ were reachable from the top at that moment,
	which means that Lemma~\ref{lem::stack::appendix::vis-proof::decreasing-stamps-in-pool} implies 
	$t_m \stamporderSymbol maxT_i$, since $chosen_i$ was the first to be found and $m$ is different from $chosen_i$
	by the previous claim.
	Hence, $t_m \stamporderSymbol maxT_i \stamporderEqSymbol maxT_n \stamporderSymbol t_m$ (contradiction).
	This proves the claim.
	
	If pool $p_m$ is in between $chPool_i$ and $chPool_{i+1}$, then $o$ must have
	picked $m$ (or some other node with even bigger timestamp), since $m$ has either infinite timestamp or finite timestamp but bigger than 
	$maxT_i$ and $maxT_{i+1}$ (since $maxT_n \stamporderSymbol t_m$). If $p_m$ is a pool appearing later than $chPool_n$, then $o$ would
	eventually pick it (or some other node with even bigger timestamp), because $m$ has either infinite timestamp or finite timestamp but bigger than 
	$maxT_n$, which contradicts that $o$ did not pick $m$. Therefore, the span of $u$ must start after the span of $o$ starts.
\end{prf}

\begin{lem}
	\label{lem::stack::appendix::vis-proof::key-invariants-atomic-case}
	All the key invariants in Figure~\ref{fig::stack::appendix::vis-proof::ts-stack-invariants} hold for the atomic timestamp TS-stack.
\end{lem}

\begin{prf}
Invariant \axiomIRef{vis-ax::stack::appendix::vis-proof::disjoint-push} holds trivially because timestamps are atomic natural numbers, meaning that the 
global counter $TS$ in Figure~\ref{fig::sub::stack::appendix::vis-proof::atomic-timestamps-gen} 
increases per push and it is unique to that push, producing only comparable timestamps. 

For invariant \axiomIRef{vis-ax::stack::vis-proof::misses-are-late}, suppose $\PEND o u$. 
Hence, we know $o,u \notin \elimSet{}$, $\AstampProp{o} \stamporderSymbol \AstampProp{u}$ (in particular $\stampProp{o} \stamporderSymbol \stampProp{u}$ 
by the overloading of $\stamporderSymbol$), and 
\begin{align}
\label{eq::stack::appendix::vis-proof::key-invariants-atomic::eq1}
\forall o' \notin \elimSet{}.\ \AstampProp{u} = \AstampProp{o'} \implies \precedesReps {\twoSpan{\spanEv {o}}} {\twoSpan{\spanEv {o'}}}
\end{align}
The result follows by the Misses Lemma~\ref{lem::stack::appendix::vis-proof::atomic-misses-lemma} if we can show the third
hypothesis in that lemma. 

So, let $o' \notin \elimSet{}$ such that $\idProp{u} = \idProp{o'}$. We can freely use the structural invariants,
since Lemma~\ref{lem::stack::appendix::vis-proof::struct-invariants-hold-atomic} already proved them.
Additionally, we can use any lemma of Section~\ref{subsect::stack::appendix::vis-proof::axioms-from-invariants}
as long as it \emph{only} uses structural invariants.

Since $o' \notin \elimSet{}$, Lemma~\ref{lem::stack::appendix::vis-proof::elim-set-basic-disj-fact} implies 
$\precedesAbs{u}{o'}$. Since $\DEF{\idProp{o'}}$, \axiomIRef{inv::stack::appendix::vis-proof::pop-ids-imply-span} implies 
$\DEF{\spanEv{o'}}$. Hence, Lemma~\ref{lem::stack::appendix::event-prec-implies-span-prec} implies 
$\spanEv{u} \precedesSpansSymbol \spanEv{o'}$, which means $\stampProp{u} = \stampProp{o'}$
by \axiomIRef{inv::stack::appendix::vis-proof::timestamps-equal}. 
In other words, $\AstampProp{u} = \AstampProp{o'}$, which implies 
$\precedesReps {\twoSpan{\spanEv {o}}} {\twoSpan{\spanEv {o'}}}$ by \eqref{eq::stack::appendix::vis-proof::key-invariants-atomic::eq1}.
\end{prf}

And we have the main theorem.

\begin{thm}
The atomic timestamp TS-stack is linearizable.
\end{thm}

\begin{prf}
By the above lemmas, all the TS-stack invariants hold for the atomic timestamp version. Therefore, by Theorem~\ref{thm::stack::appendix::vis-proof::ts-invariants-imply-vis-axioms}, the 
visibility-style axioms in Figure~\ref{fig::stack::concurrent-spec-stack} hold. 
Theorem~\ref{thm::stack::appendix::lin-proof::vis-axioms-imply-linearizability} 
then implies that the atomic timestamp TS-stack is linearizable.
\end{prf}

\subsubsection{Version: Interval Timestamps}
\label{subsubsect::stack::appendix::vis-proof::interval-case}

For the interval TS-stack, we define $\GstamporderSymbol$ over abstract timestamps as,\footnote{Recall that we overloaded the symbol $\stamporderSymbol$
	to apply to both timestamps and abstract timestamps in Definition~\ref{defn::stack::appendix::vis-proof::common-definitions}.}
\begin{align*}
	t_2 \GstamporderSymbol t_1 & \defini t_2 \stamporderSymbol t_1\ \vee\  
	\exists u_1\ u_2 \notin \elimSet{}.\ \TB {u_1} {u_2}\ \wedge\ \AstampProp {u_1}  \not\stamporderSymbol \AstampProp {u_2}\ \wedge \\ 
	& \hspace{45mm} t_2 \stamporderEqSymbol \AstampProp {u_2}\ \wedge\ \AstampProp {u_1} \stamporderEqSymbol t_1 \\
	\TB {u_1} {u_2} & \defini u_1,u_2\notin\elimSet{}\ \wedge\ \exists o_1 \notin \elimSet{}.\ \AstampProp{u_1} = \AstampProp{o_1}\ \wedge \\
	& \hspace{45mm} \forall o_2\notin \elimSet{}.\ \AstampProp{u_2} = \AstampProp{o_2} \implies \precedesReps {\twoSpan{\spanEv{o_1}}} {\twoSpan{\spanEv{o_2}}}
\end{align*}

First we prove the structural invariants, since they do not depend on $\GstamporderSymbol$.
In this way, we can freely use the structural invariants in the rest of the lemmas.

\begin{lem}
	\label{lem::stack::appendix::vis-proof::struct-invariants-hold-interval}
	All the structural invariants in Figure~\ref{fig::stack::appendix::vis-proof::ts-stack-invariants} hold for the interval timestamp TS-stack.
\end{lem}

\begin{prf}
	Identical to Lemma~\ref{lem::stack::appendix::vis-proof::struct-invariants-hold-atomic}, since
	the argument does not depend on the nature of the timestamps: it holds for atomic and
	interval timestamps.
\end{prf}

Now we prove some basic properties about the algorithm.

\begin{lem}
	\label{lem::stack::appendix::vis-proof::decreasing-stamps-in-pool-interval}
	In a pool, at any moment, all untaken nodes reachable from the top have strictly decreasing timestamps. In other words,
	let $t_n$ and $t_m$ be the timestamps of untaken nodes $n$ and $m$, respectively. If $m$ occurs after $n$ when
	traversing the linked list starting from the top, then $t_m \stamporderSymbol t_n$.
\end{lem}

\begin{prf}
Identical to the proof of Lemma~\ref{lem::stack::appendix::vis-proof::decreasing-stamps-in-pool}.
\end{prf}

\begin{lem}
	\label{lem::stack::appendix::vis-proof::untaken-nodes-are-reachable-interval}
	Untaken nodes do not get unlinked. In other words, if $n$ is an untaken node that was inserted into a pool, then it remains reachable from the top
	of the pool.
\end{lem}

\begin{prf}
Identical to the proof of Lemma~\ref{lem::stack::appendix::vis-proof::untaken-nodes-are-reachable}.
\end{prf}

\begin{lem}[Misses Lemma: Plain Timestamps]
	\label{lem::stack::appendix::vis-proof::interval-misses-lemma-plain}
If, 
\begin{itemize}
	\item $o, u \notin \elimSet{}$,
	\item $\stampProp o \stamporderSymbol \stampProp{u}$, and
	\item $\forall o' \notin \elimSet{}.\ \idProp{u} = \idProp{o'} \implies \precedesReps {\twoSpan{\spanEv {o}}} {\twoSpan{\spanEv {o'}}}$,
\end{itemize} 
then $\precedesReps {\oneSpan{\spanEv{o}}} {\oneSpan{\spanEv{u}}}$.
\end{lem}

\begin{prf}
	The proof is identical to the proof of 
	Lemma~\ref{lem::stack::appendix::vis-proof::atomic-misses-lemma}, 
	since the argument does not depend on the nature of the timestamps: it holds 
	for atomic and interval timestamps.
\end{prf}

\begin{lem}[Misses Lemma: Abstract Timestamps]
	\label{lem::stack::appendix::vis-proof::misses-lemma}
	If, 
	\begin{itemize}
		\item $o, u \notin \elimSet{}$,
		\item $\AstampProp o \stamporderSymbol \AstampProp{u}$, and
		\item $\forall o' \notin \elimSet{}.\ \AstampProp{u} = \AstampProp{o'} \implies \precedesReps {\twoSpan{\spanEv {o}}} {\twoSpan{\spanEv {o'}}}$,
	\end{itemize} 
	then $\precedesReps {\oneSpan{\spanEv{o}}} {\oneSpan{\spanEv{u}}}$.
\end{lem}

\begin{prf}
From hypothesis $\AstampProp{o} \stamporderSymbol \AstampProp{u}$, we have $\stampProp{o} \stamporderSymbol \stampProp{u}$.
The result follows by the Misses Lemma: Plain Timestamps~\ref{lem::stack::appendix::vis-proof::interval-misses-lemma-plain} if we can show the third
hypothesis in that lemma. 

So, let $o' \notin \elimSet{}$ such that $\idProp{u} = \idProp{o'}$. We can freely use the structural invariants,
since Lemma~\ref{lem::stack::appendix::vis-proof::struct-invariants-hold-interval} already proved them.
Additionally, we can use any lemma of Section~\ref{subsect::stack::appendix::vis-proof::axioms-from-invariants}
as long as it \emph{only} uses structural invariants.

Since $o' \notin \elimSet{}$, Lemma~\ref{lem::stack::appendix::vis-proof::elim-set-basic-disj-fact} implies 
$\precedesAbs{u}{o'}$. Since $\DEF{\idProp{o'}}$, \axiomIRef{inv::stack::appendix::vis-proof::pop-ids-imply-span} implies 
$\DEF{\spanEv{o'}}$. Hence, Lemma~\ref{lem::stack::appendix::event-prec-implies-span-prec} implies 
$\spanEv{u} \precedesSpansSymbol \spanEv{o'}$, which means $\stampProp{u} = \stampProp{o'}$
by \axiomIRef{inv::stack::appendix::vis-proof::timestamps-equal}. 
In other words, $\AstampProp{u} = \AstampProp{o'}$, which implies 
$\precedesReps {\twoSpan{\spanEv {o}}} {\twoSpan{\spanEv {o'}}}$ by the third hypothesis.
\end{prf}

\begin{lem}[Disjoint Generated Timestamps Lemma]
	\label{lem::stack::appendix::vis-proof::disjoint-gen-ts-lemma}
If $\precedesSpans{\spanEv{u_1}}{\spanEv{u_2}}$, then $\AstampProp{u_1} \stamporderSymbol \AstampProp{u_2}$.
\end{lem}

\begin{prf}
Since $\precedesSpans{\spanEv{u_1}}{\spanEv{u_2}}$, the invocations of $\newStampAlg$ in 
Figure~\ref{fig::sub::stack::appendix::vis-proof::interval-timestamps-gen} were disjoint in time. 
Let $[a_1,b_1]$ and $[a_2,b_2]$ be the intervals generated by $u_1$ and $u_2$, respectively.

Notice how the returned intervals at lines \ref{alg::stack::appendix::vis-proof::ts-gen::interval1},
\ref{alg::stack::appendix::vis-proof::ts-gen::interval2}, \ref{alg::stack::appendix::vis-proof::ts-gen::interval3}
have as lower bound the first read to the variable $TS$ at the start of the procedure 
(line~\ref{alg::stack::appendix::vis-proof::ts-gen::first-read}). While those intervals always have as
upper bound a value strictly smaller than the value of $TS$ at the moment the procedure returns.

This means that $b_1 \natorderSymbol TS_f \natorderSymbol TS_1 \natorderSymbol \ldots \natorderSymbol TS_m \natorderEqSymbol a_2$,
where $TS_f$ is the value of the $TS$ variable at the moment the call to $\newStampAlg$ returned for $u_1$
($TS_f$ may also be the result of calls to $\newStampAlg$ done by other threads), 
and $TS_1$, ..., $TS_m$ are increases to the global $TS$ variable done by other threads. 
Therefore, $[a_1,b_1] \stamporderSymbol [a_2,b_2]$, and so $(\idProp {u_1}, [a_1,b_1]) \stamporderSymbol (\idProp {u_2}, [a_2,b_2])$
by definition of the overloading of $\stamporderSymbol$ for abstract timestamps.
\end{prf}

\begin{lem}[Negative Transitivity Lemma: Plain Timestamps]
	\label{lem::stack::appendix::vis-proof::negative-trans-lemma-plain-ts}
	If,
	\begin{itemize}
		\item $\DEF{\stampProp{u_1}}$, $\DEF{\stampProp{u_2}}$ and $\DEF{\stampProp{u_3}}$,
		\item $\stampProp{u_1} \not\stamporderSymbol \stampProp{u_2} \not\stamporderSymbol \stampProp{u_3}$,
		\item $u_1, u_2, u_3 \notin \elimSet{}$,
		\item $\idProp{u_1} = \idProp{o_1}$ and $\idProp{u_2} = \idProp{o_2}$,
		\item $\precedesReps {\twoSpan{\spanEv{o_1}}} {\twoSpan{\spanEv{o_2}}}$,
		\item $\forall o_3\notin\elimSet{}.\ \idProp{u_3} = \idProp{o_3} \implies \precedesReps {\twoSpan{\spanEv{o_2}}} {\twoSpan{\spanEv{o_3}}}$.
	\end{itemize}
	then, $\stampProp{u_1} \not\stamporderSymbol \stampProp{u_3}$.
\end{lem}

\begin{prf}
	First, we prove the following claim.
			
			Claim: Pops $o_1$ and $o_2$ cannot take a node by reaching line~\ref{alg::stack::appendix::vis-proof::pop::try-take-elimination-pair} because $u_1$ and $u_2$ are not in an elimination pair by hypothesis (i.e., $u_1, u_2 \notin \elimSet{}$).
			
			We focus on $o_1$, since the argument for $o_2$ is similar.
			Indeed, if $o_1$ executed such line, then we must have $t_{start} \stamporderSymbol \stampProp{u_1}$ 
			(the condition at line~\ref{alg::stack::appendix::vis-proof::pop::compare-with-start-stamp}), 
			where $\stampProp{u_1}$ is the timestamp of the node $i_1$ taken by $o_1$ and $t_{start}$ is the timestamp generated by $o_1$ at 
			line~\ref{alg::stack::appendix::vis-proof::pop::create-start-stamp}. 
			Node $i_1$ was inserted by push $u_1$ by the hypothesis
			$\idProp{u_1} = \idProp{o_1}$. But since $u_1 \notin \elimSet{}$, we also have $\precedesAbs{u_1}{o_1}$,
			meaning that the calls to $\newStampAlg$ executed by $u_1$ 
			(line~\ref{alg::stack::appendix::vis-proof::push::create-stamp}) and by $o_1$ 
			(line~\ref{alg::stack::appendix::vis-proof::pop::create-start-stamp}) are disjoint, and 
			so $\stampProp{u_1} \stamporderSymbol t_{start}$ (Contradiction).
			This proves the claim.
	
	In particular, the above claim directly implies that pops $o_1$ and $o_2$ cannot prematurely end the scan of the pools
	by entering line 26. Instead, the pops are forced to scan all the pools and take a node by reaching line 31.
	
	We now focus on the lemma. Suppose for a contradiction that $\stampProp{u_1} \stamporderSymbol \stampProp{u_3}$.
	Denote by $i_1$, $i_2$, and $i_3$ the node ids (or equivalently, the nodes) inserted by pushes $u_1$, $u_2$, $u_3$, 
	respectively. Call the pool indexes where nodes $i_1$, $i_2$ and $i_3$ have been inserted, $k_1$, $k_2$ and $k_3$, respectively.
	In what follows, we will usually say that ``a pop $o$ visits pool $k$'' to mean that variable $pool$ at line 20 was set to $k$ 
	and the iteration for pool $k$ is about to start at line 21.
	
	Since $\stampProp{u_2} \not\stamporderSymbol \stampProp{u_3}$, then $\spanEv{u_2} \not\precedesSpansSymbol \spanEv{u_3}$, 
	otherwise $\stampProp{u_2} \stamporderSymbol \stampProp{u_3}$ by Lemma~\ref{lem::stack::appendix::vis-proof::disjoint-gen-ts-lemma}.
	Therefore, $\oneSpan{\spanEv{u_3}} \leq \twoSpan{\spanEv{u_2}}$.
	
	Also, since $u_2\notin\elimSet{}$ and $\idProp{u_2} = \idProp{o_2}$, 
	Lemma~\ref{lem::stack::appendix::vis-proof::eq-id-implies-eq-timestamp} implies $\spanEv{u_2} \precedesSpansSymbol \spanEv{o_2}$.\footnote{We can use 
	Lemma~\ref{lem::stack::appendix::vis-proof::eq-id-implies-eq-timestamp} because its proof only uses structural invariants, which have been proved already by Lemma~\ref{lem::stack::appendix::vis-proof::struct-invariants-hold-interval}.}
	This means that nodes $i_2$ and $i_3$ (because $\oneSpan{\spanEv{u_3}} \leq \twoSpan{\spanEv{u_2}}$) have been inserted into $k_2$ and $k_3$ before the last iteration of $o_2$ starts. 
	
	Additionally, by the 6th hypothesis, nodes $i_2$ and $i_3$ remain untaken during the entire loop of $o_2$ (up to the point when $o_2$ takes
	$i_2$). Note that the 6th hypothesis implies that $i_2 \neq i_3$, otherwise we would obtain $\precedesReps {\twoSpan{\spanEv{o_2}}} {\twoSpan{\spanEv{o_2}}}$
	as $o_2$ would take both $u_2$ and $u_3$. 
	
	We know $\stampProp{u_2} \not\stamporderSymbol \stampProp{u_3}$, but we must also have $\stampProp{u_3} \not\stamporderSymbol \stampProp{u_2}$,
	for otherwise $\stampProp{u_1} \stamporderSymbol \stampProp{u_3} \stamporderSymbol \stampProp{u_2}$ which contradicts first hypothesis.
	Since timestamps $\stampProp{u_2}$ and $\stampProp{u_3}$ cannot be compared, nodes $i_2$ and $i_3$ must have been inserted into different pools (i.e., $k_2 \neq k_3$) because nodes in the same pool have strictly increasing timestamps as they are inserted by the same thread, meaning that the push spans executed
	by the same thread are disjoint in time, producing strictly increasing timestamps by Lemma~\ref{lem::stack::appendix::vis-proof::disjoint-gen-ts-lemma}.
	With a similar argument one can conclude that $k_1 \neq k_2$. Therefore, either $k_2$ appears first than $k_3$ or the other way around. 
	
	\begin{itemize}
		\item Case $k_3 < k_2$.
		
		Since $k_3$ has an untaken node during the loop of $o_2$, denote by $i_{k_3}$ the node that $o_2$ found at $k_3$ and by $t_{k_3}$ the timestamp
		of such node obtained at line 24. Hence $\stampProp{u_3} \stamporderEqSymbol t_{k_3}$, because
		$i_{k_3}$ was found first in the pool $k_3$ and timestamps in each pool are strictly decreasing by 
		Lemma~\ref{lem::stack::appendix::vis-proof::decreasing-stamps-in-pool-interval} (the equality 
		is necessary, because $i_{k_3}$ and $i_3$ could be the same node).
		
		We must have $t_{k_3} \not\stamporderSymbol \stampProp{u_2}$, otherwise $\stampProp{u_3} \stamporderEqSymbol t_{k_3} \stamporderSymbol \stampProp{u_2}$ 
		(Contradiction).
		
		Variables $chosen$ and $maxT$ cannot have $null$ and $-\infty$ when line 24 executed, otherwise the conditional at line 27 would succeed, assigning 
		$t_{k_3}$ to $maxT$, and since $\stampProp{u_2}$ is eventually assigned to $maxT$ (since $k_3 < k_2$) we would have $t_{k_3} \stamporderSymbol \stampProp{u_2}$ (Contradiction). 
		
		Also, the conditional at line 25 must fail, otherwise $o_2$ would take $i_{k_3}$ and not $i_2$ (which is in a different pool). More specifically,
		if the CAS at line 71 succeeds, then $o_2$ would take $i_{k_3}$. If the CAS does not succeed, then $\tryRemoveAlg$ would return $\bot$, contradicting 
		that it is the last iteration of $o_2$.
		
		Hence, the $chosen$ and $maxT$ variables were assigned when the loop checked a previous pool. Denote by $k_j$ the pool where $chosen$ was found (hence $k_j < k_3$). Denote by $i_j$ and $t_j$ the node and its timestamp (as read by line 24) that $o_2$ found in $k_j$. Also, denote by $c_j$ the rep event 
		at line 48  that found $i_j$ to be not taken.
		\begin{equation}
		\label{eq::stack::appendix::vis-proof::neg-trans-1}
		\text{If $i_j$ is eventually taken by some successful CAS $x$ at line 71, then $c_j < x$}
		\end{equation}
		Also, $t_j \not\stamporderSymbol t_{k_3}$, otherwise the conditional at line 27 would have succeeded, assigning to $maxT$ $t_{k_3}$ and implying
		$t_{k_3} \stamporderSymbol \stampProp{u_2}$ (Contradiction: since $o_2$ eventually assigns to $maxT$ the timestamp $\stampProp{u_2}$
		of the taken node, and the values assigned to $maxT$ are strictly increasing).
		
		Also $\stampProp{u_1} \not\stamporderSymbol t_j$, otherwise $\stampProp{u_1} \stamporderSymbol t_j \stamporderSymbol \stampProp{u_2}$ (Contradiction).
		Also $t_j \not\stamporderSymbol \stampProp{u_3}$, otherwise $t_j \stamporderSymbol \stampProp{u_3} \stamporderEqSymbol t_{k_3}$ (Contradiction).
		Also, since $t_j \stamporderSymbol \stampProp{u_2}$, $t_j$ is finite and was created by some push span $b_{t_j}$. 
		
		Since $\stampProp{u_1} \not\stamporderSymbol t_j \not\stamporderSymbol \stampProp{u_3}$, we have
		$\oneSpan{b_{t_j}} \leq \twoSpan{\spanEv{u_1}}$ and $\oneSpan{\spanEv{u_3}} \leq \twoSpan{b_{t_j}}$ by Lemma~\ref{lem::stack::appendix::vis-proof::disjoint-gen-ts-lemma}.
		
		By a similar argument for $o_1$, we know that nodes $i_1$ and $i_2$ are present in pools $k_1$ and $k_2$ and untaken during the 
		entire loop of $o_1$.
		
		We claim that by the time the loop of $o_1$ finishes visiting pool $k_j$, node $i_3$ has been already inserted in pool $k_3$
		and will not be taken during the iteration of $o_1$.
		
		To prove the claim, we switch point of view to $o_1$. While $o_1$ visits $k_j$, denote by $c_1$ the rep event at line 22 that assigns to variable $n$ the result of calling
		$\getYoungAlg$. Since $\oneSpan{b_{t_j}} \leq \twoSpan{\spanEv{u_1}}$, $i_j$ was inserted into the pool $k_j$ before $o_1$ started, and $i_j$
		may or may not have been taken before $c_1$.
		
		\begin{itemize}
			\item If $i_j$ has been taken before $c_1$, then by \eqref{eq::stack::appendix::vis-proof::neg-trans-1}, 
			$c_j < x < c_1$, where $x$ is the CAS that took $i_j$.
			
			But this means that $i_3$ is already in the $k_3$ pool because $c_j$ is executed by $o_2$ and $i_3$ remains present and untaken during $o_2$'s loop.
			Also, since $i_3$ is taken after $i_1$, node $i_3$ will remain untaken during the rest of the iteration of $o_1$.
			
			\item $i_j$ has not been taken before $c_1$. Then, variable $n$ is not null, because $k_j$ contains at least $i_j$. 
			Denote by $i_m$ the node assigned to variable $n$ and by $t_m$ the timestamp read at line 24. Denote by $c_2$ the rep event at line 24.
			$i_j$ may or may not have been taken before $c_2$.
			
			If $i_j$ was taken before $c_2$, then by a similar argument as for $c_1$, $i_3$ is already in the $k_3$ pool and will remain there during the 
			iteration of $o_1$. 
			
			If $i_j$ was not taken before $c_2$, then $i_m$ and $i_j$ were simultaneous in the $k_j$ pool, which means
			$t_j' \stamporderEqSymbol t_m$ where $t_j'$ is the timestamp that $i_j$ had when $c_2$ executed (i.e., either $t_j' = \infty$ or
			$t_j' = t_j$). 
			
			While visiting the $k_j$ pool, the current value of the $maxT$ variable for $o_1$
			must be finite, because the maximum timestamp that $maxT$ can take for $o_1$ is $\stampProp{u_1}$ 
			(i.e., the timestamp of the node taken by $o_1$). Therefore, if $t_j' = \infty$, then $t_m = \infty$,
			which means that $maxT$ will be assigned $t_m = \infty$, contradicting that the maximum timestamp that $maxT$ is ever
			assigned is $\stampProp{u_1} \stamporderSymbol \infty$.
			So, it must be the case that $t_j' = t_j$, meaning that
			$i_j$ has already been assigned a finite timestamp, i.e., $b_{t_j}$ has finished before $c_2$. 
			
			But $\oneSpan{\spanEv{u_3}} \leq \twoSpan{b_{t_j}}$, meaning that $i_3$ has already been inserted into pool $k_3$ before $c_2$ executed.
		\end{itemize}
		This proves the claim. 
		
		Now, either $k_1 < k_3$ or $k_3 < k_1$ or $k_1 = k_3$.
		
		\begin{itemize}
		\item Case $k_1 < k_3$. 
		
		Once $o_1$ sets the $maxT$ variable to $\stampProp{u_1}$ (line 28) while visiting the pool $k_1$, the $maxT$ variable cannot change afterwards, because $o_1$ takes $u_1$. However, $o_1$ still needs to visit $k_3$. 
		Since $k_j < k_3$, before visiting $k_3$, $o_1$ will have to visit $k_j$. By the claim above, by the time $o_1$ reaches $k_3$, node $i_3$
		is already in $k_3$. But $maxT = \stampProp{u_1} \stamporderSymbol t_{i_3} \stamporderEqSymbol t_{m_3}$, where $t_{i_3}$ is the timestamp of node $i_3$ while $o_1$ visits $k_3$ (where we must have either $t_{i_3} = \infty$ or $t_{i_3} = \stampProp{u_3}$), and $t_{m_3}$ is the timestamp of the found untaken node in $k_3$ (lines 22 and 24). We have $t_{i_3} \stamporderEqSymbol t_{m_3}$
		because nodes in a single pool have strictly decreasing timestamps by Lemma~\ref{lem::stack::appendix::vis-proof::decreasing-stamps-in-pool-interval} 
		(the equality is necessary because $i_3$ could be the found node).
		Hence, variable $maxT$ is assigned the bigger timestamp $t_{m_3}$ at line 28, once $o_1$ visits $k_3$ (Contradiction).
		
		\item Case $k_3 < k_1$. 
		
		By the claim above, by the time $o_1$ reaches $k_3$, node $i_3$
		is already in $k_3$. Denote by $maxT_{k_3}$ the value of the $maxT$ variable just before $o_1$ visits the $k_3$ pool. 
		Since eventually
		$o_1$ assigns the value $\stampProp{u_1}$ to $maxT$ (since $o_1$ takes node $i_1$ from pool $k_1$), 
		we must have $maxT_{k_3} \stamporderSymbol \stampProp{u_1}$ because the timestamps assigned to the $maxT$ variable are strictly
		increasing. Denote by $t_{i_3}$ the timestamp of node $i_3$ while $o_1$ visits $k_3$ (where we must have either $t_{i_3} = \infty$ or $t_{i_3} = \stampProp{u_3}$), and by $t_{m_3}$ the timestamp of the found untaken node in $k_3$ (lines 22 and 24). So, $t_{i_3} \stamporderEqSymbol t_{m_3}$
		since nodes in a single pool have strictly decreasing timestamps by Lemma~\ref{lem::stack::appendix::vis-proof::decreasing-stamps-in-pool-interval} 
		(the equality is necessary because $i_3$ could be the found node).
		But, if $t_{i_3} = \infty$ or $t_{i_3} = \stampProp{u_3}$, then $\stampProp{u_1} \stamporderSymbol t_{i_3}$ because $\stampProp{u_1}$ is finite
		and $\stampProp{u_1} \stamporderSymbol \stampProp{u_3}$ by hypothesis. Hence, $maxT_{k_3} \stamporderSymbol \stampProp{u_1} \stamporderSymbol t_{i_3} \stamporderEqSymbol t_{m_3}$. This means that while visiting $k_3$, variable $maxT$ will be set to $t_{m_3}$ at line 28, which
		is a bigger timestamp than $\stampProp{u_1}$, even though $o_1$ has to scan pool $k_1$ still (Contradiction).
		
		\item Case $k_1 = k_3$.
		
		By the claim above, by the time $o_1$ reaches $k_1$, both nodes $i_1$ and $i_3$ have been inserted into $k_1$. 
		We must have $i_1 \neq i_3$ because $i_1$'s timestamp is smaller than $i_3$'s by hypothesis (and also, $i_3$'s timestamp could be infinite 
		at this moment). Since $i_3$'s timestamp is bigger than $i_1$'s, the $i_3$ node must appear earlier in the pool than $i_1$, because
		nodes in a pool have strictly decreasing timestamps starting from the top (Lemma~\ref{lem::stack::appendix::vis-proof::decreasing-stamps-in-pool-interval}). 
		Hence, the found node at lines 22 and 24 will be either
		$i_3$ or some node with even bigger timestamp, contradicting that $o_1$ took $i_1$.
		\end{itemize}
	
		\item Case $k_2 < k_3$. 
		
		We claim that by the time $o_1$ finishes visiting pool $k_2$, node $i_3$ is already present in pool $k_3$.
		
		Since $\stampProp{u_1} \not\stamporderSymbol \stampProp{u_2}$, then $\spanEv{u_1} \not\precedesSpansSymbol \spanEv{u_2}$, 
		otherwise $\stampProp{u_1} \stamporderSymbol \stampProp{u_2}$ by Lemma~\ref{lem::stack::appendix::vis-proof::disjoint-gen-ts-lemma}.
		Therefore, $\oneSpan{\spanEv{u_2}} \leq \twoSpan{\spanEv{u_1}}$.
		
		Since $\oneSpan{\spanEv{u_2}} \leq \twoSpan{\spanEv{u_1}}$ and $i_2$ is taken after $i_1$, we have that $i_1$ and $i_2$ are present and remain untaken during the entire last loop of $o_1$.
		In particular, $k_2$ is not empty when $o_1$ visits $k_2$.
		
		Denote by $i_m$ the node assigned to variable $n$ and by $t_m$ the timestamp read at line 24. Denote by $c_2$ the rep event at line 24.
		Denote by $t_2'$ the timestamp that $i_2$ had when $c_2$ executed (i.e., either $t_2' = \infty$ or
		$t_2' = \stampProp{u_2}$). We have $t_2' \stamporderEqSymbol t_m$ because nodes in the same pool have strictly
		decreasing timestamps (Lemma~\ref{lem::stack::appendix::vis-proof::decreasing-stamps-in-pool-interval}), 
		and the equality is needed because $i_2$ could be the found node.
		
		While visiting the $k_2$ pool, the current value of the $maxT$ variable for $o_1$
		must be finite, because the maximum timestamp that $maxT$ can take for $o_1$ is $\stampProp{u_1}$ 
		(i.e., the timestamp of the node taken by $o_1$). Therefore, if $t_2' = \infty$, then $t_m = \infty$,
		which means that $maxT$ will be assigned $t_m = \infty$, contradicting that the maximum timestamp that $maxT$ is ever
		assigned is $\stampProp{u_1} \stamporderSymbol \infty$.
		So, it must be the case that $t_2' = \stampProp{u_2}$, meaning that
		$i_2$ has already been assigned a finite timestamp, i.e., $\spanEv{u_2}$ has finished before $c_2$.
		
		But $\oneSpan{\spanEv{u_3}} \leq \twoSpan{\spanEv{u_2}}$, meaning that $i_3$ has already been inserted into pool $k_3$ before $c_2$ executed.
		In addition, $i_3$ remains untaken during $o_1$'s loop because of the 6th hypothesis.
		This proves the claim.
		
		Now, either $k_1 < k_3$ or $k_3 < k_1$ or $k_1 = k_3$. The rest of the proof is identical to the cases for
		$k_1 < k_3$ and $k_3 < k_1$ and $k_1 = k_3$ in the previous case (i.e., case for $k_3 < k_2$), but using pool $k_2$
		whenever there is a reference to pool $k_j$ in the cases $k_1 < k_3$, $k_3 < k_1$, $k_1 = k_3$.
	\end{itemize}
\end{prf}

\begin{lem}[Negative Transitivity Lemma: Abstract Timestamps]
	\label{lem::stack::appendix::vis-proof::negative-trans-lemma}
	If,
	\begin{itemize}
		\item $\DEF{\AstampProp{u_1}}$, $\DEF{\AstampProp{u_2}}$ and $\DEF{\AstampProp{u_3}}$,
		\item $\AstampProp{u_1} \not\stamporderSymbol \AstampProp{u_2} \not\stamporderSymbol \AstampProp{u_3}$,
		\item $u_1, u_2, u_3 \notin \elimSet{}$,
		\item $\AstampProp{u_1} = \AstampProp{o_1}$ and $\AstampProp{u_2} = \AstampProp{o_2}$,
		\item $\precedesReps {\twoSpan{\spanEv{o_1}}} {\twoSpan{\spanEv{o_2}}}$,
		\item $\forall o_3\notin\elimSet{}.\ \AstampProp{u_3} = \AstampProp{o_3} \implies \precedesReps {\twoSpan{\spanEv{o_2}}} {\twoSpan{\spanEv{o_3}}}$.
	\end{itemize}
	then, $\AstampProp{u_1} \not\stamporderSymbol \AstampProp{u_3}$.
\end{lem}

\begin{prf}
From the first hypothesis, we obtain $\DEF{\stampProp{u_1}}$, $\DEF{\stampProp{u_2}}$ and $\DEF{\stampProp{u_3}}$.

From the second hypothesis, $\stampProp{u_1} \not\stamporderSymbol \stampProp{u_2} \not\stamporderSymbol \stampProp{u_3}$ by overloading of $\stamporderSymbol$.

From the fourth hypothesis, $\idProp{u_1} = \idProp{o_1}$ and $\idProp{u_2} = \AstampProp{o_2}$ by definition.

So, from Lemma~\ref{lem::stack::appendix::vis-proof::negative-trans-lemma-plain-ts}, we would obtain $\stampProp{u_1} \not\stamporderSymbol \stampProp{u_3}$, and hence $\AstampProp{u_1} \not\stamporderSymbol \AstampProp{u_3}$ if we can show the sixth hypothesis of Lemma~\ref{lem::stack::appendix::vis-proof::negative-trans-lemma-plain-ts}.

So, let $o_3 \notin \elimSet{}$ such that $\idProp{u_3} = \idProp{o_3}$. We can freely use the structural invariants,
since Lemma~\ref{lem::stack::appendix::vis-proof::struct-invariants-hold-interval} already proved them.
Additionally, we can use any lemma of Section~\ref{subsect::stack::appendix::vis-proof::axioms-from-invariants}
as long as it \emph{only} uses structural invariants.

Since $o_3 \notin \elimSet{}$, Lemma~\ref{lem::stack::appendix::vis-proof::elim-set-basic-disj-fact} implies 
$\precedesAbs{u_3}{o_3}$. Since $\DEF{\idProp{o_3}}$, \axiomIRef{inv::stack::appendix::vis-proof::pop-ids-imply-span} implies 
$\DEF{\spanEv{o_3}}$. Hence, Lemma~\ref{lem::stack::appendix::event-prec-implies-span-prec} implies 
$\spanEv{u_3} \precedesSpansSymbol \spanEv{o_3}$, which means $\stampProp{u_3} = \stampProp{o_3}$
by \axiomIRef{inv::stack::appendix::vis-proof::timestamps-equal}. 
In other words, $\AstampProp{u_3} = \AstampProp{o_3}$, which implies 
$\precedesReps {\twoSpan{\spanEv {o_2}}} {\twoSpan{\spanEv {o_3}}}$ by the sixth hypothesis.
\end{prf}

And now the main lemmas.

\begin{lem}
	$\GstamporderSymbol$ is a strict partial order.
\end{lem}

\begin{prf}
	By Lemma~\ref{defn::stack::appendix::vis-proof::abstract-timestamps-order-is-strict},\footnote{Note that 
		the proof of Lemma~\ref{defn::stack::appendix::vis-proof::abstract-timestamps-order-is-strict} 
		does not make use of any invariants, because the result
		follows directly from definitions. Therefore, we can use
		Lemma~\ref{defn::stack::appendix::vis-proof::abstract-timestamps-order-is-strict} safely in our proofs.}
	$\stamporderSymbol$ is still a strict partial order when it is extended to abstract timestamps.
	
	Also, we can freely use the structural invariants,
	since Lemma~\ref{lem::stack::appendix::vis-proof::struct-invariants-hold-interval} already proved them.
	
	We now show that $\GstamporderSymbol$ is a strict partial order.
	
\begin{itemize}
	\item Irreflexivity.
	
	Suppose $t \GstamporderSymbol t$. Since $t \stamporderSymbol t$ cannot hold, we must have $\TB {u_1} {u_2}\ \wedge\ \AstampProp {u_1}  \not\stamporderSymbol \AstampProp {u_2}\ \wedge\ t \stamporderEqSymbol \AstampProp {u_2}\ \wedge\ \AstampProp {u_1} \stamporderEqSymbol t$ for some $u_1$ and $u_2$.
	Hence, $\AstampProp {u_1} \stamporderEqSymbol \stampProp {u_2}$. But $\AstampProp {u_1}  \not\stamporderSymbol \AstampProp {u_2}$, so $\AstampProp {u_1} = \AstampProp {u_2}$, which implies $\idProp {u_1} = \idProp{u_2}$ by definition of $\AstampPropName$.
	But then, $u_1 = u_2$ by \axiomIRef{inv::stack::appendix::vis-proof::prop-funcs-are-injective}, and 
	from $\TB {u_1} {u_2}$ (equivalently, $\TB {u_1} {u_1}$) we obtain $\precedesReps {\twoSpan{\spanEv{o_1}}} {\twoSpan{\spanEv{o_1}}}$ for some
	pop $o_1$ such that $\AstampProp{u_1} = \AstampProp{o_1}$ (Contradiction).

	\item Transitivity. 
	
	From $t_1 \GstamporderSymbol t_2$ and $t_2 \GstamporderSymbol t_3$, we need to consider four cases,
	\begin{itemize}
		\item Case $t_1 \stamporderSymbol t_2$ and $t_2 \stamporderSymbol t_3$. Hence $t_1 \stamporderSymbol t_3$,
		meaning $t_1 \GstamporderSymbol t_3$ by definition.
		
		\item Case $t_1 \stamporderSymbol t_2$ and $\exists u^{h_2}, u^{l_2} \notin \elimSet{}.\ \TB {u^{h_2}} {u^{l_2}}\ \wedge\ \AstampProp {u^{h_2}}  \not\stamporderSymbol \AstampProp {u^{l_2}}\ \wedge\ 
		 t_2 \stamporderEqSymbol \AstampProp {u^{l_2}}\ \wedge\ \AstampProp {u^{h_2}} \stamporderEqSymbol t_3$.
		 
		 Hence, $t_1 \stamporderSymbol t_2 \stamporderEqSymbol \AstampProp {u^{l_2}}$.
		 
		 In other words, $\TB {u^{h_2}} {u^{l_2}}\ \wedge\ \AstampProp {u^{h_2}}  \not\stamporderSymbol \AstampProp {u^{l_2}}\ \wedge\ 
		 t_1 \stamporderEqSymbol \AstampProp {u^{l_2}}$ and $\AstampProp {u^{h_2}} \stamporderEqSymbol t_3$.
		 So, $t_1 \GstamporderSymbol t_3$ by definition.
		 
		\item Case $\exists u^{h_1}, u^{l_1}\notin \elimSet{}.\ \TB {u^{h_1}} {u^{l_1}}\ \wedge\ \AstampProp {u^{h_1}}  \not\stamporderSymbol \AstampProp {u^{l_1}}\ \wedge\ 
		t_1 \stamporderEqSymbol \AstampProp {u^{l_1}}\ \wedge\ \AstampProp {u^{h_1}} \stamporderEqSymbol t_2$ and $t_2 \stamporderSymbol t_3$.
		 
		 Hence, $\AstampProp {u^{h_1}} \stamporderEqSymbol t_2 \stamporderSymbol t_3$. 
		 
		 In other words, $\TB {u^{h_1}} {u^{l_1}}\ \wedge\ \AstampProp {u^{h_1}}  \not\stamporderSymbol \AstampProp {u^{l_1}}\ \wedge\ 
		 t_1 \stamporderEqSymbol \AstampProp {u^{l_1}}\ \wedge\ \AstampProp {u^{h_1}} \stamporderEqSymbol t_3$.
		 So, $t_1 \GstamporderSymbol t_3$ by definition.
		 
		\item Case $\exists u^{h_1}, u^{l_1} \notin \elimSet{}.\ \TB {u^{h_1}} {u^{l_1}}\ \wedge\ \AstampProp {u^{h_1}}  \not\stamporderSymbol \AstampProp {u^{l_1}}\ \wedge\ 
		t_1 \stamporderEqSymbol \AstampProp {u^{l_1}}\ \wedge\ \AstampProp {u^{h_1}} \stamporderEqSymbol t_2$ and 
		$\exists u^{h_2}, u^{l_2} \notin \elimSet{}.\ \TB {u^{h_2}} {u^{l_2}}\ \wedge\ \AstampProp {u^{h_2}}  \not\stamporderSymbol \AstampProp {u^{l_2}}\ \wedge\ 
		t_2 \stamporderEqSymbol \AstampProp {u^{l_2}}\ \wedge\ \AstampProp {u^{h_2}} \stamporderEqSymbol t_3$.
		
		Either $\AstampProp {u^{h_1}} \stamporderSymbol \AstampProp {u^{h_2}}$ or not.
		If $\AstampProp {u^{h_1}} \stamporderSymbol \AstampProp {u^{h_2}}$ holds, then $\AstampProp {u^{h_1}} \stamporderSymbol \AstampProp {u^{h_2}} \stamporderEqSymbol t_3$
		and so $t_1 \GstamporderSymbol t_3$ by definition.
		
		Therefore, we can assume $\AstampProp {u^{h_1}} \not\stamporderSymbol \AstampProp {u^{h_2}}$ the rest of the proof.
		 
		Claim: $\AstampProp {u^{h_2}} \not\stamporderSymbol \AstampProp {u^{h_1}}$. Suppose $\AstampProp {u^{h_2}} \stamporderSymbol \AstampProp {u^{h_1}}$, then $\AstampProp {u^{h_2}} \stamporderSymbol \AstampProp {u^{h_1}}
		\stamporderEqSymbol t_2 \stamporderEqSymbol \AstampProp {u^{l_2}}$, which contradicts the hypothesis $\AstampProp {u^{h_2}}  \not\stamporderSymbol \AstampProp {u^{l_2}}$. This proves the claim.
		
		Since $\TB {u^{h_1}} {u^{l_1}}$ and $\TB {u^{h_2}} {u^{l_2}}$, there are pops $o^{h_1}$, $o^{h_2}$ such that the following facts
		hold,
		
		\begin{enumerate}
			\item $u^{h_1}, u^{h_2}, u^{l_1}, u^{l_2}, o^{h_1}, o^{h_2} \notin \elimSet{}$.
			\item $\AstampProp{u^{h_1}} = \AstampProp{o^{h_1}}$ and $\AstampProp{u^{h_2}} = \AstampProp{o^{h_2}}$.
			\item $\forall o^{l_1} \notin \elimSet{}.\ \AstampProp{u^{l_1}} = \AstampProp{o^{l_1}} \implies \precedesReps {\twoSpan{\spanEv{o^{h_1}}}} {\twoSpan{\spanEv{o^{l_1}}}}$.
			\item $\forall o^{l_2} \notin \elimSet{}.\ \AstampProp{u^{l_2}} = \AstampProp{o^{l_2}} \implies \precedesReps {\twoSpan{\spanEv{o^{h_2}}}} {\twoSpan{\spanEv{o^{l_2}}}}$.
		\end{enumerate}
		
		Now, we have $\AstampProp {u^{h_1}} \stamporderEqSymbol t_2 \stamporderEqSymbol \AstampProp {u^{l_2}}$, i.e.,
		$\AstampProp {u^{h_1}} \stamporderEqSymbol \AstampProp {u^{l_2}}$. So, we have two cases,
		\begin{itemize}
			\item Case $\AstampProp {u^{h_1}} \stamporderSymbol \AstampProp {u^{l_2}}$.
			
			Claim: $\twoSpan{\spanEv{o^{h_1}}} \not\precedesRepsSymbol \twoSpan{\spanEv{o^{h_2}}}$. If $\precedesReps {\twoSpan{\spanEv{o^{h_1}}}} {\twoSpan{\spanEv{o^{h_2}}}}$, then by Lemma~\ref{lem::stack::appendix::vis-proof::negative-trans-lemma} (Negative Transitivity Lemma) 
			with $\AstampProp{u^{h_1}} \not\stamporderSymbol \AstampProp{u^{h_2}} \not\stamporderSymbol \AstampProp{u^{l_2}}$ and
			facts 1,2,4 above, we obtain 
			$\AstampProp{u^{h_1}} \not\stamporderSymbol \AstampProp{u^{l_2}}$ (contradicts $\AstampProp {u^{h_1}} \stamporderSymbol \AstampProp {u^{l_2}}$). 
			This proves the claim.
			
			Claim: $\twoSpan{\spanEv{o^{h_1}}} \neq \twoSpan{\spanEv{o^{h_2}}}$. If $\twoSpan{\spanEv{o^{h_1}}} = \twoSpan{\spanEv{o^{h_2}}}$, then $o^{h_1} = o^{h_2}$ by \axiomIRef{inv::stack::appendix::vis-proof::reps-are-injective} 
			and \axiomIRef{inv::stack::appendix::vis-proof::spans-are-injective}. 
			But from fact 2 above, $\AstampProp{u^{h_2}} = \AstampProp{o^{h_2}} = \AstampProp{o^{h_1}} = \AstampProp{u^{h_1}} \stamporderSymbol \AstampProp {u^{l_2}}$, which contradicts the hypothesis $\AstampProp {u^{h_2}}  \not\stamporderSymbol \AstampProp {u^{l_2}}$.
			This proves the claim.
			
			So, by the above claims, $\precedesReps {\twoSpan{\spanEv{o^{h_2}}}} {\twoSpan{\spanEv{o^{h_1}}}}$. And from fact 3 above,
			\begin{align}
			\label{lem::stack::appendix::vis-proof::trans-proof-1}
			\forall o^{l_1}\notin\elimSet{}.\ \AstampProp{u^{l_1}} = \AstampProp{o^{l_1}} \implies \twoSpan{\spanEv{o^{h_2}}} \precedesRepsSymbol \twoSpan{\spanEv{o^{h_1}}} \precedesRepsSymbol \twoSpan{\spanEv{o^{l_1}}}
			\end{align}
			which implies $\TB{u^{h_2}}{u^{l_1}}$ by definition (we already know from facts 1,2 that $u^{h_2} \notin \elimSet{}$ is taken by
			$o^{h_2} \notin \elimSet{}$).
			
			By Lemma~\ref{lem::stack::appendix::vis-proof::negative-trans-lemma} (Negative Transitivity Lemma) 
			with $\AstampProp{u^{h_2}} \not\stamporderSymbol \AstampProp{u^{h_1}} \not\stamporderSymbol \AstampProp{u^{l_1}}$, facts 1,2 above, 
			$\precedesReps {\twoSpan{\spanEv{o^{h_2}}}} {\twoSpan{\spanEv{o^{h_1}}}}$, and
			formula \eqref{lem::stack::appendix::vis-proof::trans-proof-1}, we obtain 
			$\AstampProp{u^{h_2}} \not\stamporderSymbol \AstampProp{u^{l_1}}$.
			
			Hence, $t_1 \GstamporderSymbol t_3$ holds by definition, because $t_1 \stamporderEqSymbol \AstampProp {u^{l_1}}$ and $\AstampProp {u^{h_2}} \stamporderEqSymbol t_3$ are hypotheses,
			and we have $\TB{u^{h_2}}{u^{l_1}}$ and $\AstampProp{u^{h_2}} \not\stamporderSymbol \AstampProp{u^{l_1}}$.
			
			\item Case $\AstampProp {u^{h_1}} = \AstampProp {u^{l_2}}$.
			
			From facts 2,4 we obtain $\precedesReps {\twoSpan{\spanEv{o^{h_2}}}} {\twoSpan{\spanEv{o^{h_1}}}}$,
			since $\AstampProp{u^{l_2}} = \AstampProp{u^{h_1}} = \AstampProp{o^{h_1}}$.
			
			But from fact 3 we get \eqref{lem::stack::appendix::vis-proof::trans-proof-1}
			again, and we will obtain $t_1 \GstamporderSymbol t_3$ by following the steps after formula 
			\eqref{lem::stack::appendix::vis-proof::trans-proof-1}.
		\end{itemize}
	\end{itemize}
\end{itemize}
\end{prf}

\begin{lem}
	\label{lem::stack::appendix::vis-proof::interval-pop-totality}
$\GstamporderSymbol$ is pop-total.
\end{lem}

\begin{prf}
If $\AstampProp{u_1} \stamporderSymbol \AstampProp{u_2}$ holds, then we are done, since $\GstamporderSymbol$ contains $\stamporderSymbol$. 
Similarly if $\AstampProp{u_2} \stamporderSymbol \AstampProp{u_1}$. Hence, we can assume $\AstampProp{u_1} \not\stamporderSymbol \AstampProp{u_2}$
and $\AstampProp{u_2} \not\stamporderSymbol \AstampProp{u_1}$ in the rest of the proof.

We will focus on the case $\AstampProp{u_1} = \AstampProp{o}$, since the other case $\AstampProp{u_2} = \AstampProp{o}$ is similar.

Either there is $o_2 \notin \elimSet{}$ such that $\AstampProp{u_2} = \AstampProp{o_2}$ or not. 

If not, then $\TB{u_1}{u_2}$ holds trivially. 
Also, since $\AstampProp {u_2} \stamporderEqSymbol \AstampProp {u_2}\ \wedge\ \AstampProp {u_1} \stamporderEqSymbol \AstampProp {u_1}$ holds trivially, we have $\AstampProp {u_2} \GstamporderSymbol \AstampProp {u_1}$ by definition. 

Hence, we can assume that there is a pop $o_2 \notin \elimSet{}$ such that $\AstampProp{u_2} = \AstampProp{o_2}$. 
Since $\DEF{\stampProp{o}}$ and $\DEF{\stampProp{o_2}}$, both pops executed their spans by definition of $\stampPropName$.

\begin{itemize}
	\item Case $\precedesReps {\twoSpan{\spanEv{o}}} {\twoSpan{\spanEv{o_2}}}$.
	We claim that $\TB{u_1}{u_2}$. We already know $\AstampProp{u_1} = \AstampProp{o}$.
	Let $\AstampProp{u_2} = \AstampProp{o_2'}$ for some $o_2' \notin \elimSet{}$. 
	But $\AstampProp{o_2'} = \AstampProp{u_2} = \AstampProp{o_2}$, meaning 
	$o_2 = o_2'$ by \axiomIRef{inv::stack::appendix::vis-proof::prop-funcs-are-injective}.
	And from the case hypothesis, ${\twoSpan{\spanEv{o}}} \precedesRepsSymbol {\twoSpan{\spanEv{o_2}}} = {\twoSpan{\spanEv{o_2'}}}$ holds.
	Thus $\TB{u_1}{u_2}$.
	
	Finally, since $\AstampProp {u_2} \stamporderEqSymbol \AstampProp {u_2}\ \wedge\ \AstampProp {u_1} \stamporderEqSymbol \AstampProp {u_1}$ holds trivially, 
	and $\AstampProp{u_1} \not\stamporderSymbol \AstampProp{u_2}$, we have $\AstampProp {u_2} \GstamporderSymbol \AstampProp {u_1}$ by definition.
	
	\item Case $\precedesReps {\twoSpan{\spanEv{o_2}}} {\twoSpan{\spanEv{o}}}$.
	
	We claim that $\TB{u_2}{u_1}$. We already know $\AstampProp{u_2} = \AstampProp{o_2}$.
	Let $\AstampProp{u_1} = \AstampProp{o_1'}$ for some $o_1' \notin \elimSet{}$. But $\AstampProp{o_1'} = \AstampProp{u_1} = \AstampProp{o}$, meaning 
	$o = o_1'$ by \axiomIRef{inv::stack::appendix::vis-proof::prop-funcs-are-injective}. 
	And from the case hypothesis, ${\twoSpan{\spanEv{o_2}}} \precedesRepsSymbol {\twoSpan{\spanEv{o}}} = {\twoSpan{\spanEv{o_1'}}}$ holds.
	Thus $\TB{u_2}{u_1}$.
	
	Finally, since $\AstampProp {u_1} \stamporderEqSymbol \AstampProp {u_1}\ \wedge\ \AstampProp {u_2} \stamporderEqSymbol \AstampProp {u_2}$ holds trivially, 
	and $\AstampProp{u_2} \not\stamporderSymbol \AstampProp{u_1}$, we have $\AstampProp {u_1} \GstamporderSymbol \AstampProp {u_2}$ by definition.
		
	\item Case $\twoSpan {\spanEv{o}} = \twoSpan{\spanEv{o_2}}$. From \axiomIRef{inv::stack::appendix::vis-proof::reps-are-injective} 
	and \axiomIRef{inv::stack::appendix::vis-proof::spans-are-injective}, $o = o_2$. 
	Hence, $\AstampProp{u_1} = \AstampProp o = \AstampProp{o_2} = \AstampProp{u_2}$. Therefore, $\AstampProp{u_1} = \AstampProp{u_2}$.
\end{itemize}
\end{prf}

\begin{lem}
\label{lem::stack::appendix::vis-proof::interval-disjoint-push}
Invariant \axiomIRef{vis-ax::stack::appendix::vis-proof::disjoint-push} holds.
\end{lem}

\begin{prf}
By Lemma~\ref{lem::stack::appendix::vis-proof::disjoint-gen-ts-lemma}, $\AstampProp{u_1} \stamporderSymbol \AstampProp{u_2}$ holds.
Hence, $\AstampProp{u_1} \GstamporderSymbol \AstampProp{u_2}$ by definition.
\end{prf}

\begin{lem}
\label{lem::stack::appendix::vis-proof::interval-misses-are-late}
Invariant \axiomIRef{vis-ax::stack::vis-proof::misses-are-late} holds.
\end{lem}

\begin{prf}
From the hypothesis $\PEND o u$, we have $\AstampProp{o} \GstamporderSymbol \AstampProp{u}$. Hence, there are two cases,

\begin{itemize}
	\item Case $\AstampProp{o} \stamporderSymbol \AstampProp{u}$. The result follows by Lemma~\ref{lem::stack::appendix::vis-proof::misses-lemma} and hypothesis $\PEND o u$.
	
    \item Case there are $u_1$, $u_2 \notin \elimSet{}$, such that $\TB {u_1} {u_2}$, $\AstampProp {u_1}  \not\stamporderSymbol \AstampProp {u_2}$, $\AstampProp{o} \stamporderEqSymbol \AstampProp {u_2}$, and $\AstampProp {u_1} \stamporderEqSymbol \AstampProp{u}$.

Claim: $\AstampProp {u_1} \not\stamporderSymbol \AstampProp{o}$. If $\AstampProp {u_1} \stamporderSymbol \AstampProp{o}$, then $\AstampProp {u_1} \stamporderSymbol \AstampProp{o} \stamporderEqSymbol \AstampProp {u_2}$
(Contradicts case hypothesis $\AstampProp {u_1}  \not\stamporderSymbol \AstampProp {u_2}$). This proves the claim.

Either $\AstampProp {o} \stamporderSymbol \AstampProp{u_1}$ or not.

If $\AstampProp {o} \stamporderSymbol \AstampProp{u_1}$, then $\AstampProp {o} \stamporderSymbol \AstampProp{u_1} \stamporderEqSymbol \AstampProp {u}$. From Lemma~\ref{lem::stack::appendix::vis-proof::misses-lemma}
and hypothesis $\PEND o u$, we obtain $\precedesReps {\oneSpan{\spanEv{o}}} {\oneSpan{\spanEv{u}}}$.

Therefore, we can assume $\AstampProp {o} \not\stamporderSymbol \AstampProp{u_1}$ in the rest of the proof.

Either $\AstampProp {o} \stamporderSymbol \AstampProp{u}$ or not.

If $\AstampProp {o} \stamporderSymbol \AstampProp{u}$, then Lemma~\ref{lem::stack::appendix::vis-proof::misses-lemma} and hypothesis $\PEND o u$ 
imply $\precedesReps {\oneSpan{\spanEv{o}}} {\oneSpan{\spanEv{u}}}$. 

So, we can assume $\AstampProp {o} \not\stamporderSymbol \AstampProp{u}$ in the rest of the proof.

In other words, we have $\AstampProp {u_1} \not\stamporderSymbol \AstampProp{o}$, $\AstampProp {o} \not\stamporderSymbol \AstampProp{u_1}$ and $\AstampProp {o} \not\stamporderSymbol \AstampProp{u}$.
We now show that this leads to a contradiction (meaning that this case is impossible).

From $\AstampProp{o} \stamporderEqSymbol \AstampProp {u_2}$, and $\AstampProp {u_1} \stamporderEqSymbol \AstampProp{u}$, we need to consider four cases.
\begin{itemize}
	\item Case $\AstampProp{o} = \AstampProp {u_2}$, and $\AstampProp {u_1} = \AstampProp{u}$. 
	
	From $\AstampProp {u_1} = \AstampProp{u}$, we have $u_1 = u$ 
	by \axiomIRef{inv::stack::appendix::vis-proof::prop-funcs-are-injective}. 
	From $\TB {u_1} {u_2}$ (equivalently $\TB {u} {u_2}$), we have for some pop $o_1\notin\elimSet{}$ such that $\AstampProp{u} = \AstampProp{o_1}$,
	\[
	\forall o'\notin\elimSet{}.\ \AstampProp{u_2} = \AstampProp {o'} \implies \precedesReps {\twoSpan{\spanEv {o_1}}} {\twoSpan{\spanEv {o'}}}
	\]
	Hence, $\precedesReps {\twoSpan{\spanEv {o_1}}} {\twoSpan{\spanEv {o}}}$.
	
	But from hypothesis $\PEND o u$, we also have,
	\[
	\forall o'\notin \elimSet{}.\ \AstampProp{u} = \AstampProp {o'} \implies \precedesReps {\twoSpan{\spanEv {o}}} {\twoSpan{\spanEv {o'}}}
	\]
	Hence, $\precedesReps {\twoSpan{\spanEv {o}}} {\twoSpan{\spanEv {o_1}}}$.
	
	In other words, $\precedesReps {\twoSpan{\spanEv{o}}} {\twoSpan{\spanEv{o}}}$ (Contradiction).
	
	\item Case $\AstampProp{o} = \AstampProp {u_2}$, and $\AstampProp {u_1} \stamporderSymbol \AstampProp{u}$. 
	
	Since $\TB{u_1}{u_2}$, we have for some $o_1\notin\elimSet{}$ that $\AstampProp{u_1} = \AstampProp{o_1}$ and,
	\[
	\forall o'\notin\elimSet{}.\ \AstampProp{u_2} = \AstampProp {o'} \implies \precedesReps {\twoSpan{\spanEv {o_1}}} {\twoSpan{\spanEv {o'}}}
	\]
	Hence, $\precedesReps {\twoSpan{\spanEv{o_1}}} {\twoSpan{\spanEv{o}}}$. 
	
	Also, from $\PEND o u$, we have $\forall o'\notin\elimSet{}.\ \AstampProp{u} = \AstampProp {o'} \implies 
	\precedesReps {\twoSpan{\spanEv {o}}} {\twoSpan{\spanEv {o'}}}$.
	
	Therefore, by Lemma~\ref{lem::stack::appendix::vis-proof::negative-trans-lemma} (Negative Transitivity Lemma) 
	with $\AstampProp{u_1} \not\stamporderSymbol \AstampProp{u_2} = \AstampProp{o} \not\stamporderSymbol \AstampProp{u}$, we obtain $\AstampProp{u_1} \not\stamporderSymbol \AstampProp{u}$ (contradicts $\AstampProp {u_1} \stamporderSymbol \AstampProp{u}$).
	
	\item Case $\AstampProp{o} \stamporderSymbol \AstampProp {u_2}$ and $\AstampProp {u_1} = \AstampProp{u}$. 
	
	Since $\DEF{\AstampProp{o}}$, part 2 of 
	Lemma~\ref{lem::stack::appendix::vis-proof::eq-id-implies-eq-timestamp}\footnote{We can use 
		Lemma~\ref{lem::stack::appendix::vis-proof::eq-id-implies-eq-timestamp} because its proof
	only uses structural invariants.} imply that there is a push $u_o \notin \elimSet{}$
	such that $\AstampProp{u_o} = \AstampProp{o}$.
	From $\AstampProp {u_1} = \AstampProp{u}$ we have $u_1 = u$ by \axiomIRef{inv::stack::appendix::vis-proof::prop-funcs-are-injective}.
	
	From $\TB{u_1}{u_2}$ 
	(equivalently $\TB{u}{u_2}$) there is a pop $o_1\notin\elimSet{}$ such that $\AstampProp{u} = \AstampProp{o_1}$ and
	$\forall o_2\notin\elimSet{}.\ \AstampProp{u_2} = \AstampProp {o_2} \implies \precedesReps {\twoSpan{\spanEv {o_1}}} {\twoSpan{\spanEv {o_2}}}$.
	Hence, $\AstampProp{o_1} = \AstampProp{u} = \AstampProp{u_1}$.
	
	From $\PEND{o}{u}$, we obtain $\precedesReps {\twoSpan{\spanEv{o}}} {\twoSpan{\spanEv{o_1}}}$.
	Therefore, by Lemma~\ref{lem::stack::appendix::vis-proof::negative-trans-lemma} (Negative Transitivity Lemma) 
	with $\AstampProp{u_o} = \AstampProp{o} \not\stamporderSymbol \AstampProp{u_1} \not\stamporderSymbol \AstampProp{u_2}$, we obtain $\AstampProp{o} = \AstampProp{u_o} \not\stamporderSymbol \AstampProp{u_2}$ (contradicts $\AstampProp{o} \stamporderSymbol \AstampProp {u_2}$).
	
	\item Case $\AstampProp{o} \stamporderSymbol \AstampProp {u_2}$, and $\AstampProp {u_1} \stamporderSymbol \AstampProp{u}$.
	
	Since $\DEF{\AstampProp{o}}$, part 2 of 
	Lemma~\ref{lem::stack::appendix::vis-proof::eq-id-implies-eq-timestamp} implies that there is a push $u_o \notin \elimSet{}$
	such that $\AstampProp{u_o} = \AstampProp{o}$.
	
	From $\TB{u_1}{u_2}$ there is a pop $o_1 \notin \elimSet{}$ such that $\AstampProp{u_1} = \AstampProp{o_1}$ and
	\begin{align}
		\label{eq::stack::appendix::vis-proof::misses-are-late-eq1}
		\forall o_2\notin\elimSet{}.\ \AstampProp{u_2} = \AstampProp {o_2} \implies \precedesReps {\twoSpan{\spanEv {o_1}}} {\twoSpan{\spanEv {o_2}}}
	\end{align}
	
	From $\PEND o u$ we directly have,
	\begin{align}
		\label{eq::stack::appendix::vis-proof::misses-are-late-eq2}
		\forall o'\notin\elimSet{}.\ \AstampProp{u} = \AstampProp {o'} \implies \precedesReps {\twoSpan{\spanEv {o}}} {\twoSpan{\spanEv {o'}}}
	\end{align}
	
	We now compare the spans' end of pops $o_1$ and $o$, for which we have three cases,
	
	\begin{itemize}
		\item Case $\precedesReps {\twoSpan{\spanEv{o}}} {\twoSpan{\spanEv{o_1}}}$.
		
		By Lemma~\ref{lem::stack::appendix::vis-proof::negative-trans-lemma} (Negative Transitivity Lemma) 
		with $\AstampProp{u_o} = \AstampProp{o} \not\stamporderSymbol \AstampProp{u_1} \not\stamporderSymbol \AstampProp{u_2}$ and
		\eqref{eq::stack::appendix::vis-proof::misses-are-late-eq1}, 
		we obtain $\AstampProp{o} = \AstampProp{u_o} \not\stamporderSymbol \AstampProp{u_2}$ (contradicts $\AstampProp{o} \stamporderSymbol \AstampProp {u_2}$). 
		
		\item Case $\precedesReps {\twoSpan{\spanEv{o_1}}} {\twoSpan{\spanEv{o}}}$.
		
		By Lemma~\ref{lem::stack::appendix::vis-proof::negative-trans-lemma} (Negative Transitivity Lemma) 
		with $\AstampProp{u_1} \not\stamporderSymbol \AstampProp{u_o} = \AstampProp{o} \not\stamporderSymbol \AstampProp{u}$
		and \eqref{eq::stack::appendix::vis-proof::misses-are-late-eq2}, 
		we obtain $\AstampProp{u_1} \not\stamporderSymbol \AstampProp{u}$ (contradicts $\AstampProp {u_1} \stamporderSymbol \AstampProp{u}$). 
		
		\item Case $\twoSpan{\spanEv{o_1}} = \twoSpan{\spanEv{o}}$. So, $o_1 = o$ by 
		\axiomIRef{inv::stack::appendix::vis-proof::reps-are-injective} and 
		\axiomIRef{inv::stack::appendix::vis-proof::spans-are-injective}. 
		Therefore, $\AstampProp{u_1} = \AstampProp{o_1} = \AstampProp{o} \stamporderSymbol \AstampProp{u_2}$. In other words, 
		$\AstampProp{u_1} \stamporderSymbol \AstampProp{u_2}$ (Contradicts hypothesis $\AstampProp{u_1} \not\stamporderSymbol \AstampProp{u_2}$).
	\end{itemize}
\end{itemize}
\end{itemize}
\end{prf}

\begin{lem}
	All the key invariants in Figure~\ref{fig::stack::appendix::vis-proof::ts-stack-invariants} hold for the interval timestamp TS-stack.
\end{lem}

\begin{prf}
	Directly by Lemmas~\ref{lem::stack::appendix::vis-proof::interval-disjoint-push} and \ref{lem::stack::appendix::vis-proof::interval-misses-are-late}.
\end{prf}

And we have the main theorem.

\begin{thm}
	The interval timestamp TS-stack is linearizable.
\end{thm}

\begin{prf}
	By the above lemmas, all the TS-stack invariants hold for the interval timestamp version. Therefore, by Theorem~\ref{thm::stack::appendix::vis-proof::ts-invariants-imply-vis-axioms}, the 
	visibility-style axioms in Figure~\ref{fig::stack::concurrent-spec-stack} hold. 
	Theorem~\ref{thm::stack::appendix::lin-proof::vis-axioms-imply-linearizability} 
	then implies that the interval timestamp TS-stack is linearizable.
\end{prf}

%% file: structures.tex
\section{Visibility and Separability for Other Data Structures}

In this section we show how the methodology
laid out in Section~\ref{subsect::stack::rels::separability-relations-and-concur-specs} 
can be used to provide visibility-style concurrent axioms
for other data structures. We exemplify with
RDCSS and MCAS (Section~\ref{subsect::stack::appendix::structures::rdcss-and-mcas}),
queues (Section~\ref{subsect::stack::appendix::structures::queues}), 
and locks (Section~\ref{subsect::stack::appendix::structures::locks}).

For RDCSS and MCAS we provide a detailed 
explanation of how their visibility-style axioms
are derived and how their visibility and separability
relations are defined. For queues and locks we will
only state the sequential specifications and the result
of transforming it by following the methodology.

\subsection{RDCSS and MCAS}
\label{subsect::stack::appendix::structures::rdcss-and-mcas}

We focus on \emph{Restricted Double-Compare Single Swap (RDCSS)} and 
\emph{Multiple Compare-And-Swap (MCAS)} as presented in Harris \etal~\cite{Harris}.

\subsubsection{The algorithms}

\subparagraph*{The RDCSS algorithm}
RDCSS is
a generalization of the compare-and-swap operation CAS. Whereas
$\casRepAlg\,(pt_2,exp_2, new_2)$ updates $pt_2$ with $new_2$ if the
old value of $pt_2$ is $exp_2$, RDCSS adds another pointer $pt_1$ and
value $exp_1$ into the decision.  More precisely, RDCSS receives as
input a descriptor that is a record with five pieces of data: two
pointers $pt_1$, $pt_2$ and three values $exp_1$, $exp_2$, $new_2$.
To the invoking client, RDCSS gives the impression that it
\emph{atomically} carries out the following update: if $pt_1$ has
expected value $exp_1$ and $pt_2$ has expected value $exp_2$, then
$pt_2$ is updated to $new_2$.

\begin{figure}[t]
	\begin{multicols*}{2}
		\begin{algorithmic}[1]
			\scriptsize
			\Record{$\rdcssDesc$}
			\State $pt_1$ : $\ControlPtType$
			\State $pt_2$ : $\DataPtType$
			\State $exp_1$, $exp_2$, $new_2$ : $\ValType$
			\EndRecord
			%
			\State
			%
			\Proc{$\rdcssAlg$\,}{$desc: \rdcssDesc$}
			\State \label{alg-alloc-desc-RDCSS} $d \gets \allocRepAlg\,(\textit{desc})$ as $\DataPtType$
			\State \returnCmd $\rdcssLoopAlg\,(d, \textit{desc})$
			\EndProc
			%
			\State 
			%
			\Proc{$\rdcssLoopAlg$\,}{$d: \DataPtType$, $\textit{desc} : \rdcssDesc$}
			\State \label{alg-CAS-RDCSS} $\textit{old} \gets \casRepAlg\,(\pointTwo{desc}, \expTwo{desc}, d)$
			\If{$\isRdcssDescRepAlg\,(\textit{old})$} \label{alg-is-desc-RDCSS}
			\State \label{alg-help-complete-invoke-RDCSS} $\completeAlg\,(\textit{old})$
			\State \label{alg-help-try-again-rdcssLoop-RDCSS} \returnCmd $\rdcssLoopAlg\,(d, \textit{desc})$
			\Else
			\If {$\textit{old} = \textit{desc}.\textit{exp}_2$} \label{alg-is-cas-successful-RDCSS}
			\State \label{alg-self-complete-invoke-RDCSS} $\completeAlg\,(d)$
			\EndIf
			\State \label{alg-return-old-value-RDCSS} \returnCmd{$\textit{old}$}
			\EndIf
			\State
			\EndProc
			%
			%
			\Proc{$\completeAlg$\,}{$d: \DataPtType$}
			\State \label{alg-desc-read-RDCSS} $\textit{desc} \gets \,\readRepAlg d$
			\State \label{alg-pt1-read-RDCSS} $x \gets \,\readRepAlg \pointOne{\textit{desc}}$
			\If {$x = \expOne{\textit{desc}}$}
			\State \label{alg-pt2-write-success-RDCSS} $\casRepAlg\,(\pointTwo{\textit{desc}}, d, \newTwo{\textit{desc}})$
			\Else
			\State \label{alg-pt2-write-fail-RDCSS} $\casRepAlg\,(\pointTwo{\textit{desc}}, d, \expTwo{\textit{desc}})$
			\EndIf
			\EndProc
			%
			\State
			%
			\Proc{$\rdcssReadAlg$\,}{$\textit{pt} : \DataPtType$}
			\State \label{alg-access-Read-RDCSS} $\textit{old} \gets \,\readRepAlg \textit{pt}$
			\If {$\isRdcssDescRepAlg\,(\textit{old})$} \label{alg-is-desc-Read-RDCSS}
			\State \label{alg-help-complete-invoke-Read-RDCSS} $\completeAlg\,(\textit{old})$
			\State \label{alg-help-try-again-read-RDCSS} \returnCmd $\rdcssReadAlg\,(\textit{pt})$
			\Else
			\State \label{alg-return-old-value-Read-RDCSS} \returnCmd{$\textit{old}$}
			\EndIf
			\EndProc
			%
			\columnbreak 
			%
			\Proc{$\rdcssCasAlg$\,}{$\textit{pt} : \DataPtType$, $exp$, $new: \ValType$}
			\State \label{alg-access-CAS-RDCSS} $\textit{old} \gets \casRepAlg\,(\textit{pt}, \textit{exp}, \textit{new})$
			\If {$\isRdcssDescRepAlg\,(\textit{old})$} \label{alg-is-desc-CAS-RDCSS}
			\State \label{alg-help-complete-invoke-CAS-RDCSS} $\completeAlg\,(\textit{old})$
			\State \label{alg-help-try-again-cas-RDCSS} \returnCmd $\rdcssCasAlg\,(\textit{pt}, \textit{exp}, \textit{new})$
			\Else
			\State \label{alg-return-old-value-CAS-RDCSS} \returnCmd{$\textit{old}$}
			\EndIf
			\EndProc
			%
			\State 
			%
			\Proc{$\rdcssWriteAlg$\,}{$\textit{pt} : \DataPtType$, $v: \ValType$}
			\State \label{alg-access-Write-RDCSS} $\textit{old} \gets \,\readRepAlg \textit{pt}$
			\If {$\isRdcssDescRepAlg\,(\textit{old})$} \label{alg-is-desc-Write-RDCSS}
			\State \label{alg-help-complete-invoke-Write-RDCSS} $\completeAlg\,(\textit{old})$
			\State \label{alg-help-try-again-write-RDCSS} $\rdcssWriteAlg\,(\textit{pt}, v)$
			\Else 
			\State \label{alg-attempt-write-Write-RDCSS} $x \gets \casRepAlg\,(\textit{pt}, \textit{old}, v)$
			\If {$x \neq \textit{old}$}
			\State $\rdcssWriteAlg\,(\textit{pt}, v)$
			\EndIf 
			\EndIf
			\State
			\EndProc
			\State
			\Proc{$\rdcssReadCtlAlg$}{$pt : \ControlPtType$}
			\State \label{appendix::control-Read-RDCSS} \returnCmd{$\readRepAlg pt$}
			\EndProc
			%
			\State 
			%
			\Proc{$\rdcssCasCtlAlg$}{$pt : \ControlPtType$, $exp$, $new: \ValType$}
			\State \label{appendix::control-CAS-RDCSS} \returnCmd{$\casRepAlg(pt, exp, new)$}
			\EndProc
			%
			\State 
			%
			\Proc{$\rdcssWriteCtlAlg$}{$pt : \ControlPtType$, $v: \ValType$}
			\State \label{appendix::control-Write-RDCSS} $pt\ \writeRepAlg\ v$
			\EndProc
			%
			\State 
			%
			\Proc{$\rdcssAllocAlg$}{$v: \ValType$, $k : \ptKind$}
			\If {$k = \CONTROL$}
			\State \label{appendix::alloc-control-Alloc-RDCSS} \returnCmd{$\allocRepAlg(v)$} as $\ControlPtType$
			\Else 
			\State \label{appendix::alloc-data-Alloc-RDCSS} \returnCmd{$\allocRepAlg(v)$} as $\DataPtType$
			\EndIf
			\EndProc
		\end{algorithmic}
	\end{multicols*}
	\caption{Pseudo code of the RDCSS implementation. $\readRepAlg$,
		$\allocRepAlg$ and $\casRepAlg$ are the system calls for memory
		dereference, allocation, and compare-and-swap.
	}
	\label{alg-RDCSS}
\end{figure}

Figure~\ref{alg-RDCSS} shows the 
implementation of the RDCSS data structure ($\rdcssAlg$ and associated
methods, and the descriptor type $\rdcssDesc$). $\ValType$ denotes the
set of all possible input values; it excludes descriptors and
descriptor-storing pointers.
Every pointer exposed to the clients is classified as a control
($\ControlPtType$) or data ($\DataPtType$) pointer. Data pointers may
store $\ValType$ values, descriptors, and pointers, while control
pointers only store $\ValType$ values. Along with the $\rdcssAlg$
method, the implementation exports data pointer methods
$\rdcssReadAlg$, $\rdcssWriteAlg$, and $\rdcssCasAlg$. The latter
replace the system calls for pointer dereference, update, and
$\textit{CAS}$, which they adapt for data
pointers.  Also exported, are the methods for
dereference, update and $\textit{CAS}$ over control pointers,
$\rdcssReadCtlAlg$, $\rdcssWriteCtlAlg$, and $\rdcssCasCtlAlg$, 
respectively,
implemented simply as respective system calls. 
Methods $\rdcssLoopAlg$
and $\completeAlg$ are internal, and not exported.
The implementation further provides: (1) an exported method
$\rdcssAllocAlg\,(v,t)$ for allocating a pointer of type
$t \in \{ \CONTROL, \DATA \}$ with initial value $v$, which replaces
the system operation $\allocRepAlg$, and (2) an internal Boolean
predicate $\isRdcssDescRepAlg\,(p)$ that returns true iff pointer $p$
stores an RDCSS descriptor. 
Details for the allocation of data and control pointers are 
left unspecified.
For example, Harris \etal \cite{Harris} suggests that  
data and control pointers can be allocated in different areas
of the memory. Similarly, details for the implementation of 
$\isRdcssDescRepAlg$ are left unspecified. One possibility 
for implementing $\isRdcssDescRepAlg$ suggested
by Harris \etal, is that one bit could be reserved in the pointers
to indicate if the pointer stores a descriptor or not,
so that $\isRdcssDescRepAlg(p)$ simply reads the reserved bit
in $p$.

We next give a high-level description of $\rdcssAlg$.  Any thread $T$
invoking $\rdcssAlg$ allocates a fresh address $d$ for the input
descriptor $\textit{desc}$ (line~\ref{alg-alloc-desc-RDCSS}); $d$ will
serve as a unique identifier for the $\rdcssAlg$ invocation. Next, $T$
calls the recursive procedure $\rdcssLoopAlg$. This first CASs on
$\pointTwo{\textit{desc}}$ to read the $\textit{old}$ value and write
$d$ if the old value is the expected $\expTwo{\textit{desc}}$
(line~\ref{alg-CAS-RDCSS}). By this write, $T$ essentially requests
help with $d$, and signals to other threads that:
\begin{itemize}
	\item Descriptor $\textit{desc}$ (which is stored in $d$) is currently
	active in $\pointTwo{\textit{desc}}$.
	\item $T$ is invoking $\rdcssAlg$ in which pointer
	$\pointOne{\textit{desc}}$ still needs to be read.
	\item While $desc$ remains active in $\pointTwo{\textit{desc}}$, value
	$\expTwo{\textit{desc}}$ is stored in $\pointTwo{\textit{desc}}$
	\emph{indirectly}, as it can be reached by following the descriptor.
\end{itemize}
After the CAS in line~\ref{alg-CAS-RDCSS}, if $\textit{old}$ is a
descriptor, then $T$ first \emph{helps} by invoking
$\completeAlg{(\textit{old})}$
(line~\ref{alg-help-complete-invoke-RDCSS}), and then $T$ recurses to
reattempt its task. If $\textit{old}$ is a value not matching
$\expTwo{\textit{desc}}$, then no modification to
$\pointTwo{\textit{desc}}$ is performed and $T$ 
returns $\textit{old}$. If $\textit{old}$ is the expected value (i.e.,
the CAS succeeded), then $T$ invokes $\completeAlg{(d)}$ to help itself
(line~\ref{alg-self-complete-invoke-RDCSS}).

We now describe $\completeAlg$ from the point of view of another thread
$T'$ that reads $\pointTwo{\textit{desc}}$ and finds $d$ (lines
\ref{alg-is-desc-RDCSS}, \ref{alg-is-desc-Read-RDCSS},
\ref{alg-is-desc-CAS-RDCSS}, \ref{alg-is-desc-Write-RDCSS}). Before
doing anything else, $T'$ helps on $d$ by invoking $\completeAlg$ as
follows (lines \ref{alg-help-complete-invoke-RDCSS},
\ref{alg-help-complete-invoke-Read-RDCSS},
\ref{alg-help-complete-invoke-CAS-RDCSS},
\ref{alg-help-complete-invoke-Write-RDCSS}). $T'$ first reads pointer
$\pointOne{\textit{desc}}$ (line~\ref{alg-pt1-read-RDCSS}).  If it
finds the expected value $\expOne{\textit{desc}}$, and since pointer
$\pointTwo{\textit{desc}}$ indirectly stores the expected value, then
$\rdcssAlg$ can succeed; thus, $T'$ attempts to replace the descriptor
in $\pointTwo{\textit{desc}}$ with the new value
$\newTwo{\textit{desc}}$ (line~\ref{alg-pt2-write-success-RDCSS}). The
replacement in line~\ref{alg-pt2-write-success-RDCSS} is performed by
$\casRepAlg$, which may fail if some other thread managed to help $T$
before $T'$. Either way, $T$ has been helped after the call to
$\completeAlg$.

Otherwise, if $T'$ does not find the expected value
$\expOne{\textit{desc}}$, it attempts to undo $T$'s modification to
$\pointTwo{\textit{desc}}$ (line~\ref{alg-pt2-write-fail-RDCSS}) by
replacing $d$ in $\pointTwo{\textit{desc}}$ back to
$\expTwo{\textit{desc}}$, via $\casRepAlg$. This undoing may fail if
some other thread managed to help $T$ before $T'$.  Either way, $T$
has again been helped after the call to $\completeAlg$.

Procedures $\rdcssReadAlg$, $\rdcssCasAlg$ and $\rdcssWriteAlg$ follow
the same helping strategy, i.e., whenever they find a descriptor in
their input pointer, they invoke $\completeAlg$ to help the pending
$\rdcssAlg$ before recursing to reattempt their own task.
Procedures for control pointers $\rdcssReadCtlAlg$, 
$\rdcssCasCtlAlg$ and $\rdcssWriteCtlAlg$
directly execute their respective system calls, because
no helping is required for control pointers, as
descriptors cannot be stored in them.

\subparagraph*{The MCAS algorithm}
MCAS also generalizes
CAS, but it updates an arbitrary number of pointers at once.
More precisely, MCAS receives a list of update entries, each being a
record with three pieces of data: a pointer $\textit{pt}$ and two
values $\textit{exp}$, $\textit{new}$. To a client, MCAS gives the
impression that it atomically carries out the following conditional
multiple update: if for every update entry $i$, the pointer
$\updateEntry{\textit{pt}}{i}$ has expected value
$\updateEntry{\textit{exp}}{i}$, then for every update entry $j$, the
pointer $\updateEntry{\textit{pt}}{j}$ is updated to the new value
$\updateEntry{\textit{new}}{j}$.

\begin{figure}[t]
	\begin{multicols}{2}
		\begin{algorithmic}[1]
			\scriptsize
			\Record{$\textsc{update\_entry}$} 
			\State $\textit{pt}$ : $\DataPtType$
			\State $\textit{exp}$, $\textit{new}$ : $\ValType$
			\EndRecord
			\EnumLine{$\textsc{status}$}{$\UNDECIDED$, $\SUCCEEDED$, $\FAILED$}
			\EndEnumLine
			\Record{$\mcasDesc$}
			\State $\textit{status}$ : $\ControlPtType$ $\textsc{status}$
			\State $\textit{entries}$ : $\textsc{list}\,\textsc{update\_entry}$
			\EndRecord
			%
			\State
			\Proc{$\mcasReadAlg$}{$pt: \DataPtType$}
			\State \label{appendix::alg-access-Read-MCAS} $old \gets \rdcssReadAlg(pt)$
			\If {$\isMcasDescRepAlg(old)$} \label{appendix::alg-is-desc-Read-MCAS}
			\State \label{appendix::alg-help-complete-invoke-Read-MCAS} $\mcasHelpAlg(old)$
			\State \returnCmd $\mcasReadAlg(pt)$ \label{appendix::try-again-Read-MCAS}
			\Else 
			\State \label{appendix::alg-return-old-value-Read-MCAS} \returnCmd $old$
			\EndIf
			\EndProc
			%
			\State 
			%
			\Proc{$\mcasWriteAlg$}{$pt: \DataPtType$, $v: \ValType$}
			\State \label{appendix::alg-access-Write-MCAS} $old \gets \rdcssReadAlg(pt)$
			\If {$\isMcasDescRepAlg(old)$} \label{appendix::alg-is-desc-Write-MCAS}
			\State \label{appendix::alg-help-complete-invoke-Write-MCAS} $\mcasHelpAlg(old)$
			\State $\mcasWriteAlg(pt, v)$ \label{appendix::try-again-Write-One-MCAS}
			\Else 
			\State \label{appendix::alg-attempt-write-Write-MCAS} $x \gets \rdcssCasAlg(pt, old, v)$
			\If {$x \neq old$} \label{appendix::last-if-in-Write-MCAS}
			\State $\mcasWriteAlg(pt, v)$ \label{appendix::try-again-Write-Two-MCAS}
			\EndIf
			\EndIf
			\EndProc
			%
			\State \label{appendix::implicit-return-for-Write-MCAS}
			%
			\Proc{$\mcasAllocAlg$}{$v: \ValType$}
			\State \label{appendix::alloc-data-Alloc-MCAS} \returnCmd $\rdcssAllocAlg(v,\DATA)$
			\EndProc
			\columnbreak
			\Proc{$\mcasAlg$\,}{$\listvar u: \textsc{list}\,\textsc{update\_entry}$}
			\State \label{alloc-status-MCAS} $s \gets \rdcssNamespace{\rdcssAllocAlg\,(\UNDECIDED, \CONTROL)}$
			\State \label{create-mcas-desc-MCAS} $\textit{desc} \gets \mcasDesc\ (s, \listvar u)$
			\State \label{alloc-desc-MCAS} $d \gets \rdcssNamespace{\rdcssAllocAlg\,(\textit{desc}, \DATA)}$
			\State \returnCmd $\mcasHelpAlg\,(d)$
			\EndProc
			\State 
			\Proc{$\mcasHelpAlg$\,}{$d : \DataPtType$}
			\State \label{read-desc-MCAS} $\textit{desc} \gets \rdcssNamespace{\rdcssReadAlg\,(d)}$
			\State \label{read-phase1-status-MCAS} $\textit{phase}_1 \gets \,\rdcssReadCtlAlg(\statusProp{\textit{desc}})$
			\If{$\textit{phase}_1 = \UNDECIDED$} \label{is-phase1-still-undecided}
			\State \label{write-all-descs-MCAS} $s \gets \writeAllDescsAlg\,(d,\textit{desc})$
			\State \label{resolve-status-MCAS} $\rdcssNamespace{\rdcssCasCtlAlg\,(\statusProp{\textit{desc}}, \UNDECIDED, s)}$
			\EndIf
			\State \label{read-phase2-status-MCAS} $\textit{phase}_2 \gets \,\rdcssReadCtlAlg(\statusProp{\textit{desc}})$
			\State $b \gets (\textit{phase}_2 = \SUCCEEDED)$
			\ForEach{$e$}{$\entriesProp {\textit{desc}}$} \label{remove-all-descs-loop-MCAS}
			\State \label{remove-all-descs-MCAS} $\rdcssNamespace{\rdcssCasAlg\,(\pointGen e, d, b\ ?\ \newGen e : \expGen e)}$
			\EndForEach 
			\State \returnCmd{$b$}
			\EndProc
			\State
			\Proc{$\writeAllDescsAlg$\,}{$d : \DataPtType, \textit{desc}: \mcasDesc$}
			\ForEach{$e$}{$\entriesProp {\textit{desc}}$}
			\State \label{create-rdcss-desc-MCAS} $\textit{rd} \gets \rdcssNamespace{\rdcssDesc}\ ( \statusProp {\textit{desc}},$ 
			\State $\phantom{\textit{rd} \gets \quad}\pointGen e, \UNDECIDED, \expGen e, d )$
			\State \label{invoke-rdcss-in-MCAS} $\textit{old} \gets \rdcssNamespace{\rdcssAlg\,(\textit{rd})}$
			\If{$\isMcasDescRepAlg\,(\textit{old})$} \label{alg-is-desc-MCAS-MCAS}
			\If{$\textit{old} \neq d$} \label{is-my-desc-MCAS}
			\State \label{alg-help-complete-invoke-MCAS-MCAS} $\mcasHelpAlg\,(\textit{old})$
			\State \label{alg-help-restart-MCAS} \returnCmd{$\writeAllDescsAlg\,(d, \textit{desc})$}
			\EndIf
			\ElsIf{$\textit{old} \neq \expGen e$} \label{rdcss-failed-MCAS}
			\State \returnCmd{$\FAILED$}
			\EndIf
			\EndForEach
			\State \returnCmd{$\SUCCEEDED$}
			\EndProc
		\end{algorithmic}
	\end{multicols}
	\caption{Pseudo code of the MCAS implementation. It makes use of the exportable 
		procedures in the RDCSS implementation of Figure~\ref{alg-RDCSS}.
	}
	\label{alg-MCAS}
\end{figure}

Figure~\ref{alg-MCAS} shows the implementation of the MCAS data structure
($\mcasAlg$ and associated methods, and the descriptor type
$\mcasDesc$) atop RDCSS.

$\ValType$ is the set of input values that excludes descriptors
and descriptor-storing pointers.
We use the notation $\pointGenEntry i$, $\expGenEntry i$, and 
$\newGenEntry i$ to refer to the components of an $\textsc{update\_entry}$
$i$.

Along with the
$\mcasAlg$ method, the implementation exports data pointer methods $\mcasReadAlg$, 
$\mcasWriteAlg$, and $\mcasCasAlg$.
The latter adapt the RDCSS data pointer operations to handle the use of MCAS descriptors.
Procedures $\mcasHelpAlg$ and $\writeAllDescsAlg$ are internal and not exported.

The implementation further provides: (1) an exported method
$\mcasAllocAlg\,(v,t)$ for allocating a data pointer with initial 
value $v$, which replaces
the RDCSS operation $\rdcssAllocAlg$, and (2) an internal Boolean
predicate $\isMcasDescRepAlg\,(p)$ that returns true iff pointer $p$
stores an MCAS descriptor. 
Details for the implementation of 
$\isMcasDescRepAlg$ are left unspecified. One possibility 
for implementing $\isMcasDescRepAlg$ suggested
by Harris \etal, is that one bit could be reserved in the pointers
to indicate if the pointer stores an MCAS descriptor or not,
so that $\isMcasDescRepAlg(p)$ simply reads the reserved bit
in $p$.

A high-level description of $\mcasAlg$ is as follows.  Thread $T$
invoking $\mcasAlg$ first creates an MCAS descriptor $\textit{desc}$
containing the non-empty list of update entries and a status pointer
(line~\ref{create-mcas-desc-MCAS}).  The status pointer starts in an
undecided state ($\UNDECIDED$) and evolves into either a success
($\SUCCEEDED$) or failed ($\FAILED$) state.  $\SUCCEEDED$ indicates
that the descriptor was written into all input pointers, while
$\FAILED$ indicates the failure of at least one such write. Next, $T$
allocates a fresh address $d$ for $\textit{desc}$
(line~\ref{alloc-desc-MCAS}), which serves as the unique identifier
for the $\mcasAlg$ invocation.  Eventually, through a call to
$\mcasHelpAlg$, $T$ invokes $\writeAllDescsAlg$
(line~\ref{write-all-descs-MCAS}) to attempt storing $d$ into each
input pointer, via $\rdcssAlg$.  If any individual $\rdcssAlg$ of
$\writeAllDescsAlg$ fails, then $T$ attempts to mark the status
pointer of $desc$ as failed (line~\ref{resolve-status-MCAS}). In
general, $T$ writing $d$ into pointer $\pointGenEntry i$ of entry $i$
signals that writing $d$ into previous entries has succeeded and:
\begin{itemize}
	\item Descriptor $\textit{desc}$ (stored in $d$) is currently active
	in $\pointGenEntry j$ for every $j \leq i$.
	\item $T$ has an ongoing $\mcasAlg$ in which pointers
	$\pointGenEntry j$ for $j > i$ still need to be updated with $d$.
	\item Pointers $\pointGenEntry j$ for $j \leq i$ indirectly have
	their expected values.
\end{itemize}
If another thread $T'$ attempts to access $\pointGenEntry i$ and finds
$d$, it will \emph{help} $T$ complete $\mcasAlg$ before doing anything
else, by invoking $\mcasHelpAlg$ as follows.  $T'$ first attempts to
write the descriptor into all remaining pointers
(line~\ref{write-all-descs-MCAS}).  If $T'$ succeeds, then all
pointers had the expected values. The variable $s$ in line 
\ref{write-all-descs-MCAS} is set to
$\SUCCEEDED$, and $T'$ attempts changing the status field of $desc$
accordingly (line \ref{resolve-status-MCAS}) and replacing all the
descriptors with the new values (line~\ref{remove-all-descs-MCAS}). These
attempts utilize $\textit{CAS}$ which may fail if some other thread
already helped $T$ to either succeed or fail in its $\mcasAlg$.
Alternatively, if $T'$ fails to write $d$ into some pointer, it means
that not all pointers had the expected values. The variable $s$ in
line \ref{write-all-descs-MCAS} is bound to $\FAILED$, $T'$ attempts 
to change the status field of $desc$ accordingly 
(line~\ref{resolve-status-MCAS}), and to undo $T$'s writing of $d$
(line~\ref{remove-all-descs-MCAS}). As
before, the $\textit{CAS}$'s in these attempts may fail if some other
thread already helped $T$ to either succeed or fail in its $\mcasAlg$.

Notice that $\mcasHelpAlg$ and $\writeAllDescsAlg$ are mutually
recursive. This is necessary, because while a thread $T$ is writing
the descriptor into all the input pointers using $\writeAllDescsAlg$,
$T$ may encounter other descriptors that force it to help by invoking
$\mcasHelpAlg$ (lines
\ref{alg-is-desc-MCAS-MCAS}-\ref{alg-help-complete-invoke-MCAS-MCAS}),
after which $T$ must reattempt writing the descriptor again
(line~\ref{alg-help-restart-MCAS}).

Procedures $\mcasReadAlg$, $\mcasCasAlg$ and $\mcasWriteAlg$
follow the same helping strategy as the procedures 
$\rdcssReadAlg$, $\rdcssCasAlg$ and $\rdcssWriteAlg$ for RDCSS in
Figure~\ref{alg-RDCSS}, i.e., whenever they find an MCAS descriptor in
their input pointer, they invoke $\mcasHelpAlg$ to help the pending
$\mcasAlg$ before recursing to reattempt their own task.

\subsubsection{The specification for RDCSS}
\label{subsubsect::stack::appendix::structures::rdcss-spec}

We will focus on the specification of the $\rdcssAlg$ procedure.
The specification of the full RDCSS module, which includes 
procedures $\rdcssReadAlg$, $\rdcssReadCtlAlg$,
$\rdcssWriteAlg$, $\rdcssWriteCtlAlg$,
$\rdcssCasAlg$, $\rdcssCasCtlAlg$, 
and $\rdcssAllocAlg$, can be found in 
Dom\'inguez and Nanevski~\cite{arxiv:mcas2307.04653}.

\begin{figure}[t]
	\centering 
	\begin{subfigwrap}{State-based sequential specification.
			$\mapExt{H}{\mapEntry {p} {v}}$ is the heap (i.e., memory) obtained when
			pointer $p$ in the heap $H$ is
			mutated into $v$, while leaving the rest of $H$
			unchanged.}{subfig:atomic-spec-state-rdcss}
		\centering
		\begin{tabular}{ll}
			$(A_1)$ Successful $\rdcssAlg$ & \\
			\quad $H \xrightarrow{\rdcssAlg(d)\ \langle {\expTwo d} \rangle} \mapExt H {\mapEntry {\pointTwo d} {\newTwo d}}$ &
			if $H(\pointTwo d) = \expTwo d$ and $H(\pointOne d) = \expOne d$ \\
			$(A_{2.1})$ Failing $\rdcssAlg$ (Reads both pointers) & \\
			\quad $H \xrightarrow{\rdcssAlg(d)\ \langle \expTwo d \rangle} H$ & 
			if $H(\pointTwo d) = \expTwo d$ and $H(\pointOne d) \neq \expOne d$ \\
			$(A_{2.2})$ Failing $\rdcssAlg$ (Reads second pointer only) & \\
			\quad $H \xrightarrow{\rdcssAlg(d)\ \langle v \rangle} H$ & 
			if $H(\pointTwo d) = v$ and $v \neq \expTwo d$
		\end{tabular}
	\end{subfigwrap}
	
	\begin{subfigwrap}{History-based sequential specification. Relation $\visObsSymbol {\pointerIndx p} : \absEvent \times \absEvent$ is
			abstract.}{subfig:atomic-spec-history-rdcss}
		\centering
		\begin{tabular}{c}
			\begin{tabular}{l}
				$(B_1)$ No in-between \\
				\quad $(\visObs {\pointerIndx p} w r \wedge w' \in \writesAbs{\pointerIndx p}) \implies 
				(\precedesAbsEq {w'} w \vee \precedesAbsEq r {w'})$ \\
				$(B_2)$ Observed events are writes \\
				\quad $\visObs {\pointerIndx p} w {\_} \implies w \in \writesAbs{\pointerIndx p}$ \\
				$(B_3)$ Dependences occur in the past \\
				\quad $\visObs {\pointerIndx p} w r \implies \precedesAbs w r$\\
				$(B_{4.1})$ $\rdcssAlg$ always reads the second pointer \\
				\quad $r = \rdcssAlg(d)\left<v\right> \implies 
					\exists w_2.\ \visObs {\pointerIndx {\pointTwo d}} {w_2} {r} \wedge \inputVal {w_2} {\pointTwo d} v$ \\
				$(B_{4.2})$ $\rdcssAlg$ reads the first pointer if second pointer has the expected value\\
				\quad $r = \rdcssAlg(d)\left<\expTwo d\right> \implies 
					\exists w_1\ldot \visObs {\pointerIndx {\pointOne d}} {w_1} {r}$\\
			\end{tabular}
			\\
			\begin{tabular}{c}
				\ \\
				$\begin{aligned}
					{\inputVal x {p} {v}} \defini & \ x = \rdcssAlg(\{ \pointGenEntry{1} : \_,\ \expGenEntry{1} : \_,\ \pointGenEntry{2} : p,\ \expGenEntry{2} : \_,\ \newGenEntry{2} : v \}) \left< \_ \right> \vee {}\\
					& \ x = \rdcssWriteAlg(p,v)\left< \_ \right> \vee x = \rdcssWriteCtlAlg(p,v) \left< \_ \right> \vee x = \rdcssCasAlg(p,\_,v) \left< \_ \right> \vee {}\\ 
					& \ x = \rdcssCasCtlAlg(p,\_,v) \left< \_ \right> \vee x = \rdcssAllocAlg(v)\left< p \right> \\
				\end{aligned}$
				\\
				\ \\
				$\begin{aligned}
					x \in \writesAbs{p} \Longleftrightarrow & \ (x = \rdcssAlg(\{ \pointGenEntry{1} : p_1,\ \expGenEntry{1} : e_1,\ \pointGenEntry{2} : p,\ \expGenEntry{2} : e_2,\ \newGenEntry{2} : \_ \}) \left< e_2 \right> \wedge {} \\
					& \qquad \qquad \exists w_1.\ \visObs {\pointerIndx {p_1}} {w_1} {x} \wedge \inputVal {w_1} {p_1} {e_1}) \vee {}\\
					& \ x = \rdcssWriteAlg(p,\_)\left< \_ \right> \vee x = \rdcssWriteCtlAlg(p,\_) \left< \_ \right> \vee x = \rdcssCasAlg(p,e,\_) \left< e \right> \vee {}\\ 
					& \ x = \rdcssCasCtlAlg(p,e,\_) \left< e \right> \vee x = \rdcssAllocAlg(\_)\left< p \right>
				\end{aligned}$
			\end{tabular}
		\end{tabular}
	\end{subfigwrap}
	\caption{State-based and history-based sequential specifications for $\rdcssAlg$.}
	\label{fig::atomic-specs-rdcss}
\end{figure}

\subparagraph*{Sequential Specification}
Figure~\ref{fig::atomic-specs-rdcss} shows the
sequential state-based and history-based specifications for
$\rdcssAlg$.  

The state-based specification in
Figure~\ref{subfig:atomic-spec-state-rdcss} says in axiom $A_1$ that a
successful $\rdcssAlg(d)$ occurs when input pointers $\pointOne{d}$
and $\pointTwo{d}$ contain the
expected values; pointer $\pointTwo{d}$ is then mutated to its new
value and $\rdcssAlg$ returns the expected value for pointer
$\pointTwo{d}$. Axioms $A_{2.1}$ and $A_{2.2}$ state the failing cases for 
$\rdcssAlg$, i.e., the cases in which $\rdcssAlg$ leaves the heap
unchanged.
For $A_{2.1}$, $\rdcssAlg$ fails if pointer $\pointTwo{d}$ has the
expected value, but pointer $\pointOne{d}$ does not.
For $A_{2.2}$, $\rdcssAlg$ fails if pointer $\pointTwo{d}$ does not 
have the expected value. Note that in $A_{2.2}$, $\rdcssAlg$ is 
not required to read pointer $\pointOne{d}$, but both axioms
$A_{2.1}$ and $A_{2.2}$ imply that $\rdcssAlg$ always reads pointer 
$\pointTwo{d}$.

Figure~\ref{subfig:atomic-spec-history-rdcss} shows the history sequential
specification for $\rdcssAlg$.  The specification utilizes the
\emph{visibility relation} $\visObsSymbol {\pointerIndx p}$ to
capture a read-write causal dependence between events. In particular,
$\visObs {\pointerIndx p} w r$ means that ``event $r$ reads a value
that event $w$ wrote into pointer $p$''. 
Under this interpretation,
axioms $B_1,...,B_{4.2}$ state the following 
expected properties.

Axiom $B_1$ (No in-between) says that if $r$ reads from $w$ in $p$,
then no other successful $p$-write can occur between $w$ and $r$
(otherwise, such a write would overwrite $w$). Again,
$\precedesAbsSymbol$ denotes the \emph{returns-before} relation (with
$\precedesAbsEqSymbol$ its reflexive closure), where
$\precedesAbs x y$ means that $x$ terminated before $y$ started. 

Set $\writesAbs p$ collects the successful $p$-writes, e.g.,
writes of the form $\rdcssWriteAlg(p,\_)$ and $\rdcssWriteCtlAlg(p,\_)$, 
CASes of the form $\rdcssCasAlg(p,e,\_)$ and $\rdcssCasCtlAlg(p,e,\_)$
returning the expected value $e$ (hence, the CASes succeed),
$\rdcssAllocAlg$ events 
returning $p$ (hence, they allocate $p$ and write in $p$
some initial value), and $\rdcssAlg(d)$ events returning 
the expected value for pointer $p = \pointTwo d$.
$\rdcssAlg$ has the peculiarity that it is not possible to know if it succeeded 
even when it returns the expected value for $\pointTwo d$
(meaning that $\rdcssAlg$ read the expected value
in $\pointTwo{d}$), because the output does not say anything 
about the value in the first pointer $\pointOne{d}$. 
This is illustrated by Axioms $A_1$ and
$A_{2.1}$, where in both cases $\rdcssAlg(d)$ returns $\expTwo{d}$.
Hence, $\writesAbs p$ adds another clause for $\rdcssAlg$ regarding 
pointer $\pointOne{d}$, which states that $\rdcssAlg(d)$ observes
a writer having in its input the expected value for $\pointOne{d}$.
We now explain the predicate $\inputVal {x} {p} {v}$.

Predicate $\inputVal {x} {p} {v}$ expresses that $x$
is a writer event (independently if it succeeds to write or not)
intending to write input $v$ in pointer $p$.
It collects events of the form 
$\rdcssWriteAlg(p,v)$ and $\rdcssWriteCtlAlg(p,v)$, 
CASes of the form $\rdcssCasAlg(p,\_,v)$ and $\rdcssCasCtlAlg(p,\_,v)$,
allocs of the form $\rdcssAllocAlg(v)$ returning pointer $p$
(so that $p$ gets initialized with $v$),
and $\rdcssAlg$ events having $\newTwo{d} = v$ for pointer 
$p = \pointTwo{d}$. 
In the case of $\rdcssAlg$, definition $\inputVal {x} {p} {v}$
does not treat $\rdcssAlg$ as a writer of $\pointOne{d}$, 
because $\rdcssAlg$ never writes into
pointer $\pointOne{d}$. 

Intuitively, one should think of the statement 
$x \in \writesAbs p$
as expressing that $x$ actually modified the heap
by writing \emph{some} value in $p$, as in this diagram,

\begin{center}
	\begin{tikzcd}[ampersand replacement=\&, column sep=small, row sep=small]
		\ldots 
		\arrow[r]\&
		H
		\arrow[r, "x"] \& 
		\mapExt{H}{\mapEntry {p} {\_}}
		\arrow[r] \&
		\ldots 
	\end{tikzcd}
\end{center}

If we know the more precise statement $x \in \writesAbs p \wedge \inputVal {x} {p} {v}$,
then $x$ modified the heap
by writing $v$ in $p$,

\begin{center}
	\begin{tikzcd}[ampersand replacement=\&, column sep=small, row sep=small]
		\ldots 
		\arrow[r]\&
		H
		\arrow[r, "x"] \& 
		\mapExt{H}{\mapEntry {p} {v}}
		\arrow[r] \&
		\ldots 
	\end{tikzcd}
\end{center}

Axiom $B_2$ (Observed events are writes) says that any observed event
at $p$ must be a successful $p$-write, i.e., if $r$ reads from $w$ in
$p$, then $w$ must have actually written into $p$.  Axiom $B_3$
(Dependences occur in the past) says that if a read depends on a
write, then the write executes before the read.

Axioms $B_{4.1}$ and $B_{4.2}$ express what is true
about the output of an $\rdcssAlg$.
Since it is not possible to know if $\rdcssAlg$ succeeds or not
just by looking at its output, axioms $B_{4.1}$ and $B_{4.2}$
encode indirectly the successful and failing cases in the 
state-based sequential specification for $\rdcssAlg$ as follows.

Axiom $B_{4.1}$ states that $\rdcssAlg(d)$ always
reads pointer $\pointTwo{d}$, i.e., $\rdcssAlg(d)$
observes an event $w_2$ at $\pointTwo{d}$. Event $w_2$ executes before 
$\rdcssAlg(d)$ due to axiom $B_3$, $w_2$ actually writes $v$
due to axiom $B_2$, and pointer $\pointTwo{d}$ 
does not mutate after $w_2$ and before $\rdcssAlg(d)$
due to axiom $B_1$. Diagrammatically,

\begin{center}
	\begin{tikzcd}[ampersand replacement=\&, column sep=small, row sep=small]
		\ldots 
		\arrow[r]\&
		H_k
		\arrow[r, "{w_{2}}"{name=W,yshift=2pt}] \& 
		H_{k+1} 
		\arrow[r] \&
		\ldots 
		\arrow[r] \&
		H_n
		\arrow[r, "{\rdcssAlg(d)\left< v \right>}"{name=R}] \&[30pt] 
		\ldots
		\\
		{} \& {} \& {} \arrow[rr, mapsfrom, maps to, "\pointTwo{d} \mapsto v \text{ persists}"] \& {} \& {} \\
		\arrow[from=R, to=W, bend right, dashed, no head, "{\visObsSymbol{\pointerIndx {\pointTwo{d}}}}"{below}]
	\end{tikzcd}
\end{center}

If $v \neq \expTwo{d}$, then $\rdcssAlg(d) \notin \writesAbs{\pointTwo{d}}$ by definition,
and so $\rdcssAlg$ fails by not mutating the heap as in axiom $A_{2.2}$.

If $v = \expTwo{d}$, then Axiom $B_{4.2}$ states that $\rdcssAlg(d)$ additionally
reads $\pointOne{d}$ by observing some event $w_1$. By a similar argument,
$w_1$ will mutate the heap by writing some value $v'$ in $\pointOne{d}$ that
persists up to the execution of $\rdcssAlg(d)$.

If $v' \neq \expOne{d}$, then $\rdcssAlg(d) \notin \writesAbs{\pointTwo{d}}$
by definition, and so $\rdcssAlg$ fails by not mutating the heap as in 
axiom $A_{2.1}$.

If instead $v' = \expOne{d}$, then $\rdcssAlg(d) \in \writesAbs{\pointTwo{d}}$
by definition, and since we also have
$\inputVal{\rdcssAlg(d)\left< \_ \right>}{\pointTwo{d}}{\newTwo{d}}$ by definition, 
$\rdcssAlg$ actually mutates $\pointTwo{d}$ as in Axiom $A_1$.

\begin{figure}[t]
\begin{subfigwrap}{Concurrent specification. Relations $\visObsSymbol {\pointerIndx p}, \visSepSymbol {\pointerIndx p} : \absEvent \times \absEvent$ are
		existentially quantified.}{subfig:concurrent-spec-history-rdcss}
	\centering
	\begin{tabular}{l}
		\axiomHLabel{vis-ax::cc-no-in-between-rdcss} No in-between \\
		\quad $(\visObs {\pointerIndx p} w r \wedge w' \in \writesAbs{\pointerIndx p}) \implies 
		(\visSepEq {\pointerIndx p} {w'} w \vee \visSepEq {\pointerIndx p} r {w'})$ \\
		\axiomHLabel{vis-ax::cc-observed-are-writes-rdcss} Observed events are writes \\
		\quad $\visObs {\pointerIndx p} w {\_} \implies w \in \writesAbs{\pointerIndx p}$ \\
	\end{tabular}
	\begin{tabular}{l}
		\axiomHLabel{vis-ax::cc-no-future-dependence-rdcss} No future dependences \\
		\quad $x \transCl{\genVisSymbol} y \implies \nprecedesAbsEq y x$\\
		\axiomHLabel{vis-ax::cc-return-completion-rdcss} Return value completion \\
		\quad $\exists v.\ \postPred x v \wedge (x \in \terminatedEvent \implies v = \outputProp x)$ \\
	\end{tabular} 
\end{subfigwrap}

\begin{subfigwrap}{Defined notions.}{subfig:defined-notions-concurrent-spec-history-rdcss}
	\centering
	\begin{tabular}{c}
		\begin{tabular}{l}
			Constraint relation \\
			\quad ${\genVisSymbol} \defini \bigcup_p (\visObsSymbol {\pointerIndx p} \cup \visSepSymbol {\pointerIndx p})$\\
			Returns-before relation\\
			\quad $\precedesAbs e {e'} \defini \ETimeProp e \natorderSymbol \STimeProp {e'}$\\
		\end{tabular}
		\begin{tabular}{l}
			Set of terminated events\\
			\quad $\terminatedEvent \defini \{ e \mid \ETimeProp e \neq \bot \}$\\
			Closure of terminated events\\
			\quad $\closedEvent \defini \{ e \mid \exists t \in \terminatedEvent.\ e \refleTransCl{\genVisSymbol} t \} $\\
		\end{tabular}
		\\
		\begin{tabular}{c}
			\\
			$\postPred {\rdcssAlg(d)} v \defini 
				\begin{aligned}[t]
					\exists w_2\ldot
					\visObs {\pointerIndx {\pointTwo d}} {w_2} {\rdcssAlg(d)} \wedge 
					\inputVal {w_2} {\pointTwo d} v \wedge {} \\
					(v = \expTwo d \implies \exists w_1\ldot \visObs {\pointerIndx {\pointOne d}} {w_1} {\rdcssAlg(d)})
				\end{aligned}$
			\\
			\ \\
			$\rdcssAlg(d) \in \writesAbs{p} \Longleftrightarrow
			\begin{aligned}[t]
				p = \pointTwo d \wedge \exists w_1, w_2.\ \visObs
				{\pointerIndx {\pointOne d}} {w_1} {\rdcssAlg(d)} \wedge 
				\visObs {\pointerIndx {\pointTwo d}} {w_2} {\rdcssAlg(d)} \wedge {} \\
				\inputVal {w_1} {\pointOne d} {\expOne d} \wedge 
				\inputVal {w_2} {\pointTwo d} {\expTwo d}
			\end{aligned}$
		\end{tabular}
	\end{tabular}
\end{subfigwrap}
\caption{Concurrent history-based specification for $\rdcssAlg$. Variables $w$, $w'$, $r$, $x$, $y$ range over $\closedEvent$.
	Variables $e$, $e'$ range over $\absEvent$.}
\label{fig::concurrent-spec-rdcss}
\end{figure}

\subparagraph*{Concurrent Specification}
As was explained in Section~\ref{subsect::stack::rels::separability-relations-and-concur-specs},
concurrent execution histories do not satisfy sequential history-based specifications 
for two main reasons.  First,
concurrent events can \emph{overlap} in real time. As a consequence,
the axioms $B_1$ (No in-between) and $B_3$ (Dependencies occur in the
past) are too restrictive, as they force events to be disjoint, due to
the use of the returns-before relation $\precedesAbsSymbol$.
Second, events can no longer be treated as atomic; thus event's start
and end times (if the event is terminated) must be taken into
account. As a consequence, axioms $B_{4.1}$ and $B_{4.2}$ and set
$\writesAbs p$ must be modified to account for the output of an
unfinished event not being available yet. 

Figure~\ref{fig::concurrent-spec-rdcss} shows the concurrent
specification that addresses the above issues. As expected, in
addition to the visibility relation, the specification utilizes the
\emph{separability relation} $\visSepSymbol {\pointerIndx p}$.
We now explain how the concurrent specification of
Figure~\ref{fig::concurrent-spec-rdcss} is obtained from the sequential
one in Figure~\ref{subfig:atomic-spec-history-rdcss}.

Axiom \axiomHRef{vis-ax::cc-no-in-between-rdcss} is obtained from $B_1$ by
carrying out the replacement method described in 
Section~\ref{subsect::stack::rels::separability-relations-and-concur-specs}
where we replace $\precedesAbsSymbol$ with $\visSepSymbol p$.\footnote{Note
that since $\precedesAbsSymbol$ occurs in positive position in $B_1$,
the replacement method states that we do not need to flip the order of 
the variables nor apply a negation.}  
The intuition
is that we want to relax the real-time strong separation imposed by
$\precedesAbsSymbol$ into a more permissive separation
$\visSepSymbol p$. We shall see in
Section~\ref{subsubsect::stack::appendix::structures::rdcss-and-mcas} 
the concrete definition of
$\visSepSymbol p$ for the
$\rdcssAlg$ implementation of Figure~\ref{alg-RDCSS}.

Axiom \axiomHRef{vis-ax::cc-observed-are-writes-rdcss} is unchanged compared to $B_2$.

Axiom \axiomHRef{vis-ax::cc-no-future-dependence-rdcss} is obtained from $B_3$ 
identically as in Section~\ref{subsect::stack::rels::separability-relations-and-concur-specs}
for obtaining \axiomSRef{vis-ax::stack::cc-no-future-dependence}. 
Here, the \emph{constraint
relation} $\genVisSymbol$ is again the union of the 
visibility and separability relations 
${\genVisSymbol} \defini {\cup_p {(\visObsSymbol p \cup \visSepSymbol
	p)}}$.

Axiom \axiomHRef{vis-ax::cc-return-completion-rdcss} has an identical explanation
as that for axiom \axiomSRef{vis-ax::stack::cc-return-completion} in 
Section~\ref{subsect::stack::rels::separability-relations-and-concur-specs}. 
Axiom \axiomHRef{vis-ax::cc-return-completion-rdcss} simply coalesces 
axioms $B_{4.1}$ and $B_{4.2}$ into the
postcondition predicate $\postPredSymbol$.

We further emphasize that in the concurrent setting we also need to
redefine the set $\writesAbs p$ of successful writes into pointer $p$.
It does not suffice to state that $\rdcssAlg$ 
returns $\expTwo{d}$; we need a criterion for an unfinished $\rdcssAlg$. 
Thus, we define in
Figure~\ref{subfig:defined-notions-concurrent-spec-history-rdcss} that an
$\rdcssAlg(d)$ is a successful $p$-write if $p = \pointTwo{d}$ 
and $\rdcssAlg(d)$ observes events writing
the expected values for both pointers $\pointOne{d}$ and
$\pointTwo{d}$.

\subsubsection{The specification for MCAS}

We will focus on the specification of the $\mcasAlg$ procedure.
The specification of the full MCAS module, which includes 
procedures $\mcasReadAlg$, $\mcasWriteAlg$,
and $\mcasAllocAlg$, can be found in 
Dom\'inguez \etal~\cite{arxiv:mcas2307.04653}.

\begin{figure}[t]
	\centering 
	\begin{subfigwrap}{State-based sequential specification.
			$\mapExtThree{H}{\mapEntry {\pointGenEntry i} {\newGenEntry
					i}}{i\in\listvar{u}}$ is the heap (i.e., memory) obtained when
			pointers $\pointGenEntry i$ ($i \in \listvar u$) in the heap $H$ are
			mutated into $\newGenEntry i$, while leaving the rest of $H$
			unchanged.}{subfig:atomic-spec-state}
		\centering
		\begin{tabular}{ll}
			$(A_1)$ Successful $\mcasAlg$ & \\
			\quad $H \xrightarrow{\mcasAlg(\listvar{u})\ \langle true \rangle} \mapExtThree{H}{\mapEntry {\pointGenEntry i} {\newGenEntry i}}{i\in\listvar{u}}$ &
			if $\forall i \in \listvar{u}.\ H(\pointGenEntry i) = \expGenEntry i$\\
			$(A_2)$ Failing $\mcasAlg$ & \\
			\quad $H \xrightarrow{\mcasAlg(\listvar{u})\ \langle false \rangle} H$ &
			if $\exists i \in \listvar{u}.\ H(\pointGenEntry i) \neq \expGenEntry i$ \\
		\end{tabular}
	\end{subfigwrap}
	
	\begin{subfigwrap}{History-based sequential specification. Relation $\visObsSymbol {\pointerIndx p} : \absEvent \times \absEvent$ is
			abstract.}{subfig:atomic-spec-history}
		\centering
		\begin{tabular}{c}
			\begin{tabular}{l}
				$(B_1)$ No in-between \\
				\quad $(\visObs {\pointerIndx p} w r \wedge w' \in \writesAbs{\pointerIndx p}) \implies 
				(\precedesAbsEq {w'} w \vee \precedesAbsEq r {w'})$ \\
				$(B_2)$ Observed events are writes \\
				\quad $\visObs {\pointerIndx p} w {\_} \implies w \in \writesAbs{\pointerIndx p}$ \\
				$(B_3)$ Dependences occur in the past \\
				\quad $\visObs {\pointerIndx p} w r \implies \precedesAbs w r$\\
				$(B_{4.1})$ Successful $\mcasAlg$ \\
				\quad $r = \mcasAlg(\listvar{u})\left<true\right> \implies 
				\forall i \in \listvar{u}.\ \exists w.\ \visObs {\pointerIndx {\pointGenEntry i}} w r \wedge
				\inputVal w {\pointGenEntry i} {\expGenEntry i}$ \\
				$(B_{4.2})$ Failing $\mcasAlg$ \\
				\quad $r = \mcasAlg(\listvar{u})\left<false\right> \implies 
				\exists i \in \listvar{u}.\ \exists w.\ \exists v'.\ \visObs {\pointerIndx {\pointGenEntry i}} w r \wedge {}
				\inputVal w {\pointGenEntry i} {v'} \wedge v' \neq \expGenEntry i$\\
			\end{tabular}
			\\
			\begin{tabular}{c}
				\ \\
				$\begin{aligned}
					{\inputVal x {p} {v}} \defini & \ x = \mcasAlg([\ldots,\{ \pointGenEntry{} : p,\ \expGenEntry{} : \_,\ \newGenEntry{} : v \},\ldots]) \left< \_ \right> \vee {}\\
					& \ x = \mcasWriteAlg(p,v)\left< \_ \right> \vee x = \mcasAllocAlg(v)\left< p \right>\\
				\end{aligned}$
				\\
				\ \\
				$\begin{aligned}
					x \in \writesAbs{p} \Longleftrightarrow & \ x = \mcasAlg([\ldots,\{ \pointGenEntry{} : p,\ \ldots\},\ldots]) \left< true \right> \vee
					x = \mcasWriteAlg(p,\_)\left< \_ \right> \vee x = \mcasAllocAlg(\_)\left< p \right>\\
				\end{aligned}$
			\end{tabular}
		\end{tabular}
	\end{subfigwrap}
	\caption{State-based and history-based sequential specifications for $\mcasAlg$.}
	\label{fig::atomic-specs-mcas}
\end{figure}

\subparagraph*{Sequential Specification}
Figure~\ref{fig::atomic-specs-mcas} shows the
sequential state-based and history-based specifications for
$\mcasAlg$.  

The state-based specification in
Figure~\ref{subfig:atomic-spec-state} says in axiom $A_1$ that a
successful $\mcasAlg$ occurs when all input pointers contain the
expected values; the pointers are then mutated to their new
values. Axiom $A_2$ says that a failing $\mcasAlg$ occurs when some
input pointer does not have the expected value, leaving the heap
unchanged.

Figure~\ref{subfig:atomic-spec-history} shows the history sequential
specification for $\mcasAlg$.  Similarly to the specification of 
$\rdcssAlg$, Figure~\ref{subfig:atomic-spec-history} utilizes the
\emph{visibility relation} $\visObsSymbol {\pointerIndx p}$ to
capture a read-write causal dependence between events. In particular,
$\visObs {\pointerIndx p} w r$ means that ``event $r$ reads a value
that event $w$ wrote into pointer $p$''. 
Under this interpretation,
axioms $B_1$, $B_2$, and $B_3$ are identical to those
for $\rdcssAlg$ in Figure~\ref{fig::atomic-specs-rdcss}.
This should not be surprising, since both $\rdcssAlg$
and $\mcasAlg$ are read-write pointer procedures,
which means that there must exist some common properties
among them.

Similarly to the specification of $\rdcssAlg$, 
set $\writesAbs p$ collects the successful $p$-writes, e.g.,
writes of the form $\mcasWriteAlg(p,\_)$, $\mcasAllocAlg$ events 
returning $p$, and $\mcasAlg(\listvar u)$ events returning $true$ and having $p$ in
$\listvar u$.

Axioms $B_{4.1}$ (Successful $\mcasAlg$) and $B_{4.2}$ (Failing
$\mcasAlg$) essentially encode the state-based sequential
specification for $\mcasAlg$.  For example, axiom $B_{4.1}$ directly
says that a successful $\mcasAlg(\listvar{u})\left<true\right>$ event
$r$ observes---for each of its input entries $\listvar u$---a successful write
event ($\mcasAlg$, $\mcasWriteAlg$ or $\mcasAllocAlg$, as per axiom $B_2$) that
wrote the expected value into the
appropriate pointer. Because the write events are
\emph{observed} by $r$, axiom $B_3$ ensures that they execute before $r$, while 
axiom $B_1$ guarantees the none of them is
overwritten before $r$ executes. Thus, diagrammatically, 
for each entry $i \in \listvar u$, the execution
looks as follows, where the write into pointer $\pointGenEntry {i}$
persists until the heap $H_n$ at which $r$
executes.

\begin{center}
	\begin{tikzcd}[ampersand replacement=\&, column sep=small, row sep=small]
		\ldots 
		\arrow[r]\&
		H_k
		\arrow[r, "{w_{i}}"{name=Wi,yshift=2pt}] \& 
		H_{k+1} 
		\arrow[r] \&
		\ldots 
		\arrow[r] \&
		H_n
		\arrow[r, "{\mcasAlg(\listvar{u})\left< true \right>}"{name=MOne}] \&[30pt] 
		\ldots
		\\
		{} \& {} \& {} \arrow[rr, mapsfrom, maps to, "\pointGenEntry{i} \mapsto \expGenEntry{i} \text{ persists}"] \& {} \& {} \\
		\arrow[from=MOne, to=Wi, bend right, dashed, no head, "{\visObsSymbol{\pointerIndx {\pointGenEntry {i}}}}"{below}]
	\end{tikzcd}
\end{center}

\begin{figure}[t]
	\begin{subfigwrap}{Concurrent specification. Relations $\visObsSymbol {\pointerIndx p}, \visSepSymbol {\pointerIndx p} : \absEvent \times \absEvent$ are
			existentially quantified.}{subfig:concurrent-spec-history}
		\centering
		\begin{tabular}{l}
			\axiomHLabel{vis-ax::cc-no-in-between} No in-between \\
			\quad $(\visObs {\pointerIndx p} w r \wedge w' \in \writesAbs{\pointerIndx p}) \implies 
			(\visSepEq {\pointerIndx p} {w'} w \vee \visSepEq {\pointerIndx p} r {w'})$ \\
			\axiomHLabel{vis-ax::cc-observed-are-writes} Observed events are writes \\
			\quad $\visObs {\pointerIndx p} w {\_} \implies w \in \writesAbs{\pointerIndx p}$ \\
		\end{tabular}
		\begin{tabular}{l}
			\axiomHLabel{vis-ax::cc-no-future-dependence} No future dependences \\
			\quad $x \transCl{\genVisSymbol} y \implies \nprecedesAbsEq y x$\\
			\axiomHLabel{vis-ax::cc-return-completion} Return value completion \\
			\quad $\exists v.\ \postPred x v \wedge (x \in \terminatedEvent \implies v = \outputProp x)$ \\
		\end{tabular} 
	\end{subfigwrap}
	
	\begin{subfigwrap}{Defined notions.}{subfig:defined-notions-concurrent-spec-history}
		\centering
		\begin{tabular}{c}
			\begin{tabular}{l}
				Constraint relation \\
				\quad ${\genVisSymbol} \defini \bigcup_p (\visObsSymbol {\pointerIndx p} \cup \visSepSymbol {\pointerIndx p})$\\
				Returns-before relation\\
				\quad $\precedesAbs e {e'} \defini \ETimeProp e \natorderSymbol \STimeProp {e'}$\\
			\end{tabular}
			\begin{tabular}{l}
				Set of terminated events\\
				\quad $\terminatedEvent \defini \{ e \mid \ETimeProp e \neq \bot \}$\\
				Closure of terminated events\\
				\quad $\closedEvent \defini \{ e \mid \exists t \in \terminatedEvent.\ e \refleTransCl{\genVisSymbol} t \} $\\
			\end{tabular}
			\\
			\begin{tabular}{c}
				\\
				$\postPred {\mcasAlg(\listvar{u})} v \defini v \in \BoolType \wedge
				\begin{cases}
					\forall i \in \listvar{u}.\ \exists w.\ \visObs {\pointerIndx {\pointGenEntry i}} w {\mcasAlg(\listvar{u})} \wedge
					\inputVal w {\pointGenEntry i} {\expGenEntry i} &
					\text{if $v = true$} \\
					\exists i \in \listvar{u}.\ \exists w.\ \exists v'.\ \visObs {\pointerIndx {\pointGenEntry i}} w {\mcasAlg(\listvar{u})} \wedge {} & \text{if $v = false$} \\
					\hspace{30pt} \inputVal w {\pointGenEntry i} {v'} \wedge v' \neq \expGenEntry i & \\
				\end{cases}$
				\\
				\ \\
				$\begin{aligned}
					\mcasAlg(\listvar{u}) \in \writesAbs{p} \Longleftrightarrow
					(\exists j \in \listvar{u}.\ p = \pointGenEntry j) \wedge 
					\forall i \in \listvar{u}.\ \exists w.\ \visObs {\pointerIndx {\pointGenEntry i}} w {\mcasAlg(\listvar{u})} \wedge
					\inputVal w {\pointGenEntry i} {\expGenEntry i}
				\end{aligned}$
			\end{tabular}
		\end{tabular}
	\end{subfigwrap}
	\caption{Concurrent history-based specification for $\mcasAlg$. Variables $w$, $w'$, $r$, $x$, $y$ range over $\closedEvent$.
		Variables $e$, $e'$ range over $\absEvent$.}
	\label{fig::concurrent-spec-mcas}
\end{figure}

\subparagraph*{Concurrent Specification}
Axioms \axiomHRef{vis-ax::cc-no-in-between}, \axiomHRef{vis-ax::cc-observed-are-writes},
and \axiomHRef{vis-ax::cc-no-future-dependence} are obtained identically as 
the corresponding ones in Section~\ref{subsubsect::stack::appendix::structures::rdcss-spec}. 

Also, axiom \axiomHRef{vis-ax::cc-return-completion} has an identical explanation
as in Section~\ref{subsubsect::stack::appendix::structures::rdcss-spec}.
The axiom simply coalesces 
axioms $B_{4.1}$ and $B_{4.2}$ into the
postcondition predicate $\postPredSymbol$.

Similarly, we need to redefine the set $\writesAbs p$ of successful writes into pointer $p$.
It does not suffice to consider an $\mcasAlg$ as a successful write if
it returns $true$; we need a criterion when an unfinished $\mcasAlg$
is a successful write as well. Thus, we define in
Figure~\ref{subfig:defined-notions-concurrent-spec-history} that an
$\mcasAlg(\listvar u)$ is a successful $p$-write if $p$ is a pointer
in $\listvar u$, and $\mcasAlg(\listvar u)$ observes events writing
the expected values for each entry in $\listvar u$.

\subsubsection{Defining the Visibility and Separability Relations}
\label{subsubsect::stack::appendix::structures::rdcss-and-mcas}

\begin{figure}[t]
	\includegraphics[scale=0.5]{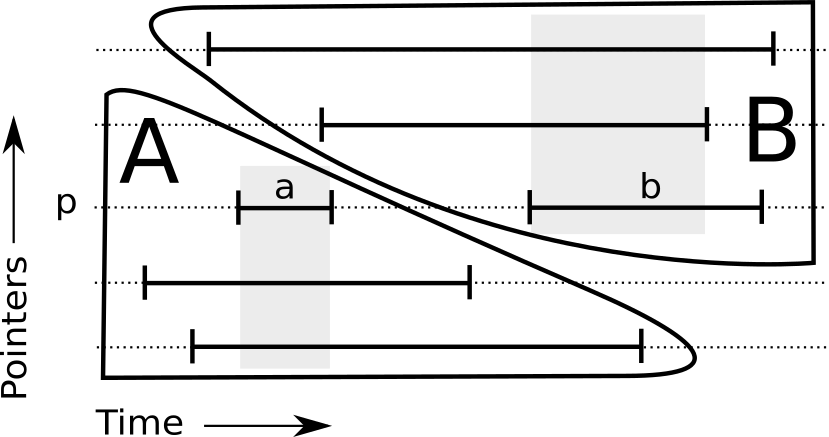}
	\centering
	\caption{An execution of $\mcasAlg$ events $A$ and $B$.  Each event
		encircles the spans belonging to it (i.e., helping it).  The
		intersection of spans for each event are shown as gray
		columns. Spans $a$ and $b$, both accessing pointer $p$, are named
		for later reference.}
	\label{fig::spans-example}
\end{figure}

\subparagraph{Defining the relations for $\mcasAlg$}
We now proceed to explain how relations $\visObsSymbol p$ and 
$\visSepSymbol p$ are defined for the implementation of $\mcasAlg$ in Figure 
\ref{alg-MCAS}. We focus first on $\mcasAlg$ because it illustrates 
the most general case, as it turns out that $\rdcssAlg$ is 
a special case of the analysis we do in this section.

The key idea is to analyze the behavior of the descriptors during the
execution.  Figure~\ref{fig::spans-example} shows the execution of two
overlapping invocations of $\mcasAlg$, which we refer to as events $A$
and $B$, respectively.\footnote{We focus on the successful
	$\mcasAlg$ case. Dominguez \etal~\cite{arxiv:mcas2307.04653} also describe the 
	failing $\mcasAlg$ case. Nevertheless, the successful case suffices 
to discover the definitions for $\visObsSymbol p$ and 
$\visSepSymbol p$, since the failing case also
produces the same definitions.} During the execution of
$\mcasAlg$, threads write the descriptor into a pointer
(line~\ref{invoke-rdcss-in-MCAS} in Figure~\ref{alg-MCAS}), and then
remove it (line~\ref{remove-all-descs-MCAS}). We term \emph{span} the
time interval between writing the descriptor and removing it. We also
say that the removed descriptor is
\emph{resolved}. Figure~\ref{fig::spans-example} depicts spans as
bounded line segments. 

There are two key invariants of $\mcasAlg$ that give rise to 
its visibility and separability relations.
The first key
invariant is that all the spans of an event \emph{must overlap} (we
call this the \emph{bunching invariant}). Concretely for $\mcasAlg$
events, any descriptor write must occur before any descriptor
removal. This is so because of the following two properties related to
how control flows through $\mcasAlg$. First, threads can reach
line~\ref{remove-all-descs-MCAS} (descriptor removal) only if some
thread changed the descriptor status at
line~\ref{resolve-status-MCAS}.\footnote{Recall that all descriptors
	are $\UNDECIDED$ initially, and can evolve to $\SUCCEEDED$ or
	$\FAILED$ only.}  This holds because
line~\ref{remove-all-descs-MCAS} (descriptor removal) is in the
$\mathit{false}$ branch of the conditional at
line~\ref{is-phase1-still-undecided} that checks if the descriptor
status is $\UNDECIDED$. Therefore, if no thread has changed the
descriptor status at line~\ref{resolve-status-MCAS}, threads starting
the $\mcasHelpAlg$ procedure will keep entering the $\mathsf{true}$
branch at line~\ref{is-phase1-still-undecided}.  The second property
is that any descriptor write at line~\ref{invoke-rdcss-in-MCAS} must
occur before some thread changes the descriptor status at
line~\ref{resolve-status-MCAS}.  This holds because the $\rdcssAlg$ at
line~\ref{invoke-rdcss-in-MCAS} (descriptor write attempt) fails if
the descriptor status is no longer $\UNDECIDED$.  These two properties
imply that any descriptor write must occur before the status change,
and any descriptor removal must occur after the status change, i.e.,
any span starts before any span finishes in the event; hence, span
intersection is non-empty. Figure~\ref{fig::spans-example} depicts the
spans' intersection for each event as gray columns. Since the presence
of a descriptor signals that a pointer indirectly has the expected
value, inside the gray column all the event's pointers simultaneously
have the expected values. Therefore, the event can be considered as
abstractly occurring (i.e., having its LP) anywhere inside the gray
column.

The second key invariant is that spans writing into \emph{the same
	pointer} (e.g., spans $a$ and $b$ in
Figure~\ref{fig::spans-example}) \emph{must be disjoint} (we call this
the \emph{disjointness invariant}).
This is so because $\mcasAlg$ uses helping; thus an event writes into
a pointer only if there is no descriptor currently present. More
specifically, the $\rdcssAlg$ at line \ref{invoke-rdcss-in-MCAS} in
Figure~\ref{alg-MCAS} writes the descriptor only if the expected value
is present directly (i.e., not via a descriptor). Since spans start 
by writing a descriptor, no span can start while there is a descriptor present, or
equivalently, when another span is already active in the pointer. The span
can start only after the currently present descriptor is removed.

These two invariants imply, as also apparent from
Figure~\ref{fig::spans-example}, that gray columns of
different events $A$ and $B$ accessing a common pointer must be
disjoint. Thus, we can \emph{separate} event $A$ before event $B$ in
time, because $A$'s gray column
executes before $B$'s. In the figure, we can describe the separation
(i.e., the gap between the gray columns) by saying that
there is a span in $A$ (namely $a$) and a span in $B$ (namely $b$),
both accessing a common pointer, and $a$ completes before $b$ starts.
Spans $a$ and $b$, being disjoint, induce a gap in the
gray columns. 

We define the separability relation $\visSepSymbol {\pointerIndx p}$ so that 
it directly formalizes the above description of the separation between $A$ and $B$,
\begin{align}
	\label{eq::overview::helpers-separability}
	\visSep{\pointerIndx p} A B \defini \exists x \in \hspans p A, y \in \hspans p B.\ \precedesSpans x y 
\end{align}
In the definition, $x \in \hspans p A$ means that $x$ is a span
accessing pointer $p$ in event $A$ (similarly for
$y \in \hspans p B$), and $\precedesSpans x y$ means that span $x$
terminates before $y$ starts.
Relation $\precedesSpansSymbol$ is a partial order on spans that
totally orders spans accessing the same pointer due to span
disjointness. $\precedesSpansEqSymbol$ denotes its reflexive closure.

We now focus on the visibility relation
$\visObsSymbol {\pointerIndx p}$.  We previously informally explained
that $\visObs {\pointerIndx p} A B$ intuitively captures that event
$B$ reads a value written by the $p$-write $A$, with no other
intervening writes between.
To illustrate how this intuition can be expressed using spans, consider the
span $b$ in Figure~\ref{fig::spans-example}.
When this span starts, it reads the input pointer (this is so because the $\rdcssAlg$ at
line \ref{invoke-rdcss-in-MCAS} in Figure \ref{alg-MCAS} obtains the
pointer's value, writes the descriptor and returns the read
value). This obtained value must have been written by the most recent
span that successfully wrote into the pointer, which, in
Figure~\ref{fig::spans-example} is span $a$.

To formally capture the described situation, we say that event $A$ is
\emph{observed} by $B$ at pointer $p$ if there are spans $x$ and $y$ in
$A$ and $B$, respectively, such that $x$ is the most recent span that terminated before
$y$ started, and that successfully wrote into $p$,
\begin{align}
	\label{eq::overview::helpers-observation}
	\visObs {\pointerIndx p} A B \defini \exists x \in 
	\hspans p A, y \in \hspans p B.\  x = \max_{\precedesSpansEqSymbol} \{ z \in \writesSpans p \mid \precedesSpans z y \}
\end{align}
In the definition, the maximum is taken under
$\precedesSpansEqSymbol$, and set $\writesSpans p$ collects the
spans that \emph{successfully} write into $p$. We differentiate
successful writes because some spans do not carry out changes. For
example, line \ref{remove-all-descs-MCAS} in Figure
\ref{alg-MCAS} produces a span that writes the new value only if the
descriptor's status was set to $\SUCCEEDED$, but produces a span that
undoes changes (as if the descriptor was not written) if the status
was set to $\FAILED$.  
Notice that it makes sense 
to take the maximum under $\precedesSpansEqSymbol$
because spans accessing the same pointer are disjoint, hence,
linearly ordered under $\precedesSpansEqSymbol$.

\subparagraph*{Spans and the Visibility-Style Axioms} We
briefly illustrate how the definitions satisfy the no in-between axiom
\axiomHRef{vis-ax::cc-no-in-between} using Figure~\ref{fig::spans-example}.  In
Figure~\ref{fig::spans-example}, assertion
$\visObs {\pointerIndx p} A B$ means that span $b$ reads a value
written by span $a$ in pointer $p$.  Since $a$ is the most recent
$p$-writing span before $b$ in real-time, and due to span
disjointness, any other $p$-writing span $c$ must execute either
completely before $a$ or completely after $b$. Therefore, if $c$
belongs to some $p$-write $C$, there will be a gap between the gray
columns of either $C$ and $A$ or $B$ and $C$, depending on whether $c$
executes before $a$ or after $b$, meaning that either 
$\visSep {\pointerIndx p} C A$ or $\visSep {\pointerIndx p} B C$.

\subparagraph*{Alternative Span Definitions}

As we just saw, the definitions
\eqref{eq::overview::helpers-separability} and
\eqref{eq::overview::helpers-observation} satisfy axiom \axiomHRef{vis-ax::cc-no-in-between}. 
Notice, however, that the argument relied solely on the span disjointness
invariant, and not on a specific definition of spans.  This is actually
an overarching property that we shall utilize when proving each of the
visibility axioms from Figure~\ref{subfig:concurrent-spec-history}.
Neither proof will depend on the particular definition of spans, but
only on high-level abstract span invariants such as bunching,
disjointness and a small number of additional structural 
ones~\cite{arxiv:mcas2307.04653}.

This abstraction affords some freedom to deviate from the operational
definition of spans as the time intervals between writing the descriptor
and resolving it,
so long as the high-level span invariants remain satisfied by the
programs.

To illustrate, bunching and disjointness remain satisfied if the
successful CAS at line~\ref{resolve-status-MCAS} is chosen as the
ending point of a span, instead of line~\ref{remove-all-descs-MCAS}
that we previously considered. Recall that while explaining the bunching
invariant,
we stated that any descriptor write must occur before the status change
at line~\ref{resolve-status-MCAS}. Therefore, the successful CAS at line~\ref{resolve-status-MCAS} is perfectly 
fine to end spans, because any span must start before it, i.e., bunching 
holds because any span starts before the CAS, which is the ending point for all the spans. 
Disjointness still holds because of the same reason: no descriptor can be 
written if there is currently a descriptor present. Therefore, if we end the span
prematurely at line~\ref{resolve-status-MCAS}, still no span can start until the descriptor
is removed at line~\ref{remove-all-descs-MCAS} much later.

Additionally, single reads that return a $\ValType$ value can be treated as 
spans that write a descriptor and instantaneously resolve it (i.e. a ``collapsed span'').
For example, the read at line~\ref{appendix::alg-access-Read-MCAS}.
It is essential that we treat as spans only those reads that return $\ValType$ values
to ensure the span disjointness invariant: if we treat reads that
return descriptors as spans, then the read will occur while a descriptor is present,
i.e., inside another span, violating disjointness.

\subparagraph*{Defining the relations for $\rdcssAlg$}

Perhaps surprisingly, definitions
\eqref{eq::overview::helpers-separability} and
\eqref{eq::overview::helpers-observation} apply without change 
to $\rdcssAlg$ as well.

The reason is that $\rdcssAlg$ events can be seen as a special case of
Figure \ref{fig::spans-example}, in which each $\rdcssAlg$ event
executes at most two spans. The first span of $\rdcssAlg$ is generated
by writing the descriptor at line \ref{alg-CAS-RDCSS} in Figure
\ref{alg-RDCSS}, and resolved at either line
\ref{alg-pt2-write-success-RDCSS} or line
\ref{alg-pt2-write-fail-RDCSS}.  The second span consists of a single
read of the control pointer at line \ref{alg-pt1-read-RDCSS} (thus, it
is a ``collapsed'' span). Trivially, $\rdcssAlg$ satisfies the bunching
invariant because the thread that resolves the descriptor at either
line \ref{alg-pt2-write-success-RDCSS} or line
\ref{alg-pt2-write-fail-RDCSS}, must have also previously read the
control pointer at line \ref{alg-pt1-read-RDCSS} while the descriptor
was present. In other words, the collapsed span overlaps the first
span.  Also, $\rdcssAlg$ satisfies span disjointness: an $\rdcssAlg$
invocation cannot write a new descriptor into a pointer if there is
currently another descriptor present.

A full proof that these definitions satisfy the visibility-style axioms
of Figures \ref{fig::concurrent-spec-rdcss} and 
\ref{fig::concurrent-spec-mcas} can be found in 
Dominguez \etal~\cite{arxiv:mcas2307.04653}.

\subsection{Queues}
\label{subsect::stack::appendix::structures::queues}

\begin{figure}[t]
	\centering 
	\begin{subfigwrap}{State-based sequential specification. The queue is represented as a list, with
		head the first element of the list. $\concat$ denotes list concatenation.}{subfig::stack::appendix::structures::atomic-spec-state-queues}
		\centering
		\begin{tabular}{l}
			$(A_1)$ Non-empty $\deqAlg$ \\
			\quad $\queueFull v Q \xrightarrow{\deqAlg()\ \langle v \rangle} Q$ \\
		\end{tabular}
		\quad
		\begin{tabular}{l}
			$(A_2)$ Empty $\deqAlg$ \\
			\quad $\queueEmpty \xrightarrow{\deqAlg()\ \langle \EMPTY \rangle} \queueEmpty$ \\
		\end{tabular}
		\quad
		\begin{tabular}{l}
			$(A_3)$ $\enqAlg$ \\
			\quad $Q \xrightarrow{\enqAlg(v)\ \langle \unitValue \rangle} Q \concat \queueList{v}$ \\
		\end{tabular}
	\end{subfigwrap}
	
	\begin{subfigwrap}{History-based sequential specification. Relation $\visObsSYMBOL : \absEvent \times \absEvent$ is
			abstract.}{subfig::stack::appendix::structures::atomic-spec-history-queues}
		\centering
		\begin{tabular}{c}
			\begin{tabular}{l}
				$(B_1)$ FIFO \\
				\quad $\visObs {} {e_1} {d_1} \wedge \precedesAbs {e_2} {e_1} \implies 
				\exists d_2.\ \visObs{} {e_2} {d_2} \wedge \precedesAbs {d_2} {d_1}$ \\
				$(B_2)$ Dequeue uniqueness \\
				\quad $\visObs {} {e} {d_1} \wedge \visObs {} e {d_2} \implies d_1 = d_2$ \\
				$(B_3)$ Dependences occur in the past \\
				\quad $\visObs {} e d \implies \precedesAbs e d$\\
				$(B_{4.1})$ Non-empty $\deqAlg$ \\
				\quad $d = \deqAlg()\left<v\right> \wedge v \neq \EMPTY \implies 
				\exists e.\ \visObs{} e d \wedge v = \inProp e$ \\
				$(B_{4.2})$ Empty $\deqAlg$ \\
				\quad $d_1 = \deqAlg()\left<\EMPTY\right> \implies 
				\forall e.\ \precedesAbs e {d_1} \implies \exists d_2.\ \visObs{} e {d_2} \wedge \precedesAbs {d_2} {d_1}$ \\
				$(B_{4.3})$ $\enqAlg$ \\
				\quad $e = \enqAlg(\_)\left<v\right> \implies v = \unitValue$ \\
			\end{tabular}
		\end{tabular}
	\end{subfigwrap}
	\caption{State-based and history-based sequential specifications for queues. Variables $e$, $d$, and their 
		indexed variants, range over enqueues and dequeues, respectively.}
	\label{fig::stack::appendix::structures::atomic-specs-queues}
\end{figure}

\subparagraph*{Sequential Specification}
Figure~\ref{fig::stack::appendix::structures::atomic-specs-queues} shows the
sequential state-based and history-based specifications for
queues.

In the state-based specification, 
axiom $A_1$ says that a $\deqAlg$ removes the head element $v$ from a non-empty
queue and returns $v$.  Axiom $A_2$ says that $\deqAlg$ returns
$\EMPTY$ when the queue is empty, leaving the queue unchanged. Axiom
$A_3$ says $\enqAlg(v)$ inserts $v$ to the end of the queue, 
returning the trivial value $\unitValue$.

The history-based specification utilizes the \emph{visibility relation} 
$\visObsSYMBOL$ to capture a
enqueue-dequeue causal dependence between events. In particular,
$\visObs {} e d$ means that ``event $d$ dequeues a value that event $e$
added to the queue''. Under this interpretation,
axioms $B_1,...,B_{4.3}$ state the following expected
properties.

Axiom $B_1$
encodes the FIFO discipline. More specifically, 
if $e_2$ enqueues a value before enqueue $e_1$
(i.e., $e_2 \precedesAbsSymbol e_1$),
but we know that $e_1$ has already been dequeued
by $d_1$ (i.e., $\visObs {} {e_1} {d_1}$),
then $e_2$ must have been dequeued before $d_1$,
because $e_2$ enqueued first, i.e.,
the first to enqueue is dequeued first.

Axiom $B_2$ (Dequeue uniqueness) says that a enqueue is observed by at most
one dequeue. 

Axiom $B_3$ (Dependences occur in the past) says that if a
dequeue depends on an enqueue, then the enqueue executes before the dequeue.

Axioms $B_{4.1}$ (Non-empty $\deqAlg$), $B_{4.2}$ (Empty $\deqAlg$),
and $B_{4.3}$ essentially are the counterparts of the state-based
sequential axioms $A_1$-$A_3$, respectively, as we show next.

Axiom $B_{4.1}$ says that a $\deqAlg()\left<v\right>$ event $d$
observes an enqueue $e$ that enqueued $v$.
This axiom, along with $B_1$-$B_3$, ensures that $d$ relates to $e$ as
in the following diagram.
\begin{center}
	\begin{tikzcd}[ampersand replacement=\&, column sep=small, row sep=small] 
		Q 
		\arrow[r, "{e}"{name=E,yshift=2pt}] \&
		Q \concat \queueList{v}
		\arrow[r, "{x_i}"] \& 
		\dots 
		\arrow[r, "{x_j}"] \& 
		Q_1 \concat (\queueFull{v}{Q_2})
		\arrow[r, "{x_k}"] \& 
		\ldots 
		\arrow[r, "{x_l}"] \& 
		\queueFull{v}{Q_3}
		\arrow[r, "{d = \deqAlg()\left< v \right>}"{name=D}] \&[30pt] 
		Q_3 
		\\
		\arrow[from=D, to=E, bend right, dashed, no head, "{\visObsSYMBOL}"{below}]
	\end{tikzcd}
\end{center}
In particular: (i) $e$ executes before $d$ (by axiom $B_3$, because
$\visObs{} {e}{d}$), (ii) every enqueue before $e$ is dequeued
before $d$ (by axiom $B_1$), and each enqueue is dequeued exactly once (by
axiom $B_2$). Thus, the subqueue $Q$ contains values enqueued before
$e$, and since each of these values is dequeued before $d$, the subqueue
$Q$ is empty when $d$ executes, which means that $v$ is at the head
of the queue when $d$ executes. Between $e$ and $d$, other enqueues 
could enqueue values after $v$, so that subqueue $Q_2$ has a smaller or equal 
length than $Q_3$. Between $e$ and $d$, dequeues will remove the values before
$v$, so that subqueue $Q$ has a bigger or equal length than $Q_1$. 
When $d$ executes the queue is modified
from $\queueFull{v}{Q_3}$ to $Q_3$. 
This explains that $B_{4.1}$ is essentially a
history-based version of $A_1$.

Similarly, axiom $B_{4.2}$ states that if a
$\deqAlg()\left<\EMPTY\right>$ event $d$ occurs, then every enqueue
before $d$ must have been dequeued before $d$, as this ensures that the
queue is empty when $d$ is reached. Hence, the axiom is counterpart to
$A_2$.

Finally, axiom $B_{4.3}$ says that the output of a enqueue event is the
trivial value $\unitValue$. The axiom imposes no conditions on the
queue, as a value can always be enqueued. In this, $B_{4.3}$ is the
counterpart to $A_3$ which also imposes no conditions on the input
queue, and posits that push's output value is trivial. However, unlike
$A_3$, $B_{4.3}$ does not directly says that the value is enqueued 
at the end of the queue, as that aspect is captured by the relationships
between enqueues and dequeues described by $B_{4.1}$.

\begin{figure}[t]
	\begin{subfigwrap}{Concurrent specification. Relations $\visObsSYMBOL, \visSepSYMBOL : \absEvent \times \absEvent$ are abstract.}{subfig::stack::appendix::structures::concurrent-spec-history-queues}
		\centering
		\begin{tabular}{l}
			\axiomQLabel{vis-ax::stack::appendix::structures::cc-concurrent-fifo} Concurrent FIFO \\
			\quad $\visObs {} {e_1} {d_1} \wedge {e_1} \nVisSepEqSYMBOL {e_2} \implies 
			\exists d_2.\ \visObs{} {e_2} {d_2} \wedge \visSep {} {d_2} {d_1}$ \\
			\axiomQLabel{vis-ax::stack::appendix::structures::cc-deq-uniqueness} Dequeue uniqueness \\
			\quad $\visObs {} {e} {d_1} \wedge \visObs {} e {d_2} \implies d_1 = d_2$ \\
		\end{tabular}
		\begin{tabular}{l}
			\axiomQLabel{vis-ax::stack::appendix::structures::cc-no-future-dependence-queues} No future dependences \\
			\quad $x \transCl{\genVisSymbol} y \implies \nprecedesAbsEq y x$\\
			\axiomQLabel{vis-ax::stack::appendix::structures::cc-return-completion-queues} Return value completion \\
			\quad $\exists v.\ \postPred x v \wedge (x \in \terminatedEvent \implies v = \outputProp x)$ \\
		\end{tabular} 
	\end{subfigwrap}
	
	\begin{subfigwrap}{Defined notions.}{subfig::stack::appendix::structures::defined-notions-concurrent-spec-history-queues}
		\centering
		\begin{tabular}{c}
			\begin{tabular}{l}
				Constraint relation \\
				\quad ${\genVisSymbol} \defini \visObsSYMBOL \cup \visSepSYMBOL$\\
				Returns-before relation\\
				\quad $\precedesAbs {k_1} {k_2} \defini \ETimeProp {k_1} \natorderSymbol \STimeProp {k_2}$\\
			\end{tabular}
			\begin{tabular}{l}
				Set of terminated events\\
				\quad $\terminatedEvent \defini \{ k \mid \ETimeProp k \neq \bot \}$\\
				Closure of terminated events\\
				\quad $\closedEvent \defini \{ k \mid \exists t \in \terminatedEvent.\ k \refleTransCl{\genVisSymbol} t \} $\\
			\end{tabular}
			\\
			\begin{tabular}{c}
				\\
				$\postPred {d_1} v \defini
				\begin{cases}
					\exists e.\ \visObs{} {e} {d_1} \wedge v = \inProp {e} &
					\text{if $v \neq \EMPTY$} \\
					\forall e.\ {d_1} \nVisSepEqSYMBOL {e} \implies \exists d_2.\ \visObs{} e {d_2} 
					\wedge \visSep{} {d_2} {d_1} & \text{if $v = \EMPTY$} \\
				\end{cases}$
				\qquad $\postPred {e} v \defini v = \unitValue$
			\end{tabular}
		\end{tabular}
	\end{subfigwrap}
	\caption{Concurrent history-based specification for queues. Variables $e$, $d$, and their 
		indexed variations, range over enqueues and dequeues in $\closedEvent$, respectively.
		Variables $x$, $y$ range over $\closedEvent$.
		Variable $k$, and its indexed variations, range over $\absEvent$.}
	\label{fig::stack::appendix::structures::concurrent-spec-queues}
\end{figure}

\subparagraph*{Concurrent Specification}
Figure~\ref{fig::stack::appendix::structures::concurrent-spec-queues}
shows the result of transforming the sequential specification of
Figure~\ref{subfig::stack::appendix::structures::atomic-spec-history-queues}.
The transformation procedure is identical to the one in 
Section~\ref{subsect::stack::rels::separability-relations-and-concur-specs}.

Axiom \axiomQRef{vis-ax::stack::appendix::structures::cc-concurrent-fifo}
is obtained from $B_1$ by replacing $\precedesAbsSymbol$ for $\visSepSYMBOL$
following the replacement rules described in 
Section~\ref{subsect::stack::rels::separability-relations-and-concur-specs}:
negative occurrences of $a \precedesAbsSymbol b$ are replaced
with $b \not\visSepEqSYMBOL a$; positive occurrences of 
$a \precedesAbsSymbol b$ are replaced with $a \visSepSYMBOL b$.

Axiom \axiomQRef{vis-ax::stack::appendix::structures::cc-deq-uniqueness}
is $B_2$ unchanged, i.e., there is no need to replace, since $B_2$ does 
not have occurrences of $\precedesAbsSymbol$.

Axiom \axiomQRef{vis-ax::stack::appendix::structures::cc-no-future-dependence-queues}
replaces $B_3$ with an identical explanation as for
axiom \axiomSRef{vis-ax::stack::cc-no-future-dependence} in
Section~\ref{subsect::stack::rels::separability-relations-and-concur-specs}.

Axiom \axiomQRef{vis-ax::stack::appendix::structures::cc-return-completion-queues}
coalesces $B_{4.1}$, $B_{4.2}$, and $B_{4.3}$ into the postcondition 
predicate $\postPred{x}{v}$ following the replacement rules for $\precedesAbsSymbol$
in $B_{4.1}$, $B_{4.2}$, and $B_{4.3}$.
Axiom \axiomQRef{vis-ax::stack::appendix::structures::cc-return-completion-queues}
has an identical explanation to axiom \axiomSRef{vis-ax::stack::cc-return-completion} in
Section~\ref{subsect::stack::rels::separability-relations-and-concur-specs}.

\subsection{Locks}
\label{subsect::stack::appendix::structures::locks}

\begin{figure}[t]
	\centering 
	\begin{subfigwrap}{State-based sequential specification. The initial state is $(false,0)$.}{subfig::stack::appendix::structures::atomic-spec-state-locks}
		\centering
		\begin{tabular}{l}
			$(A_1)$ 
			\quad $(false, 0) \xrightarrow{\lockAlg()\ \langle \unitValue \rangle} (true,0)$ \\
			$(A_2)$ 
			\quad $(true,c) \xrightarrow{\unlockAlg()\ \langle \unitValue \rangle} (false,c)$ \\
		\end{tabular}
		\quad
		\begin{tabular}{l}
			$(A_3)$ 
			\quad $(false,n) \xrightarrow{\regAlg()\ \langle \unitValue \rangle} (false,n+1)$ \\
		$(A_4)$ 
		\quad $(b,n+1) \xrightarrow{\deregAlg()\ \langle \unitValue \rangle} (b,n)$ \\
	\end{tabular}
	\end{subfigwrap}

\begin{subfigwrap}{Alternative equivalent state-based sequential specification. It restricts $A_2$ with $c = 0$ and $A_4$ with $b=false$. The initial state is $(false,0)$.}{subfig::stack::appendix::structures::alternative-atomic-spec-state-locks}
	\centering
	\begin{tabular}{l}
		$(A_1')$ 
		\quad $(false, 0) \xrightarrow{\lockAlg()\ \langle \unitValue \rangle} (true,0)$ \\
		$(A_2')$ 
		\quad $(true,0) \xrightarrow{\unlockAlg()\ \langle \unitValue \rangle} (false,0)$ \\
	\end{tabular}
	\quad
	\begin{tabular}{l}
		$(A_3')$ 
		\quad $(false,n) \xrightarrow{\regAlg()\ \langle \unitValue \rangle} (false,n+1)$ \\
		$(A_4')$ 
		\quad $(false,n+1) \xrightarrow{\deregAlg()\ \langle \unitValue \rangle} (false,n)$ \\
	\end{tabular}
\end{subfigwrap}
	
	\begin{subfigwrap}{History-based sequential specification. Relation $\visObsSYMBOL : \absEvent \times \absEvent$ is
			abstract.}{subfig::stack::appendix::structures::atomic-spec-history-locks}
		\centering
		\begin{tabular}{c}
			\begin{tabular}{l}
				$(B_1)$ No locks in-between \\
				\quad $\visObs {} {u_1} {l_1} \implies \forall l_2.\ \precedesAbsEq{l_2} {u_1} \vee \precedesAbsEq{l_1}{l_2}$ \\
				$(B_2)$ No unlocks in-between \\
				\quad $\visObs {} {l_1} {u_1} \implies \forall u_2.\ \precedesAbsEq{u_2} {l_1} \vee \precedesAbsEq{u_1}{u_2}$ \\
				$(B_3)$ No locks in-between pending reg \\
				\quad $\visObs {} {u} {r} \implies \forall l.\ \precedesAbsEq{l} {u} \vee \precedesAbsEq{r}{l}$ \\
				$(B_4)$ No pending regs before lock \\
				\quad $\precedesAbs{r}{l} \implies \exists d.\ \visObs{} r {d} \wedge \precedesAbs {d} {l}$ \\
				$(B_5)$ Reg uniqueness \\
				\quad $\visObs {} {r_1} {d} \wedge \visObs {} {r_2} {d} \implies r_1 = r_2$ \\
				$(B_6)$ Dereg uniqueness \\
				\quad $\visObs {} {r} {d_1} \wedge \visObs {} {r} {d_2} \implies d_1 = d_2$ \\
				$(B_7)$ Dependences occur in the past \\
				\quad $\visObs {} x y \implies \precedesAbs x y$\\
				$(B_{8.1})$ Case $\lockAlg$ \\
				\quad $l_1 = \lockAlg()\ \langle v \rangle \implies v = \unitValue \wedge  
				[(\forall l_2.\ \precedesAbsEq{l_1}{l_2}) \vee (\exists u.\ \visObs{} u {l_1})]$ \\
				$(B_{8.2})$ Case $\unlockAlg$ \\
				\quad $u = \unlockAlg()\ \langle v \rangle \implies v = \unitValue \wedge
				\exists l.\ \visObs{} l {u}$ \\
				$(B_{8.3})$ Case $\regAlg$ \\
				\quad $r = \regAlg()\ \langle v \rangle \implies v = \unitValue \wedge
				[(\forall l.\ \precedesAbs{r}{l}) \vee (\exists u.\ \visObs{} u {r})]$ \\
				$(B_{8.4})$ Case $\deregAlg$ \\
				\quad $d = \deregAlg()\ \langle v \rangle \implies v = \unitValue \wedge
				\exists r.\ \visObs{} r {d}$ \\
			\end{tabular}
		\end{tabular}
	\end{subfigwrap}
	\caption{State-based and history-based sequential specifications for readers/writer locks. Variables $l$, $u$, $r$, $d$, and their 
		indexed variants, range over locks, unlocks, regs, and deregs, respectively. Variables $x$ and $y$ range over arbitrary events.}
	\label{fig::stack::appendix::structures::atomic-specs-locks}
\end{figure}

\subsubsection{Readers/Writer Locks}

A readers/writer lock allows concurrent access for read-only operations (readers lock), 
whereas it imposes exclusive access to write operations (writer lock). 
This means that multiple readers can acquire the readers lock and execute in parallel, 
but only one writer can acquire the writer lock and modify the data. 
When a writer acquires the writer lock, all other writers and readers are blocked until the writer releases the
writer lock. 
A writer cannot acquire the writer lock if there is at least one reader that has not released the 
readers lock. 

The abstract state of a readers/writer lock is a tuple $(b,c)$ where $b$ is a boolean
that indicates if the writer lock is acquired ($true$) or not ($false$), and
$c$ is a natural number, counting the number of readers that have
acquired the readers lock. The readers lock is not acquired if $c = 0$
and acquired if $c > 0$.

The data structure for a readers/writer lock consists on four methods.
$\lockAlg()$ acquires the writer lock if both the writer and readers lock are currently not acquired;
$\unlockAlg()$ releases the writer lock if it is currently locked;
$\regAlg()$ (i.e., ``register'') acquires the readers lock by increasing the readers count
only if the writer lock is currently not acquired;
$\deregAlg()$ (i.e., ``deregister'') decrements by one the readers lock only if the
readers lock is acquired. All the above methods have
the trivial output $\unitValue$. 

\subparagraph*{Sequential Specification}
The methods are better described by the state-based sequential specification in
Figure~\ref{subfig::stack::appendix::structures::atomic-spec-state-locks}. 
In axiom $A_1$, $\lockAlg()$ changes the writer lock to $true$ (acquired) 
only if the writer lock is $false$ (not acquired) and the readers count is $0$ 
(i.e., readers lock is not acquired).
In axiom $A_2$, $\unlockAlg()$ changes the writer lock to $false$ (not acquired) 
only if the writer lock is $true$ (acquired).
In axiom $A_3$, $\regAlg()$ increases the readers count by one, but only if
the writer lock is $false$ (not acquired). In axiom $A_4$, $\deregAlg()$ decreases 
the readers count by one, but only if the readers count is bigger than $0$ (i.e.,
readers lock is acquired).

Figure~\ref{subfig::stack::appendix::structures::alternative-atomic-spec-state-locks} 
shows an alternative state-based specification which restricts 
the readers counter $c$ in $A_2$ to $0$ and the boolean
$b$ in $A_4$ to $false$. One can easily show that both
state-based specifications in 
Figures~\ref{subfig::stack::appendix::structures::atomic-spec-state-locks} and 
\ref{subfig::stack::appendix::structures::alternative-atomic-spec-state-locks} 
are equivalent: (1) we must have $c = 0$ in
$A_2$ because once a lock executes, no register or deregister can execute 
because the lock set the boolean to $true$ and the counter to $0$; therefore,
when the unlock executes, the counter will be $0$; (2) we must have $b=false$
in $A_4$ because once a register executes, no lock or unlock can execute 
because the register increased the counter and left the boolean in the $false$ state;
therefore, when the deregister executes, the boolean will still be $false$.

The axioms in 
Figure~\ref{subfig::stack::appendix::structures::atomic-spec-state-locks} 
impose less constraints on the states, because $A_2$ does not require any
constraint on the readers counter, and $A_4$ does not require any constraint
on the boolean. This will make easier to explain the connection between
the visibility relation and the state constraints in $A_1$-$A_4$ once we move 
to the history-based specification.
Nevertheless, when illustrating executions in a history, we will
use the axioms in 
Figure~\ref{subfig::stack::appendix::structures::alternative-atomic-spec-state-locks}.

Figure~\ref{subfig::stack::appendix::structures::atomic-spec-history-locks}
shows the corresponding history-based sequential specification.
Again, the specification utilizes the \emph{visibility relation} 
$\visObsSYMBOL$ to capture a
causal dependence between events. In particular,
$\visObs {} x y$ means that ``event $y$ can execute because of event $x$''. 
The axioms $B_1$-$B_{8.4}$ make use of four cases for the relation 
$\visObsSYMBOL$, each case corresponding to a state constraint
in the axioms $A_1$-$A_4$, as we explain next.

Case $\visObs {} u l$ for unlock $u$ and lock $l$. The statement 
$\visObs {} u l$ means
that $l$ can execute because $u$ released the writer lock, i.e., 
the writer lock is not acquired, which is a constraint in axiom $A_1$. 
Note that axiom $A_1$ also requires that the readers lock be 
not acquired, but we will see later that axiom $B_4$ models this 
constraint, so that there is no need to introduce a case in 
the visibility relation that relates locks with registers 
and deregisters.

Case $\visObs {} l u$ for lock $l$ and unlock $u$. 
The statement $\visObs {} l u$ means
that $u$ can execute because $l$ acquired the writer lock, i.e., 
the writer lock is acquired, which is the only constraint in axiom $A_2$. 

Case $\visObs{}{u}{r}$ for unlock $u$ and register $r$. 
The statement $\visObs {} u r$ means
that $r$ can execute because $u$ released the writer lock, i.e., 
the writer lock is not acquired, which is the only constraint in axiom $A_3$. 

Case $\visObs{}{r}{d}$ for register $r$ and deregister $d$. 
The statement $\visObs {} r d$ means
that $d$ can execute because $r$ increased the readers count, i.e., 
the readers lock is acquired, which is the only constraint in axiom $A_4$. 

Under this interpretation of the $\visObsSYMBOL$ relation,
axioms $B_1,...,B_{8.4}$ state the following properties.

$B_1$ states that if lock $l_1$ can execute because unlock $u_1$
released the writer lock ($\visObs {} {u_1} {l_1}$), then any other lock
$l_2$ cannot execute between $u_1$ and $l_1$, otherwise either: (1) the 
writer lock is already acquired by $l_2$ when $l_1$ tries to execute, or (2) 
$l_2$ was actually released by some other unlock $u_2$, but then this would contradict
that $u_1$ is the \emph{cause} for $l_1$.

$B_2$ states something identical to $B_1$ but for unlocks: 
if unlock $u_1$ can execute because lock $l_1$
acquired the writer lock ($\visObs {} {l_1} {u_1}$), then any other unlock
$u_2$ cannot execute between $l_1$ and $u_1$.

$B_3$ states that if a register $r$ can execute because unlock 
$u$ released the writer lock ($\visObs {} {u} {r}$) then no lock $l$
can start between $u$ and $r$, otherwise either: (1) the 
writer lock is already acquired by $l$ when $r$ tries to execute, or (2) 
$l$ was actually released by some other unlock $u'$, but then this would contradict
that $u$ is the \emph{cause} for $r$.

$B_4$ states that any register $r$ before a lock $l$ must have been deregistered by
some $d$ before the lock executes, otherwise, the readers counter would not be zero
when lock $l$ tries to execute. Note that $B_4$ encodes the state constraint 
in axiom $A_1$ that states $\lockAlg()$ cannot acquire the writer lock if the readers lock 
is acquired.

$B_5$ and $B_6$ state that the visibility relation pairs each deregister
with a unique register, and each register with a unique deregister. In other words,
each decrease in the readers counter comes from a unique increase, and 
each increase in the readers counter can only be paired with a unique decrease. 

$B_7$ is the standard axiom stating that any dependence of an event $y$ must execute before $y$.

Axioms $B_{8.1}$-$B_{8.4}$ are the history-based version of axioms $A_1$-$A_4$.
As we explain next.

Axiom $B_{8.1}$ states that either lock $l_1 = \lockAlg()$ is the first lock in the history
or there is an unlock $u$ observed by $l_1$ (which happens when $l_1$ is not the first lock in the history).

In case $l_1$ is the first lock in the history, the execution looks as follows,
\begin{center}
	\begin{tikzcd}[ampersand replacement=\&, column sep=small, row sep=small] 
		(false,0) 
		\arrow[r, "{x_h}"] \&
		\dots
		\arrow[r, "{x_i}"] \& 
		(false,c)
		\arrow[r, "{x_j}"] \& 
		\dots
		\arrow[r, "{x_k}"] \&
		(false,0)
		\arrow[r, "{l_1 = \lockAlg()\langle \unitValue \rangle}"] \&[35pt] 
		(true,0)
	\end{tikzcd}
\end{center}
We must have that the boolean does not change from $false$ to $true$ before $l_1$, because there is no lock before $l_1$.
Also, registers and deregisters do not change the boolean. The readers counter could increase and decrease
arbitrarily before $l_1$, but (1) $B_4$ ensures that every register is deregistered before $l_1$, and (2) 
each increase of the readers counter is decreased exactly once by $B_5$ and $B_6$. Therefore, just before
$l_1$ executes, the readers counter must be zero and the boolean is $false$, and $l_1$ executes as in $A_1$.

In case $l_1$ observes an unlock $u$, the execution looks as follows,
\begin{center}
	\begin{tikzcd}[ampersand replacement=\&, column sep=small, row sep=small] 
		(true,0)
		\arrow[r, "{u}"{name=U,yshift=2pt}] \&
		(false,0)
		\arrow[r, "{x_i}"] \& 
		\dots 
		\arrow[r, "{x_j}"] \& 
		(false,m)
		\arrow[r, "{x_k}"] \& 
		\ldots 
		\arrow[r, "{x_l}"] \& 
		(false,0)
		\arrow[r, "{l_1 = \lockAlg()\langle \unitValue \rangle}"{name=L}] \&[35pt] 
		(true,0) 
		\\
		\arrow[from=L, to=U, bend right, dashed, no head, "{\visObsSYMBOL}"{below}]
	\end{tikzcd}
\end{center}
First, $u$ executes before $l_1$ by $B_7$. $u$ changes the boolean to $false$. 
We must have that the boolean does not change from $false$ to $true$ in between $u$ and $l_1$, 
because $B_1$ implies that there is no lock in between $u$ and $l_1$.
Also, registers and deregisters do not change the boolean.
The readers counter could increase and decrease
arbitrarily in between $u$ and $l_1$, but (1) $B_4$ ensures that every register is deregistered before $l_1$, and (2) 
each increase of the readers counter is decreased exactly once by $B_5$ and $B_6$. Therefore, just before
$l_1$ executes, the readers counter must be zero and the boolean is $false$, and $l_1$ executes as in $A_1$.

Axiom $B_{8.2}$ states that any unlock $u=\unlockAlg()$ must observe a lock $l$. The execution will look as follows,
\begin{equation}
	\label{eq::stack::appendix::structures::b-eight-two-exec}
	\begin{tikzcd}[ampersand replacement=\&, column sep=small, row sep=small] 
		(false,0)
		\arrow[r, "{l}"{name=L,yshift=2pt}] \&
		(true,0)
		\arrow[r, "{u = \unlockAlg()\langle \unitValue \rangle}"{name=U}] \&[40pt] 
		(false,0) 
		\\
		\arrow[from=U, to=L, bend right, dashed, no head, "{\visObsSYMBOL}"{below}]
	\end{tikzcd}
\end{equation}
First, $l$ executes before $u$ by $B_7$. $l$ changes the state to $(true,0)$. We now show that there cannot exist 
an event between $l$ and $u$. Axiom $B_2$ states that there cannot exist an unlock between $l$ and $u$. It remains to show
that there cannot exist a lock, register, and deregister between $l$ and $u$. 

Suppose there is a lock $l_2$ in between $l$ and $u$. Since $l_2$ is not the first lock (i.e., $l$ executes before $l_2$), 
axiom $B_{8.1}$ states that $l_2$ must be observing an unlock $u_2$ that executes before $l_2$ by $B_7$.
If $u_2$ executes before $l$, i.e.,
\begin{center}
	\begin{tikzcd}[ampersand replacement=\&, column sep=small, row sep=small] 
		\dots
		\arrow[r, "{u_2}"{name=U2,yshift=2pt}] \&
		\dots 
		(false,0)
		\arrow[r, "{l}"{name=L,yshift=2pt}] \&
		(true,0)
		\dots 
		\arrow[r, "{l_2}"{name=L2,yshift=2pt}] \&
		\dots
		\arrow[r, "{u = \unlockAlg()\langle \unitValue \rangle}"{name=U}] \&[40pt] 
		(false,0) 
		\\
		\arrow[from=U, to=L, bend right, dashed, no head, "{\visObsSYMBOL}"{above}]
		\arrow[from=L2, to=U2, bend right, dashed, no head, "{\visObsSYMBOL}"{below}]
	\end{tikzcd}
\end{center}
then $l$ is in between $u_2$ and $l_2$, which contradicts $B_1$.

If $u_2$ executes after $l$, i.e.,
\begin{center}
	\begin{tikzcd}[ampersand replacement=\&, column sep=small, row sep=small] 
		(false,0)
		\arrow[r, "{l}"{name=L,yshift=2pt}] \&
		(true,0)
		\dots
		\arrow[r, "{u_2}"{name=U2,yshift=2pt}] \&
		\dots 
		\arrow[r, "{l_2}"{name=L2,yshift=2pt}] \&
		\dots
		\arrow[r, "{u = \unlockAlg()\langle \unitValue \rangle}"{name=U}] \&[40pt] 
		(false,0) 
		\\
		\arrow[from=U, to=L, bend right, dashed, no head, "{\visObsSYMBOL}"{above}]
		\arrow[from=L2, to=U2, bend right, dashed, no head, "{\visObsSYMBOL}"{below}]
	\end{tikzcd}
\end{center}
then $u_2$ is in between $l$ and $u$, which contradicts $B_2$.

Suppose there is a register $r$ in between $l$ and $u$. Since $l$ executes before $r$, 
axiom $B_{8.3}$ states that $r$ must be observing an unlock $u_2$ that executes before $r$ by $B_7$.
If $u_2$ executes before $l$, i.e.,
\begin{center}
	\begin{tikzcd}[ampersand replacement=\&, column sep=small, row sep=small] 
		\dots
		\arrow[r, "{u_2}"{name=U2,yshift=2pt}] \&
		\dots 
		(false,0)
		\arrow[r, "{l}"{name=L,yshift=2pt}] \&
		(true,0)
		\dots 
		\arrow[r, "{r}"{name=R,yshift=2pt}] \&
		\dots
		\arrow[r, "{u = \unlockAlg()\langle \unitValue \rangle}"{name=U}] \&[40pt] 
		(false,0) 
		\\
		\arrow[from=U, to=L, bend right, dashed, no head, "{\visObsSYMBOL}"{above}]
		\arrow[from=R, to=U2, bend right, dashed, no head, "{\visObsSYMBOL}"{below}]
	\end{tikzcd}
\end{center}
then $l$ is in between $u_2$ and $r$, which contradicts $B_3$.

If $u_2$ executes after $l$, i.e.,
\begin{center}
	\begin{tikzcd}[ampersand replacement=\&, column sep=small, row sep=small] 
		(false,0)
		\arrow[r, "{l}"{name=L,yshift=2pt}] \&
		(true,0)
		\dots
		\arrow[r, "{u_2}"{name=U2,yshift=2pt}] \&
		\dots 
		\arrow[r, "{r}"{name=R,yshift=2pt}] \&
		\dots
		\arrow[r, "{u = \unlockAlg()\langle \unitValue \rangle}"{name=U}] \&[40pt] 
		(false,0) 
		\\
		\arrow[from=U, to=L, bend right, dashed, no head, "{\visObsSYMBOL}"{above}]
		\arrow[from=R, to=U2, bend right, dashed, no head, "{\visObsSYMBOL}"{below}]
	\end{tikzcd}
\end{center}
then $u_2$ is in between $l$ and $u$, which contradicts $B_2$.

Suppose there is a deregister $d$ between $l$ and $u$. Axiom $B_{8.4}$ states that $d$ must be observing 
a register $r$ that executes before $d$ by $B_7$. If $r$ executes after $l$, i.e.,
\begin{center}
	\begin{tikzcd}[ampersand replacement=\&, column sep=small, row sep=small] 
		(false,0)
		\arrow[r, "{l}"{name=L,yshift=2pt}] \&
		(true,0)
		\dots
		\arrow[r, "{r}"{name=R,yshift=2pt}] \&
		\dots 
		\arrow[r, "{d}"{name=D,yshift=2pt}] \&
		\dots
		\arrow[r, "{u = \unlockAlg()\langle \unitValue \rangle}"{name=U}] \&[40pt] 
		(false,0) 
		\\
		\arrow[from=U, to=L, bend right, dashed, no head, "{\visObsSYMBOL}"{above}]
		\arrow[from=D, to=R, bend right, dashed, no head, "{\visObsSYMBOL}"{below}]
	\end{tikzcd}
\end{center}
it would contradict 
the previous case which states that no register can execute between $l$ and $u$. Hence, $r$ must execute 
before $l$. But $B_4$ implies that $r$ must be deregistered by some $d_2$ executing before $l$, i.e.,
\begin{center}
	\begin{tikzcd}[ampersand replacement=\&, column sep=small, row sep=small] 
		\dots
		\arrow[r, "{r}"{name=R,yshift=2pt}] \&
		\dots
		\arrow[r, "{d_2}"{name=D2,yshift=2pt}] \&
		\dots
		(false,0)
		\arrow[r, "{l}"{name=L,yshift=2pt}] \&
		(true,0)
		\dots 
		\arrow[r, "{d}"{name=D,yshift=2pt}] \&
		\dots
		\arrow[r, "{u = \unlockAlg()\langle \unitValue \rangle}"{name=U}] \&[40pt] 
		(false,0) 
		\\
		\arrow[from=U, to=L, bend right, dashed, no head, "{\visObsSYMBOL}"{above}]
		\arrow[from=D, to=R, bend right, dashed, no head, "{\visObsSYMBOL}"{below}]
		\arrow[from=D2, to=R, bend right, dashed, no head, "{\visObsSYMBOL}"{below}]
	\end{tikzcd}
\end{center}
and so, $d = d_2$ by $B_6$, which is impossible.

Therefore, the execution must look as in \eqref{eq::stack::appendix::structures::b-eight-two-exec},
and $u$ executes as in $A_2'$ (which is equivalent to $A_2$).

Axiom $B_{8.3}$ states that either register $r = \regAlg()$ executes before any lock in the history
or there is an unlock $u$ observed by $r$ (which happens when there is some lock executing before $r$).

In case every lock executes after $r$, the execution looks as follows,
\begin{center}
	\begin{tikzcd}[ampersand replacement=\&, column sep=small, row sep=small] 
		(false,0) 
		\arrow[r, "{x_h}"] \&
		\dots
		\arrow[r, "{x_i}"] \& 
		(false,c)
		\arrow[r, "{x_j}"] \& 
		\dots
		\arrow[r, "{x_k}"] \&
		(false,n)
		\arrow[r, "{r = \regAlg()\langle \unitValue \rangle}"] \&[30pt] 
		(false,n+1)
	\end{tikzcd}
\end{center}
We must have that the boolean does not change from $false$ to $true$ before $r$, because there is no lock before $r$.
Also, registers and deregisters do not change the boolean. Therefore, just before
$r$ executes, the boolean is $false$, and $r$ executes as in $A_3$.

In case $r$ observes an unlock $u$, the execution looks as follows,
\begin{center}
	\begin{tikzcd}[ampersand replacement=\&, column sep=small, row sep=small] 
		(true,0)
		\arrow[r, "{u}"{name=U,yshift=2pt}] \&
		(false,0)
		\arrow[r, "{x_i}"] \& 
		\dots 
		\arrow[r, "{x_j}"] \& 
		(false,c)
		\arrow[r, "{x_k}"] \& 
		\ldots 
		\arrow[r, "{x_l}"] \& 
		(false,n)
		\arrow[r, "{r = \regAlg()\langle \unitValue \rangle}"{name=R}] \&[30pt] 
		(false,n+1) 
		\\
		\arrow[from=R, to=U, bend right, dashed, no head, "{\visObsSYMBOL}"{below}]
	\end{tikzcd}
\end{center}
First, $u$ executes before $r$ by $B_7$. $u$ changes the boolean to $false$. 
We must have that the boolean does not change from $false$ to $true$ in between $u$ and $r$, 
because $B_3$ implies that there is no lock in between $u$ and $r$.
Also, registers and deregisters do not change the boolean.
Therefore, just before
$r$ executes, the boolean is $false$, and $r$ executes as in $A_3$.

Axiom $B_{8.4}$ states that any deregister $d=\deregAlg()$ must observe a register $r$. The execution
will look as follows,
\begin{center}
	\begin{tikzcd}[ampersand replacement=\&, column sep=small, row sep=small] 
		(false,c)
		\arrow[r, "{r}"{name=R,yshift=2pt}] \&
		(false,c+1)
		\arrow[r, "{x_i}"] \&  
		\dots 
		\arrow[r, "{x_l}"] \& 
		(false,n+1)
		\arrow[r, "{d = \deregAlg()\langle \unitValue \rangle}"{name=D}] \&[35pt] 
		(false,n) 
		\\
		\arrow[from=D, to=R, bend right, dashed, no head, "{\visObsSYMBOL}"{below}]
	\end{tikzcd}
\end{center}
We claim that before $d$ executes, the readers counter must be positive (i.e., $n+1$, for some $n$).
If the counter is zero, it means that the increase produced by $r$ must have been decreased \emph{before}
$d$ executes. But axioms $B_5$ and $B_6$ state that increases and decreases are uniquely paired,
meaning that the number of increases must equal the number of decreases before $d$, 
because the counter is zero. Hence, $r$ must have been deregistered before $d$ executed.
But this is impossible, because $d$ observes $r$ and registers and 
deregisters are paired uniquely by $B_5$ and $B_6$.

Also, we claim that there cannot be a lock between $r$ and $d$ changing the boolean from $false$
to $true$. Suppose for a contradiction that there is such lock $l$.
$B_4$ implies that $r$ must be deregistered by some $d_2$ executing before $l$, i.e.,
\begin{center}
	\begin{tikzcd}[ampersand replacement=\&, column sep=small, row sep=small] 
		(false,c)
		\arrow[r, "{r}"{name=R,yshift=2pt}] \&
		(false,c+1)
		\dots 
		\arrow[r, "{d_2}"{name=D2,yshift=2pt}] \& 
		\dots
		\arrow[r, "{l}"] \& 
		\ldots  
		(false,n+1)
		\arrow[r, "{d = \deregAlg()\langle \unitValue \rangle}"{name=D}] \&[35pt] 
		(false,n) 
		\\
		\arrow[from=D, to=R, bend right, dashed, no head, "{\visObsSYMBOL}"{below}]
		\arrow[from=D2, to=R, bend right, dashed, no head, "{\visObsSYMBOL}"{below}]
	\end{tikzcd}
\end{center}
So, $d_2 = d$ by $B_6$, which is impossible.

Therefore, just before $d$ executes, the boolean is $false$ and the readers counter is positive,
so that $d$ executes as in $A_4'$ (which is equivalent to $A_4$).

\begin{figure}[t]
	\begin{subfigwrap}{Concurrent specification. Relations $\visObsSYMBOL, \visSepSYMBOL : \absEvent \times \absEvent$ are abstract.}{subfig::stack::appendix::structures::concurrent-spec-history-locks}
		\centering
		\begin{tabular}{l}
			\axiomLLabel{vis-ax::stack::appendix::structures::no-lock-in-between} No locks in-between \\
			\quad $\visObs {} {u_1} {l_1} \implies \forall l_2.\ \visSepEq{}{l_2} {u_1} \vee \visSepEq{}{l_1}{l_2}$ \\
			\axiomLLabel{vis-ax::stack::appendix::structures::no-unlock-in-between} No unlocks in-between \\
			\quad $\visObs {} {l_1} {u_1} \implies \forall u_2.\ \visSepEq{}{u_2} {l_1} \vee \visSepEq{}{u_1}{u_2}$ \\
			\axiomLLabel{vis-ax::stack::appendix::structures::no-lock-in-between-pending-reg} No locks in-between pending reg \\
			\quad $\visObs {} {u} {r} \implies \forall l.\ \visSepEq{}{l} {u} \vee \visSepEq{}{r}{l}$ \\
			\axiomLLabel{vis-ax::stack::appendix::structures::no-pending-regs-before-lock} No pending regs before lock \\
			\quad $l \not\visSepEqSYMBOL r \implies \exists d.\ \visObs{} r {d} \wedge \visSep{} {d} {l}$
		\end{tabular}
		\begin{tabular}{l}
			\axiomLLabel{vis-ax::stack::appendix::structures::reg-uniqueness} Reg uniqueness \\
			\quad $\visObs {} {r_1} {d} \wedge \visObs {} {r_2} {d} \implies r_1 = r_2$ \\
			\axiomLLabel{vis-ax::stack::appendix::structures::dereg-uniqueness} Dereg uniqueness \\
			\quad $\visObs {} {r} {d_1} \wedge \visObs {} {r} {d_2} \implies d_1 = d_2$ \\
			\axiomLLabel{vis-ax::stack::appendix::structures::cc-no-future-dependence-locks} No future dependences \\
			\quad $x \transCl{\genVisSymbol} y \implies \nprecedesAbsEq y x$\\
			\axiomLLabel{vis-ax::stack::appendix::structures::cc-return-completion-locks} Return value completion \\
			\quad $\exists v.\ \postPred x v \wedge (x \in \terminatedEvent \implies v = \outputProp x)$ \\
		\end{tabular} 
	\end{subfigwrap}
	
	\begin{subfigwrap}{Defined notions.}{subfig::stack::appendix::structures::defined-notions-concurrent-spec-history-locks}
		\centering
		\begin{tabular}{c}
			\begin{tabular}{l}
				Constraint relation \\
				\quad ${\genVisSymbol} \defini \visObsSYMBOL \cup \visSepSYMBOL$\\
				Returns-before relation\\
				\quad $\precedesAbs {k_1} {k_2} \defini \ETimeProp {k_1} \natorderSymbol \STimeProp {k_2}$\\
			\end{tabular}
			\begin{tabular}{l}
				Set of terminated events\\
				\quad $\terminatedEvent \defini \{ k \mid \ETimeProp k \neq \bot \}$\\
				Closure of terminated events\\
				\quad $\closedEvent \defini \{ k \mid \exists t \in \terminatedEvent.\ k \refleTransCl{\genVisSymbol} t \} $\\
			\end{tabular}
			\\
			\\
			\begin{tabular}{ll}
				$\postPred {l_1} {v}$ & \kern-1em $\defini v = \unitValue \wedge [(\forall l_2.\ \visSepEq{}{l_1}{l_2}) \vee (\exists u.\ \visObs{} u {l_1})]$ \\
				$\postPred {u} {v}$ & \kern-1em $\defini v = \unitValue \wedge \exists l.\ \visObs{} l {u}$ \\
			\end{tabular}
		\begin{tabular}{ll}
				$\postPred {r} {v}$ & \kern-1em $\defini v = \unitValue \wedge [(\forall l.\ \visSep{}{r}{l}) \vee (\exists u.\ \visObs{} u {r})]$ \\
				$\postPred {d} {v}$ & \kern-1em $\defini v = \unitValue \wedge \exists r.\ \visObs{} r {d}$ \\
			\end{tabular}
		\end{tabular}
	\end{subfigwrap}
	\caption{Concurrent history-based specification for readers/writer locks. Variables $l$, $u$, $r$, $d$, and their 
		indexed variants, range over locks, unlocks, regs, and deregs in $\closedEvent$, respectively. 
		Variables $x$, $y$ range over $\closedEvent$.
		Variable $k$, and its indexed variations, range over $\absEvent$.}
	\label{fig::stack::appendix::structures::concurrent-spec-locks}
\end{figure}

\subparagraph*{Concurrent Specification}
Figure~\ref{fig::stack::appendix::structures::concurrent-spec-locks} shows the result of 
transforming the sequential specification of
Figure~\ref{subfig::stack::appendix::structures::atomic-spec-history-locks} into a concurrent 
specification. As expected, the transformation generalizes
the returns-before relation by replacing it with an abstract separability relation $\visSepSYMBOL$.

Axioms \axiomLRef{vis-ax::stack::appendix::structures::no-lock-in-between}-\axiomLRef{vis-ax::stack::appendix::structures::dereg-uniqueness}
are the direct result of transforming axioms $B_1$-$B_6$ by applying the 
replacement rules described in Section~\ref{subsect::stack::rels::separability-relations-and-concur-specs}: 
positive occurrences of the subformula $\precedesAbs{a}{b}$ are replaced with 
$\visSep{}{a}{b}$; negative occurrences with $b \not\visSepEqSYMBOL a$.

Axiom \axiomLRef{vis-ax::stack::appendix::structures::cc-no-future-dependence-locks}
replaces $B_7$. Axiom \axiomLRef{vis-ax::stack::appendix::structures::cc-no-future-dependence-locks} 
has an identical explanation as axiom \axiomSRef{vis-ax::stack::cc-no-future-dependence}
in Section~\ref{subsect::stack::rels::separability-relations-and-concur-specs}.

Axiom \axiomLRef{vis-ax::stack::appendix::structures::cc-return-completion-locks}
coalesces $B_{8.1}$-$B_{8.4}$ into the postcondition predicate $\postPred{x}{v}$
after replacing the occurrences of $\precedesAbsSymbol$ in $B_{8.1}$-$B_{8.4}$ with $\visSepSYMBOL$.
Axiom \axiomLRef{vis-ax::stack::appendix::structures::cc-return-completion-locks}
has an identical explanation as axiom \axiomSRef{vis-ax::stack::cc-return-completion}
in Section~\ref{subsect::stack::rels::separability-relations-and-concur-specs}.

\begin{figure}[t]
		\begin{tabular}{l}
			$(A_1)$ 
			\quad $false \xrightarrow{\lockAlg()\ \langle \unitValue \rangle} true$ \\
		\end{tabular}
		\quad
		\begin{tabular}{l}
			$(A_2)$ 
			\quad $true \xrightarrow{\unlockAlg()\ \langle \unitValue \rangle} false$ \\
		\end{tabular}
	\caption{State-based sequential specification for simple locks. The initial state is $false$.}
	\label{fig::stack::appendix::structures::atomic-specs-simple-locks}
\end{figure}

\subsubsection{Simple Locks}

A simple lock is a lock that only has the $\lockAlg()$ and 
$\unlockAlg()$ methods. The abstract state of a simple lock is a boolean, where
$true$ indicates that the lock is acquired, and $false$ otherwise. 
Figure~\ref{fig::stack::appendix::structures::atomic-specs-simple-locks} 
shows the state-based sequential specification of simple locks.

We get a history-based sequential specification for simple locks
by removing from Figure~\ref{subfig::stack::appendix::structures::atomic-spec-history-locks} 
the axioms $B_3$, $B_4$, $B_5$, $B_6$, $B_{8.3}$, and $B_{8.4}$,
all of which talk about registers and deregisters.

Accordingly, we get a concurrent specification for simple locks
simply by removing from Figure~\ref{fig::stack::appendix::structures::concurrent-spec-locks}
the axioms 
\axiomLRef{vis-ax::stack::appendix::structures::no-lock-in-between-pending-reg},
\axiomLRef{vis-ax::stack::appendix::structures::no-pending-regs-before-lock},
\axiomLRef{vis-ax::stack::appendix::structures::reg-uniqueness},
\axiomLRef{vis-ax::stack::appendix::structures::dereg-uniqueness};
and also, removing the definitions $\postPred {r} {v}$ and $\postPred {d} {v}$
for registers and deregisters in the post-condition predicate.